\documentclass[graybox,envcountchap]{svmult}

\usepackage{mathptmx}
\usepackage{helvet}
\usepackage{courier}
\usepackage{braket} 
\usepackage{algorithmicx}
\usepackage{algpseudocode} 
\usepackage{amsfonts}
\usepackage{amsmath}
\usepackage{simplewick}
\usepackage{type1cm}         
\usepackage{exercise}
\usepackage{makeidx}         
\usepackage{rotating}
\usepackage{graphicx}        
\usepackage{multicol}        
\usepackage[bottom]{footmisc}
\usepackage{xcolor}
\usepackage{listings}
\usepackage{epic}
\usepackage{eepic}
\usepackage{a4wide}
\usepackage{amsmath}
\usepackage{amssymb}
\usepackage[T1]{fontenc}
\usepackage{cite} 
\usepackage{shadow}
\usepackage{hyperref}                     
\usepackage{bookmark}                     
\usepackage{bezier}
\usepackage{pstricks}
\usepackage{textcomp,type1ec,pdfpages}
\usepackage{bera}
\usepackage{slashed}
\usepackage{environ}
\usepackage{esvect}
\usepackage{xspace}
\usepackage{bm}
\usepackage{array}

\usepackage{chapterbib}
\definecolor{dkgreen}{rgb}{0,0.6,0}
\definecolor{gray}{rgb}{0.5,0.5,0.5}
\definecolor{mauve}{rgb}{0.58,0,0.82}
\definecolor{green}{rgb}{0,0.5,0}

\lstnewenvironment{Python}[1]{
\lstset{
language=python,
basicstyle=\footnotesize\setstretch{1},
stringstyle=\color{red},
showstringspaces=false,
alsoletter={1234567890},
otherkeywords={\ , \}, \{},
keywordstyle=\color{black},
emph={access,and,break,class,continue,def,del,elif ,else,%
except,exec,finally,for,from,global,if,import,in,is,%
lambda,not,or,pass,print,raise,return,try,while},
emphstyle=\color{black}\bfseries,
emph={[2]True, False, None, self},
emphstyle=[2]\color{red},
emph={[3]from, import, as},
emphstyle=[3]\color{blue},
upquote=true,
morecomment=[s]{"""}{"""},
commentstyle=\color{dkgreen}\slshape, 
emph={[4]1, 2, 3, 4, 5, 6, 7, 8, 9, 0},
emphstyle=[4]\color{blue},
framexleftmargin=1mm, framextopmargin=1mm, rulesepcolor=\color{blue},
breakatwhitespace=true,
tabsize=2
}}{}

\lstnewenvironment{C++}[1]{
\lstset{
language=c++,
basicstyle=\footnotesize\setstretch{1},
stringstyle=\color{red},
showstringspaces=false,
alsoletter={1234567890},
otherkeywords={\ , \}, \{},
keywordstyle=\color{black},
emph={access,and,break,class,continue,def,del,elif ,else,%
except,exec,finally,for,from,global,if,import,in,is,%
lambda,not,or,pass,print,raise,return,try,while},
emphstyle=\color{black}\bfseries,
emph={[2]True, False, None, self},
emphstyle=[2]\color{red},
emph={[3]from, import, as},
emphstyle=[3]\color{blue},
upquote=true,
morecomment=[s]{"""}{"""},
commentstyle=\color{dkgreen}\slshape, 
emph={[4]1, 2, 3, 4, 5, 6, 7, 8, 9, 0},
emphstyle=[4]\color{blue},
framexleftmargin=1mm, framextopmargin=1mm, rulesepcolor=\color{blue},
breakatwhitespace=true,
tabsize=2
}}{}
\usepackage{calrsfs}

\makeindex             

\newcommand{\trace}{\mathrm{tr}\,}
\newcommand{\tr}{\trace}
\newcommand{\sgn}{\mathrm{sgn}\,}

\newcommand{\AC}{\ensuremath{\mathcal{A}}}
\newcommand{\OC}{\ensuremath{\mathcal{O}}}
\newcommand{\PC}{\ensuremath{\mathcal{P}}}

\newcommand{\nuc}[2]{\ensuremath{^{#2}\mathrm{#1}}}

\newcommand{\nn}{\ensuremath{\bar{n}}}

\newcommand{\NNNLO}{N$^3$LO}
\newcommand{\NNLO}{NNLO}
\newcommand{\NNLOsat}{$\text{NNLO}_\text{sat}$}
\newcommand{\Vlowk}{\ensuremath{V_{\text{low-k}}}}

\newcommand{\Tint}{\ensuremath{\TO_\text{int}}}
\newcommand{\Tcm}{\ensuremath{\TO_\text{cm}}}

\newcommand{\Hint}{\ensuremath{\HO_\text{int}}}

\newcommand{\Rch}{\ensuremath{R_\text{ch}}}
\newcommand{\Hfinal}{\ensuremath{\mkern 3mu\overline{\mkern-3muH}}}

\newcommand{\lambdaSRG}{\ensuremath{\lambda}}

\newcommand{\hw}{\ensuremath{\hbar\omega}}
\newcommand{\eMax}{\ensuremath{e_{\text{max}}}}

\newcommand{\EMax}{\ensuremath{E_{3\text{max}}}}
\newcommand{\Nmax}{\ensuremath{N_\text{max}}}

\newcommand{\fm}{\ensuremath{\,\text{fm}}}
\newcommand{\fmi}{\ensuremath{\,\text{fm}^{-1}}}
\newcommand{\keV}{\ensuremath{\,\text{keV}}}
\newcommand{\MeV}{\ensuremath{\,\text{MeV}}}
\newcommand{\GeV}{\ensuremath{\,\text{GeV}}}

\renewcommand{\ket}[1]{\ensuremath{\,|{#1}\rangle}}
\renewcommand{\braket}[2]{\ensuremath{\langle{#1}|{#2}\rangle}}
\newcommand{\expect}[1]{\ensuremath{\langle{#1}\rangle}}
\newcommand{\matrixe}[3]{\ensuremath{\langle{#1}|\,{#2}\,|{#3}\rangle}}
\newcommand{\dmatrixe}[2]{\ensuremath{\langle{#1}|\,{#2}\,|{#1}\rangle}}

\newcommand{\totd}[2]{\ensuremath{ \frac{d {#1}} {d {#2}} }}

\newcommand{\op}[1]{\ensuremath{\hat{#1}}}
\newcommand{\adj}[1]{\ensuremath{#1}^\dag}
\renewcommand{\vec}[1]{\ensuremath{\mathbf{#1}}}

\newcommand{\aO}{\ensuremath{a}}
\newcommand{\aaO}{\ensuremath{\adj{a}}}

\newcommand{\hO}{\ensuremath{h}}
\newcommand{\hhO}{\ensuremath{\adj{h}}}

\newcommand{\tO}{\op{t}}
\newcommand{\vO}{\op{v}}

\newcommand{\AO}{\op{A}}
\newcommand{\BO}{\op{B}}
\newcommand{\CO}{\op{C}}
\newcommand{\HO}{\op{H}}
\newcommand{\OO}{\op{O}}
\newcommand{\PO}{\op{P}}

\newcommand{\TO}{\op{T}}
\newcommand{\UO}{\op{U}}
\newcommand{\VO}{\op{V}}

\newcommand{\ZO}{\op{Z}}

\newcommand{\etaO}{\op{\eta}}
\newcommand{\sigmaO}{\op{\sigma}}

\newcommand{\OmegaO}{\op{\Omega}}

\newcommand{\QQO}{\adj{\op{Q}}}
\newcommand{\XXO}{\adj{\op{X}}}
\newcommand{\UUO}{\adj{\op{U}}}
\newcommand{\ZZO}{\adj{\op{Z}}}

\newcommand{\eetaO}{\adj{\op{\eta}}}

\newcommand{\idO}{\ensuremath{\op{I}}}

\newcommand{\pOV}{\op{\vec{p}}}
\newcommand{\qOV}{\op{\vec{q}}}
\newcommand{\rOV}{\op{\vec{r}}}

\newcommand{\sigmaOV}{\op{\vec{\sigma}}}

\newcommand{\nord}[1]{\big\{#1\big\}}

\newcommand{\comm}[2]{\ensuremath{[{#1},{#2}]}}
\newcommand{\acomm}[2]{\ensuremath{ \big\{ {#1}, {#2} \big\} }}

\begin{document}

\title{In-Medium Similarity Renormalization Group Approach to the Nuclear Many-Body Problem}

\author{Heiko Hergert, Scott K.~Bogner, Justin G.~Lietz, Titus D.~Morris, Samuel Novario, Nathan M.~Parzuchowski, and Fei Yuan}
\institute{
Heiko Hergert  \at Department of Physics and Astronomy and National Superconducting Cyclotron Laboratory, Michigan State University, East Lansing, Michigan USA, \email{hergert@nscl.msu.edu}, \and 
Scott Bogner  \at Department of Physics and Astronomy and National Superconducting Cyclotron Laboratory, Michigan State University, East Lansing, Michigan USA, \email{bogner@nscl.msu.edu}, \and 
Justin G.~Lietz \at Department of Physics and Astronomy and National Superconducting Cyclotron Laboratory, Michigan State University, East Lansing, Michigan,  USA, \email{lietz@nscl.msu.edu}, \and 
Titus Morris \at Department of Physics and Astronomy, University of Tennessee, Knoxville, Tennessee, USA,
and Physics Division, Oak Ridge National Laboratory, Oak Ridge, Tennessee, USA, \email{titusmorris@gmail.com}, \and 
Samuel Novario \at Department of Physics and Astronomy and National Superconducting Cyclotron Laboratory, Michigan State University, East Lansing, Michigan,  USA, \email{novarios@nscl.msu.edu},\and 
Nathan Parzuchowski  \at Department of Physics and Astronomy and National Superconducting Cyclotron Laboratory, Michigan State University, East Lansing, Michigan USA, \email{parzuchowski@frib.msu.edu}, \and 
Fei Yuan \at Department of Physics and Astronomy and National
Superconducting Cyclotron Laboratory, Michigan State University, East
Lansing, Michigan USA, \email{yuan@nscl.msu.edu}} 
\maketitle
\abstract{
We present a pedagogical discussion of Similarity Renormalization
Group (SRG) methods, in particular the In-Medium SRG (IMSRG) approach for 
solving the nuclear many-body problem. These methods use continuous unitary
transformations to evolve the nuclear Hamiltonian to a desired shape. The
IMSRG, in particular, is used to decouple the ground state from all excitations
and solve the many-body Schr\"odinger equation. We discuss the IMSRG formalism 
as well as its numerical implementation, and use the method to study the pairing model 
and infinite neutron matter. We compare our results with those of Coupled cluster 
theory (Chapter 8), Configuration-Interaction Monte Carlo 
(Chapter 9), and the Self-Consistent Green's Function approach discussed 
in Chapter 11. The chapter concludes with an expanded overview of 
current research directions, and a look ahead at upcoming developments.
}

\section{Introduction}
Effective Field Theory (EFT) and Renormalization Group (RG) methods
have become important tools of modern (nuclear) many-body theory ---
one need only look at the table of contents of this book to see the
veracity of this claim.

Effective Field Theories allow us to systematically take into account the separation of
scales when we construct theories to describe natural phenomena. One
of the earliest examples that every physics student encounters is the 
effective force law of gravity near the surface of the Earth: For a
mass $m$ at a height $h$ above ground, Newton's force law becomes
\begin{equation}
  F(R+h) = G\frac{m M}{(R+h)^2} = m \frac{GM}{R^2} \frac{1}{1+\left(\frac{h}{R}\right)^2}
  =m \underbrace{\frac{GM}{R^2}}_{\equiv g} + \OC\left(\frac{h^2}{R^2}\right)\,,
\end{equation}
where $M$ and $R$ are the mass and radius of the Earth, respectively.
Additional examples are the multipole expansion of electric fields 
\cite{Jackson:1999yg}, which shows that only the moments of an electric
charge distribution with characteristic length scale $R$ are resolved
by probes with long wave lengths $\lambda\gg R$, or Fermi's theory of 
beta decay \cite{Fermi:1934eu}, which can nowadays be derived from the
Standard Model by expanding the propagator of the $W^\pm$ bosons that
mediate weak processes for small momenta $q\ll M_W=80\,\GeV$ (in units
where $\hbar=c=1$).

The strong-interaction sector of the Standard Model is provided by
Quantum Chromodynamics (QCD), but the description of nuclear
observables on the level of quarks and gluons is not feasible, except
in the lightest few-nucleon systems (see, e.g., \cite{Detmold:2015xw}
and the chapters on Lattice QCD in this book). The main issue is that
QCD is an asymptotically free theory
\cite{Gross:1973pd,Politzer:1973lq}, i.e., it is weak and amenable to
perturbative methods from Quantum Field Theory at large momentum
transfer, but highly non-perturbative in the low-momentum regime which
is relevant for nuclear structure physics. A consequence of the latter
property is that quarks are confined in baryons and mesons at low
momentum or energy scales, which makes those confined particles
suitable degrees of freedom for an EFT approach. Chiral EFT, in
particular, is constructed in terms of nucleon and pion fields, with
some attention now also being given to the lowest excitation mode of
the nucleon, namely the $\Delta$ resonance.  It provides provides a
constructive framework and organizational hierarchy for $NN$, $3N$,
and higher many-nucleon forces, as well as consistent electroweak
operators (see, e.g.,
\cite{Epelbaum:2009ve,Machleidt:2011bh,Epelbaum:2015gf,Entem:2015qf,Gezerlis:2014zr,Lynn:2016ec,Pastore:2009zr,Pastore:2011dq,Piarulli:2013vn,Kolling:2009yq,Kolling:2011bh}).
Since Chiral EFT is a low-momentum expansion, high-momentum
(short-range) physics is not explicitly resolved by the theory, but
parametrized by the so-called low-energy constants (LECs).  In
principle, the LECs can be determined by matching calculations of the
same observables in chiral EFT and (Lattice) QCD in the overlap region
of the two theories. Since such a calculation is currently not
feasible, they are fit to experimental data for low-energy QCD
observables, typically in the $\pi{}N$, $NN$, and $3N$ sectors
\cite{Epelbaum:2009ve,Machleidt:2011bh,Ekstrom:2015fk,Shirokov:2016wo}.

RG methods are natural companions for
EFTs, because smoothly connect theories with different resolution scales 
and degrees of freedom \cite{Lepage:1989hf,Lepage:1997py}. Since they 
were introduced in low-energy nuclear physics around the start 
of the millennium \cite{Bogner:2003os,Bogner:2007od,Bogner:2010pq,Furnstahl:2013zt}, 
they have provided a systematic framework for formalizing many ideas on the 
renormalization of nuclear interactions and many-body effects that had been 
discussed in the nuclear structure community since the 1950s. For instance,
soft and hard-core $NN$ interactions can reproduce scattering data equally
well, but have significantly different saturation properties, which caused
the community to all but abandon the former in the 1970s (see, e.g., \cite{Bethe:1971qf}). 
What was missing at that time was the recognition 
of the intricate link between the off-shell $NN$ interaction and $3N$ forces 
that was formally demonstrated for the first time by Polyzou and Gl\"ockle in 1990 \cite{Polyzou:1990fk}.
From the modern RG perspective, soft- and hard-core interactions emerge
as representations of low-energy QCD at different resolution scales, 
and the dialing of the resolution scale necessarily leads to induced
$3N$ forces, in such a way that observables (including saturation
properties) remain invariant under the RG flow (see Sec.~\ref{sec:srg_induced} 
and \cite{Bogner:2010pq,Furnstahl:2013zt}). In conjunction, chiral EFT
and nuclear RG applications demonstrate that one cannot treat the $NN, 3N, \ldots$
sectors in isolation from each other.

Brueckner introduced the idea of renormalizing the $NN$ interaction in the
nuclear medium by summing correlations due to the scattering of nucleon
pairs to high-energy states into the so-called $G$-matrix, which became
the basis of an attempted perturbation expansion of nuclear observables 
\cite{Brueckner:1954qf,Brueckner:1955rw,Bethe:1957qv,Goldstone:1957zz,Day:1967zl,Brandow:1967tg}.
Eventually, the nuclear structure community uncovered severe issues with
this approach, like a lack of order-by-order convergence 
\cite{Barrett:1970jl,Kirson:1971la,Barrett:1972bs,Kirson:1974oq,Goode:1974pi},
and a strong model space dependence in summations over virtual excitations
\cite{Vary:1973dn}. One of the present authors led a study that revisited 
this issue, and demonstrated that the $G$ matrix retains significant coupling 
between low- and high-momentum nodes of the underlying interaction 
\cite{Bogner:2010pq}, so the convergence issues are not surprising from a 
modern perspective. In the Similarity Renormalization Group \cite{Glazek:1993il,Wegner:1994dk}
and other modern RG approaches, low- and high-momentum physics are decoupled 
\emph{properly}, which results in low-momentum interactions that can be
treated successfully in finite-order many-body perturbation theory (MBPT) 
\cite{Bogner:2006qf,Bogner:2010pq,Roth:2010ys,Tichai:2016vl}. These 
interactions are not just suited as input for MBPT, but for all
methods that work with momentum- or energy-truncated configuration 
spaces. The decoupling of low- and high-momentum modes greatly improves 
the convergence behavior of such methods, which can then be applied to 
heavier and heavier nuclei
\cite{Barrett:2013oq,Jurgenson:2013fk,Hergert:2013ij,Roth:2014fk,Binder:2014fk,Hagen:2014ve,Hagen:2016rb}.

The idea of decoupling energy scales can also be used to directly
tackle the nuclear many-body problem. We implement it in the so-called
In-Medium SRG \cite{Tsukiyama:2011uq,Hergert:2013mi,Hergert:2016jk},
which is discussed at length in this chapter. In a nutshell, we will 
use SRG-like flow equations to decouple physics at different excitation 
energy scales of the nucleus, and render the Hamiltonian matrix in 
configuration space block or band diagonal in the process. We will see
that this can be achieved while working on the operator level, freeing
us from the need to construct the Hamiltonian matrix in a factorially 
growing basis of configurations. We will show that the IMSRG evolution
can also be viewed as a re-organization of the many-body perturbation
series, in which correlations that are described explicitly by the 
configuration space are absorbed into an \emph{RG-improved} Hamiltonian. 
With an appropriately chosen decoupling strategy, it is even possible to 
extract specific eigenvalues and eigenstates of the Hamiltonian, which is
why the IMSRG qualifies as an \emph{ab initio} (first principles) method 
for solving quantum many-body problems.

The idea of using flow equations to solve quantum many-body problems
dates back (at least) to Wegner's initial work on the SRG \cite{Wegner:1994dk} 
(also see \cite{Kehrein:2006kx} and references therein). In the 
solid-state physics literature, the approach is also known as 
continuous unitary transformation theory  
\cite{Heidbrink:2002kx,Drescher:2011kx,Krull:2012bs,Fauseweh:2013zv,Krones:2015ft}.
When we discuss our decoupling strategies for the nuclear many-body
problem, it will become evident that the IMSRG is related to Coupled Cluster theory (CC), see also chapter 8,  
and a variety of other many-body methods that are used heavily in 
quantum chemistry (see, e.g., \cite{Shavitt:2009,Hagen:2014ve,White:2002fk,
Yanai:2007kx,Nakatsuji:1976yq,Mukherjee:2001uq,Mazziotti:2006fk,Evangelista:2014rq}).
What sets the IMSRG apart from these methods is that the Hamiltonian 
instead of the wave function is at the center of attention, in the
spirit of RG methodology. This seems to be a trivial distinction, but
there are practical advantages of this viewpoint, like the ability to
simultaneously decouple ground and a number of excited states from the
rest of the spectrum (see Secs.~\ref{sec:imsrg_decoupling} and 
\ref{sec:current_hamiltonians}).

%
%
\subsubsection*{Organization of this Chapter}
We conclude our introduction by looking ahead at the remainder of this
chapter. In Sec.~\ref{sec:srg}, we will introduce the basic concepts of
the SRG, and apply it to a pedagogical toy model
(Sec.~\ref{sec:srg_toy}), the pairing Hamiltonian that is also discussed 
in Chapters 8, 9, and 11 
(Sec.~\ref{sec:srg_pairing}), and last but not least, we will discuss the 
SRG evolution of modern nuclear interactions (Sec.~\ref{sec:srg_interactions}). 

The issue of SRG-induced operators (Sec.~\ref{sec:srg_induced}) will serve 
as our launching point for the discussion of the IMSRG in Sec.~\ref{sec:imsrg}.
First, we will introduce normal-ordering techniques as a means to control the 
size of induced interaction terms (Sec.~\ref{sec:nord}). This is followed
by the derivation of the IMSRG flow equations, determination of decoupling
conditions, and the construction of generators in Secs.~\ref{sec:imsrg_flow}--\ref{sec:imsrg_generator}.
We discuss the essential steps of an IMSRG implementation through the 
example of a symmetry-unconstrained Python code (Sec.~\ref{sec:imsrg_implementation}), 
and use this code to revisit the pairing Hamiltonian in Sec.~\ref{sec:imsrg_pairing}.
In Sec.~\ref{sec:imsrg_neutron_matter}, we compute the neutron matter 
equation-of-state in the IMSRG(2) truncation scheme, and compare our result
to that of corresponding Coupled Cluster, Quantum Monte Carlo, and Self-Consistent 
Green's Function results with the same interaction. 

Section \ref{sec:current} introduces the three major directions of
current research: First, we present the Magnus formulation of the
IMSRG (Sec.~\ref{sec:current_magnus}), which is the key to the
efficient computation of observables and the construction of
approximate version of the IMSRG(3). Second, we give an overview of
the multireference IMSRG (MR-IMSRG), which generalizes our framework
to arbitrary reference states, and gives us new freedom to manipulate
the correlation content of our many-body calculations
\ref{sec:current_mrimsrg}.  Third, we will discuss applications of
IMSRG-evolved, RG-improved Hamiltonians as input to many-body
calculations, in particular for the nuclear (interacting) Shell model
and Equation-of-Motion (EOM) methods
(Sec.~\ref{sec:current_hamiltonians}). An outlook on how these three
research thrusts interweave concludes the section
(Sec.~\ref{sec:current_remarks}).

In Sec.~\ref{sec:conclusions}, we make some final remarks and close the main 
body of the chapter in Sec.~\ref{sec:conclusions}.
Section \ref{sec:exercises_projects} contains exercises that further flesh 
out subjects discussed in the preceding sections, as well as 
outlines for computational projects. Formulas for products and commutators 
of normal-ordered operators are collected in an Appendix.

\section{\label{sec:srg}The Similarity Renormalization Group}

\subsection{Concept}
The basic idea of the Similarity Renormalization Group (SRG) method is 
quite general: We want to ``simplify'' our system's Hamiltonian
$\HO(s)$ by means of a continuous unitary transformation that is parametrized
by a one-dimensional parameter $s$,
\begin{equation}\label{eq:cut}
  \HO(s)=\UO(s)\HO(0)\UUO(s)\,.
\end{equation}
By convention, $\HO(s=0)$ is the starting Hamiltonian. To specify what we mean by 
simplifying $\HO$, it is useful to briefly think of it as a matrix rather than an 
operator. As in any quantum-mechanical problem, we are primarily interested in 
finding the eigenstates of $\HO$ by diagonalizing its matrix representation. This task 
is made easier if we can construct a unitary transformation that renders the Hamiltonian 
more and more diagonal as $s$ increases. Mathematically, we want to split the Hamiltonian 
into suitably defined diagonal and off-diagonal parts,
\begin{equation}
  \HO(s) = \HO_d(s) + \HO_{od}(s)\,,
\end{equation}
and find $\UO(s)$ so that
\begin{equation}\label{eq:trafo_goal}
  \HO(s)\underset{s\to\infty}{\longrightarrow}\HO_d(s)\,,\quad \HO_{od}(s)\underset{s\to\infty}{\longrightarrow} 0\,.
\end{equation}

To implement the continuous unitary transformation, we take the derivative of
Eq.~\eqref{eq:cut} with respect to $s$ to obtain 
\begin{equation}
  \totd{\HO(s)}{s} = \totd{\UO(s)}{s}\HO(0)\UUO(s) + \UO(s)\HO(0)\totd{\UUO(s)}{s}
                   = \totd{\UO(s)}{s}\UUO(s)\HO(s) + \HO(s)\UO(s)\totd{\UUO(s)}{s}\,.
\end{equation}
Since $\UO(s)$ is unitary, we also have
\begin{equation}
  \totd{}{s}\left(\UO(s)\UUO(s)\right) = \totd{}{s} \left(\idO\right) = 0 \quad \Longrightarrow 
  \quad \totd{\UO(s)}{s}\UUO(s) = - \UO(s)\totd{\UUO(s)}{s}\,.
\end{equation}
Defining the anti-Hermitian operator
\begin{equation}\label{eq:def_eta}
  \etaO(s) \equiv \totd{\UO(s)}{s} \UUO(s) = - \eetaO(s)\,,
\end{equation}
we can write the differential equation for the $s$-dependent Hamiltonian as
\begin{equation} \label{eq:opflow}
  \totd{}{s} \HO(s) = \comm{\etaO(s)}{\HO(s)}\,.
\end{equation}
This is the SRG \emph{flow equation} for the Hamiltonian, which describes the 
evolution of $\HO(s)$ under the action of a dynamical generator $\etaO(s)$. 
Since we are considering a \emph{unitary} transformation, the spectrum of the 
Hamiltonian is preserved\footnote{There are mathematical subtleties due to 
$\HO(s)$ being an operator that is only bounded from below, and having a 
spectrum that is part discrete, part continuous (see, e.g., \cite{Bach:2010zr,Boutin:2016ef})
. In practice, we are forced
to work with approximate, finite-dimensional matrix representations of $\HO(s)$
in any case.}. Thus, the SRG is related to so-called isospectral flows, a class 
of transformations that has been studied extensively in the mathematics literature 
(see for example Refs.~\cite{Brockett:1991kx,Chu:1994vn,Chu:1995ys,Bach:2010zr,Boutin:2016ef}).

The flow equation \eqref{eq:opflow} is the most practical way of implementing
an SRG evolution: We can obtain $\HO(s)$ by integrating Eq.~\eqref{eq:opflow} 
numerically, without explicitly constructing the unitary transformation itself. 
Formally, we can also obtain $\UO(s)$ by rearranging Eq.~\eqref{eq:def_eta} into
\begin{equation}\label{eq:Uflow}
  \totd{}{s}\UO(s) = \etaO(s)\UO(s)\,.
\end{equation}
The solution to this differential equation is given by the $S$-ordered 
exponential
\begin{align}
  U(s) &= \mathcal{S}\exp \int^s_0 ds'\,\eta(s') \label{eq:def_U_pathexp}\,,
\end{align}
because the generator changes dynamically during the flow. This expression
is defined equivalently either as a product of infinitesimal unitary transformations,
\begin{align}       \label{eq:def_U_expds}
   U(s) &= \lim_{N\to\infty}\prod^{N}_{i=0} e^{\eta(s_i)\delta s_i}\,,\quad s_{i+1}=s_i+\delta s_i\,,\quad \sum_{i}\delta s_i=s\,,
\end{align}
or through a series expansion:
\begin{align}     \label{eq:def_U_series}
   U(s) &= \sum_n \frac{1}{n!}\int^s_0 ds_1 \int^s_0 ds_2 \ldots 
          \int^s_0 ds_n \mathcal{S}\{\eta(s_1)\ldots\eta(s_n)\}\,.
\end{align}
Here, the $S$-ordering operator $\mathcal{S}$ ensures that the flow parameters 
appearing in the integrands are always in descending order,
$s_1 > s_2 > \ldots$. Note that neither Eq.~\eqref{eq:def_U_expds} nor Eq.~\eqref{eq:def_U_series} 
can be written as a single proper exponential, so we do not obtain
a simple Baker-Campbell-Hausdorff expansion of the transformed
Hamiltonian. Instead, we would have to use these complicated expressions 
to construct $\HO(s)=\UO(s)\HO(0)\UUO(s)$, and to make matters even worse, Eqs.~\eqref{eq:def_U_expds}
and \eqref{eq:def_U_series} depend on the generator at \emph{all intermediate points} 
of the flow trajectory. The associated storage needs would make numerical applications
impractical or entirely unfeasible.

Let us focus on the flow equation \eqref{eq:opflow}, then, and specify a
generator that will transform the Hamiltonian to the desired structure 
(Eq.~\eqref{eq:trafo_goal}). Inspired by the work of Brockett \cite{Brockett:1991kx} 
on the so-called double-bracket flow, Wegner \cite{Wegner:1994dk} proposed the generator
\begin{equation}\label{eq:def_Wegner_general}
  \etaO(s) \equiv \comm{H_d(s)}{H_{od}(s)}\,.
\end{equation}
A fixed point of the SRG flow is reached when $\etaO(s)$ vanishes. At finite $s$, this 
can occur if $\HO_d(s)$ and $\HO_{od}(s)$ happen to commute, e.g., due to a degeneracy 
in the spectrum of $\HO(s)$. A second fixed point at $s\to\infty$ exists if $\HO_{od}(s)$ 
vanishes as required. 

Going back over the discussion, you may notice that we never specified in
detail \emph{how} we split the Hamiltonian into diagonal and off-diagonal
parts. By ``diagonal'' we really mean the desired structure of the Hamiltonian, 
and ``off-diagonal'' labels the contributions we have to suppress in the
limit $s\to\infty$ to obtain that structure. The basic concepts described
here are completely general, and we will discuss two examples in which we
apply them to the diagonalization of matrices in the following.
The renormalization of Hamiltonians (or other operators) is a more specific 
application of continuous unitary transformations. We make contact with 
renormalization ideas by imposing a block or band-diagonal structure on
the representation of operators in bases that are organized by momentum
or energy. This implies a decoupling of low and high momenta or energies
in the renormalization group sense. We will conclude this section with a
brief discussion of how this SRG decoupling of scales is used to render  
nuclear Hamiltonians more suitable for \emph{ab initio} many-body 
calculations \cite{Bogner:2007od,Bogner:2010pq,Morris:2015ve,Hergert:2016jk,Hergert:2017kx}.

\subsection{\label{sec:srg_toy}A Two-Dimensional Toy Problem}

In order to get a better understanding of the SRG method, we first consider  
a simple $2\times 2$ matrix problem that can be solved analytically, and 
compare the flow generated by Eq.~\eqref{eq:opflow} with standard 
diagonalization algorithms like Jacobi's rotation method (see, e.g., 
Ref.~\cite{Golub:2013le}).

Let us consider a symmetric matrix $H$, 
\begin{equation} 
  H \equiv \begin{pmatrix} H_{11} & H_{12} \\ H_{12} & H_{22}\end{pmatrix}. 
\end{equation}
and an orthogonal (i.e., unitary and real) matrix $U$,
\begin{equation}
  U = \begin{pmatrix} \cos\gamma & \sin\gamma \\ -\sin\gamma & \cos\gamma \end{pmatrix}, 
\end{equation}
that parameterizes a rotation of the basis in which $H$ and $U$ are
represented. We want to find an angle $\gamma$ so that $H' = UHU^T$ is diagonal, 
and to achieve this, we need to solve
\begin{equation}
(H_{22} - H_{11})\cos\gamma\,\sin\gamma + H_{12}(\cos^2\gamma - \sin^2\gamma) = 0\,.
\end{equation}
Using the addition theorems $\cos^2\gamma-\sin^2\gamma = \cos(2\gamma)$ and 
$\cos\gamma\,\sin\gamma = \sin(2\gamma)/2$, we can rewrite this equation as
\begin{equation}
  \tan(2\gamma) = \frac{2 H_{12}}{H_{11}-H_{22}}
\end{equation}
and obtain
\begin{equation} 
\gamma = \frac{1}{2} \tan^{-1} \left( \frac{2H_{12}}{H_{11}-H_{22}}
\right) + \frac{k\pi}{2}, \quad k \in \mathbb{Z}, \label{eq:0} 
\end{equation}
where $k\pi/2$ is added due to the periodicity of the $\tan$ function.
Note that  $k=0$ gives a diagonal matrix of the form
\begin{equation} 
H'_{k=0} = \begin{pmatrix} E_1 & 0 \\ 0 & E_2 \end{pmatrix},
\label{eq:1} 
\end{equation}
while  $k=1$ interchanges the diagonal elements:  
\begin{equation} 
H'_{k=1} = \begin{pmatrix} E_2 & 0 \\ 0 & E_1 \end{pmatrix}.
\label{eq:2}
\end{equation}

Let us now solve the same problem with an SRG flow. We parameterize the Hamiltonian
as
\begin{equation}
  H(s) = H_d(s) + H_{od}(s) \equiv \begin{pmatrix}E_1(s) & 0 \\ 0 & E_2(s)\end{pmatrix} + 
         \begin{pmatrix} 0 & V(s) \\ V(s) & 0\end{pmatrix}\,,
\end{equation}
working in the eigenbasis of $H_d(s)$. We can express $H(s)$ in terms of the $2\times2$ 
identity matrix and the Pauli matrices:
\begin{equation}
  H(s) = E_{+}(s) \idO + E_{-}(s)\sigma_3 
         + V(s)\sigma_1\,,
\end{equation}
where we have introduced
\begin{equation}
  E_{\pm}(s) \equiv \frac{1}{2}\left(E_1(s) \pm E_2(s)\right)\,.
\end{equation}

The Wegner generator can be determined readily using the algebra of
the Pauli matrices,
$\comm{\sigma_j}{\sigma_k}=2i\varepsilon_{jkl}\sigma_l$:
\begin{equation}
  \eta(s) = \comm{H_d(s)}{H_{od}(s)} = 2 i E_{-}(s) V(s) \sigma_2\,.
\end{equation}
By evaluating both $\dot{H}(s)$ as well as $\comm{\etaO(s)}{\HO(s)}$
and comparing the coefficients of the $2\times2$ matrices, we obtain the
following system of flow equations:
\begin{align}
  \dot{E}_+ & = 0 \label{eq:flow_Eplus}\\
  \dot{E}_{-} &= 4 V^2 E_{-} \label{eq:flow_Eminus}\\
  \dot{V}     &= -4 E_{-}^2V \label{eq:flow_V}\,,
\end{align}
where we have suppressed the flow parameter dependence for brevity.
The first flow equation reflects the conservation of the Hamiltonian's
trace under unitary transformations,
\begin{equation}
  \tr (E_1(s) + E_2(s)) = \tr (E_1(0) + E_2(0)) = \text{const.}
\end{equation} 
The third flow equation can be rearranged into
\begin{equation}
  \frac{1}{V}dV = - 4 E_{-}^2 ds
\end{equation}
and integrated, which yields
\begin{equation}
  V(s) = V(0)\exp\left(-4\int_0^sds'\,E^2_{-}(s') \right)\,.
\end{equation}
Since $E_{-}(s)$ is real, the integral is positive for all values
of $s$, and this means that the off-diagonal matrix element will 
be suppressed exponentially as we evolve $s\to\infty$ (barring
singular behavior in $E_{\pm}(s)$).

To proceed, we introduce new variables $\Omega$ and $\theta$:
\begin{equation}
  \Omega \equiv \sqrt{E_{-}^2 + V^2}\,,\quad \tan\frac{\theta}{2} \equiv \frac{V}{E_{-}}\,.
\end{equation}
Using Eqs.~\eqref{eq:flow_Eminus} and \eqref{eq:flow_V}, we can show 
that $\Omega$ is a flow invariant:
\begin{equation}
  \dot\Omega = \frac{1}{2\Omega}(2 E_{-}\dot{E}_{-} + 2 V \dot{V}) = 0\,.
\end{equation}
Rewriting $E_{-}(s)$ and $V(s)$ in terms of $\Omega$ and $\theta(s)$,
we then have
\begin{equation}\label{eq:srg_toy_parameters}
  E_{-}(s) = \Omega \cos\frac{\theta(s)}{2}\,,\quad V(s) = \Omega \sin\frac{\theta(s)}{2}\,.
\end{equation}
Using these expressions, we find that Eqs.~\eqref{eq:flow_Eminus} and \eqref{eq:flow_V}
reduce to a single differential equation for $\theta(s)$:
\begin{equation}
  \dot\theta = - 8\Omega^2\sin\frac{\theta(s)}{2}\cos\frac{\theta(s)}{2} = -4\Omega^2\sin\theta\,.
\end{equation}
Bringing $\sin\theta$ to the left-hand side and using
\begin{equation}
  \totd{}{x}\ln\tan\frac{x}{2}=\frac{1}{\sin x}\,,
\end{equation}
we can integrate the ordinary differential equation (ODE) and obtain
\begin{equation}\label{eq:srg_toy_theta}
  \tan\frac{\theta(s)}{2} = \tan\frac{\theta(0)}{2} e^{-4\Omega^2 s}\,.
\end{equation}
At $s=0$ we have
\begin{equation}
  \theta(0) = 2 \tan^{-1} \frac{V(0)}{E_{-}(0)} + 2k\pi= 2 \tan^{-1}\frac{2 V(0)}{E_{1}(0) - E_{2}(0)}+2k\pi\,,
  \quad k \in \mathbb{Z}\,,
\end{equation}
which is just four times the angle of the Jacobi rotation that diagonalizes our initial
matrix, Eq.~\eqref{eq:0}. Likewise $\theta(s)$ is (up to the prefactor) the 
angle of the Jacobi rotation that will diagonalize the evolved $H(s)$ for $s>0$. In 
the limit $s\to\infty$,
$\theta(s)\to 0$ because the SRG flow has driven the off-diagonal matrix element to 
zero and the Hamiltonian is already diagonal.

\begin{figure}[t]
  \begin{center}
    \includegraphics[width=0.7\textwidth]{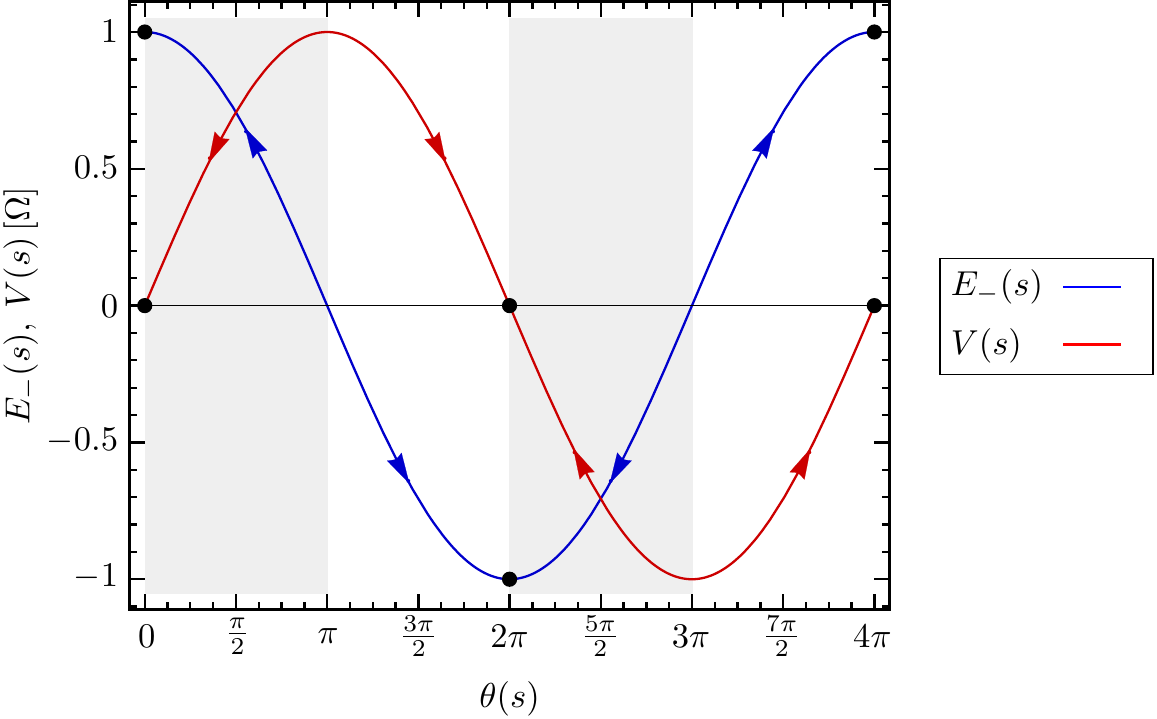}
  \end{center}
  \caption{\label{fig:srg_toy}
    $E_{-}(s)$ and $V(s)$ as a function of the flowing angle $\theta(s)$, in units of the
    flow invariant $\Omega$. Black dots and
    arrows the fixed points at $s\to\infty$ and the directions of the SRG flow, respectively,
    in the domains corresponding to specific sign combinations for $E_{-}(s)$ and $V(s)$ (see
    text).
  }
\end{figure}

When we introduced the parameterization \eqref{eq:srg_toy_parameters}, we chose $\Omega$ 
to be positive, which means that $\theta(s)$ must encode all information on the signs of 
$E_{-}(s)$ and $V(s)$. In Fig.~\ref{fig:srg_toy}, we show these quantities as a function
of $\theta(s)$ over the interval $[0,4\pi]$. We see that the four possible sign 
combinations correspond to distinct regions in the interval. We can map any set of initial
values --- or any point of the flow, really --- to a distinct value $\theta(s)$ 
in this figure, even in limiting cases: For instance, if the diagonal matrix elements 
are degenerate, $E_1=E_2$, we have $E_{-}=0$, the angle $\theta$ will approach $\pm\pi$ and 
$V(s)/\Omega\to\pm 1$. From this point the SRG flow will drive $E_{-}(s)$ and $V(s)$ to 
the nearest fixed point at a multiple of $2\pi$ according to the trajectory for $\theta(s),
$Eq.~\eqref{eq:srg_toy_theta}. The fixed points and flow directions are indicated in the 
figure.

\subsection{\label{sec:srg_pairing}The Pairing Model}

%
%
\subsubsection{Preliminaries}
\begin{table}[t]
  \begin{center}
      \begin{tabular*}{0.25\textwidth}{@{\extracolsep\fill}|c|c|r|r|}
      \hline 
      state & $p$ & $2s_z$ & $\varepsilon$
      \\ \hline 
      $0$ & $1$ & $ 1$ & $     0 $ \\ 
      $1$ & $1$ & $-1$ & $     0 $ \\ 
      $2$ & $2$ & $ 1$ & $ \delta$ \\ 
      $3$ & $2$ & $-1$ & $ \delta$ \\ 
      $4$ & $3$ & $ 1$ & $2\delta$ \\ 
      $5$ & $3$ & $-1$ & $2\delta$ \\ 
      $6$ & $4$ & $ 1$ & $3\delta$ \\ 
      $7$ & $4$ & $-1$ & $3\delta$ \\ \hline
      \end{tabular*}
  \end{center}
  \caption{\label{tab:srg_pairing_sp}Single-particle states and their quantum numbers and their energies from Eq.~\eqref{eq:def_pairing_hamiltonian}. The degeneracy for every quantum number $p$ is equal to 2 due to the two possible spin values.} 
\end{table}

As our second example for diagonalizing matrices by means of SRG flows, we will
consider the pairing model that was discussed in the context of Hartree-Fock and
beyond-HF methods in chapter 8. In second quantization, the 
pairing Hamiltonian is given by
\begin{equation}\label{eq:def_pairing_hamiltonian}
  \HO = \delta \sum_{p \sigma} (p-1) a^{\dagger}_{p \sigma} a_{p\sigma}
        -\frac{1}{2}g \sum_{pq} a^{\dagger}_{p+}a^{\dagger}_{p-}a_{q-}a_{q+}\,,
\end{equation}
where $\delta$ controls the spacing of single-particle levels that are indexed
by a principal quantum number $p=0,\ldots,3$ and their spin projection $\sigma$ (see 
Tab.~\ref{tab:srg_pairing_sp}), and $g$ the strength of the pairing interaction. 

We will consider the case of four particles in eight single-particle states. 
Following the Full Configuration Interaction (FCI) approach discussed in 
Chap.~8, we can construct a many-body basis of Slater 
determinants by placing our particles into the available single-particle 
basis states. Since each single-particle state can only be occupied by one 
particle, there are
\begin{equation}
  \begin{pmatrix} 8 \\ 4\end{pmatrix} = 70
\end{equation}
unique configurations. The specific form of the pairing interaction
implies that the total spin projection $S_z = \sum_{n=1}^4 s_z^{(n)}$ 
is conserved, and the Hamiltonian will have a block diagonal structure.
The possible values for $S_z$ are $-2,-1,0,1,2$, depending on the number
of particles in states with spin up ($N_+$) or spin down ($N_-$). The 
dimension can be calculated via
\begin{equation}
 d_{S_z} = \begin{pmatrix} 4 \\ N_{+}\end{pmatrix} \times 
           \begin{pmatrix} 4 \\ N_{-}\end{pmatrix}\,,
\end{equation}
which yields 
\begin{equation}
 d_{\pm 2} = 1\,, \quad d_{\pm 1} = 16\,, \quad d_{0} = 36\,.
\end{equation}
Since the pairing interaction only couples pairs of single-particle 
states with the same principal quantum number but opposite spin
projection, it does not break pairs of particles that occupy such 
states --- in other words, the number of particle pairs is another
conserved quantity, which allows us to decompose the $S_z$ blocks
into even smaller sub blocks. As in chapter 8, we 
consider the $S_z=0$
sub block with two particle pairs. In this block, the Hamiltonian 
is represented by the six-dimensional matrix (suppressing block 
indices)
\begin{equation}\label{eq:def_h_matrix}
  H = \begin{pmatrix}
  2\delta -g  &      -g/2  &       -g/2 &      -g/2 &      -g/2 &        0 \\ 
         -g/2 & 4\delta -g &       -g/2 &      -g/2 &        -0 &     -g/2 \\ 
         -g/2 &       -g/2 & 6\delta -g &         0 &      -g/2 &     -g/2 \\ 
         -g/2 &       -g/2 &          0 & 6\delta-g &      -g/2 &     -g/2 \\ 
         -g/2 &          0 &       -g/2 &      -g/2 & 8\delta-g &     -g/2 \\ 
            0 &       -g/2 &       -g/2 &      -g/2 &      -g/2 & 10\delta -g
  \end{pmatrix}\,.
\end{equation}

%
%
\subsubsection{\label{sec:srg_pairing_flow}SRG Flow for the Pairing Hamiltonian}
As in earlier sections, we split the Hamiltonian matrix \eqref{eq:def_h_matrix}
into diagonal and off-diagonal parts:
\begin{equation}
  H_d(s) = \mathrm{diag}(E_0(s)\,,\ldots,E_5(s))\,,\quad H_{od}(s) = H(s) - H_{d}(s)\,,
\end{equation}
with initial values defined by Eq.~\eqref{eq:def_h_matrix}. Since $H_d(s)$
is diagonal throughout the flow per construction, the Slater determinants
that span our specific subspace are the eigenstates of this matrix. In our 
basis representation, Eq.~\eqref{eq:opflow} can be written as
\begin{align}
  \totd{}{s}\matrixe{i}{\HO}{j}
  &=\sum_k\left(\matrixe{i}{\etaO}{k}\matrixe{k}{\HO}{j}-\matrixe{i}{\HO}{k}\matrixe{k}{\etaO}{j}\right)\notag\\
  &=-\left(E_i-E_j\right)\matrixe{i}{\etaO}{j}
    +\sum_k\left(\matrixe{i}{\etaO}{k}\matrixe{k}{\HO_{od}}{j}-\matrixe{i}{\HO_{od}}{k}\matrixe{k}{\etaO}{j}\right),
\end{align}
where $\matrixe{i}{\HO_{od}}{i}=0$ and block indices as well as the $s$-dependence
have been suppressed for brevity. The Wegner generator, Eq.~\eqref{eq:def_Wegner_general}, 
is given by
\begin{equation}
  \matrixe{i}{\etaO}{j}=\matrixe{i}{\comm{\HO_d}{\HO_{od}}}{j}=(E_i-E_j)\matrixe{i}{\HO_{od}}{j}\,,
\end{equation}
and inserting this into the flow equation, we obtain
\begin{align}
  \totd{}{s}\matrixe{i}{\HO}{j}
  =-\left(E_i-E_j\right)^2\matrixe{i}{\HO_{od}}{j}
  +\sum_k\left(E_i+E_j-2E_k\right)\matrixe{i}{\HO_{od}}{k}\matrixe{k}{\HO_{od}}{j}\,.\label{eq:schema_wegner}
\end{align}
Let us assume that the transformation generated by $\etaO$ truly suppresses 
$\HO_{od}$, and consider the asymptotic behavior for large flow parameters 
$s>s_0\gg0$. If $||\HO_{od}(s_0)||\ll 1$ in some suitable norm, the second 
term in the flow equation can be neglected compared to the first one. For
the diagonal and off-diagonal matrix elements, this implies
\begin{align}
  \totd{E_i}{s} &= \totd{}{s} \matrixe{i}{\HO_d}{i} = 2\sum_k (E_i - E_k) \matrixe{i}{\HO_{od}}{k}\matrixe{k}{\HO_{od}}{i}
  \approx0
\end{align}
and
\begin{align}
  \totd{}{s}\matrixe{i}{\HO}{j}
  \approx-\left(E_i-E_j\right)^2\matrixe{i}{\HO_{od}}{j}\,,
\end{align}
respectively. Thus, the diagonal matrix elements will be (approximately)
constant in the asymptotic region,
\begin{equation}
  E_i(s) \approx E_i(s_0)\,,\quad s>s_0\,,
\end{equation}
which in turn allows us to integrate the flow equation for the off-diagonal
matrix elements. We obtain
\begin{equation}\label{eq:def_hod_general}
  \matrixe{i}{\HO_{od}(s)}{j} \approx \matrixe{i}{\HO_{od}(s_0)}{j}\,e^{-(E_i-E_j)^2 (s-s_0)}\,, \quad s>s_0\,,
\end{equation}
i.e., the off-diagonal matrix elements are suppressed exponentially, as for the
$2\times2$ matrix toy problem discussed in the previous section. If the pairing
strength $g$ is sufficiently small so that $\HO^2_{od}(0)\sim \mathcal{O}(g^2)$ can 
be neglected, we expect to see the exponential suppression of the off-diagonal 
matrix elements from the very onset of the flow. 

Our solution for the off-diagonal matrix elements, Eq.~\eqref{eq:def_hod_general},
shows that the characteristic decay scale for each matrix element is determined 
by the square of the energy difference between the states it couples, 
$(\Delta E_{ij})^2 = (E_i-E_j)^2$. Thus, states with larger energy differences
are decoupled before states with small energy differences, which means that 
the Wegner generator generates a proper renormalization group transformation.
Since we want to diagonalize \eqref{eq:def_h_matrix} in the present example,
we are only interested in the limit $s\to\infty$, and it does not really matter
whether the transformation is an RG or not. Indeed, there are alternative choices
for the generator which are more efficient in achieving the desired diagonalization
(see Sec.~\ref{sec:imsrg_generator} and Refs.~\cite{Hergert:2016jk,Hergert:2017kx}).
The RG property will matter in our discussion of SRG-evolved nuclear interactions 
in the next section.

%
%
\subsubsection{\label{sec:srg_pairing_implementation}Implementation}
We are now ready to discuss the implementation of the SRG flow for the pairing
Hamiltonian. The main numerical task is the integration of the flow equations,
a system of first order ODEs. Readers who are interested in learning the 
nuts-and-bolts details of implementing an ODE solver are referred to the excellent 
discussion in \cite{Press:2007vn}, while higher-level discussions of the algorithms
can be found, e.g., in \cite{Shampine:1975qq,Landau:2012zr,Hjorth-Jensen:2015mz}.
A number of powerful ODE solvers have been developed over the past decades and 
integrated into readily available libraries, and we choose to rely on one of these 
here, namely ODEPACK (see \url{www.netlib.org/odepack} and 
\cite{Hindmarsh:1983pd,Radhakrishnan:1993fk,Brown:1989qd}). The SciPy package
provides convenient Python wrappers for the ODEPACK solvers.

The following source code shows the essential part of our Python implementation of
the SRG flow for the pairing model. The full program with additional features can 
be downloaded from 
\url{https://github.com/ManyBodyPhysics/LectureNotesPhysics/tree/master/Programs/Chapter10-programs/python/srg_pairing}.

\begin{lstlisting}
import numpy as np
from numpy import array, dot, diag, reshape
from scipy.linalg import eigvalsh
from scipy.integrate import odeint

# Hamiltonian for the pairing model
def Hamiltonian(delta,g):

    H = array(
      [[2*delta-g,    -0.5*g,     -0.5*g,     -0.5*g,    -0.5*g,          0.],
       [   -0.5*g, 4*delta-g,     -0.5*g,     -0.5*g,        0.,     -0.5*g ], 
       [   -0.5*g,    -0.5*g,  6*delta-g,         0.,    -0.5*g,     -0.5*g ], 
       [   -0.5*g,    -0.5*g,         0.,  6*delta-g,    -0.5*g,     -0.5*g ], 
       [   -0.5*g,        0.,     -0.5*g,     -0.5*g, 8*delta-g,     -0.5*g ], 
       [       0.,    -0.5*g,     -0.5*g,     -0.5*g,    -0.5*g, 10*delta-g ]]
    )

  return H

# commutator of matrices
def commutator(a,b):
    
    return dot(a,b) - dot(b,a)

# right-hand side of the flow equation
def derivative(y, t, dim):

    # reshape the solution vector into a dim x dim matrix
    H = reshape(y, (dim, dim))

    # extract diagonal Hamiltonian...
    Hd  = diag(diag(H))

    # ... and construct off-diagonal the Hamiltonian
    Hod = H-Hd

    # calculate the generator
    eta = commutator(Hd, Hod)

    # dHdt is the derivative in matrix form 
    dH  = commutator(eta, H)

    # convert dH into a linear array for the ODE solver
    dy = reshape(dH, -1)
      
    return dy


#------------------------------------------------------------------------------
# Main program
#------------------------------------------------------------------------------
def main():
    g     = 0.5
    delta = 1

    H0    = Hamiltonian(delta, g)
    dim   = H0.shape[0]

    # calculate exact eigenvalues
    eigenvalues = eigvalsh(H0)

    # turn initial Hamiltonian into a linear array
    y0  = reshape(H0, -1)                 

    # flow parameters for snapshot images
    flowparams = array([0.,0.001,0.01,0.05,0.1, 1., 5., 10.])

    # integrate flow equations - odeint returns an array of solutions,
    # which are 1d arrays themselves
    ys  = odeint(derivative, y0, flowparams, args=(dim,))

    # reshape individual solution vectors into dim x dim Hamiltonian
    # matrices
    Hs  = reshape(ys, (-1, dim,dim))
\end{lstlisting}

The routine \texttt{Hamiltonian} sets up the Hamiltonian matrix \eqref{eq:def_h_matrix}
for general values of $\delta$ and $g$. The right-hand side of the flow equation
is implemented in the routine \texttt{derivative}, which splits $H(s)$
in diagonal and off-diagonal parts, and calculates both the generator and
the commutator $\comm{\etaO}{\HO}$ using matrix products. Since the
interface of essentially all ODE libraries require the ODE system and
derivative to be passed as a one-dimensional array, NumPy's \texttt{reshape}
functionality used to rearrange these arrays into $6\times 6$ matrices 
and back again.

The main routine calls the ODEPACK wrapper \texttt{odeint}, passing 
the initial Hamiltonian as a one-dimensional array \texttt{y0} as well
as a list of flow parameters $s$ for which a solution is requested. The 
routine \texttt{odeint} returns these solutions as a two-dimensional arrays.

%
%
\subsubsection{\label{sec:srg_pairing_example}A Numerical Example}
\begin{figure*}[t]
  \begin{center}
    \includegraphics[width=0.95\textwidth]{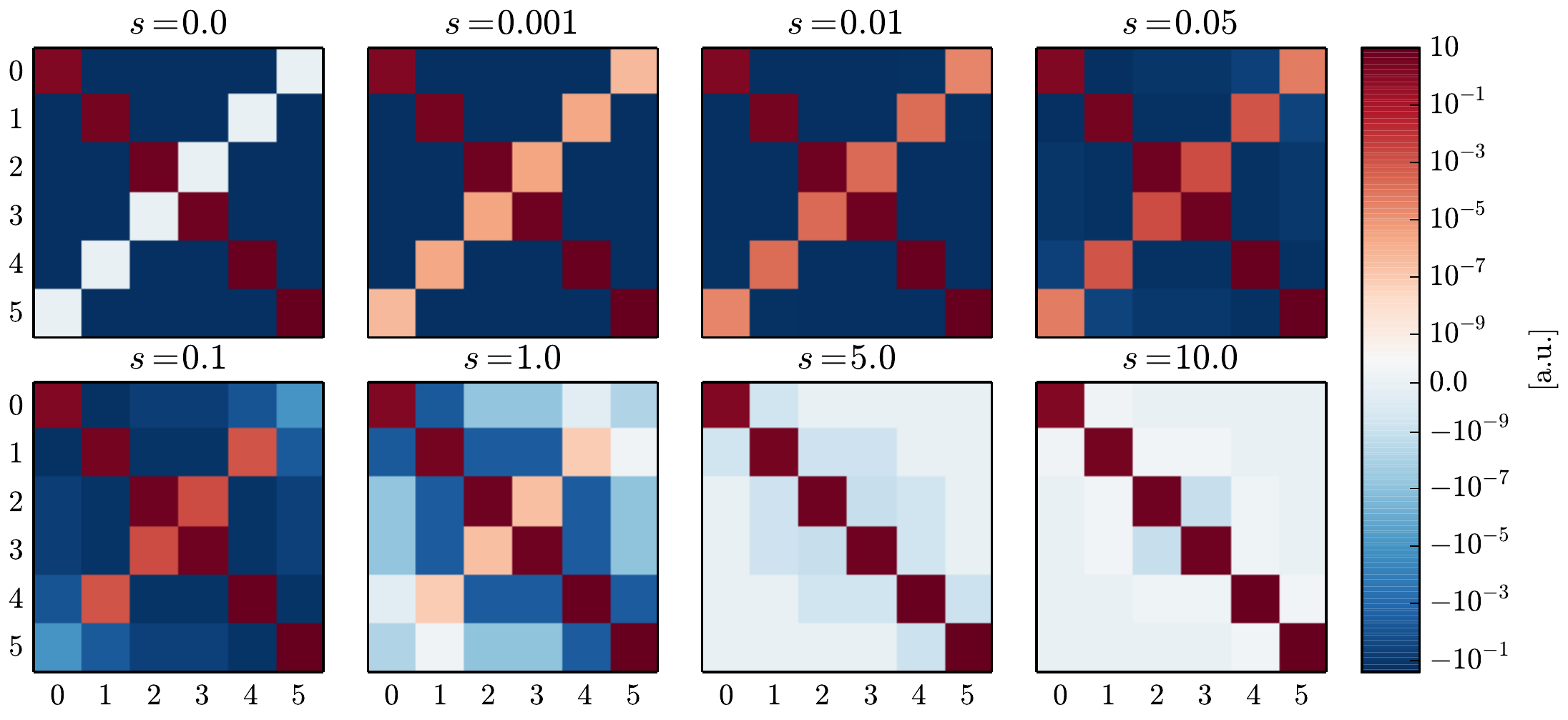}
  \end{center}    
  \caption{\label{fig:srg_pairing}SRG evolution of the pairing Hamiltonian with
  $\delta=1, g=0.5$. The 
  figures show snapshots of the Hamiltonian's matrix representation at various 
  stages of the flow, indicated by the flow parameters $s$. Note the essentially 
  logarithmic scales of the positive and negative matrix elements, which are bridged
  by a linear scale in the vicinity of 0.}
\end{figure*}

As a numerical example, we solve the pairing Hamiltonian for $\delta=1.0$
and $g=0.5$. In Fig.~\ref{fig:srg_pairing}, we show snapshots of the matrix
$H(s)$ at different stages of the flow. We can nicely see how the SRG evolution
drives the off-diagonal matrix elements to zero. The effect becomes noticeable 
on our logarithmic color scale around $s=0.05$, where the outermost off-diagonal 
matrix elements start to lighten. At $s=0.1$, $H_{05}(s)$ has been reduced by
four to five orders of magnitude, and at $s=1.0$, essentially all of the 
off-diagonal matrix elements have been affected to some extent. Note that the
strength of the suppression depends on the distance from the diagonal, aside from 
$H_{05}$ itself, which has a slightly larger absolute value than $H_{04}$ and
$H_{15}$. The overall behavior is as expected from our approximate solution 
\eqref{eq:def_hod_general}, and the specific deviations can be explained by 
the approximate nature of that result. Once we have reached $s=10$, the matrix
is essentially diagonal, with off-diagonal matrix elements reduced to  
$10^{-10}$ or less. Only the $2\times2$ block spanned by the states labeled 2
and 3 has slightly larger off-diagonal matrix elements remaining, which is 
due to the degeneracy of the corresponding eigenvalues.

\begin{figure*}[t]
  \setlength{\unitlength}{\textwidth}
  \begin{picture}(1.0000,0.4000)
    \put(0.0300,0.0300){\includegraphics[width=0.45\unitlength]{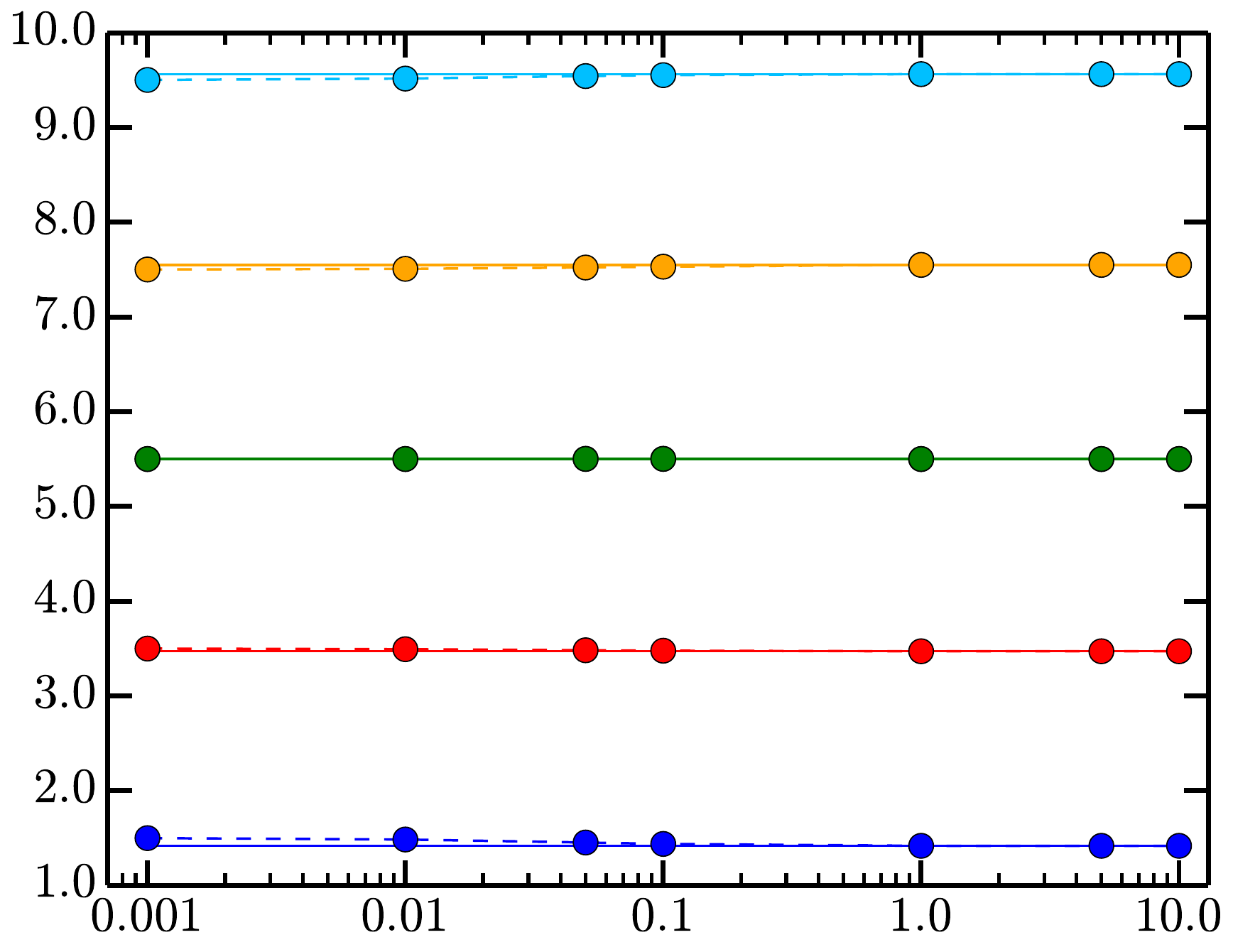}}
    \put(0.0100,0.0400){\begin{sideways}\parbox{0.3500\unitlength}{\centering$ E_i, H_{ii}(s)$ }\end{sideways}}
    \put(0.0200,0.0150){\parbox{0.500\unitlength}{\centering$s$}}
    \put(0.5200,0.0300){\includegraphics[width=0.46\unitlength]{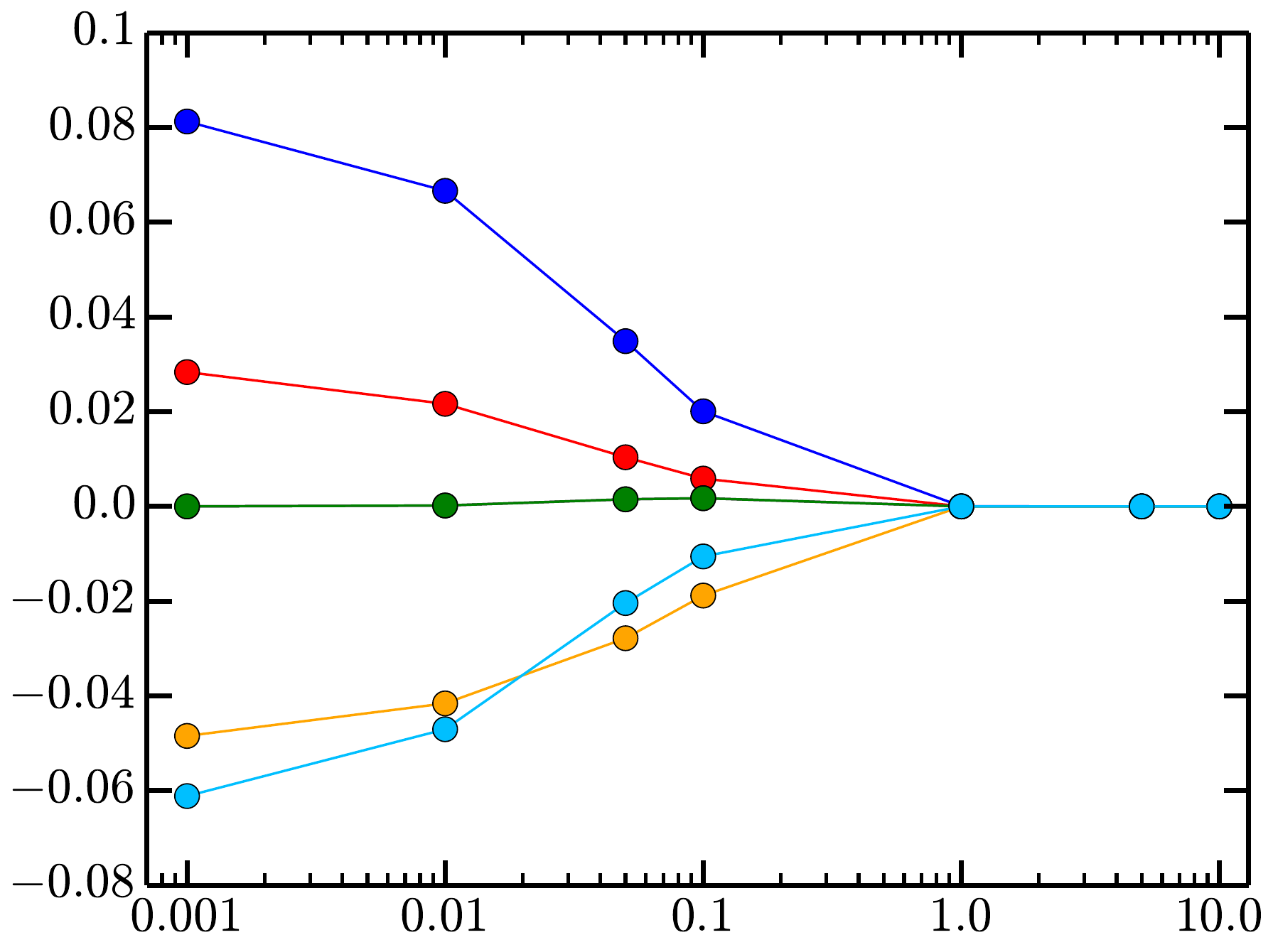}}    
    \put(0.5000,0.0400){\begin{sideways}\parbox{0.3500\unitlength}{\centering$ H_{ii}(s) - E_i$ }\end{sideways}}
    \put(0.5200,0.0150){\parbox{0.500\unitlength}{\centering$s$}}

  \end{picture}

  \caption{\label{fig:srg_pairing_diags}SRG evolution of the pairing Hamiltonian
  with $\delta=1, g=0.5$. The left panel shows the diagonal matrix elements $H_{ii}(s)$ as a function of
  the flow parameter $s$ (dashed lines)  and
  the corresponding eigenvalues (solid lines), the right panel the difference of
  the two numbers. The color coding is the same in both panels.}
\end{figure*}
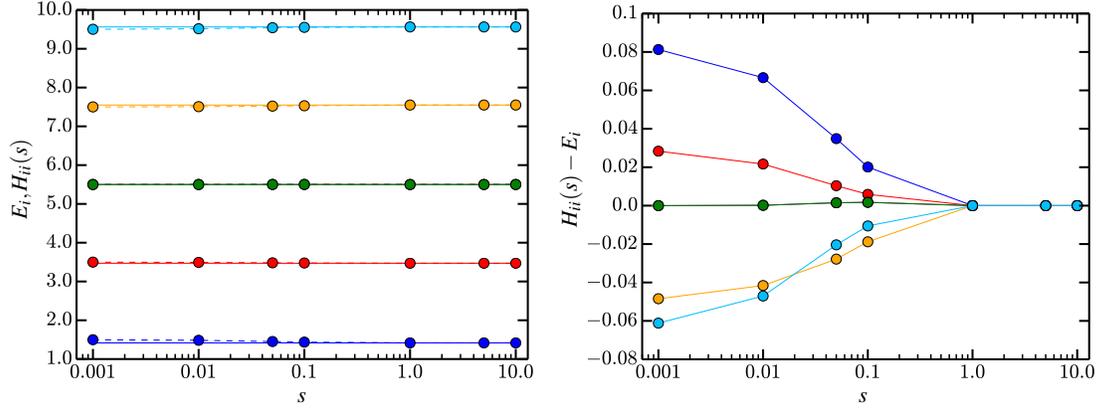

In Fig.~\ref{fig:srg_pairing_diags}, we compare the flowing diagonal matrix
elements $H_{ii}(s)$ to the eigenvalues of the pairing Hamiltonian. As
we have just mentioned, the pairing Hamiltonian has a doubly degenerate 
eigenvalue $E_2=E_3=6\delta-g$, which is why we see only five curves
in these plots. For our choice of parameters, the diagonal matrix elements
are already in fairly good agreement with the eigenvalues to begin with. Focusing
on the right-hand panel of the figure in particular, we see that $H_{00}$ (blue) 
and $H_{11}$ (red) approach their eigenvalues from above, while 
$H_{44}$ (orange) and $H_{55}$ (light blue) approach from below as we evolve
to large $s$. It is interesting that the diagonal matrix elements 
are already practically identical to the eigenvalues once we have 
integrated up to $s=1.0$, despite the non-vanishing off-diagonal matrix elements 
that are visible in the $s=1.0$ snapshot shown in Fig.~\ref{fig:srg_pairing}.

\subsection{\label{sec:srg_interactions}Evolution of Nuclear Interactions}

%
%
\subsubsection{\label{sec:srg_opflow}Matrix and Operator Flows}
In our discussion of the schematic pairing model in the previous section, we 
have used SRG flows to solve the eigenvalue problem arising from a four-body 
Schr\"odinger equation, so we may want to use the same method for the more 
realistic case of $A$ nucleons interacting by nuclear $NN$, $3N$, etc. 
interactions (see chapter 8). However, we quickly realize the
main problem of such an approach: Working in a full configuration interaction (FCI) 
picture
and assuming even a modest single-particle basis size, e.g., 50 proton and neutron 
states each, a basis for the description of a nucleus like $\nuc{O}{16}$ would 
naively have
\begin{equation}
 d(\nuc{O}{16}) = \begin{pmatrix} 50 \\ 8\end{pmatrix} \times 
           \begin{pmatrix} 50 \\ 8\end{pmatrix}
                \approx 2.88\times10^{17}
\end{equation}
configurations, i.e., we would need about 2 exabytes (EB) of memory to store all 
the coefficients of just one eigenvector (assuming double precision floating-point 
numbers), and $7\times10^{17}$ EB to construct
the complete Hamiltonian matrix! State-of-the-art methods for large-scale 
diagonalization are able to reduce the memory requirements and computational
effort significantly by exploiting matrix sparseness, and using modern versions of
Lanczos-Arnoldi \cite{Lanczos:1950sp,Arnoldi:1951kk} or Davidson algorithms 
\cite{Davidson:1989pi}, but nuclei in the vicinity of the oxygen isotopic 
chain are among the heaviest accessible with today's computational
resources (see, e.g., \cite{Yang:2013ly,Barrett:2013oq} and references therein).
A key feature of Lanczos-Arnoldi and Davidson methods is that the
Hamiltonian matrix only appears in the calculation
of matrix-vector products. In this way, an explicit construction of the Hamiltonian
matrix in the CI basis is avoided, because the matrix-vector product can be
calculated from the input $NN$ and $3N$ interactions that only require $\OC(n^4)$ 
and $\OC(n^6)$ storage, respectively, where $n$ is the size of the single-particle
basis (see Sec.~\ref{sec:imsrg}). However, the SRG flow of the previous section 
clearly forces us to construct
and store the Hamiltonian matrix in its entirety --- at best, we could save some
storage by resizing the matrix once its off-diagonal elements have been 
sufficiently suppressed.

Instead of trying to evolve the many-body Hamiltonian matrix, we therefore focus
on the Hamiltonian operator itself instead. Let us consider a nuclear Hamiltonian 
with a two-nucleon interaction for simplicity:
\begin{equation}
  \Hint = \Tint + \VO^{[2]}\,.
\end{equation}
Since nuclei are self-bound objects, we have to consider the intrinsic form of the 
kinetic energy, 
\begin{equation}
  \Tint \equiv \TO - \Tcm\,.
\end{equation}
It is straightforward to show that $\Tint$ can be written either as a sum of
one- and two-body operators,
\begin{equation}\label{eq:def_Tint_1B2B}
  \Tint= \left(1-\frac{1}{\AO}\right)\sum_{i}\frac{\pOV_i^2}{2m} - \frac{1}{\AO}\sum_{i<j}\frac{\pOV_{i}\cdot \pOV_{j}}{m}
\end{equation}
or as a pure two-body operator
\begin{equation}\label{eq:def_Tint_2B}
  \Tint= \frac{2}{\AO}\sum_{i<j}\frac{\qOV_{ij}^2}{2\mu}\,,\quad \qOV_{ij}\equiv\pOV_{i}-\pOV_{j}\,.
\end{equation}
Here, $\AO$ should be treated as a particle-number \emph{operator} (see \cite{Hergert:2009wh}),
and $\mu=m/2$ is the reduced nucleon mass (neglecting the proton-neutron mass difference).
Using Eq.~\eqref{eq:def_Tint_2B} for the present discussion, we can write the intrinsic 
Hamiltonian as
\begin{equation}\label{eq:def_Hint_2B}
  \Hint = \frac{2}{\AO}\sum_{i<j}\frac{\qOV_{ij}^2}{2\mu} + \sum_{i<j}\vO^{[2]}_{ij}\,,
\end{equation}
and directly consider the evolution of this operator via the flow equation
\eqref{eq:opflow}. It is customary to absorb the flow-parameter dependence 
completely into the interaction part of the Hamiltonian, and leave the kinetic
energy invariant --- in our previous examples, this simply amounts to moving the
$s-$dependent part of $\HO_d(s)$ into $\HO_{od}(s)$. We end up with a flow equation 
for the two-body interaction:
\begin{align}\label{eq:opflow_NN}
  \totd{}{s}v^{[2]}_{ij} &= \comm{\eta}{\vO^{[2]}_{ij}}\,.
\end{align}
In cases where we can expand the two-body interaction in terms of a finite algebra of
``basis'' operators, Eq.~\eqref{eq:opflow_NN} becomes a system of ODEs for the expansion 
coefficients, the so-called running couplings of the Hamiltonian, as explained
in earlier chapters of this book. An example is the toy problem discussed in Sec.~
\ref{sec:srg_toy}: We actually expanded our $2\times2$ in terms of the algebra 
$\{\idO,\sigmaO_{1},\sigmaO_{2},\sigmaO_{3}\}$,
and related the matrix elements to the coefficients in this expansion. While 
the representation of the basis operators of our algebra would force us to use
extremely large matrices when we deal with an $A-$body system, we may be able
to capture the SRG flow completely with a small set of ODEs for the couplings 
of the Hamiltonian!

If we cannot identify a set of basis operators for the two-body interaction, we 
can still resort to representing it as a matrix between two-body states. For a given
choice of single-particle basis with size $n$, $\vO^{[2]}$ is then represented
by $\OC(n^4)$ matrix elements, as mentioned above. In general, we will then have
to face the issue of induced many-body forces, as discussed in Sec.~
\ref{sec:srg_induced}.

%
%
\subsubsection{\label{sec:srg_nn}SRG in the Two-Nucleon System}

\begin{figure*}[t]
  \setlength{\unitlength}{\textwidth}
  \begin{center}
  \begin{picture}(0.9000,0.5100)
    \put(0.0000,0.0400){\includegraphics[width=0.45\unitlength]{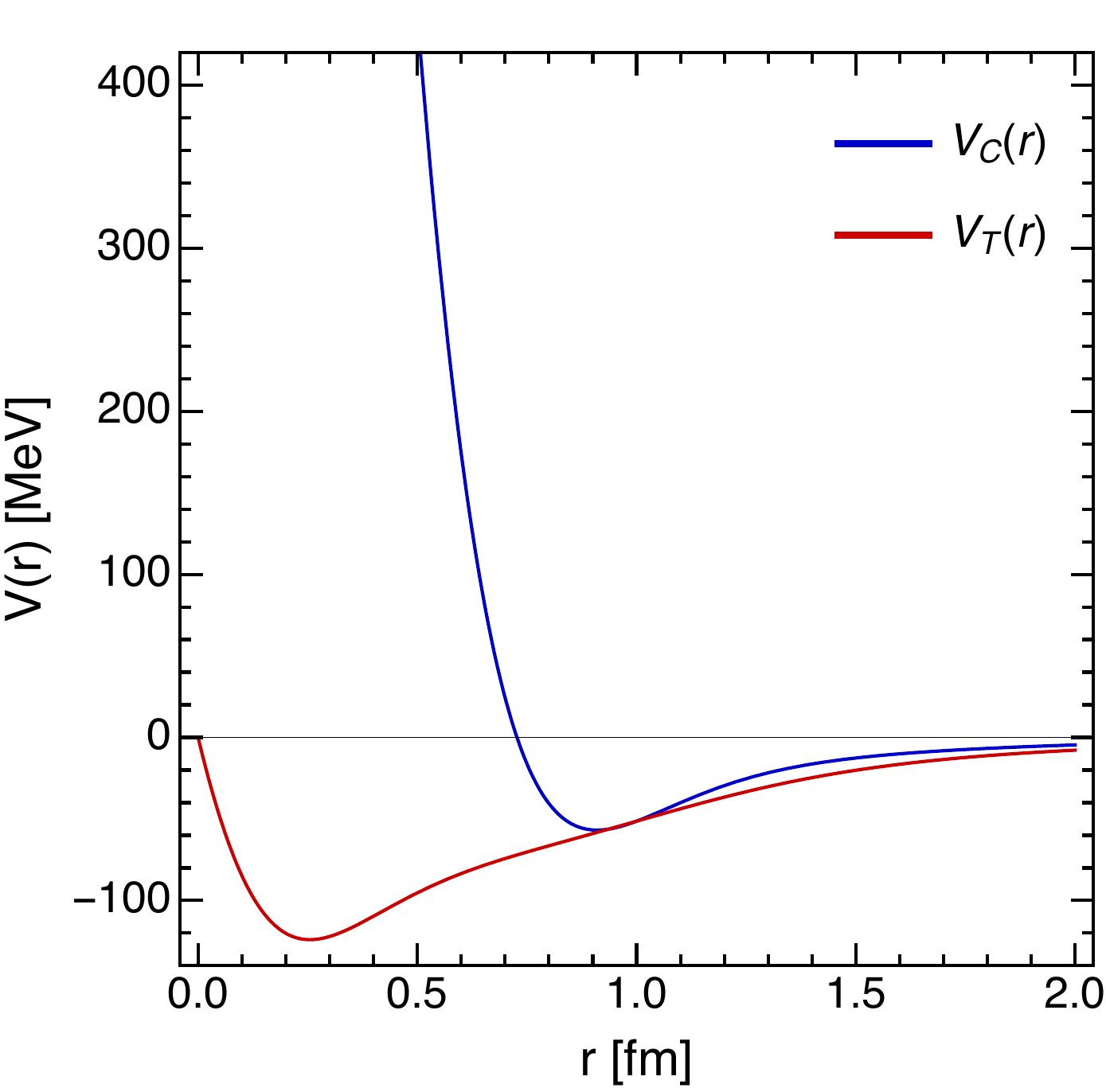}}
    \put(0.5000,0.2550){\includegraphics[width=0.40\unitlength]{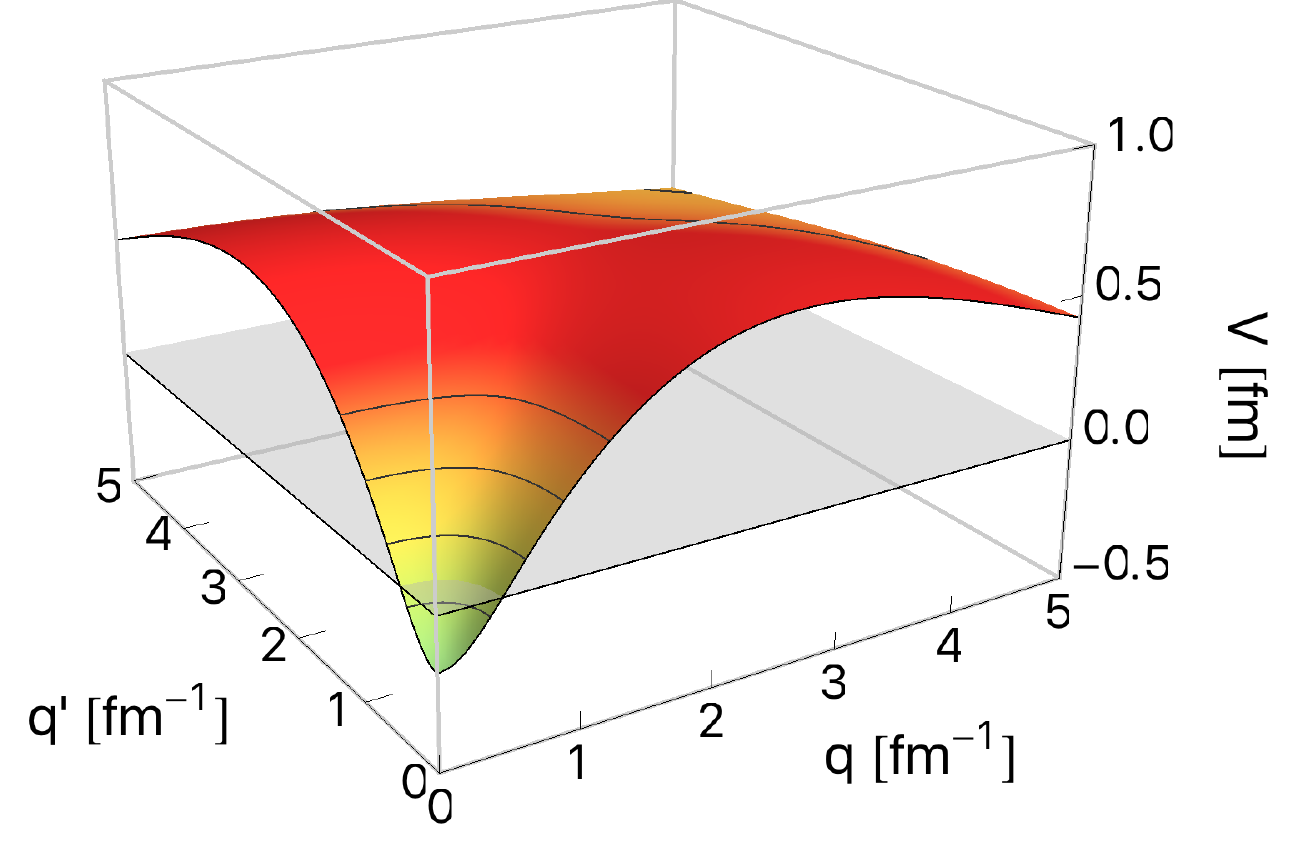}}
    \put(0.5000,0.0000){\includegraphics[width=0.40\unitlength]{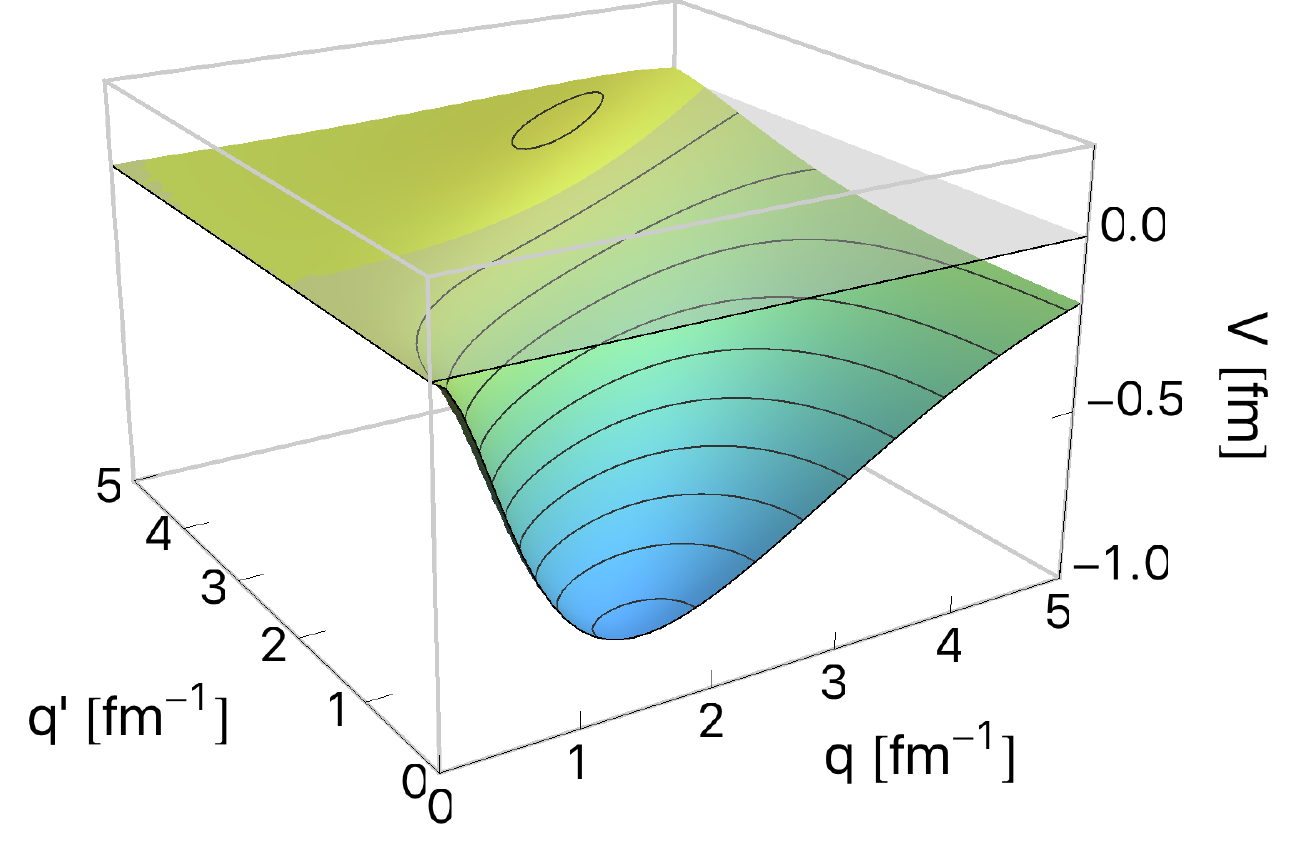}}
    \put(0.8700,0.4800){\fbox{\parbox{0.04\unitlength}{\raggedleft\large${}^3S_1$}}}
    \put(0.7900,0.2400){\fcolorbox{black}{white}{\parbox{0.12\unitlength}{\raggedleft\large${}^3S_1-{}^3D_1$}}}
  \end{picture}
  \end{center}
  \caption{\label{fig:av18}
    Repulsive core  and tensor force of the Argonne
    V18 $NN$ interaction \cite{Wiringa:1995or} in the $(S,T)=(1,0)$
    channel. In the left panel, the radial dependencies of the central ($V_C(r)$)  and tensor 
    components ($V_T(r)$) of Argonne V18 are shown, while the right panel shows its momentum 
    space matrix elements in the deuteron partial waves.
  }
\end{figure*}

Let us now consider the operator flow of the $NN$ interaction
in the two-nucleon system, Eq.~\eqref{eq:opflow_NN}. Since the nuclear Hamiltonian is invariant
under translations and rotations, it is most convenient to work in
momentum and angular momentum eigenstates of the form
\begin{equation}
  \ket{q(LS)JMTM_T}\,.
\end{equation}
Because of the rotational symmetry, the $NN$ interaction conserves
the total angular momentum quantum number $J$, and it is easy to show 
that the total spin $S$ of the nucleon pair is a conserved quantity as 
well. The
orbital angular momentum is indicated by the quantum number $L$,
and we remind our readers that $L$ is \emph{not} conserved, because
the nuclear tensor operator
\begin{equation}
  S_{ij}(\rOV,\rOV) = \frac{3}{\rOV^2}(\sigmaOV_i\cdot\rOV)(\sigmaOV_j\cdot\rOV)-\sigmaOV_i\cdot\sigmaOV_{j}
\end{equation}
can couple states with $\Delta L = \pm 2$. We
assume that the interaction is charge-dependent in the isospin channel
$T=1$, i.e., matrix elements will depend on the projection $M_T=-1,0,1$, 
which indicates the neutron-neutron, neutron-proton, and proton-proton 
components of the nuclear Hamiltonian.

In Fig.~\ref{fig:av18} we show features of the central 
and tensor forces of the Argonne V18 (AV18) interaction \cite{Wiringa:1995or}
in the $(S,T)=(1,0)$ channel, which has the quantum numbers of the deuteron. This 
interaction belongs to a group of so-called
realistic interactions that describe nucleon-nucleon scattering data
with high accuracy, but precede the modern chiral forces (see
chapter 8, \cite{Epelbaum:2009ve,Machleidt:2011bh}). AV18
is designed to be maximally local in order to be a suitable input for
nuclear Quantum Monte Carlo calculations \cite{Carlson:2015lq,Gezerlis:2014zr,
Lynn:2016ec}. Because of the required locality, AV18 has a strong repulsive
core in the central part of the interaction. Like all $NN$ interactions,
it also has a strong tensor force that results from pion exchange. The
radial dependencies of these interaction components are shown in the left
panel of Fig.~\ref{fig:av18}. 

When we switch to the momentum representation,
we see that the ${}^3S_1$ partial wave\footnote{We use the conventional partial wave
notation ${}^{2S+1}L_J$, where $L=0,1,2,\ldots$ is indicated by the letters 
$S,P,D,\ldots$. The isospin channel is fixed by requiring the antisymmetry
of the $NN$ wavefunction, leading to the condition $(-1)^{L+S+T}=-1$.} which
gives the dominant contribution to the deuteron wave function has strong
off-diagonal matrix elements, with tails extending over the entire shown
range and as high as $|\qOV| \sim 20\,\fmi$.
The matrix elements of the ${}^3S_1-{}^3D_1$ mixed partial wave, which
are generated exclusively by the tensor force, are sizable as well. The
strong coupling between states with low and high relative momenta forces
us to use large Hilbert spaces in few- and many-body calculations, even
if we are only interested in the lowest eigenstates. Methods like the 
Lanczos algorithm (see chapter 8 and \cite{Lanczos:1950sp})
extract eigenvalues and eigenvectors by repeatedly acting with the Hamiltonian
on an arbitrary starting vector in the many-body space, i.e., by repeated
matrix-vector products. Even if that vector only has low-momentum or low-energy 
components in the beginning, an interaction like AV18 will mix in high-momentum 
components even after a single matrix-vector multiplication, let alone tens 
or hundreds as in typical many-body calculations. Consequently, the 
eigenvalues and eigenstates of the nuclear Hamiltonian converge very slowly
with respect to the basis size of the Hilbert space (see, e.g., 
\cite{Barrett:2013oq}). To solve this problem, we perform an RG evolution of
the $NN$ interaction.

\begin{figure*}[t]
  \begin{center}
    \setlength{\unitlength}{\textwidth}
    \begin{picture}(0.9000,0.4100)
      \put(0.0000,0.0000){\includegraphics[width=0.9\textwidth]{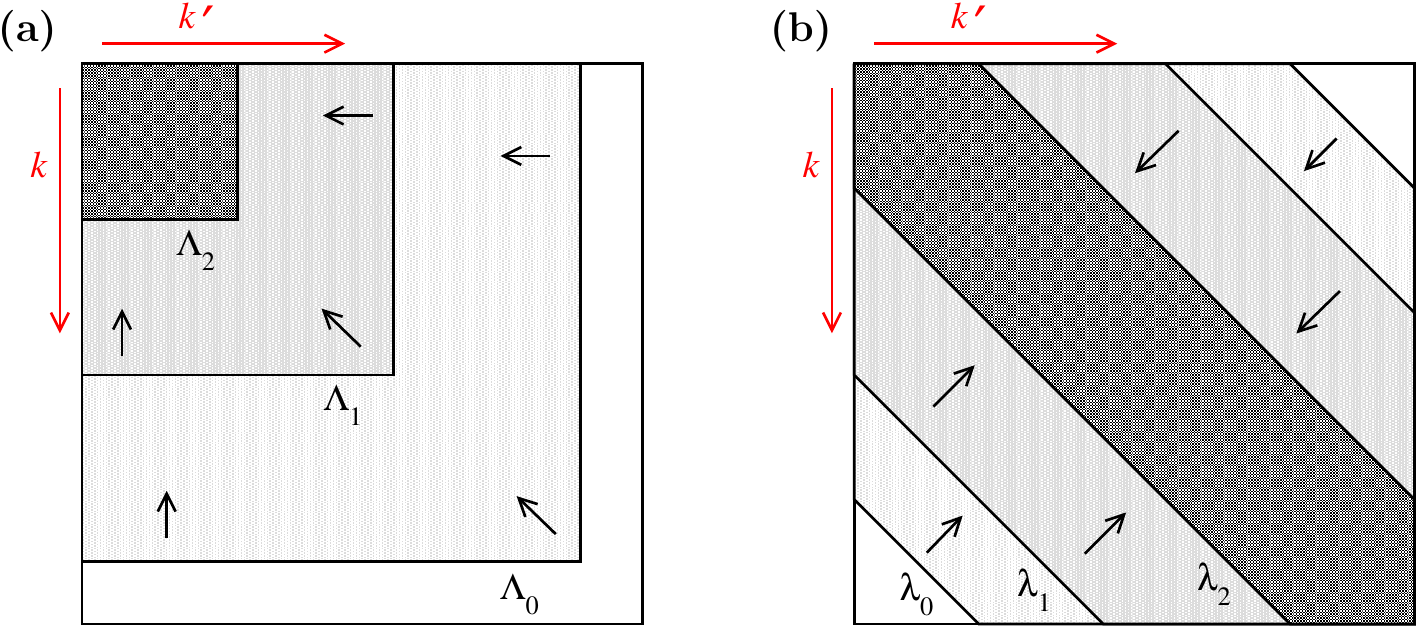}}
      \put(0.0000,0.2850){\parbox{0.1\unitlength}{\colorbox{white}{\color{red}\large$q'$}}}
      \put(0.4900,0.2850){\parbox{0.1\unitlength}{\colorbox{white}{\color{red}\large$q'$}}}
      \put(0.1100,0.3860){\parbox{0.1\unitlength}{\colorbox{white}{\color{red}\large$q$}}}
      \put(0.6000,0.3860){\parbox{0.1\unitlength}{\colorbox{white}{\color{red}\large$q$}}}
    \end{picture}
  \end{center}

  \caption{\label{fig:schematic}Schematic illustration of two types of RG evolution 
    for $NN$ potentials in momentum space: (a) \Vlowk{} running in $\Lambda$, and 
    (b) SRG running in $\lambdaSRG$ (see main text). Here, $q$ and $q'$ denote the 
    relative momenta of the initial and final state, respectively. At each $\Lambda_i$ 
    or $\lambdaSRG_i$, the matrix elements outside of the corresponding 
    blocks or bands are negligible, implying that high- and low-momentum 
    states are decoupled.
  }
\end{figure*}
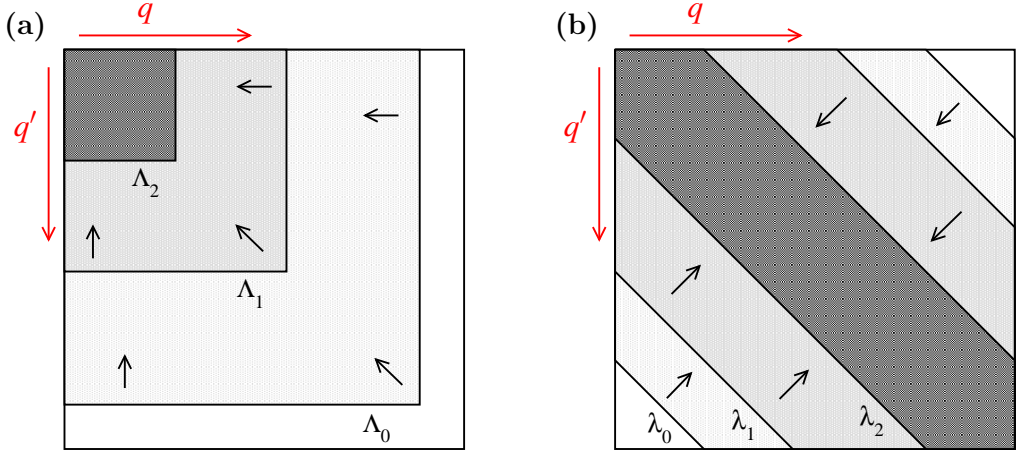

In Fig.~\ref{fig:schematic}, we show examples for two types of RG evolution
that decouple the low- and high-momentum pieces of $NN$ interactions. The first
example, Fig.~\ref{fig:schematic}(a), is a so-called  \emph{RG decimation}, 
in which the interaction is evolved to decreasing cutoff scales $\Lambda_0 > \Lambda_1 > \Lambda_2$, 
and high-momentum modes are ``integrated out''. This is the so-called \Vlowk{}
approach, which was first used in nuclear physics in the early 2000s 
\cite{Bogner:2003os,Bogner:2010pq}. Note that the resulting low-momentum 
interaction is entirely confined to states with relative momentum $q\leq\Lambda$.
In contrast, Fig.~\ref{fig:schematic}(b) shows the SRG evolution of the $NN$ interaction
to a band-diagonal shape via the flow equation \eqref{eq:opflow_NN}, using a 
generator built from the relative kinetic energy in the two-nucleon system \cite{Bogner:2007od,Bogner:2010pq}:
\begin{equation}\label{eq:def_srg_generator}
  \eta(\lambda) \equiv \comm{\frac{\qOV^2}{2\mu}}{v(\lambda)}\,.
\end{equation}
Instead of the flow parameter $s$, we have parameterized the evolution by 
$\lambdaSRG=s^{-1/4}$, which has the dimensions of a momentum in natural units. 
Note that the generator \eqref{eq:def_srg_generator} would vanish if the
interaction were diagonal in momentum space. As suggested by Fig.~
\ref{fig:schematic}(b), $\lambdaSRG$ is a measure for the ``width'' of the 
band in momentum space. Thus, momentum transfers between nucleons are limited 
according to
\begin{equation}\label{eq:momentum_transfer}
  Q\equiv |\qOV' - \qOV| \lesssim \lambdaSRG\,,
\end{equation} 
and low- and high-lying momenta are decoupled as $\lambdaSRG$ is decreased. 

Equation \eqref{eq:momentum_transfer} implies that the spatial resolution scale
of an SRG-evolved interaction (or a \Vlowk{} if we determine the maximum
momentum transfer in the low-momentum block) is $\sim 1/Q \geq 1/\lambdaSRG$,
i.e., only long-ranged components of the $NN$ interaction are resolved explicitly
and short-range components of the interaction can just as well be replaced 
by contact interactions \cite{Lepage:1997py,Bogner:2003os,Holt:2004ux,Bogner:2010pq}.
This is the reason why the realistic $NN$ interactions that accurately 
describe $NN$ scattering data collapse to a universal long-range interaction
when RG-evolved, namely one-pion exchange (OPE). This universal behavior emerges
in the range $1.5\,\fmi\leq\lambdaSRG\leq2.5\,\fmi$. Any further evolution to
lower $\lambdaSRG$ starts to remove pieces of OPE, and eventually generates a 
pion-less theory that is essentially parameterized in terms of contact 
interactions. While it is possible to implement such an evolution in
the two-body system without introducing pathological behavior \cite{Wendt:2011ys},
such an interaction must be complemented by strong induced many-nucleon
forces once it is applied in finite nuclei, as discussed in Sec.~\ref{sec:srg_induced}.
For this reason, $NN$ interactions are only ever evolved to the aforementioned
range of $\lambdaSRG$ values.

Nowadays, SRG evolutions are 
preferred over \Vlowk{} style decimations in nuclear many-body theory, because 
they can be readily extended to $3N,\ldots$ interactions and to general 
observables 
\cite{Bogner:2010pq,Anderson:2010br,Schuster:2014oq,More:2015bx,Jurgenson:2009bs,Hebeler:2012ly,Wendt:2013ys}. 
Moreover, we could easily achieve a block decoupling as in 
Fig.~\ref{fig:schematic}(a) by using a generator like \cite{Anderson:2008hx}
\begin{equation}\label{eq:def_srg_generator_PQ}
  \eta(\lambda) \equiv \comm{\underbrace{P_{\Lambda}H(\lambda)P_{\Lambda}+Q_{\Lambda}H(\lambda)Q_{\Lambda}}_{\equiv H_d(\lambda)}}{H(\lambda)}\,.
\end{equation}
where the projection operators $P_\Lambda$ and $Q_\Lambda$ partition the
relative momentum basis in states with $|\qOV|\leq\Lambda$ and $|\qOV|>\Lambda$,
respectively. In this case, $\lambda$ is an auxiliary parameter that is eliminated
by evolving $\lambda\to0$, just like we evolved $s\to\infty$ in Secs.~\ref{sec:srg_toy}
and \ref{sec:srg_pairing}. 

%
%
\subsubsection{\label{sec:srg_nn_flow}Implementation of the Flow Equations}
We are now ready to implement the flow equations for the $NN$ interaction in
the momentum-space partial-wave representation. Using basis states that 
satisfy the orthogonality and completeness relations
\begin{equation}
  \braket{qLSJMTM_T}{q'L'S'J'M'T'M'_T} = 
    \frac{\delta(q-q')}{qq'}\delta_{LL'}\delta_{SS'}\delta_{JJ'}\delta_{MM'}\delta_{TT'}\delta_{M_TM'_T}
\end{equation}
and
\begin{equation}
  \idO = \sum_{LSJMTM_T}\int_0^\infty dq\,q^2 \ket{qLSJMTM_T}\bra{qLSJMTM_T}\,,
\end{equation}
respectively, we obtain \cite{Bogner:2007od,Bogner:2010pq}
\begin{align}
  \left(-\frac{\lambda^5}{4}\right)\frac{d\matrixe{qL}{\vO}{q'L'}}{d\lambda} =&- (q^2 - q'^{2})^2
  \matrixe{qL}{\vO}{q'L'}
  \notag\\
    &+ \sum_{\bar{L}}\int_0^{\infty}dp p^2\,
      (q^2 + q'{}^2 - 2p^2)\matrixe{qL}{\vO}{p\bar{L}}
      \matrixe{p\bar{L}}{\vO}{q'L'}\,,\label{eq:flow_nn}
\end{align}
where we have used scattering units ($\hbar^2/m=1$) and suppressed
the $\lambda$-dependence of $\vO$ as well as the conserved quantum 
numbers for brevity. Note that a prefactor $-\lambda^5/4$ 
appears due to our change of variables from $s$ to $\lambda$. 

We can turn this integro-differential equation back into a matrix 
flow equation by discretizing the relative momentum variable, e.g.,
on uniform or Gaussian quadrature meshes. The matrix elements of the 
relative kinetic energy operator are then simply given by
\begin{equation}
  \matrixe{q_iL}{\tO}{q_jL'}=q_i^2 \delta_{q_iq_j}\delta_{LL'}\,
\end{equation}
(with $\hbar^2/m=1$). The discretization turns the integration
into a simple summation,
\begin{equation}
  \int_0^\infty dq\,q^2\;\rightarrow\;\sum_{i}w_i q^2_i\,,
\end{equation}
where the weights $w_i$ depend on our choice of mesh. For a uniform
mesh, all weights are identical and correspond to the mesh spacing,
while for Gaussian quadrature rules the mesh points and weights have
to be determined numerically \cite{Press:2007vn}. For convenience,
we absorb the weights and $q^2$ factors from the integral measure into 
the interaction matrix element,
\begin{equation}
  \matrixe{q_iL}{\overline{v}}{q_jL'} \equiv \sqrt{w_iw_j}q_i q_j\matrixe{q_iL}{\vO}{q_jL'}\,.
\end{equation}
The discretized flow equation can then be written as
\begin{equation}\label{eq:flow_nn_discrete}
  \totd{}{\lambda} \matrixe{q_iL}{\overline{v}}{q_jL'} = - \frac{4}{\lambda^5}
    \matrixe{q_iL}{\comm{\comm{\tO}{\overline{v}}}{\tO+\overline{v}}}{q_jL'}\,.
\end{equation}

We can solve Eq.~\eqref{eq:flow_nn_discrete} using a modified version of our 
Python code for the pairing model, discussed in Sec.~\ref{sec:srg_pairing}.
The Python code and sample inputs can be downloaded from 
\url{https://github.com/ManyBodyPhysics/LectureNotesPhysics/tree/master/Programs/Chapter10-programs/python/srg_nn}.
Let us briefly discuss the most important modifications.

First, we have a set of functions that read the momentum mesh
and the input matrix elements from a file:
\begin{lstlisting}
def uniform_weights(momenta):
    weights = np.ones_like(momenta)
    weights *= abs(momenta[1]-momenta[0])
    return weights

def read_mesh(filename):
    data = np.loadtxt(filename, comments="#")  
    dim  = data.shape[1]
    
    momenta = data[0,:dim]

    return momenta
  
def read_interaction(filename):
    data = np.loadtxt(filename, comments="#")  
    dim  = data.shape[1]
    V = data[1:,:dim]
    return V
\end{lstlisting}

The matrix element files have the following format:
\begin{lstlisting}
# momentum space matrix elements 
# partial wave J=1, L=0/0, S=1, T=0, MT=0 
# 
# momentum grid [fm^-1]
0.000000 0.050000 0.100000 0.150000 0.200000 0.250000 0.300000 0.350000 0.400000 0.450000 
0.500000 0.550000 0.600000 0.650000 0.700000 0.750000 0.800000 0.850000 0.900000 0.950000 
...
6.300000 6.350000 6.400000 6.450000 6.500000 6.550000 6.600000 6.650000 6.700000 6.750000 
6.800000 6.850000 6.900000 6.950000 7.000000 
#
# matrix elements [MeV fm^3] 
-36.94918 -36.83554 -36.49896 -35.95649 -35.23306 -34.35875 -33.36536 -32.28337 -31.13985 
-29.95722 -28.75281 -27.53904 -26.32400 -25.11217 -23.90513 -22.70231 -21.50158 -20.29973 
...
0.00000   0.00000   0.00000   0.00000   0.00000   0.00000   0.00000   0.00000   0.00000   
0.00000   0.00000   0.00000   0.00000   0.00000   0.00000   0.00000   0.00000   0.00000
\end{lstlisting}
Comments, indicated by the \texttt{\#} character, are ignored. The first set
of data is a row containing the mesh points. Here, we have $141$ points in total,
ranging from $0$ to $7\,\fmi$ with a spacing of $0.05\,\fmi$. The range of momenta
is sufficient for the chiral $NN$ interaction we use in our example, the \NNNLO{} 
potential by Entem and Machleidt with cutoff $\Lambda=500\,\MeV$ \cite{Entem:2003th,Machleidt:2011bh}, 
which is considerably softer than the AV18 interaction discussed above. This is followed
by a simple $141 \times 141$ array of matrix elements. It is straightforward
to adapt the format and I/O routines to Gaussian quadrature meshes by including
mesh points (i.e., the abscissas) and weights in the data file.

The derivative routine is almost unchanged, save for the prefactor due to the
use of $\lambda$ instead of $s$ to parameterize the flow, and the treatment of
the kinetic energy operator as explicitly constant:
\begin{lstlisting}
def derivative(lam, y, T):
    dim = T.shape[0]

    # reshape the solution vector into a dim x dim matrix
    V = reshape(y, (dim, dim))

    # calculate the generator
    eta = commutator(T, V)

    # dV is the derivative in matrix form 
    dV  = -4.0/(lam**5) * commutator(eta, T+V)

    # convert dH into a linear array for the ODE solver
    dy = reshape(dV, -1)
      
    return dy
\end{lstlisting}

In the main routine of the program, we first set up the mesh and 
then proceed to read the interaction matrix elements for the 
different partial waves.
We are dealing with a coupled-channel problem because
the tensor forces connects partial waves with $\Delta L=2$
in all $S=1$ channels. In our example, we restrict ourselves
to the partial waves that contribute to the deuteron bound
state, namely ${}^3S_1$, ${}^3D_1$,
and ${}^3S_1-{}^3D_1$. Indicating the orbital angular momenta
of these partial waves by indices, we have
\begin{equation}
  \mathbf{T}=
  \begin{pmatrix} 
      \;t\; &   \\
        &  \;t\;
  \end{pmatrix}\,,
  \quad
  \mathbf{V}=
  \begin{pmatrix} 
      \;\overline{v}_{00}\;       &  \;\overline{v}_{02}\; \\
      \;\overline{v}^\dag_{02}\;  &  \;\overline{v}_{22}\;
  \end{pmatrix}\,,
\end{equation}
where 
\begin{equation}
  t = \mathrm{diag}\left(q_0^2,\ldots,q_\text{max}^2\right)\,,
\end{equation}
since the kinetic energy is independent of $L$. As soon as we pass 
from the $S-$ into the $D-$ wave in either the
rows or the columns, the momentum mesh simply starts from the
lowest mesh point again. We use \texttt{NumPy}'s \texttt{hstack} 
and \texttt{vstack} functions to assemble the interaction matrix
from the partial-wave blocks:

\begin{lstlisting}
def main():
...  

    # read individual partial waves
    partial_waves=[]
    for filename in ["n3lo500_3s1.meq", "n3lo500_3d1.meq", "n3lo500_3sd1.meq"]:
      partial_waves.append(read_interaction(filename))
      # print partial_waves[-1].shape

    # assemble coupled channel matrix
    V = np.vstack((np.hstack((partial_waves[0], partial_waves[2])), 
                   np.hstack((np.transpose(partial_waves[2]), partial_waves[1]))
                  ))

    # switch to scattering units
    V = V/hbarm

...
\end{lstlisting}

As discussed earlier, we work in scattering units with $\hbar^2/m=1$. Thus,
we have to divide the input matrix elements by this factor. We also need to
absorb the weights and explicit momentum factors into the interaction matrix.
It is convenient to define a conversion matrix for this purpose, which can
be multiplied element-wise with the entries of $\mathbf{V}$ using the 
regular \texttt{*} operator (recall that the matrix product is implemented
by the \texttt{NumPy} function \texttt{dot}).

% ...

Since we changed variables from $s$ to $\lambda$, we now start the
integration at $\lambda=\infty$, or $\lambda\gg 1\fmi$ in practice.
As discussed above, we do not evolve all the way to $\lambda=0\,\fmi$,
but typically stop before we start integrating out explicit pion
physics, e.g., at $\lambda=1.5\,\fmi$. For typical $NN$ interactions,
especially those with a hard core like AV18, the flow equations tend
to become stiff because they essentially depend on cubic products of
the kinetic energy and interaction. For this reason, we use \texttt{SciPy}'s
\texttt{ode} class, which provides access to a variety of solvers
and greater control over the parameters of the integration process.
Specifically, we choose the VODE solver package and its 5th-order 
Backward Differentiation method \cite{Brown:1989wj}, which is 
efficient and works robustly for a large variety of input interactions.

\begin{lstlisting}
...

    lam_initial = 20.0
    lam_final = 1.5

    # integrate using scipy.ode instead of scipy.odeint - this gives
    # us more control over the solver
    solver = ode(derivative,jac=None)

    # equations may get stiff, so we use VODE and Backward Differentiation
    solver.set_integrator('vode', method='bdf', order=5, nsteps=1000)
    solver.set_f_params(T)
    solver.set_initial_value(y0, lam_initial)

...
\end{lstlisting}

Finally, we reach the loop that integrates the ODE system. We request
output from the solver in regular intervals, reducing these intervals
as we approach the region of greatest practical interest, 
$1.5\,\fmi\leq \lambda \leq 2.5\,\fmi$: 

\begin{lstlisting}
...

    while solver.successful() and solver.t > lam_final:
        # adjust the step size in different regions of the flow parameter
        if solver.t >= 6.0:
          ys = solver.integrate(solver.t-1.0)
        elif solver.t < 6.0 and solver.t >= 2.5:
          ys = solver.integrate(solver.t-0.5)
        elif solver.t < 2.5 and solver.t >= lam_final:
          ys = solver.integrate(solver.t-0.1)

        # add evolved interactions to the list
        flowparams.append(solver.t)
        Vtmp = reshape(ys,(dim,dim))
        Vs.append(Vtmp)

        print("%8.5f %14.8f"%(solver.t, eigvalsh((T + Vtmp)*hbarm)[0]))

...
\end{lstlisting}

Of course, the ODE solver will typically take several hundred adaptive 
steps to propagate the solution with the desired accuracy between successive 
requested values of $\lambda$. At the end of each external 
step, we diagonalize the evolved Hamiltonian
and check whether the lowest eigenvalue, i.e., the deuteron binding
energy, remains invariant within the numerical tolerances we use for
the ODE solver. To illustrate the evolution of the $NN$ interaction, 
the code will also generate a sequence of matrix plots at the
desired values of $\lambda$, similar to Fig.~\ref{fig:srg_pairing}
for the pairing Hamiltonian.

%
%
\subsubsection{\label{sec:srg_n3lo}Example: Evolution of a Chiral $NN$ Interaction}

\begin{figure*}[t]
  \setlength{\unitlength}{\textwidth}
  \begin{picture}(1.0000,1.0500)
    \put(0.0000,0.0000){\includegraphics[width=0.5\unitlength]{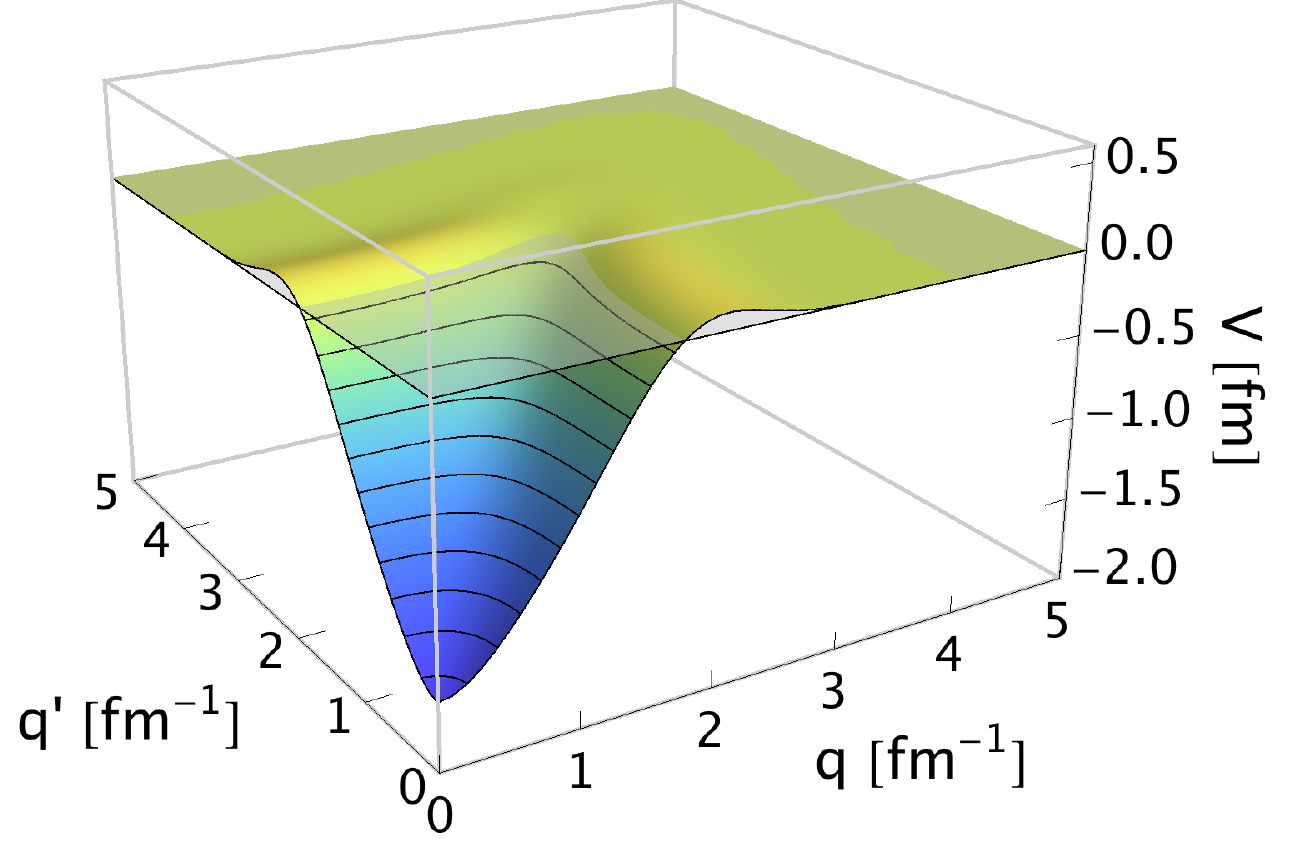}}
    \put(0.5500,0.0000){\includegraphics[width=0.45\unitlength]{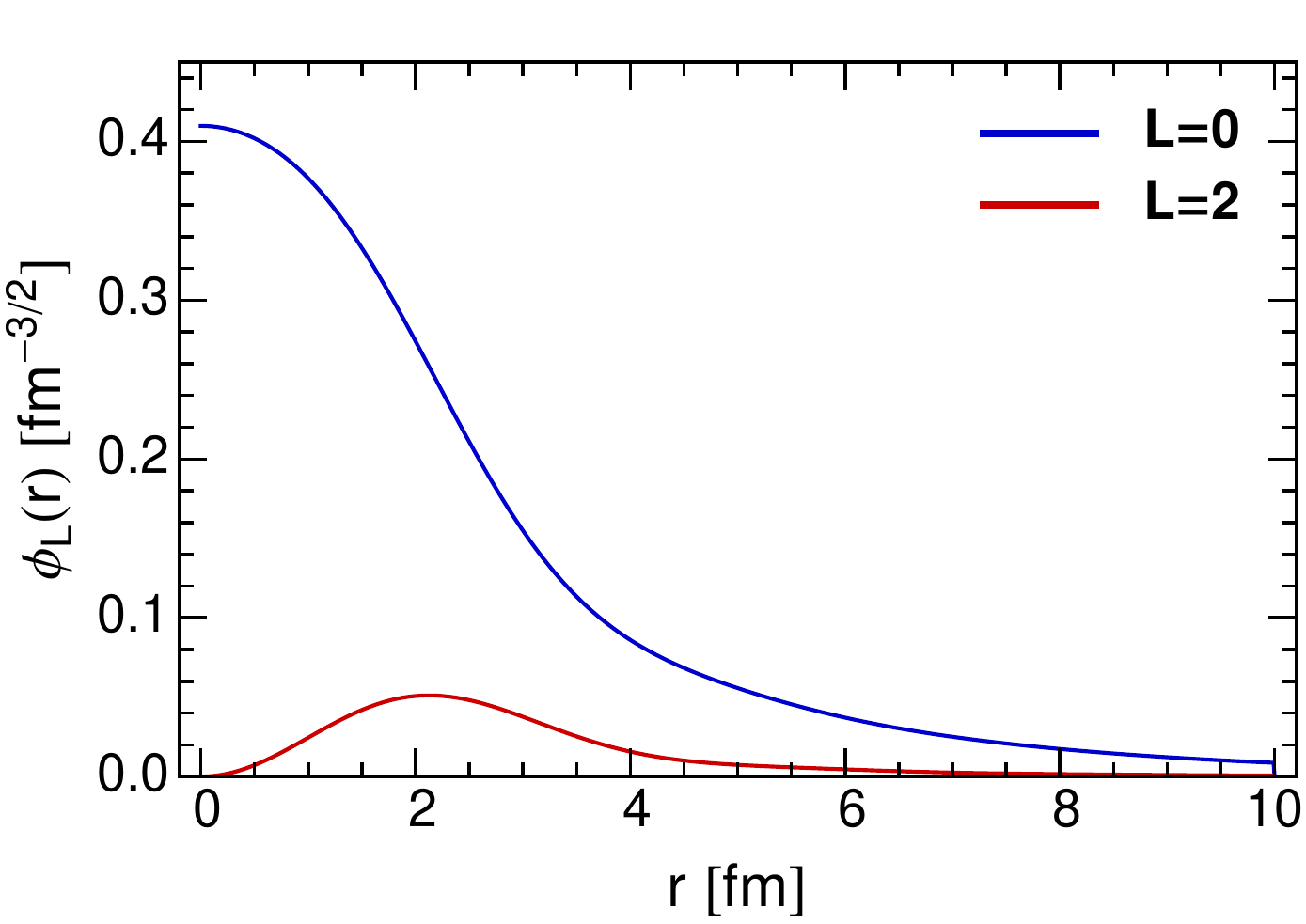}}
    \put(0.0000,0.3500){\includegraphics[width=0.5\unitlength]{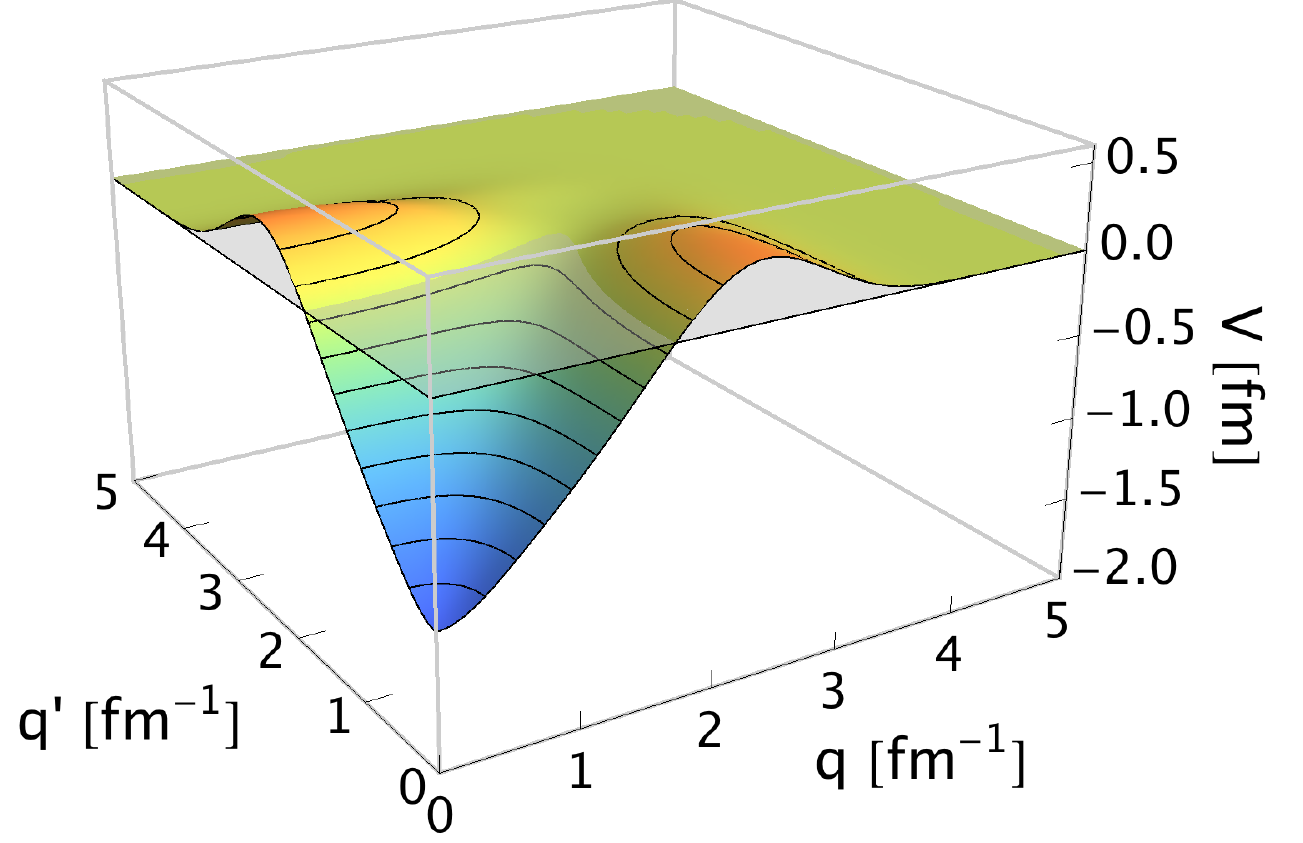}}
    \put(0.5500,0.3500){\includegraphics[width=0.45\unitlength]{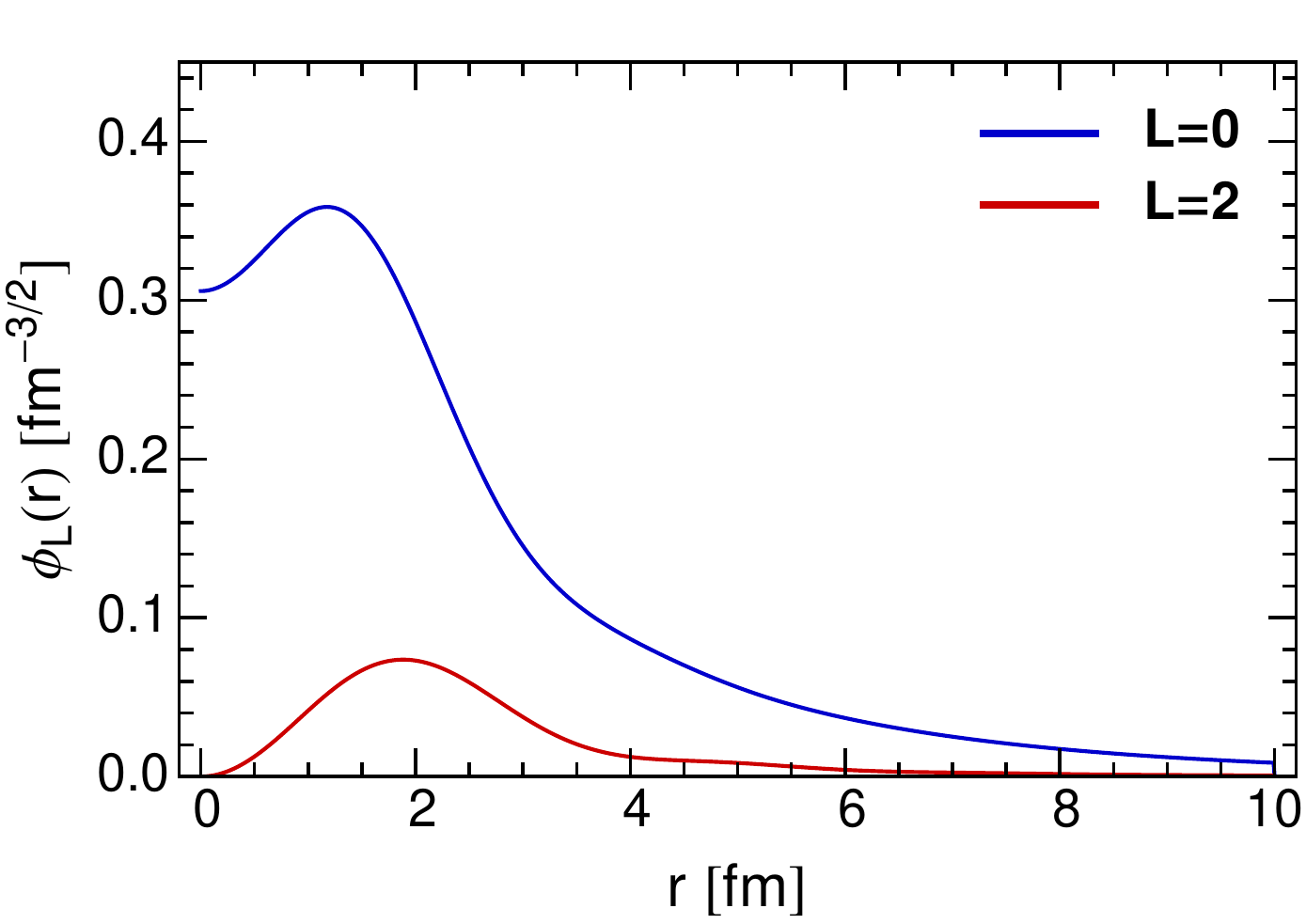}}
    \put(0.0000,0.7000){\includegraphics[width=0.5\unitlength]{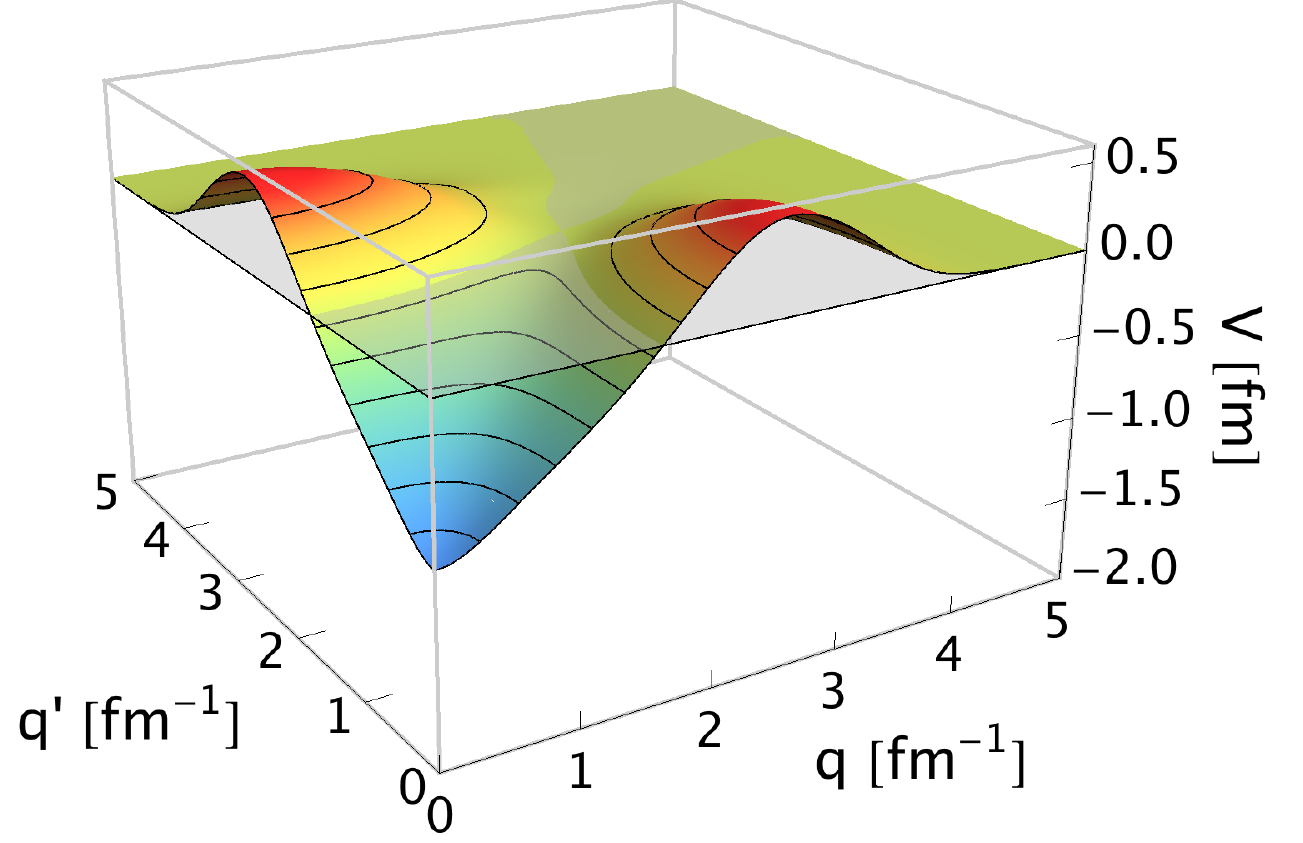}}
    \put(0.5500,0.7000){\includegraphics[width=0.45\unitlength]{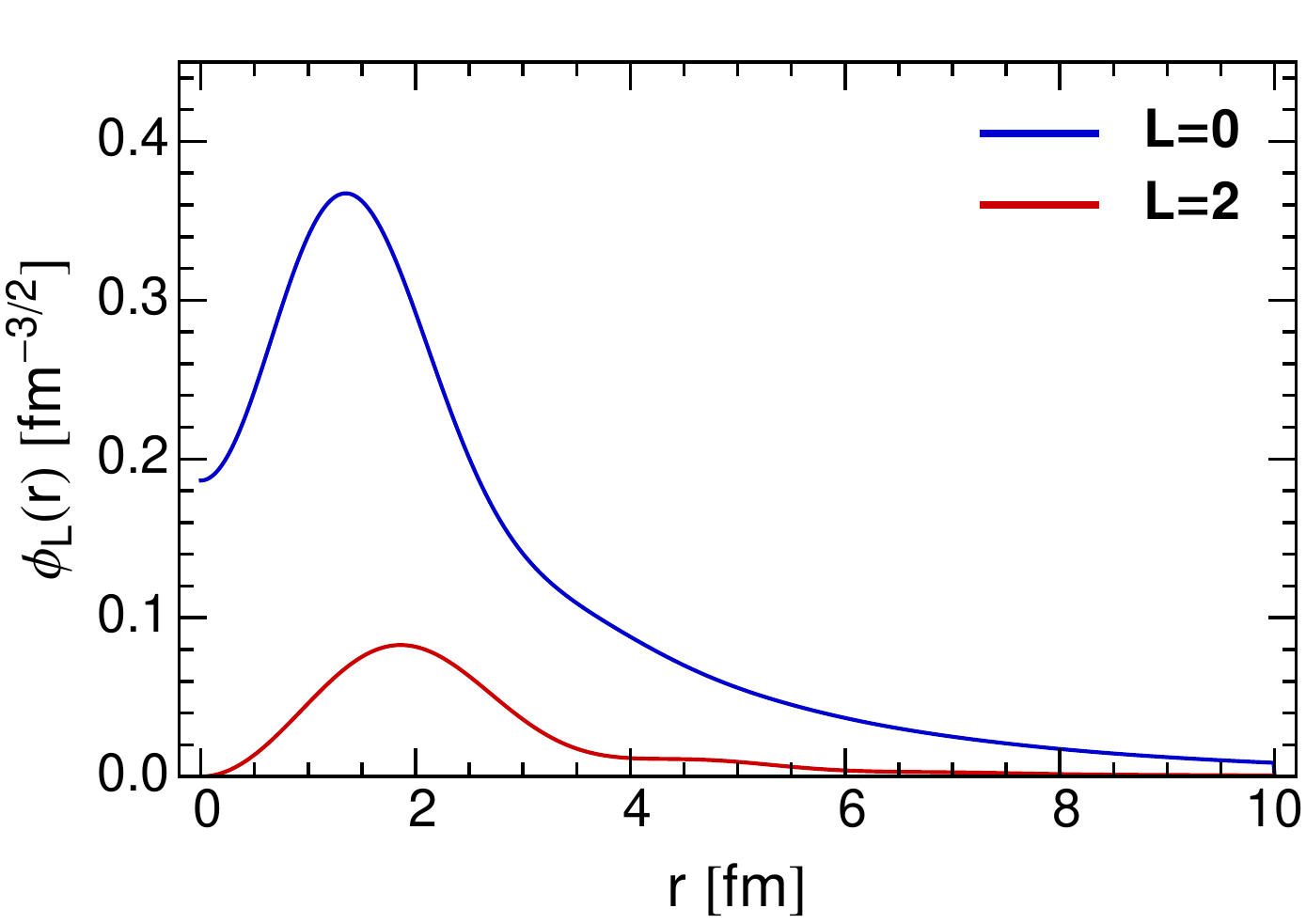}}
    \put(0.0000,1.0000){\fcolorbox{black}{white}{\parbox{0.15\unitlength}{\centering\large``$\lambda = \infty$''}}}
    \put(0.0000,0.6500){\fcolorbox{black}{white}{\parbox{0.15\unitlength}{\centering\large$\lambda = 3\,\fmi$}}}
    \put(0.0000,0.3000){\fcolorbox{black}{white}{\parbox{0.15\unitlength}{\centering\large$\lambda = 2\,\fmi$}}}
  \end{picture}
  \\[10pt]

  \caption{\label{fig:vsrg_momentum}SRG evolution of the chiral \NNNLO{} nucleon-nucleon interaction
  by Entem and Machleidt, with initial cutoff $\Lambda=500\,\MeV$ \cite{Entem:2003th,Machleidt:2011bh}. 
  In the left column, we show the
  momentum-space matrix elements of the interaction in the ${}^3S_1$ partial wave for different values
  of the SRG resolution scale $\lambdaSRG$. The top-most row shows the initial interaction at $s=0\,\fm^4$\,,
  i.e., ``$\lambda=\infty$''. In the right column, we show the $S-$ and $D-$wave components
  of the deuteron wave function that is obtained by solving the Schr\"odinger equation with the corresponding
  SRG-evolved interaction.}
\end{figure*}
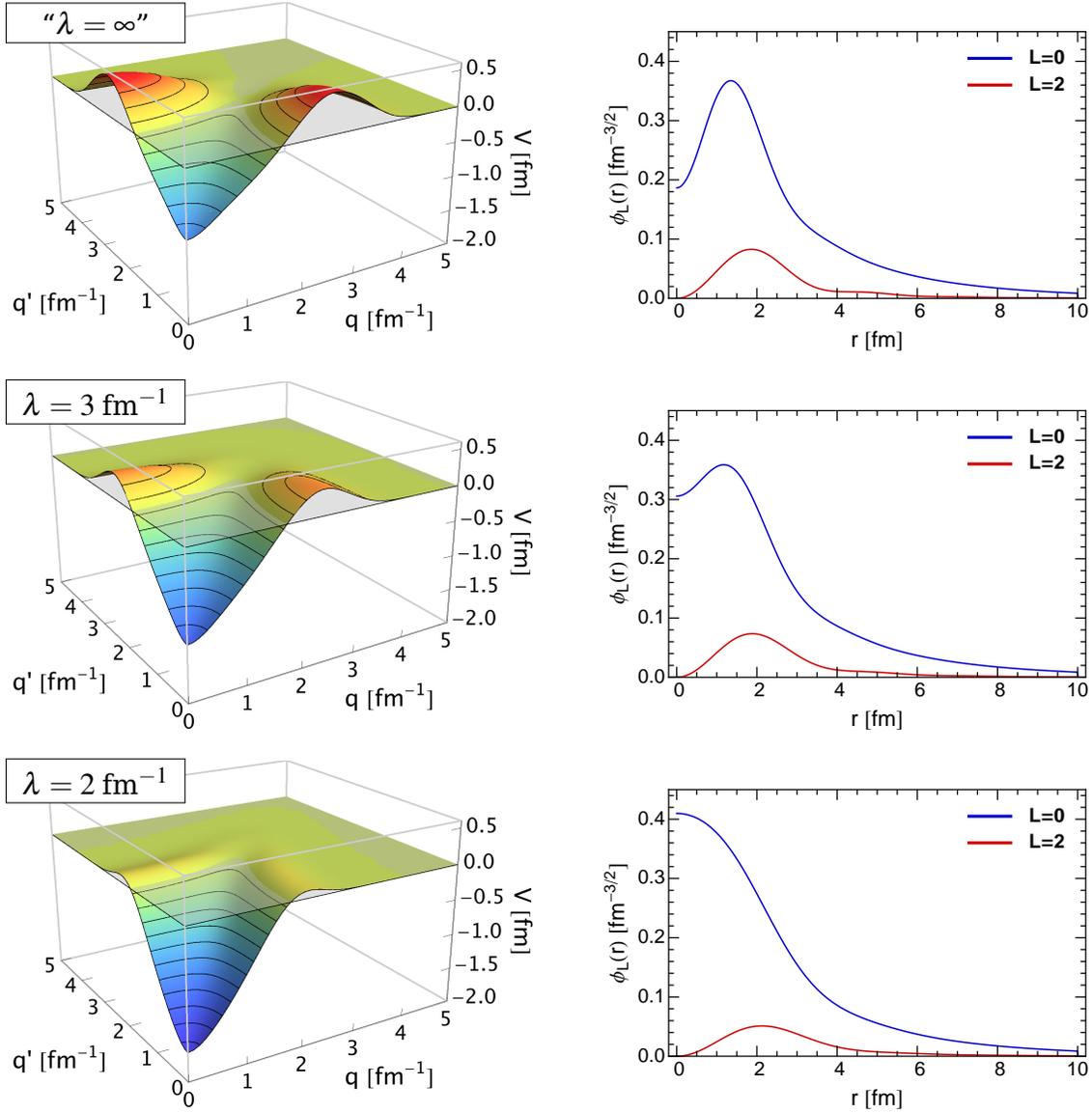

As an example of a realistic application, we discuss the SRG evolution of the
chiral \NNNLO{} nucleon-nucleon interaction by Entem and Machleidt with initial 
cutoff $\Lambda=500\,\MeV$ \cite{Entem:2003th,Machleidt:2011bh}. The momentum-space
matrix elements of this interaction in the deuteron partial waves are distributed
with the Python code discussed in the previous section.

In the top row of Fig.~\ref{fig:vsrg_momentum}, we show the matrix elements
of the initial interaction in the ${}^3S_1$ partial wave; the ${}^3S_1-{}^3D_1$
and ${}^3D_1$ are not shown to avoid clutter. Comparing the matrix elements to
those of the AV18 interaction we discussed in Sec.~\ref{sec:srg_nn}, shown
in Fig.~\ref{fig:av18}, we note that the chiral interaction has much weaker
off-diagonal matrix elements to begin with. While the AV18 matrix elements
extend as high as $|\qOV|\sim 20\,\fmi$, the chiral interaction has no appreciable
strength in states with $|\qOV|\sim 4.5\,\fmi$. In nuclear physics jargon, 
AV18 is a much harder interaction than the chiral interaction because of 
the former's strongly repulsive core. By evolving the initial interaction
to $3\,\fmi$ and then to $2\,\fmi$, the offdiagonal matrix elements are 
suppressed, and the interaction is almost entirely contained in a block
of states with $|\qOV|\sim 2\,\fmi$, except for a weak diagonal ridge.

Next to the matrix elements, we also show the deuteron wave functions
that we obtain by solving the Schr\"odinger equation with the initial
and SRG-evolved chiral interactions. For the unevolved $NN$ interaction,
the $S-$wave ($L=0$) component of the wave function is 
suppressed at small relative distances, which reflects short-range 
correlations between the nucleons. (For AV18, the $S-$wave component of 
the deuteron wave function vanishes at $r=0\,\fm$ due to the hard core.)
There is also a significant $D-$wave ($L=2$) admixture due to the tensor interaction.
As we lower the resolution scale, the ``correlation hole'' in the wave
function is filled in, and all but eliminated once we reach 
$\lambdaSRG=2.0\,\fmi$. The $D-$wave admixture is reduced significantly,
as well, because the evolution suppresses the 
matrix elements in the ${}^3S_1-{}^3D_1$ wave, which are responsible
for this mixing \cite{Bogner:2010pq}. Focusing just on the $S-$wave,
the wave function is extremely simple and matches what we would expect
for two almost independent, uncorrelated nucleons. The Pauli principle
does not affect the coordinate-space part of the wave function here 
because the overall antisymmetry of the deuteron wave function is ensured 
by its spin and isospin parts.

Let us dwell on the removal of short-range correlations from
the wave function for another moment, and consider the exact eigenstates of
the initial $NN$ Hamiltonian, 
\begin{equation}
  \HO(0) \ket{\psi_n} = E_n \ket{\psi_n}\,.
\end{equation}
The eigenvalues are invariant under a unitary transformation, e.g.,
an SRG evolution,
\begin{equation}
  \HO(\lambda)\UO(\lambda)\ket{\psi_n} \equiv \UO(\lambda) \HO(0) \UUO(\lambda) \UO(\lambda)\ket{\psi_n} 
    = E_n \UO(\lambda) \ket{\psi_n}\,.
\end{equation}
We can interpret this equation as a shift of correlations from the
wave function into the effective, \emph{RG-improved} Hamiltonian. When
we solve the Schr\"odinger equation numerically, we can usually only
obtain an approximation $\ket{\phi_n}$ of the exact eigenstate. In 
the ideal case, this is merely due to finite-precision arithmetic on
a computer, but more often, we also have systematic approximations,
e.g., mesh discretizations, finite basis sizes, many-body truncations 
(think of the cluster operator in Coupled Cluster, for instance, cf.~
chapter 8), etc.
If we use the evolved Hamiltonian $\HO(\lambda)$, we only need 
to approximate the transformed eigenstate,
\begin{equation}\label{eq:approx_eigenstate}
  \ket{\phi_n}\approx\UO(\lambda)\ket{\psi_n}\,
\end{equation}
instead of $\ket{\psi_n}$, which is often a less demanding task.
This is certainly true for our deuteron example at $\lambda=2.0\,\fmi$,
where we no longer have to worry about short-range correlations.

%
%
\subsubsection{\label{sec:srg_induced}Induced Interactions}

As discussed earlier in this section, our motivation for using the SRG to
decouple the low- and high-lying momentum components of $NN$ interactions 
is to improve the convergence of many-body calculations. The decoupling 
prevents the Hamiltonian from scattering nucleon pairs from low to high 
momenta or energies, which in turn allows configuration-space based methods 
to achieve convergence in much smaller Hilbert spaces than for a ``bare'',
unevolved interaction. This makes it possible to extend the reach of these
methods to heavier nuclei \cite{Roth:2011kx,Barrett:2013oq,Jurgenson:2013fk,Roth:2014fk,Hergert:2013ij,Hergert:2013mi,Hergert:2014vn,Hergert:2016jk,Hagen:2010uq,Roth:2012qf,Binder:2013zr,Binder:2014fk,Soma:2011vn,Soma:2013ys,Soma:2014fu,Soma:2014eu}.

In practical applications, we pay a price for the improved convergence.
To illustrate the issue, we consider the Hamiltonian in a second-quantized 
form, assuming only a two-nucleon interaction for simplicity (cf.~Eq.~\eqref{eq:def_Hint_2B}):
\begin{equation}
  \Hint = \frac{1}{4}\sum_{pqrs} \matrixe{pq}{\frac{\qOV^2}{2\mu}+\vO}{rs}\aaO_p\aaO_q\aO_s\aO_r\,.
\end{equation}
If we plug the kinetic energy and interaction into the commutators in 
Eqs.~\eqref{eq:def_srg_generator} and \eqref{eq:opflow}, we obtain
\begin{equation}\label{eq:comm2B_vac}
  \comm{\aaO_i\aaO_j\aO_l\aO_k}{\aaO_p\aaO_q\aO_s\aO_r}=
  \delta_{lp}\aaO_i\aaO_j\aaO_q\aO_s\aO_r\aO_k + \aaO\aaO\aaO\aO\aO\aO
  -\delta_{lp}\delta_{kq}\aaO_i\aaO_j\aO_s\aO_r + \aaO\aaO\aO\aO\,, 
\end{equation}
where the terms with suppressed indices schematically stand for additional 
two- and three-body operators. Even if we start from a pure $NN$
interaction, the SRG flow will induce operators of higher rank, i.e., 
$3N$, $4N$, and in general up to $A$-nucleon interactions. Of course, 
these induced interactions are only probed if we study an $A$-nucleon 
system. If we truncate the SRG flow equations at the two-body level, 
the properties of the two-nucleon system are preserved, in particular
the $NN$ scattering phase shifts and the deuteron binding energy. A 
truncation at the three-body level ensures the invariance of observables 
in $A=3$ nuclei, e.g.~$\nuc{H}{3}$ and $\nuc{He}{3}$ ground-state energies, 
and so on. 

\begin{figure*}[t]
  \setlength{\unitlength}{\textwidth}
  \begin{center}
    \includegraphics[width=0.55\unitlength]{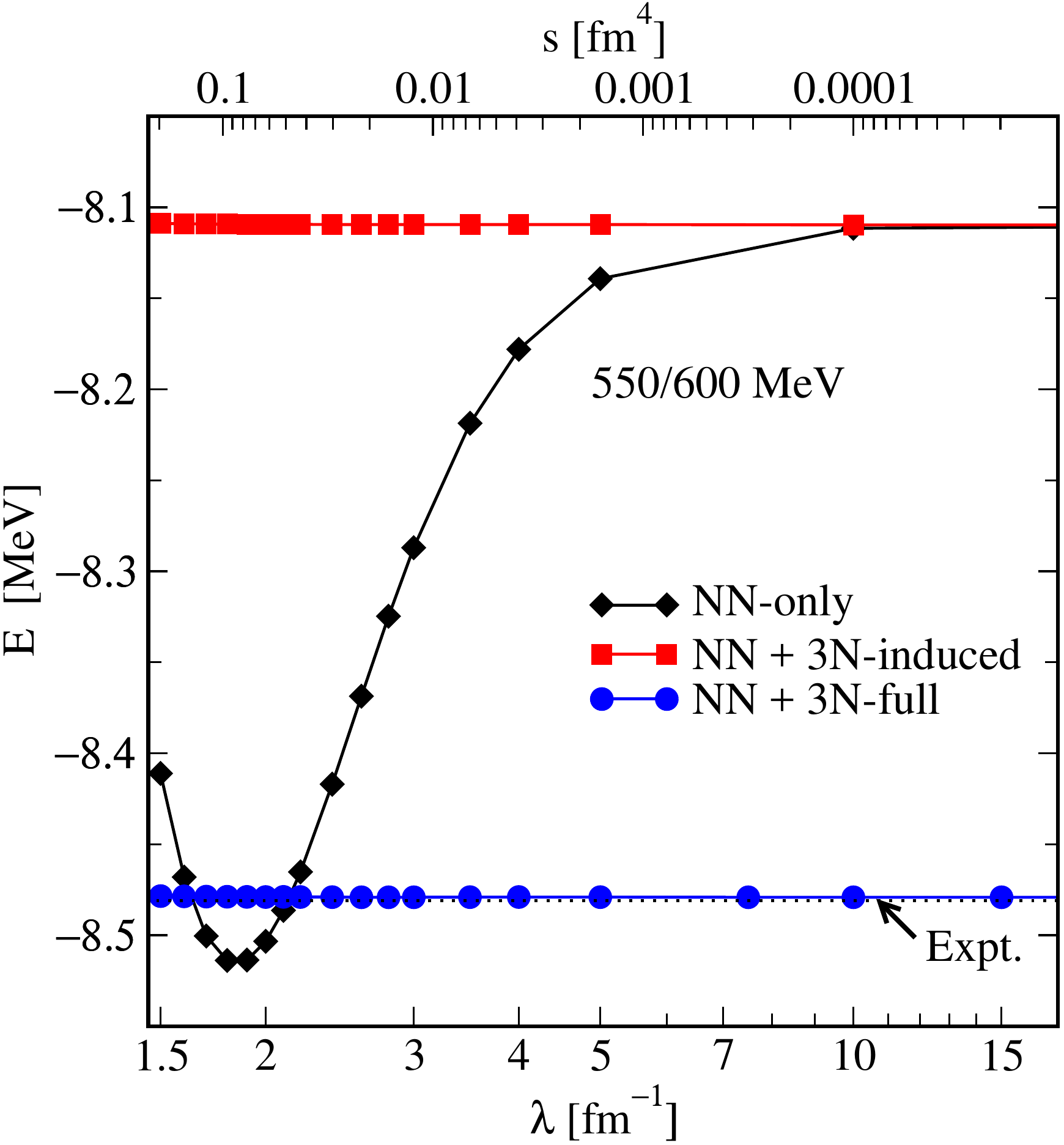}
  \end{center}  
  \vspace{-10pt}

  \caption{\label{fig:triton}Ground state energy of $\nuc{H}{3}$ as a function 
  of the flow parameter $\lambdaSRG$ for chiral \NNLO{} $NN$ and $NN\!+\!3N$ 
  interactions (see \cite{Hebeler:2012ly} for details). $NN$-only means initial and 
  induced $3N$ interactions are discarded, $NN\!+\!3N$-induced takes only 
  induced $3N$ interactions into account, and $3N$-full contains initial
  $3N$ interactions as well. The black dotted line shows the experimental 
  binding energy \cite{Wang:2012uq}. Data for the figure courtesy of 
  K.~Hebeler.}
\end{figure*}

Nowadays, state-of-the-art SRG evolutions of nuclear interactions are 
performed in the three-body system \cite{Jurgenson:2009bs,Jurgenson:2011zr,Jurgenson:2013fk, 
Hebeler:2012ly,Wendt:2013uq}. In Fig.~\ref{fig:triton}, we show 
$\nuc{H}{3}$ ground-state energies that have been calculated with
a family of SRG-evolved interactions that is generated from a 
chiral \NNLO{} $NN$ interaction by Epelbaum, Gl\"ockle, and Mei\ss{}ner 
\cite{Epelbaum:2002nr,Epelbaum:2006mo},
and a matching $3N$ interaction (see \cite{Hebeler:2012ly} for
full details). As mentioned above, the SRG evolution is not unitary in 
the three-body system if we truncate the evolved interaction and
the SRG generator at the two-body level (\emph{N$N$-only}). The 
depends strongly on $\lambdaSRG$, varying by 5--6\% over the 
typical range that we consider here (cf.~Sec.~\ref{sec:srg_nn}). 
If we truncate the operators at the three-body level instead, induced 
$3N$ interactions are properly included and the unitarity of the 
transformation is restored (\emph{$NN\!+\!3N$-induced}): 
The energy does not change as $\lambdaSRG$ is varied. Finally, the 
curve \emph{$NN\!+\!3N$-full} shows results of calculations in
which a $3N$ force was included in the initial Hamiltonian and
evolved consistently to lower resolution scale as well. Naturally,
the triton ground-state energy is invariant under the SRG flow, and 
it closely reproduces the experimental value because the $3N$ interaction's 
low-energy constants are usually fit to give the correct experimental 
$\nuc{H}{3}$ ground-state energy (see, e.g., 
\cite{Epelbaum:2009ve,Machleidt:2011bh,Gazit:2009qf}).

Our example shows that it is important to track induced interactions,
especially when we want to use evolved nuclear Hamiltonians beyond
the few-body systems we have focused on here. The nature of the SRG 
as a continuous evolution works at least somewhat in our favor: As
discussed above, truncations of the SRG flow equations lead to a 
violation of unitarity that manifests as a (residual) dependence 
of our calculated few- and many-body observables on the resolution 
scale $\lambdaSRG$. We can use this dependence as a tool to assess 
the size of missing contributions, although one has to take great
care to disentangle them from the effects of many-body truncations,
unless one uses quasi-exact methods like the NCSM (see, e.g., 
\cite{Bogner:2010pq,Jurgenson:2009bs,Hebeler:2012ly,Roth:2011kx,Hergert:2013mi,Hergert:2013ij,Binder:2014fk,
Soma:2014eu,Griesshammer:2015dp}). If we want more detailed information,
then we cannot avoid to work with $3N, 4N, \ldots$ or higher many-nucleon 
forces. The empirical observation that SRG evolutions down to $\lambdaSRG\sim1.5\fmi$
appear to preserve the natural hierarchy of nuclear forces, i.e.,  
$NN > 3N > 4N > \ldots$, suggests that we can truncate induced
forces whose contributions would be smaller than the desired accuracy 
of our calculations. 

While we may not have to go all the way to the treatment of induced
$A-$nucleon operators, which would be as expensive as implementing 
the matrix flow in the $A-$body system (cf.~Sec.~\ref{sec:srg_nn}),
dealing with induced $3N$ operators is already computationally 
expensive enough. Treating induced $4N$ forces explicitly is out 
of the question, except in schematic cases. However, there is a
way of accounting for effects of induced $3N,\ldots$ forces in
an implicit manner, by performing SRG evolutions in the nuclear
medium. 


\section{\label{sec:imsrg}The In-Medium SRG}
As discussed in the previous section, we now want to carry out the
operator evolution \eqref{eq:opflow} in the nuclear medium. The idea
is to decompose a given $N$-body operator into in-medium contributions
of lower rank and residual components that can be truncated safely. To
this end, we first have to lay the groundwork by reviewing the
essential elements of normal ordering, as well as Wick's theorem.

\subsection{\label{sec:nord}Normal Ordering and Wick's Theorem}

%
%
\subsubsection{\label{sec:nord_ops}Normal-Ordered Operators}
To construct normal-ordered operators, we start from the usual Fermionic
creation and annihilation operators, $\aaO_i$ and $\aO_i$, which satisfy
the canonical anticommutation relations
\begin{equation}
  \acomm{\aaO_i}{\aaO_j} = \acomm{\aO_i}{\aO_j} = 0\,,\quad \acomm{\aaO_i}{\aO_j}=\delta_{ij}\,.
\end{equation}
The indices are collective labels for the quantum numbers of our single-particle 
states. Using the creators and annihilators, we can express any given $A$-body 
operator in second quantization. Moreover, we can construct a complete basis 
for a many-body Hilbert space by acting with products of $\aaO_i$ on the particle 
vacuum,
\begin{equation}\label{eq:vacbasis}
  \ket{\Phi\{i_1\ldots i_A\}}=\prod_{k=1}^A\aaO_{i_k}\ket{\text{vac}}\,,
\end{equation}
and letting the indices $i_1,\ldots,i_A$ run over all single-particle states.
The states $\ket{\Phi\{i_1\ldots i_A\}}$ are, of course, nothing but 
antisymmetrized product states, i.e., Slater determinants.

Of course, not all of the Slater determinants in our basis are created equal. 
We can usually find a Slater determinant that is a fair approximation 
to the nuclear ground state, and use it as a \emph{reference state} for the 
construction and organization of our many-body basis. By simple energetics, the 
ground state and low-lying excitation spectrum of an $A$-body nucleus are usually 
dominated by excitations of particles in the vicinity of the reference state's 
Fermi energy. This is especially true for $NN$ interactions that have
been evolved to a low resolution scale $\lambdaSRG$ (see Sec.~
\ref{sec:srg_nn}). For such forces, the coupling between basis states whose
energy expectation values differ by much more than the characteristic energy 
$\hbar^2\lambda^2/m$ is suppressed. 

Slater determinants that are variationally optimized through a Hartree-Fock 
(HF) calculation have proven to be reasonable reference states for interactions 
with $\lambda\approx2.0\fm^{-1}$ (see, e.g., Refs.~\cite{Bogner:2010pq,Roth:2010vp,Barrett:2013oq,Hagen:2014ve,Hergert:2016jk,Tichai:2016vl} and references therein), allowing post-HF methods like MBPT, 
CC, or the IMSRG discussed below to converge rapidly to the exact result. 
Starting from such a HF reference state $\ket{\Phi}$, we can obtain 
a basis consisting of the state itself and up to $A$-particle, $A$-hole ($ApAh$) 
excitations:
\begin{equation}
  \ket{\Phi},\,\aaO_{p_1}\aO_{h_1}\ket{\Phi},\;\ldots\;,\,\aaO_{p_1}\ldots\aaO_{p_A}\aO_{h_A}\ldots\aO_{h_1}\ket{\Phi}\,.
\end{equation}
Here, indices $p_i$ and $h_i$ run over all one-body basis states with energies above 
(\emph{particle} states) and below the Fermi level (\emph{hole} states), respectively.
Such bases work best for systems with large gaps in the single-particle 
spectrum, e.g., closed-shell nuclei. If the gap is small, excited basis 
states can be nearly degenerate with the reference state, which usually 
results in spontaneous symmetry breaking and strong configuration mixing.

We can now introduce a one-body operator that is normal-ordered with respect
to the reference state $\ket{\Phi}$ by defining
\begin{equation}\label{eq:def_no}
  \aaO_i\aO_j \equiv \nord{\aaO_i\aO_j} +\, \contraction[1.5ex]{}{\aO}{{}^\dag_i}{\aO}\aaO_i\aO_j\,,
\end{equation}
where the brackets $\{\ldots\}$ indicate normal ordering, and the brace 
over a pair of creation and annihilation operators means that
they have been \emph{contracted}. The contraction itself is merely the 
expectation value of the operator in the reference state $\ket{\Phi}$:
\begin{equation}\label{eq:def_particle_contraction}
  \contraction[1.5ex]{}{\aO}{{}^\dag_i}{\aO}\aaO_i\aO_j \equiv \matrixe{\Phi}{\aaO_i\aO_j}{\Phi} \equiv \rho_{ji} \,.
\end{equation}
By definition, the contractions are identical to the elements of the one-body 
density matrix of $\ket{\Phi}$ \cite{Ring:1980bb}. Starting from the one-body
case, we can define normal-ordered $A$-body operators recursively by evaluating 
all contractions between creation and annihilation operators, e.g., 
\begin{align}\label{eq:def_no_nbody}
   &\aaO_{i_1}\ldots\aaO_{i_A}\aO_{j_A}\ldots\aO_{j_1} \notag\\
    &\equiv\, \nord{\aaO_{i_1}\ldots\aaO_{i_A}\aO_{j_{A}}\ldots\aO_{j_{1}}} \notag\\
    &\hphantom{\equiv}
       + \contraction[1.5ex]{}{\aO}{{}^\dag_{i_1}}{\aO}\aaO_{i_1}\aO_{j_1} 
        \nord{\aaO_{i_2}\ldots\aaO_{i_A}\aO_{j_{A}}\ldots\aO_{j_{2}}} 
      -\; \contraction[1.5ex]{}{\aO}{{}^\dag_{i_1}}{\aO}\aaO_{i_1}\aO_{j_2} 
        \nord{\aaO_{i_2}\ldots\aaO_{i_A}\aO_{j_A}\ldots\aO_{j_{3}}\aO_{j_{1}}}
      + \text{\,singles}\notag\\
    &\hphantom{\equiv}
       + \left(
          \contraction[1.5ex]{}{\aO}{{}^\dag_{i_1}}{\aO}\aaO_{i_1}\aO_{j_1}
          \contraction[1.5ex]{}{\aO}{{}^\dag_{i_1}}{\aO}\aaO_{i_2}\aO_{j_2}
          -
          \contraction[1.5ex]{}{\aO}{{}^\dag_{i_1}}{\aO}\aaO_{i_1}\aO_{j_2}
          \contraction[1.5ex]{}{\aO}{{}^\dag_{i_1}}{\aO}\aaO_{i_2}\aO_{j_1}
        \right) 
        \nord{\aaO_{i_3}\ldots\aaO_{i_A}\aO_{j_A}\ldots\aO_{j_{3}}} + \text{\,doubles} \notag\\
    &\hphantom{\equiv}
  +\,\ldots\,+ \text{\,full contractions}\,.
\end{align}
Here, we have followed established quantum chemistry jargon (singles, doubles, 
etc.) for the number of contractions in a term (cf.~chapter 8).
Note that the double contraction shown in the next-to-last line is identical
to the factorization formula for the two-body density matrix of a Slater 
determinant,
\begin{equation}
  \rho_{j_1j_2i_1i_2}
  \equiv
  \matrixe{\Phi}{\aaO_{i_1}\aaO_{i_2}\aO_{j_2}\aO_{j_1}}{\Phi}
  = \rho_{i_1j_1}\rho_{i_2j_2}- \rho_{i_1j_2}\rho_{i_2j_1}\,.
\end{equation}

From Eq.~\eqref{eq:def_no}, it is evident that $\matrixe{\Phi}{\nord{\aaO_i\aO_j}}{\Phi}$ 
must vanish, and this is readily generalized to expectation values of 
arbitrary normal-ordered operators in the reference state $\ket{\Phi}$,
\begin{equation}\label{eq:nord_ex}
  \matrixe{\Phi}{\nord{\aaO_{i_1}\ldots\aO_{i_1}}}{\Phi} = 0\,.
\end{equation}
This property of normal-ordered operators greatly facilitates calculations 
that require the evaluation of matrix elements in a space spanned by excitations 
of $\ket{\Phi}$. Another important property is that we can freely anticommute 
creation and annihilation operators within a normal-ordered string
(see problem \ref{problem:nord}):
\begin{equation}\label{eq:nord_acomm}
  \nord{\ldots\aaO_i\aO_j\ldots} = -\nord{\ldots\aO_j\aaO_i\ldots}\,.
\end{equation}

As an example, we consider an intrinsic nuclear $A$-body Hamiltonian 
containing both $NN$ and $3N$ interactions,
\begin{equation}\label{eq:def_Hint}
  H = \left(1-\frac{1}{\AO}\right)\TO^{[1]} + \frac{1}{\AO}\TO^{[2]} + \VO^{[2]} +\VO^{[3]}\,,
\end{equation}
where the one- and two-body kinetic energy terms are
\begin{align}
  \TO^{[1]} &\equiv \sum\frac{\pOV^2_i}{2m}\,,\\
  \TO^{[2]} &\equiv -\frac{1}{m}\sum_{i<j}\pOV_i\cdot\pOV_j
\end{align}
(see Sec.~\ref{sec:srg_nn} and \cite{Hergert:2009wh}). Choosing a 
single Slater determinant $\ket{\Phi}$ as the reference state, we can 
rewrite the Hamiltonian \emph{exactly} in terms of normal-ordered operators,
\begin{align}\label{eq:Hno}
  \HO &= E + \sum_{ij}f_{ij}\nord{\aaO_{i}\aO_{j}} + \frac{1}{4}\sum_{ijkl}\Gamma_{ijkl}\nord{\aaO_{i}\aaO_{j}\aO_{l}\aO_{k}}
    + \frac{1}{36}\sum_{ijklmn}W_{ijklmn}\nord{\aaO_{i}\aaO_{j}\aaO_{k}\aO_{n}\aO_{m}\aO_{l}}\,,
\end{align}
where the labels for the individual contributions have been chosen for 
historical reasons. For convenience, we will work in the eigenbasis of 
the one-body density matrix in the following, so that
\begin{equation}\label{eq:def_natorb}
  \rho_{ab}=n_{a}\delta_{ab}\,,\quad n_{a}\in\{0,1\}\,.
\end{equation}
The individual normal-ordered contributions in Eq.~\eqref{eq:Hno} are then 
given by
\begin{align}
  E &= \left(1-\frac{1}{A}\right)\sum_{a}\matrixe{a}{\tO^{[1]}}{a}n_{a}
      + \frac{1}{2}\sum_{ab}\matrixe{ab}{\frac{1}{A}\tO^{[2]}\!+\!\vO^{[2]}}{ab}n_{a}n_{b}
      + \frac{1}{6}\sum_{abc}\matrixe{abc}{\vO^{[3]}}{abc}n_{a}n_{b}n_{c}\,,\label{eq:E0}\\
  f_{ij} &= \left(1-\frac{1}{A}\right)\matrixe{i}{\tO^{[1]}}{j} 
      + \sum_{a}\matrixe{ia}{\frac{1}{A}\tO^{[2]}\!+\!\vO^{[2]}}{ja}n_{a}
      + \frac{1}{2}\sum_{ab}\matrixe{iab}{\vO^{[3]}}{jab}n_{a}n_{b}\,,\label{eq:f}   
      \\
  \Gamma_{ijkl} &= \matrixe{ij}{\frac{1}{A}\tO^{(2)}\!+\!\vO^{[2]}}{kl} + \sum_{a}\matrixe{ija}{\vO^{[3]}}{kla}n_{a}\,,\label{eq:Gamma}\\
  W_{ijklmn}&=\matrixe{ijk}{\vO^{[3]}}{lmn}\,.
\end{align}
Due to the occupation number factors in Eqs.~\eqref{eq:E0}--\eqref{eq:Gamma}, the sums 
run only over states that are occupied in the reference state. This means that the 
zero-, one-, and two-body parts of the Hamiltonian all contain in-medium contributions 
from the free-space 3N interaction.

For low-momentum interactions, it has been shown empirically that the omission 
of the normal-ordered three-body piece of the Hamiltonian causes a deviation of merely 
1--2\% in ground-state and (absolute) excited state energies of light and medium-mass 
nuclei \cite{Hagen:2007zc,Roth:2011kx,Roth:2012qf,Binder:2013fk,Gebrerufael:2016fe}. 
This \emph{normal-ordered two-body approximation} (NO2B) to the Hamiltonian is
useful for practical calculations, because it provides an efficient means to 
account for $3N$ force effects in nuclear many-body calculations without incurring 
the computational expense of explicitly treating three-body operators. In Sec.~ 
\ref{sec:imsrg_flow}, we will see that the NO2B approximation also meshes in a natural 
way with the framework of the IMSRG, which makes it especially appealing for our 
purposes.

%
%
\subsubsection{Wick's Theorem}
The normal-ordering formalism has additional benefits for the evaluation
of products of normal-ordered operators. Wick's theorem (see, e.g., \cite{Shavitt:2009}), 
which is a direct consequence of Eq.~\eqref{eq:def_no_nbody}, allows us to 
expand such products in the following way:
\begin{align}\label{eq:def_wick}
   & \nord{\aaO_{i_1}\ldots\aaO_{i_N}\aO_{j_{N}}\ldots\aO_{j_{1}}} 
     \nord{\aaO_{k_1}\ldots\aaO_{k_M}\aO_{l_{M}}\ldots\aO_{l_{1}}}
   \notag\\
    &=(-1)^{M\cdot N}  
     \nord{\aaO_{i_1}\ldots\aaO_{i_N}\aaO_{k_1}\ldots\aaO_{k_M}\aO_{j_{N}}\ldots\aO_{j_{1}}\aO_{l_{M}}\ldots\aO_{l_{1}}} 
    \notag\\
    &\hphantom{=}
       + (-1)^{M\cdot N}\contraction[1.5ex]{}{\aO}{{}^\dag_{i_1}}{\aO}\aaO_{i_1}\aO_{l_1} 
        \nord{\aaO_{i_2}\ldots\aaO_{k_M}\aO_{j_{N}}\ldots\aO_{l_{2}}} \notag\\
    &\hphantom{\equiv}
      + (-1)^{(M-1)(N-1)}\contraction[1.5ex]{}{\aO}{{}^\dag_{j_N}}{\aO}\aO_{j_N}\aaO_{k_1} 
        \nord{\aaO_{i_1}\ldots\aaO_{k_M}\aO_{j_{N}}\ldots\aO_{j_2}}
      \notag\\
    &\hphantom{=}
      + \text{\,singles}+ \text{\,doubles} + \ldots \,.
\end{align}
The phase factors appear because we anti-commute the creators and 
annihilators until they are grouped in the canonical order, i.e.,
all $\aaO$ appear to the left of the $\aO$. In the process, we also 
encounter a new type of contraction,
\begin{equation}\label{eq:def_hole_contraction}
  \contraction[1.5ex]{}{\aO}{{}^\dag_{i}}{\aO}\aO_{i}\aaO_{j} 
  \equiv \matrixe{\Phi}{\aO_{i}\aaO_{j}}{\Phi}
  = \delta_{ij} - \rho_{ij} \equiv \overline{\rho}_{ij}\,,
\end{equation}
as expected from the canonical anti-commutator algebra. $\overline\rho$
is the so-called \emph{hole density matrix}. 

The defining feature of Eq.~\eqref{eq:def_wick} is that only contractions 
between one index from each of the two strings of creation and annihilation
operators appear in the expansion, because contractions between indices 
within a single operator string have already been subtracted when we normal 
ordered it initially. In practical calculations, this leads to a substantial 
reduction of terms. An immediate consequence of Eq.~\eqref{eq:def_wick} is 
that a product of normal-ordered $M$ and $N$-body operators has the general 
form
\begin{equation}\label{eq:wick_product_schematic}
  \AO^{[M]}\BO^{[N]} = \sum_{k=|M-N|}^{M+N}\CO^{[k]}\,.
\end{equation}
Note that zero-body contributions, i.e., plain numbers, can only be generated 
if both operators have the same particle rank.

\subsection{\label{sec:imsrg_flow}In-Medium SRG Flow Equations}

%
%
\subsubsection{Induced Forces Revisited}
In Sec.~\ref{sec:srg_induced}, we discussed how SRG evolutions naturally
induce $3N$ and higher many-nucleon forces, because every evaluation of 
the commutator on the right-hand side of the operator flow equation 
\eqref{eq:opflow} increases the particle rank of $\HO(s)$, e.g.,
\begin{equation}\label{eq:induced_3N}
  \sum_{ijklpqrs}\eta_{ijkl}H_{pqrs}\comm{\aaO_{i}\aaO_{j}\aO_{l}\aO_{k}}{\aaO_{p}\aaO_{q}\aO_{s}\aO_{r}}
    = -\sum_{ijkqrs}\eta_{ijkl}H_{kqrs} \aaO_{i}\aaO_{j}\aaO_{q}\aO_{s}\aO_{r}\aO_{l} + \text{3N terms} + \text{2N terms}\,.
\end{equation}
Note that there are no induced $4N$ interactions, and that commutators
involving at least one one-body operator do not change the particle rank
(see problem \ref{problem:nord}). In the free-space evolution, we 
found that the truncation of $3N$ forces in the flowing Hamiltonian caused
a significant flow-parameter dependence of observables in $A\geq 3$ systems.

Working in the medium and using normal-ordered operators, we can expand the 
induced $3N$ operators:
\begin{align}\label{eq:induced_3N_no}
  \sum_{ijkqrs}\eta_{ijkl}H_{kqrs} \aaO_{i}\aaO_{j}\aaO_{q}\aO_{s}\aO_{r}\aO_{l}
  =\sum_{ijkqrs}\eta_{ijkl}H_{kqrs} 
    \Big(&
      \nord{\aaO_{i}\aaO_{j}\aaO_{q}\aO_{s}\aO_{r}\aO_{l}} 
       + n_q\delta_{qs}\nord{\aaO_{i}\aaO_{j}\aO_{r}\aO_{l}} 
       + n_j n_q\delta_{jr}\delta_{qs}\nord{\aaO_{i}\aO_{l}} \notag\\
  &
       + n_in_j n_q\delta_{il}\delta_{jr}\delta_{qs}+ \text{permutations}
    \Big)\,.
\end{align}
If we now truncate operators to the normal-ordered two-body level, we keep
all the in-medium contributions of the induced $3N$ terms, and retain 
information that we would have lost in the free-space evolution. These
in-medium contributions continuously feed into the 0B, 1B, and 2B matrix 
elements of the flowing Hamiltonian as we integrate Eq.~\eqref{eq:opflow}.

%
%
\subsubsection{The IMSRG(2) Scheme}
The evolution of the Hamiltonian or any other observable by means of
the flow equation \eqref{eq:opflow} is a continuous unitary transformation
in $A-$nucleon space only if we keep up to induced $A$-nucleon 
forces. Because an explicit treatment of induced contributions up to the 
$A$-body level is simply not feasible, we have to introduce a truncation 
to close the system of flow equations.

As explained in the previous subsection, we can make such truncations 
more robust if we normal order all operators with respect to a reference
state that is a fair approximation to the ground state of our system
(or another exact eigenstate we might want to target). Here, we choose
to truncate operators at the two-body level, to avoid the computational
expense of treating explicit three-body operators. For low-momentum $NN+3N$
Hamiltonians, the empirical success of the NO2B approximation mentioned 
at the end of Sec.~\ref{sec:nord_ops} seems to support this truncation: 
The omission of the normal-ordered $3N$ term in exact calculations causes 
deviations of only $\sim1\%$ in the oxygen, calcium, and nickel isotopes
\cite{Roth:2012qf,Binder:2013fk,Binder:2014fk}.

Following this line of reasoning,  we demand that for all values of the
flow parameter $s$
\begin{align} 
  \etaO(s) &\approx\etaO^{(1)}(s)+\etaO^{(2)}(s)\,,\label{eq:imsrg2_eta}\\
  \HO(s) &\approx E(s) + f(s) + \Gamma(s)\,,\label{eq:imsrg2_H}\\
  \totd{}{s}\HO(s) &\approx \totd{}{s}E(s) + \totd{}{s}f(s) + \totd{}{s}\Gamma(s)\label{eq:imsrg2_dH}\,.
\end{align}
This is the so-called IMSRG(2) truncation, which has been our primary
workhorse in past applications 
\cite{Tsukiyama:2011uq,Tsukiyama:2012fk,Hergert:2013mi,Hergert:2013ij,Hergert:2014vn,Morris:2015ve,Hergert:2016jk}. It is the basis for all results that we will discuss in the remainder of this chapter. 
The IMSRG(2) is a cousin to Coupled Cluster with Singles and Doubles
(CCSD) and the ADC(3) scheme in Self-Consistent Green's Function Theory 
(see chapters 8 and 11). Since all three
methods (roughly) aim to describe the same type and level of many-body 
correlations, we expect to obtain similar results for observables.

Let us introduce the permutation symbol $P_{ij}$ to interchange the 
indices of any expression, i.e.,
\begin{equation}\label{eq:def_Pij}
  P_{ij} g(\ldots,i,\ldots,j) \equiv g(\ldots,j,\ldots,i)\,,
\end{equation}
Plugging equations \eqref{eq:imsrg2_eta}--\eqref{eq:imsrg2_dH} into 
the operator flow equation \eqref{eq:opflow} and evaluating the 
commutators with the expressions from the appendix, we obtain the 
following system of IMSRG(2) flow equations:
\begin{align}
  \totd{E}{s}&= \sum_{ab}(n_a-n_b)\eta_{ab} f_{ba} 
    + \frac{1}{2} \sum_{abcd}\eta_{abcd}\Gamma_{cdab} n_a n_b\bar{n}_c\bar{n}_d
    \label{eq:imsrg2_m0b}\,,\\[5pt]
%
  \totd{f_{ij}}{s} &= 
  \sum_{a}(1+P_{ij})\eta_{ia}f_{aj} +\sum_{ab}(n_a-n_b)(\eta_{ab}\Gamma_{biaj}-f_{ab}\eta_{biaj}) \notag\\ 
  &\quad +\frac{1}{2}\sum_{abc}(n_an_b\bar{n}_c+\bar{n}_a\bar{n}_bn_c) (1+P_{ij})\eta_{ciab}\Gamma_{abcj}
  \label{eq:imsrg2_m1b}\,,\\[5pt]
%
  \totd{\Gamma_{ijkl}}{s}&= 
  \sum_{a}\left\{ 
    (1-P_{ij})(\eta_{ia}\Gamma_{ajkl}-f_{ia}\eta_{ajkl} )
    -(1-P_{kl})(\eta_{ak}\Gamma_{ijal}-f_{ak}\eta_{ijal} )
    \right\}\notag \\
  &\quad+ \frac{1}{2}\sum_{ab}(1-n_a-n_b)(\eta_{ijab}\Gamma_{abkl}-\Gamma_{ijab}\eta_{abkl})
    \notag\\
  &\quad+\sum_{ab}(n_a-n_b) (1-P_{ij})(1-P_{kl})\eta_{aibk}\Gamma_{bjal}
    \label{eq:imsrg2_m2b}\,.
\end{align}
Here, $\nn_i=1-n_i$, and the $s$-dependence has been suppressed for brevity. 
To obtain ground-state energies, we integrate Eqs.~\eqref{eq:imsrg2_m0b}--\eqref{eq:imsrg2_m2b} 
from $s=0$ to $s\to\infty$, starting from the initial components of the 
normal-ordered Hamiltonian (Eqs.~\eqref{eq:E0}--\eqref{eq:Gamma})
(see Secs.~\ref{sec:imsrg_pairing} and \ref{sec:imsrg_neutron_matter} 
for numerical examples).

By integrating the flow equations, we absorb many-body correlations into
the flowing normal-ordered Hamiltonian, summing certain classes of terms 
in the many-body expansion to all orders \cite{Hergert:2016jk}. We can
identify specific structures by looking at the occupation-number dependence 
of the terms in Eqs.~\eqref{eq:imsrg2_m0b}--\eqref{eq:imsrg2_m2b}: for 
instance, $\nn_i$ and $n_i$ restrict summations to particle and hole states, 
respectively (cf.~equation \eqref{eq:def_natorb}). Typical IMSRG generators 
(see Sec.~\ref{sec:imsrg_generator}) are proportional to the (offdiagonal) 
Hamiltonian, which means that the two terms in the zero-body flow equation 
essentially have the structure of second-order energy corrections, but 
evaluated for the \emph{flowing} Hamiltonian $\HO(s)$. Thus, we can express
the equation in terms of Hugenholtz diagrams as
%
\newcommand{\diagEa}{
  \setlength{\unitlength}{0.12\textwidth}
  \begin{picture}(0.7000,1.0000)
    \put(0.0000,0.0000){\includegraphics[height=\unitlength]{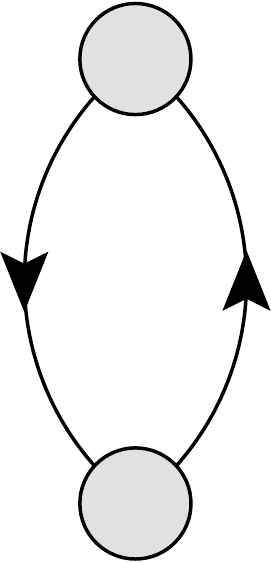}}
  \end{picture}
}
\newcommand{\diagEb}{
  \setlength{\unitlength}{0.12\textwidth}
  \begin{picture}(0.7000,1.0000)
    \put(0.0000,0.0000){\includegraphics[height=\unitlength]{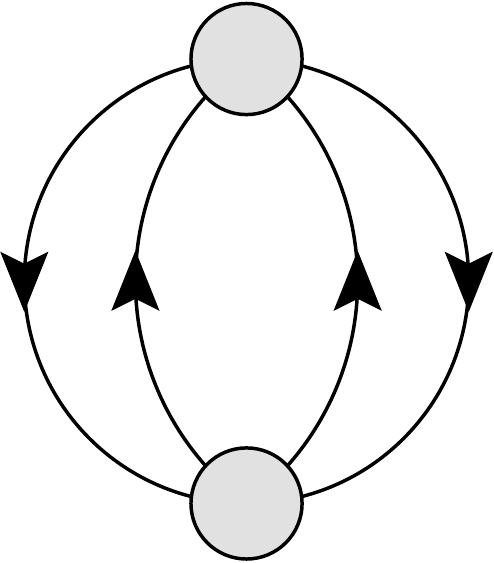}}
  \end{picture}
}
\begin{equation}
  \totd{}{s} E = \vcenter{\hbox{\diagEa}} + \vcenter{\hbox{\diagEb}}\,.
\end{equation}
Note that the energy denominators associated with the propagation of the 
intermediate state are consistently calculated with $\HO(s)$ here. 

\newcommand{\diagA}{
  \setlength{\unitlength}{0.1\textwidth}
  \begin{picture}(0.7000,1.0000)
    \put(0.0000,0.0000){\includegraphics[height=\unitlength]{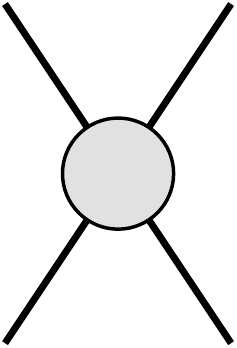}}
  \end{picture}
}

\newcommand{\diagB}{
  \setlength{\unitlength}{0.1\textwidth}
  \begin{picture}(0.7000,1.0000)
    \put(0.0000,0.0000){\includegraphics[height=\unitlength]{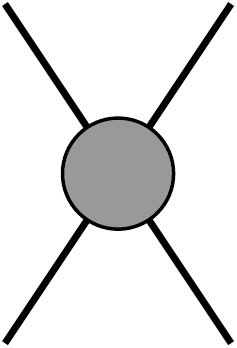}}
  \end{picture}
}
\newcommand{\diagBpp}{
  \setlength{\unitlength}{0.1\textwidth}
  \begin{picture}(0.7000,1.5000)
    \put(0.0000,0.0000){\includegraphics[height=1.5\unitlength]{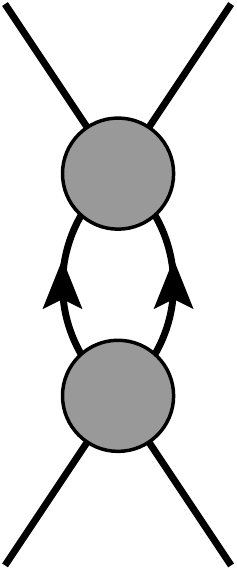}}
  \end{picture}
}
\newcommand{\diagBhh}{
  \setlength{\unitlength}{0.1\textwidth}
  \begin{picture}(0.7000,1.5000)
    \put(0.0000,0.0000){\includegraphics[height=1.5\unitlength]{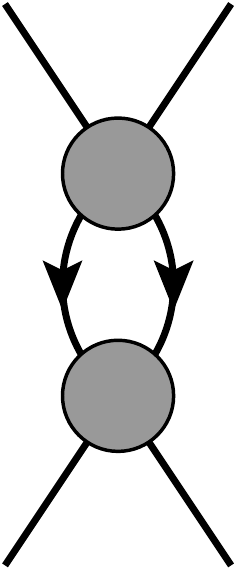}}
  \end{picture}
}
\newcommand{\diagBph}{
  \setlength{\unitlength}{0.1\textwidth}
  \begin{picture}(1.3000,1.0000)
    \put(0.0000,0.0000){\includegraphics[height=\unitlength]{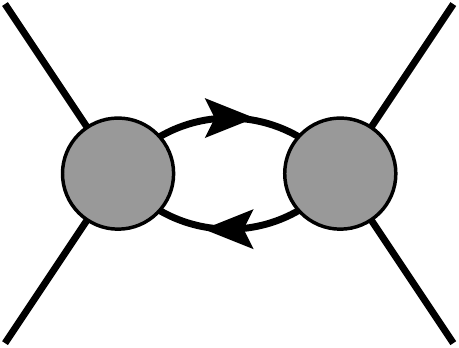}}
  \end{picture}
}

\newcommand{\diagC}{
  \setlength{\unitlength}{0.1\textwidth}
  \begin{picture}(0.7000,1.0000)
    \put(0.0000,0.0000){\includegraphics[height=\unitlength]{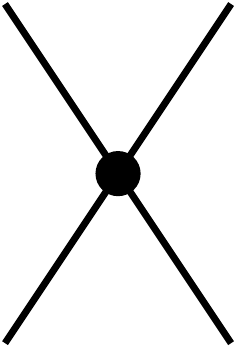}}
  \end{picture}
}
\newcommand{\diagCppa}{
  \setlength{\unitlength}{0.1\textwidth}
  \begin{picture}(0.7000,1.5000)
    \put(0.0000,0.0000){\includegraphics[height=1.5\unitlength]{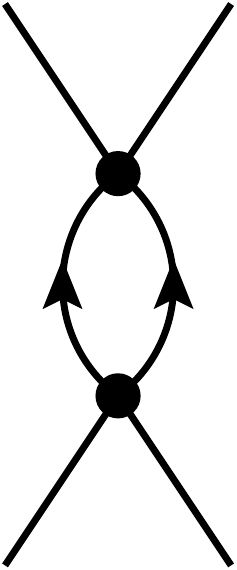}}
  \end{picture}
}
\newcommand{\diagCppb}{
  \setlength{\unitlength}{0.1\textwidth}
  \begin{picture}(0.7000,2.0000)
    \put(0.0000,0.0000){\includegraphics[height=2.0\unitlength]{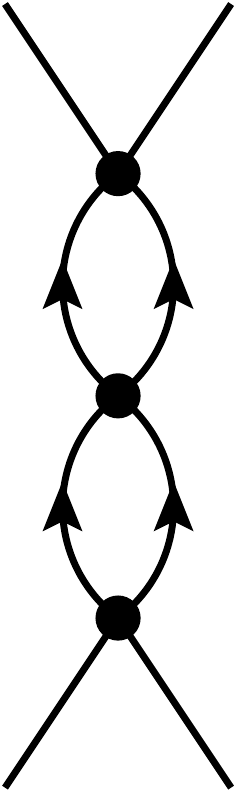}}
  \end{picture}
}
\newcommand{\diagCppc}{
  \setlength{\unitlength}{0.1\textwidth}
  \begin{picture}(0.7000,2.0000)
    \put(0.0000,0.0000){\includegraphics[height=2.0\unitlength]{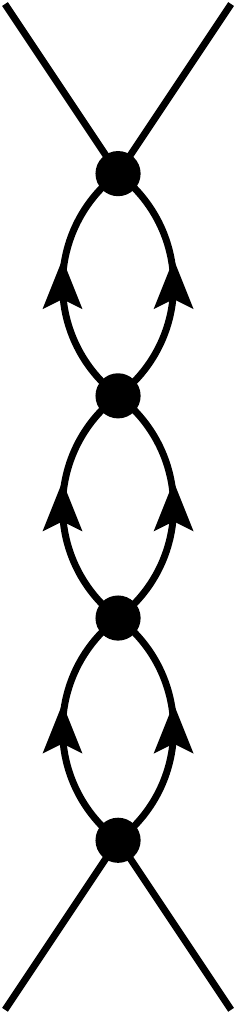}}
  \end{picture}
}
\newcommand{\diagCpha}{
  \setlength{\unitlength}{0.1\textwidth}
  \begin{picture}(1.3000,1.0000)
    \put(0.0000,0.0000){\includegraphics[height=\unitlength]{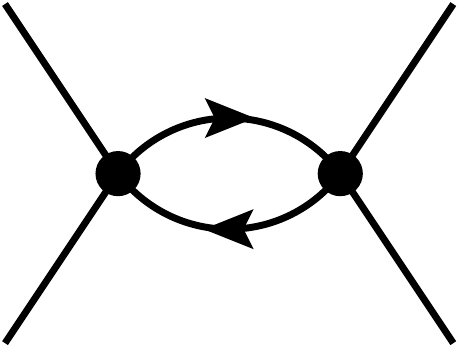}}
  \end{picture}
}
\newcommand{\diagCphb}{
  \setlength{\unitlength}{0.1\textwidth}
  \begin{picture}(1.9000,1.0000)
    \put(0.0000,0.0000){\includegraphics[height=\unitlength]{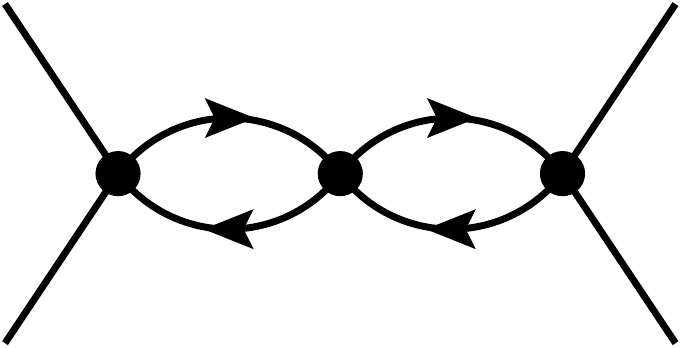}}
  \end{picture}
}
\newcommand{\diagCphc}{
  \setlength{\unitlength}{0.1\textwidth}
  \begin{picture}(2.5000,1.0000)
    \put(0.0000,0.0000){\includegraphics[height=\unitlength]{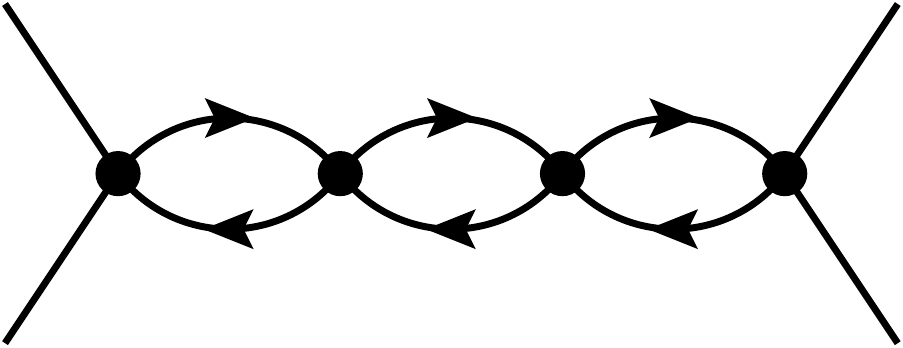}}
  \end{picture}
}
\newcommand{\diagCpppha}{
  \setlength{\unitlength}{0.1\textwidth}
  \begin{picture}(1.2000,1.5000)
    \put(0.0000,0.0000){\includegraphics[height=1.5\unitlength]{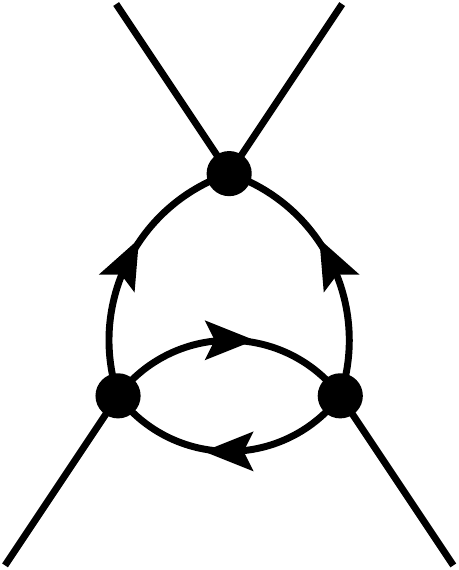}}
  \end{picture}
}
\newcommand{\diagCppphb}{
  \setlength{\unitlength}{0.1\textwidth}
  \begin{picture}(0.9000,1.5000)
    \put(0.1000,0.0000){\includegraphics[height=1.5\unitlength]{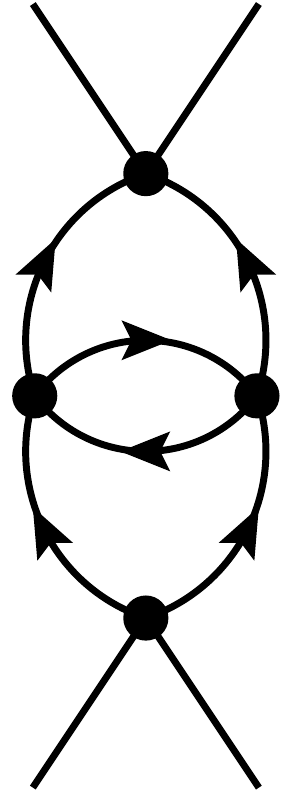}}
  \end{picture}
}

In the flow equation for the two-body vertex $\Gamma$, terms that
are proportional to 
\begin{equation}
  1-n_a-n_b = \nn_a\nn_b - n_a n_b
\end{equation}
will build up a summation of particle-particle and hole-hole \emph{ladder 
diagrams} as we integrate the flow equations $s\to\infty$. Similarly, 
the terms proportional to 
\begin{equation}
  n_a-n_b = n_a \nn_b - \nn_a n_b\,,
\end{equation}
will give rise to a summation of \emph{chain diagrams} representing 
particle-hole terms at all orders. We can illustrate this by expanding 
the vertex we obtain after two integration steps,
$\Gamma(2\delta s)$, in terms of the prior vertices $\Gamma(\delta s)$ and
$\Gamma(0)$. Indicating these vertices by light gray, dark gray, and black
circles, we schematically have
\begin{align}\label{eq:vertex_schematic}
  \vcenter{\hbox{\diagA}} 
  &= \vcenter{\hbox{\diagB}} + \vcenter{\hbox{\diagBpp}} + 
      \vcenter{\hbox{\diagBhh}} + \vcenter{\hbox{\diagBph}} + \;\ldots\notag\\
  &= \vcenter{\hbox{\diagC}} + 
      \vcenter{\hbox{\diagCppa}} + \vcenter{\hbox{\diagCppb}} + \vcenter{\hbox{\diagCppc}} + \;\ldots\notag\\
  &\hphantom{=}
      +\vcenter{\hbox{\diagCpha}} + \vcenter{\hbox{\diagCphb}} + \vcenter{\hbox{\diagCphc}} + \;\ldots\notag\\
  &\hphantom{=}
      +\vcenter{\hbox{\diagCpppha}} + \vcenter{\hbox{\diagCppphb}} + \;\ldots
\end{align}
In the first line, we see that $\Gamma(2\delta s)$ is given by
the vertex of the previous step, $\Gamma(\delta s)$, plus second-order
corrections. As in the energy flow equation, it is assumed that the
energy denominators associated with the propagation of the intermediate
states are calculated with $\HO(\delta s)$. For brevity, we have suppressed
additional permutations of the shown diagrams, as well as the diagrams
that result from contracting one- and two-body operators in Eq.~\eqref{eq:imsrg2_m2b}.

In the next step, we expand each of the $\Gamma(\delta s)$ vertices in
terms of $\Gamma(0)$, and assume that energy denominators are now
expressed in terms of $\HO(0)$. In the second line of Eq.~\eqref{eq:vertex_schematic},
we explicitly show the ladder-type diagrams with intermediate 
particle-particle states that are generated by expanding
the first two diagrams for $\Gamma(\delta s)$, many additional diagrams
are suppressed. Likewise, the third line illustrates the emergence of 
the chain summation via the particle-hole diagrams that 
are generated by expanding the fourth diagram for $\Gamma(\delta s)$.
In addition to the ladder and chain summations, the IMSRG(2) will also
sum interference diagrams like the ones shown in the last row of 
Eq.~\eqref{eq:vertex_schematic}.
Such terms are not included in traditional summation methods, like
the $G$-matrix approach for ladders, or the Random Phase Approximation
(RPA) for chains \cite{Day:1967zl,Brandow:1967tg,Fetter:2003ve}. We
conclude our discussion at this point, and refer interested readers
to the much more detailed analysis in Ref.~\cite{Hergert:2016jk}.

%
%
\subsubsection{Computational Scaling}
Let us briefly consider the computational scaling of the IMSRG(2) scheme,
ahead of the discussion of an actual implementation in Sec.~\ref{sec:imsrg_implementation}.
When performing a single integration step, the computational effort is 
dominated by the two-body flow equation \eqref{eq:imsrg2_m2b}, which 
naively requires $\OC(N^6)$ operations, where $N$ denotes the size of
the single-particle basis. This puts the IMSRG(2) in the same category 
as CCSD and ADC(3) (see chapters 8 and 11).
Fortunately, large portions of the flow equations can be expressed in 
terms of matrix products, allowing us to use optimized linear algebra 
libraries provided by high-performance computing vendors. 

Moreover, we can further reduce the computational cost by distinguishing 
particle and hole states, because the number of hole states $N_h$ is 
typically much smaller than the number of particle states $N_p\sim N$. 
The best scaling we can achieve in the IMSRG(2) depends on the choice of 
generator (see Sec.~\ref{sec:imsrg_generator}). If the one- and two-body 
parts of the 
generator only consist of $ph$ and $pphh$ type matrix elements and their 
Hermitian conjugates, the scaling is reduced to $\OC(N_h^2N_p^4)$, 
which matches the cost of solving the CCSD amplitude equations.

\subsection{\label{sec:imsrg_decoupling}Decoupling}

%
%
\subsubsection{\label{sec:imsrg_offidag}The Off-Diagonal Hamiltonian}

\begin{figure}[t]
  \setlength{\unitlength}{\textwidth}
  \begin{center}
  \begin{picture}(0.9000,0.3700)
    \put(0.1100,0.0400){\includegraphics[width=0.3\unitlength]{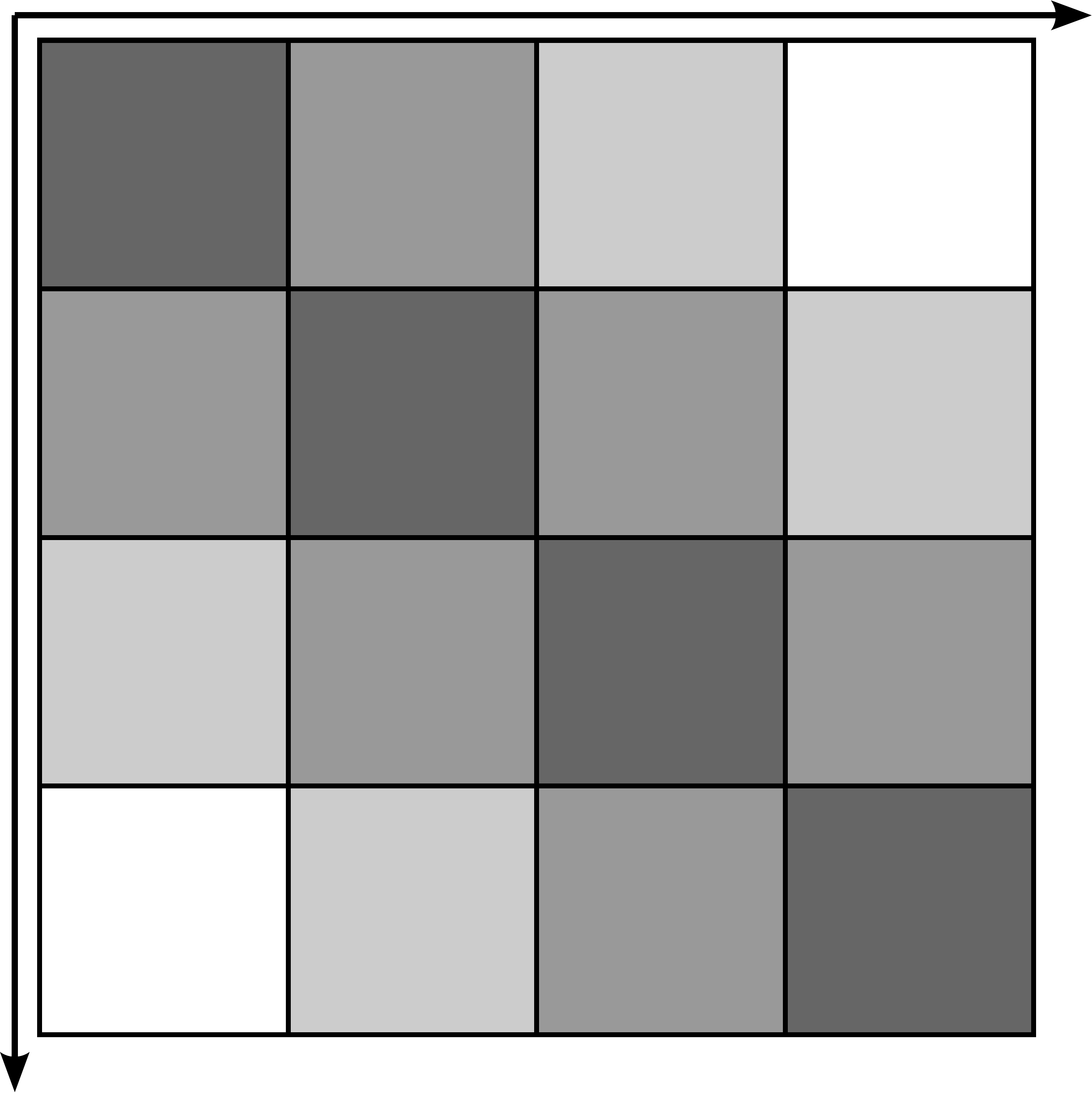}}
    \put(0.5500,0.0400){\includegraphics[width=0.3\unitlength]{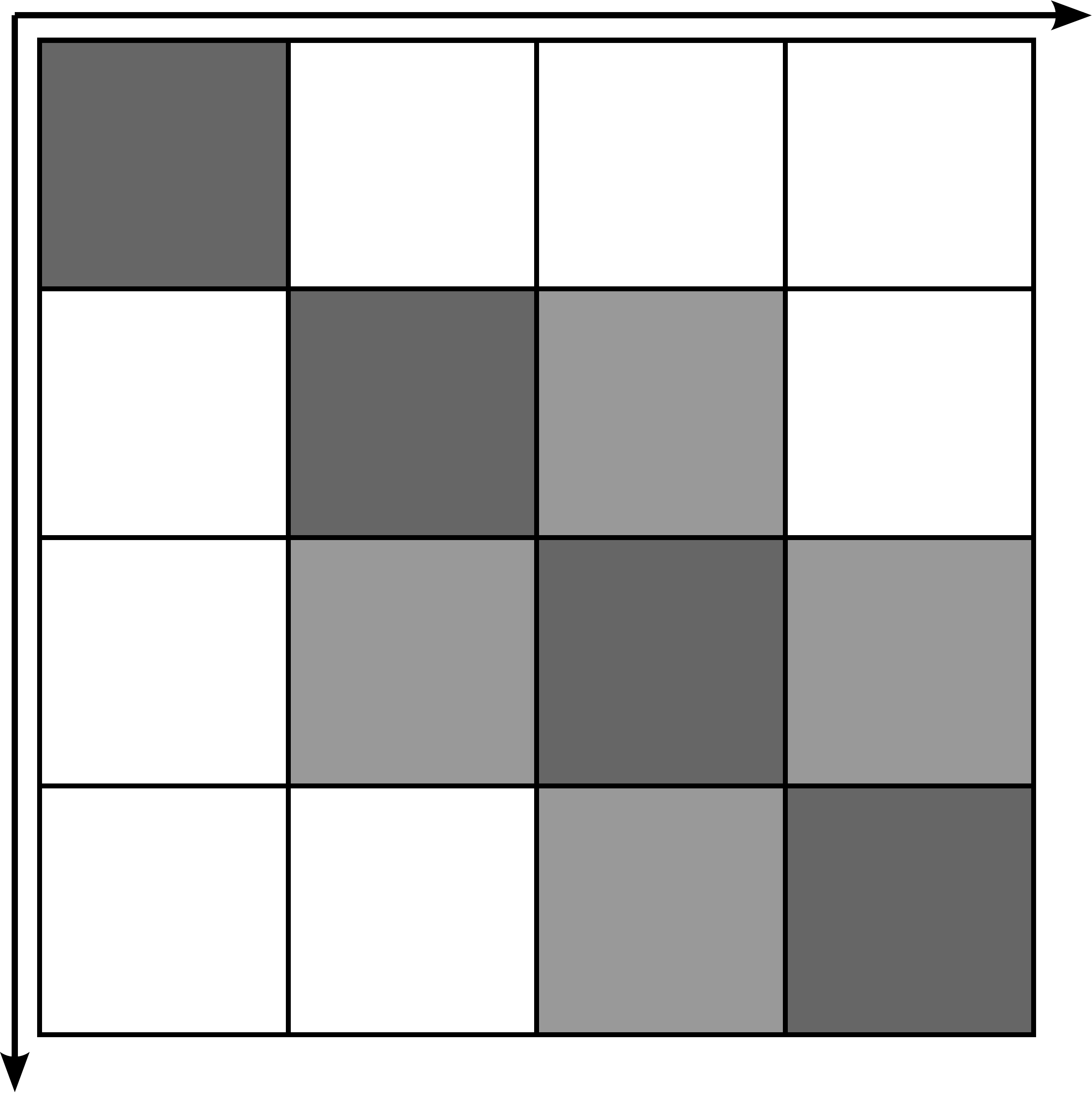}}
    \put(0.1100,0.0200){\parbox{0.3\unitlength}{\centering$\matrixe{i}{\HO(0)}{j}$}}
    \put(0.5500,0.0200){\parbox{0.3\unitlength}{\centering$\matrixe{i}{\HO(\infty)}{j}$}}
    \put(0.4300,0.1900){\parbox{0.1\unitlength}{\centering\Large$\stackrel{s\to\infty}{\longrightarrow}$}}
    \put(0.0800,0.0500){\rotatebox{90}{\parbox{0.075\unitlength}{\small\centering$\ket{\Phi^{pp'p''}_{\!hh'\!h''}\!}$}}}
    \put(0.0800,0.1200){\rotatebox{90}{\parbox{0.075\unitlength}{\small\centering$\ket{\Phi^{pp'}_{hh'}}$}}}
    \put(0.0800,0.1900){\rotatebox{90}{\parbox{0.075\unitlength}{\small\centering$\ket{\Phi^{p}_{h}}$}}}
    \put(0.0800,0.2600){\rotatebox{90}{\parbox{0.075\unitlength}{\small\centering$\ket{\Phi}$}}}
    \put(0.1150,0.3500){\parbox{0.075\unitlength}{\small\centering$\ket{\Phi}$}}
    \put(0.1850,0.3500){\parbox{0.075\unitlength}{\small\centering$\ket{\Phi^{p}_{h}}$}}
    \put(0.2550,0.3500){\parbox{0.075\unitlength}{\small\centering$\ket{\Phi^{pp'}_{hh'}}$}}
    \put(0.3200,0.3500){\parbox{0.075\unitlength}{\small\centering$\ket{\Phi^{pp'p''}_{\!hh'\!h''}\!}$}}
    \put(0.8450,0.1300){\rotatebox{-90}{\parbox{0.075\unitlength}{\small\centering$\ket{\Phi^{pp'p''}_{\!hh'\!h''}\!}$}}}
    \put(0.8450,0.2000){\rotatebox{-90}{\parbox{0.075\unitlength}{\small\centering$\ket{\Phi^{pp'}_{hh'}}$}}}
    \put(0.8450,0.2700){\rotatebox{-90}{\parbox{0.075\unitlength}{\small\centering$\ket{\Phi^{p}_{h}}$}}}
    \put(0.8450,0.3400){\rotatebox{-90}{\parbox{0.075\unitlength}{\small\centering$\ket{\Phi}$}}}
    \put(0.5550,0.3500){\parbox{0.075\unitlength}{\small\centering$\ket{\Phi}$}}
    \put(0.6250,0.3500){\parbox{0.075\unitlength}{\small\centering$\ket{\Phi^{p}_{h}}$}}
    \put(0.6950,0.3500){\parbox{0.075\unitlength}{\small\centering$\ket{\Phi^{pp'}_{hh'}}$}}
    \put(0.7600,0.3500){\parbox{0.075\unitlength}{\small\centering$\ket{\Phi^{pp'p''}_{\!hh'\!h''}\!}$}}
  \end{picture}
  \end{center}
  \vspace{-10pt}

  \caption{\label{fig:imsrg}
  Schematic view of single-reference IMSRG decoupling in a many-body Hilbert space spanned by
  a Slater determinant reference $\ket{\Phi}$ and its particle-hole excitations $\ket{\Phi^{p\ldots}_{h\ldots}}$.
  }
\end{figure}
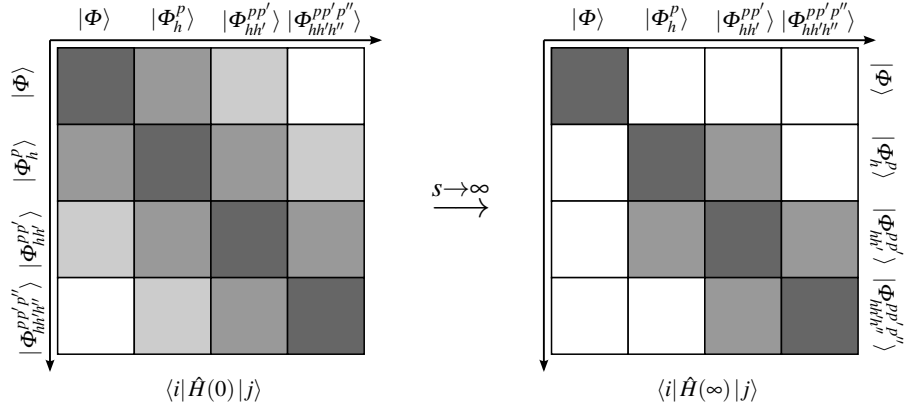

Having set up the IMSRG flow equations, we now need to specify our decoupling 
strategy, i.e., how we split the Hamiltonian into diagonal parts we 
want to keep, and off-diagonal parts we want to suppress (cf.~Sec.~
\ref{sec:srg}). To this end, we refer to the matrix representation of 
the Hamiltonian in a basis of $A$-body Slater determinants, but let
us stress that we never actually construct the Hamiltonian matrix in 
this representation.

Our Slater determinant basis consists of a reference determinant and 
all its possible particle-hole excitations (cf.~Sec.~\ref{sec:nord}):
\begin{equation}
  \ket{\Phi},\,\nord{\aaO_{p}\aO_h}\ket{\Phi},\,\nord{\aaO_{p}\aaO_{p'}\aO_{h'}\aO_{h}}\ket{\Phi},\ldots\,.
\end{equation}
Note that
\begin{equation}
  \nord{\aaO_{p_1}\ldots\aaO_{p_i}\aO_{h_i}\ldots\aO_{h_1}}=\aaO_{p_1}\ldots\aaO_{p_i}\aO_{h_i}\ldots\aO_{h_1}
\end{equation}
because contractions of particle and hole indices vanish by construction. 
Using Wick's theorem, one can show that the particle-hole excited Slater 
determinants are orthogonal to the reference state as well as each other
(see problem \ref{problem:ph_orthogonality}).
In the Hilbert space spanned by this basis, the matrix representation of
our initial Hamiltonian in the NO2B approximation (cf.~Sec.~\ref{sec:nord_ops}) 
has the structure shown in the left panel of Fig.~\ref{fig:imsrg}, i.e., it is 
band-diagonal, and can at most couple $npnh$ and $(n\pm2)p(n\pm2)h$ 
excitations. 

We now have to split the Hamiltonian into appropriate diagonal and off-diagonal 
parts on the operator level \cite{Kutzelnigg:1982ly,Kutzelnigg:1983ve,Kutzelnigg:1984qf}. 
Using a broad definition of diagonality is ill-advised because we must 
avoid inducing strong in-medium $3N, \ldots$ interactions to maintain the
validity of the IMSRG(2) truncation. For this reason, we choose a so-called 
\emph{minimal decoupling scheme} that only aims to decouple the one-dimensional 
block spanned by the reference state from all particle-hole excitations, as
shown in the right panel of Fig.~\ref{fig:imsrg}. 

If we could implement the minimal decoupling without approximations, 
we would extract a single eigenvalue and eigenstate of the many-body Hamiltonian
for the nucleus of interest in the limit $s\to\infty$. The eigenvalue
would simply be given by the zero-body piece of $H(\infty)$, while the eigenstate
is obtained by applying the unitary IMSRG transformation to the reference
state, $\UUO(\infty)\ket{\Phi}$. In practice, truncations cannot be
avoided, of course, and we only obtain an approximate eigenvalue and 
mapping. We will explicitly demonstrate in Sec.~\ref{sec:imsrg_pairing}
that the chosen reference state plays an important role in determining
which eigenvalue and eigenstate of the Hamiltonian we end up extracting
in our minimal decoupling scheme. An empirical rule of thumb is that the 
IMSRG flow will connect the reference state to the eigenstate with which 
it has the highest overlap. If we are interested in the exact ground state, 
this is typically the case for a HF Slater determinant, because it minimizes 
both the absolute energy and the correlation energy. 

Analyzing the matrix elements between the reference state and its
excitations with the help of Wick's theorem, we first see that the 
Hamiltonian couples the $0p0h$ block to $1p1h$ excitations through 
the matrix elements
\begin{align}
\matrixe{\Phi}{\HO\nord{\aaO_{p}\aO_{h}}}{\Phi}
  &=E\matrixe{\Phi}{\nord{\aaO_{p}\aO_{h}}}{\Phi}
    +\sum_{ij}f_{ij}\matrixe{\Phi}{\nord{\aaO_{i}\aO_{j}}\nord{\aaO_{p}\aO_{h}}}{\Phi}\notag\\
  &\hphantom{=}+\frac{1}{4}\sum_{ijkl}\Gamma_{ijkl}
      \matrixe{\Phi}{\nord{\aaO_{i}\aaO_{j}\aO_{l}\aO_{k}}\nord{\aaO_{p}\aO_{h}}}{\Phi}\notag\\
  &=\sum_{ij}f_{ij}\delta_{ih}\delta_{pj}n_i\nn_j = f_{hp} \label{eq:coupl_1p1h}
\end{align}
and their Hermitian conjugates. The contributions from the zero-body and 
two-body pieces of the Hamiltonian vanish because they are expectation
values of normal-ordered operators in the reference state (cf.~Eq.~\eqref{eq:nord_ex}).
Likewise, the $0p0h$ and $2p2h$ blocks are coupled by the matrix elements
\begin{equation}
\matrixe{\Phi}{\HO\nord{\aaO_{p}\aaO_{p'}\aO_{h'}\aO_{h}}}{\Phi} = \Gamma_{hh'pp'}
\end{equation}
and their conjugates. It is precisely these two-body matrix elements 
that couple $npnh$ and $(n\pm2)p(n\pm2)h$ states and generate the
outermost side diagonals of the Hamiltonian matrix. This suggests that 
we can transform the Hamiltonian to the shape shown in the top right panel
of Fig.~\ref{fig:imsrg} by defining its offdiagonal part as
\begin{equation}
  \HO_{od} \equiv \sum_{ph}f_{ph}\nord{\aaO_{p}\aO_{h}} + 
              \frac{1}{4}\sum_{pp'hh'}\Gamma_{pp'hh'}\nord{\aaO_{p}\aaO_{p'}\aO_{h'}\aO_{h}} + \text{H.c.}\,.
\end{equation}
In Sec.~\ref{sec:imsrg_pairing}, we will show that the IMSRG flow does indeed
exponentially suppress the matrix elements of $\HO_{od}$ and 
achieve the desired decoupling in the limit $s\to\infty$. 

%
%
\subsubsection{\label{sec:imsrg_variational}Variational Derivation of Minimal Decoupling}
Our minimal decoupling scheme is very reminiscent of the strategy 
followed in Coupled Cluster approaches \cite{Shavitt:2009,Hagen:2014ve}, except 
that we specifically use a unitary transformation instead of a general 
similarity transformation. It is also appealing for a different reason: As 
we will discuss now, it can be derived from a variational approach, tying
the seemingly unrelated ideas of energy minimization and renormalization 
in the many-body system together. 

Consider the energy expectation value of the final IMSRG evolved Hamiltonian, 
\begin{equation}
  \Hfinal \equiv \HO(\infty)\,,
\end{equation}
in the reference state (which is assumed to be normalized):
\begin{equation}
  E = \dmatrixe{\Phi}{\Hfinal}\,.
\end{equation}
We can introduce a \emph{unitary} variation, which we are free to apply 
either to the reference state\,,
\begin{equation}
  \ket{\Phi}\rightarrow e^{\ZO}\ket{\Phi}\,,\quad\ZZO=-\ZO\,,
\end{equation}
or, equivalently, to the Hamiltonian:
\begin{equation}
  e^{\ZZO}\Hfinal e^{\ZO} = e^{-\ZO}\Hfinal e^{\ZO}\,.
\end{equation}
The variation of the energy is
\begin{align}
  \delta E = \dmatrixe{\Phi}{e^{-\ZO}(\Hfinal - E) e^{\ZO}} 
           = \dmatrixe{\Phi}{\Hfinal-E} + \dmatrixe{\Phi}{\comm{\Hfinal-E}{\ZO}} + O(||\ZO||^2)\,,
\end{align}
where $||\cdot||$ is an appropriate operator norm. The first term obviously 
vanishes, as does the commutator of $\ZO$ with the energy, because the latter
is a mere number. Thus, the energy is stationary if
\begin{equation}
  \delta E = \dmatrixe{\Phi}{\comm{\Hfinal}{\ZO}} = 0\,. 
\end{equation}
Expanding 
\begin{equation}
  \ZO =\sum_{ph} Z_{ph}\nord{\aaO_{p}\aO_{h}} + \frac{1}{4}\sum_{pp'hh'} Z_{pp'hh'}\nord{\aaO_{p}\aaO_{p'}\aO_{h'}\aO_{h}} 
        + \text{H.c.} + \ldots\,,
\end{equation}
and using the independence of the expansion coefficients (save for the unitarity
conditions), we obtain the system of equations
\begin{align}
  \dmatrixe{\Phi}{\comm{\Hfinal}{\nord{\aaO_{p}\aO_{h}}}} &= 0 \,,\\
  \dmatrixe{\Phi}{\comm{\Hfinal}{\nord{\aaO_{h}\aO_{p}}}} &= 0 \,,\\
  \dmatrixe{\Phi}{\comm{\Hfinal}{\nord{\aaO_{p}\aaO_{p'}\aO_{h'}\aO_{h}}}} &= 0 \,,\\
  \dmatrixe{\Phi}{\comm{\Hfinal}{\nord{\aaO_{h}\aaO_{h'}\aO_{p'}\aO_{p}}}} &= 0 \,,\\
  \ldots\notag
\end{align}
which are special cases of the so-called \emph{irreducible Brillouin conditions (IBCs)}
\cite{Mukherjee:2001uq,Kutzelnigg:2002kx,Kutzelnigg:2004vn,Kutzelnigg:2004ys}. 
Writing out the commutator in the first equation, we obtain
\begin{equation}
  \dmatrixe{\Phi}{\comm{\Hfinal}{\nord{\aaO_{p}\aO_{h}}}} 
    = \dmatrixe{\Phi}{\Hfinal\nord{\aaO_{p}\aO_{h}}} - \dmatrixe{\Phi}{\nord{\aaO_{p}\aO_{h}}\Hfinal}
    = \dmatrixe{\Phi}{\Hfinal\nord{\aaO_{p}\aO_{h}}} 
    = 0\,,
\end{equation}
where the second term vanishes because it is proportional to $n_p \nn_h = 0$.
The remaining equations can be evaluated analogously, and we find that the
energy is stationary if the IMSRG evolved Hamiltonian $\Hfinal$ no longer 
couples the reference state and its particle-hole excitations, as discussed 
above. However, we need to stress that the IMSRG is \emph{not} variational,
because any truncation of the flow equation breaks the unitary equivalence
of the initial and evolved Hamiltonians. Thus, the final IMSRG(2) energy
cannot be understood as an upper bound for the true eigenvalue in a strict
sense, although the qualitative behavior might suggest so in numerical
applications.

\subsection{\label{sec:imsrg_generator}Choice of Generator}
In the previous section, we have identified the matrix elements of the
Hamiltonian that couple the ground state to excitations, and collected
them into a definition of the off-diagonal Hamiltonian that we want to 
suppress with an IMSRG evolution. While we have decided on a decoupling 
``pattern'' in this way, we have a tremendous amount of freedom in 
implementing this decoupling. As long as we use the same off-diagonal
Hamiltonian, many different types of generators will drive the Hamiltonian
to the desired shape in the limit $s\to\infty$, and some of these generators
stand out when it comes to numerical efficiency \cite{Hergert:2016jk}. 

%
%
\subsubsection{\label{sec:imsrg_generator_general}Construction of Generators for Single-Reference Applications}
A wide range of suitable generators for the single-reference case
is covered by the ansatz
\begin{equation}
  \eta= \sum_{ph} \eta_{ph}\nord{\aaO_{p}\aO_{h}}
        +\frac{1}{4}\sum_{pp'hh'} \eta_{pp'hh'}\nord{\aaO_{p}\aaO_{p'}\aO_{h'}\aO_{h}} - \text{H.c.}\,,
\end{equation}
constructing the one- and two-body matrix elements directly from those
of the offdiagonal Hamiltonian and a tensor $G$ that ensures the 
anti-Hermiticity of $\eta$:
\begin{align}
  \eta_{ph}     &\equiv G_{ph} f_{ph}\,,\\
  \eta_{pp'hh'} &\equiv G_{pp'hh'} \Gamma_{pp'hh'}\,.
\end{align}
To identify possible options for $G$, we consider the flow
equations in perturbation theory (see Ref.~\cite{Hergert:2016jk}
for a detailed discussion). We assume a Hartree-Fock reference state,
and partition the Hamiltonian as
\begin{equation}
  \HO = \HO_0 + \HO_I\,,
\end{equation}
with 
\begin{align}
  \HO_0 &\equiv E + \sum_{i} f_{ii}\nord{\aaO_{i}\aO_{i}}
                 + \frac{1}{4} \sum_{ij} \Gamma_{ijij}\nord{\aaO_{i}\aaO_{j}\aO_{j}\aO_{i}}\,,\\
  \HO_I &\equiv \sum_{ij}^{i\neq j} f_{ij}\nord{\aaO_{i}\aO_{j}} 
                 + \frac{1}{4} \sum_{ijkl}^{(ij)\neq (kl)} \Gamma_{ijkl}\nord{\aaO_{i}\aaO_{j}\aO_{l}\aO_{k}}\,.
\end{align}
We introduce a power counting in terms of the auxiliary parameter
$g$, and count the diagonal Hamiltonian $\HO_0$ as unperturbed ($\OC(1)$)
, while the perturbation $\HO_I$ is counted as $\OC(g)$. In the space of 
up to $2p2h$ excitations, our partitioning is a second-quantized form of 
the one used by Epstein and Nesbet \cite{Epstein:1926fp,Nesbet:1955lq}.

We now note that the one-body piece of the initial Hamiltonian is diagonal in 
the HF orbitals, which implies
\begin{equation}
  f_{ph}=0,\quad \eta_{ph}=0\,.
\end{equation} 
Inspecting the Eq.~\eqref{eq:imsrg2_m1b}, we see that corrections 
to $f$ that are induced by the flow are at least of order $\OC(g^2)$, 
because no diagonal matrix elements of $\Gamma$ appear:
\begin{align}
  \left.\totd{}{s}f_{ij}\right|_{s=0} &=
    \frac{1}{2}\sum_{abc}
    \left(\eta_{iabc}\Gamma_{bcja}-\Gamma_{iabc}\eta_{bcja}\right)\left(n_{a}\bar{n}_{b}\bar{n}_{c}+\bar{n}_{a}n_{b}n_{c}\right)=\OC(g^2)\,.\label{eq:flow1b_pert}
\end{align}
Using this knowledge, the two-body flow equation for the $pphh$ matrix elements 
of the off-diagonal Hamiltonian reads
\begin{align}
  \totd{}{s}\Gamma_{pp'hh'}&=  
  -\left(f_{pp}+f_{p'p'}-f_{hh}-f_{h'h'}\right)\eta_{pp'hh'}
  -\left(\Gamma_{hh'hh'}+\Gamma_{pp'pp'}\right)\eta_{pp'hh'}
  \notag\\
  &\hphantom{=}
    +\left(\Gamma_{p'h'p'h'}+\Gamma_{phph}+\Gamma_{ph'ph'}+\Gamma_{p'hp'h}
     \right)\eta_{pp'hh'} + \OC(g^2)
  \notag\\
  &=-\Delta_{pp'hh'}\eta_{pp'hh'} + \OC(g^2)\,.\label{eq:flow2b_pert}
\end{align}
Note that $\etaO$, which is of order $\OC(g)$, is multiplied
by unperturbed, diagonal matrix elements of the Hamiltonian 
in the leading term. Because of this restriction, the sums
in the particle-particle and hole-hole ladder terms (line 2 of Eq.~\eqref{eq:imsrg2_m2b})
collapse, and the pre-factors $\tfrac{1}{2}$ are canceled by 
factors $2$ from the unrestricted summation over indices, e.g.,
\begin{equation}
  \frac{1}{2}\sum_{h_1h_2}\eta_{pp'h_1h_2}\Gamma_{h_1h_2hh'}(1-n_{h_1}-n_{h_2})
  =-\frac{1}{2}\eta_{pp'hh'}\Gamma_{hh'hh'}-\frac{1}{2}\eta_{pp'h'h}\Gamma_{h'hhh'}
  =-\eta_{pp'hh'}\Gamma_{hh'hh'}\,.
\end{equation}
In equation \eqref{eq:flow2b_pert}, we have introduced the quantity 
\begin{align}
  \Delta_{pp'hh'}&\equiv
    f_{pp}+f_{p'p'}-f_{hh}-f_{h'h'} + \Gamma_{hh'hh'}+\Gamma_{pp'pp'}
    -\Gamma_{phph}-\Gamma_{p'h'p'h'}-\Gamma_{ph'ph'}-\Gamma_{p'hp'h} \notag\\
    &=\dmatrixe{\Phi}{\nord{\aaO_{h}\aaO_{h'}\aO_{p'}\aO_{p}}\HO\nord{\aaO_{p}\aaO_{p'}\aO_{h'}\aO_{h}}}-\dmatrixe{\Phi}{\HO}\notag\\
    &=\dmatrixe{\Phi}{\nord{\aaO_{h}\aaO_{h'}\aO_{p'}\aO_{p}}\HO_0\nord{\aaO_{p}\aaO_{p'}\aO_{h'}\aO_{h}}}-\dmatrixe{\Phi}{\HO_0}\,,
    \label{eq:def_epsteinnesbet_2b}
\end{align}
i.e., the unperturbed energy difference between the two states
that are coupled by the matrix element $\Gamma_{pp'hh'}$, namely
the reference state $\ket{\Phi}$ and the excited state $\nord{\aaO_{p}\aaO_{p'}\aO_{h'}\aO_{h}}\ket{\Phi}$.
Since it is expressed in terms of diagonal matrix elements, $\Delta_{pp'hh'}$
would appear in precisely this form in appropriate energy denominators
of Epstein-Nesbet perturbation theory. 

Plugging our ansatz for $\eta$ into equation \eqref{eq:flow2b_pert}, we obtain
\begin{equation}
  \totd{}{s}\Gamma_{pp'hh'} 
  =-\Delta_{pp'hh'}G_{pp'hh'}\Gamma_{pp'hh'} + \OC(g^2)\,,
\end{equation}
Neglecting $\OC(g^2)$ terms in the flow equations, the one-body part of $\HO$ 
remains unchanged, and assuming that $G$ itself is independent of $s$
at order $\OC(g)$, we can integrate equation \eqref{eq:flow2b_pert}:
\begin{equation}\label{eq:Gamma_pert}
  \Gamma_{pp'hh'}(s) = \Gamma_{pp'hh'}(0) e^{-\Delta_{pp'hh'}G_{pp'hh'} s}\,.
\end{equation}
Clearly, the offdiagonal matrix elements of the Hamiltonian will be
suppressed for $s\to\infty$ if the product $\Delta_{pp'hh'} G_{pp'hh'}$ 
is positive. $G_{pp'hh'}$ also allows us
to control the details of this suppression, e.g., the decay scales.
To avoid misconceptions, we stress that we do not impose perturbative
truncations in practical applications, and treat all matrix elements 
and derived quantities, including the $\Delta_{pp'hh'}$, as 
$s$-dependent.

%
%
\subsubsection{\label{sec:imsrg_generator_white}White's Generators}
A generator that is particularly powerful in numerical applications is inspired
by the work of White on canonical transformation theory in quantum chemistry 
\cite{White:2002fk,Tsukiyama:2011uq,Hergert:2016jk}. In the language we have
set up above, it uses $G_{pp'hh'}$ to \emph{remove} the scale dependence
of the IMSRG flow. This so-called White generator is defined as
\begin{align}
  \etaO^\text{W}(s)
  &\equiv\sum_{ph}\frac{f_{ph}(s)}{\Delta_{ph}(s)}\nord{\aaO_{p}\aO_{h}}
  +\frac{1}{4}\sum_{pp'hh'}\frac{\Gamma_{pp'hh'}(s)}{\Delta_{pp'hh'}(s)}
    \nord{\aaO_{p}\aaO_{p'}\aO_{h'}\aO_{h}}-\;\text{H.c.}\label{eq:eta_white}\,,
\end{align}
where the Epstein-Nesbet denominators use the energy differences defined in 
equations \eqref{eq:def_epsteinnesbet_2b} and \eqref{eq:def_epsteinnesbet_1b}.

For the White generator, we find
\begin{equation}
  \Gamma_{pp'hh'}(s) = \Gamma_{pp'hh'}(0) e^{-s}\,,
\end{equation}
i.e., \emph{all} off-diagonal matrix 
elements are suppressed \emph{simultaneously} with a decay scale identical (or close to) 1 
\cite{Hergert:2016jk}. 
While this means that $\eta^\text{W}$ does \emph{not} generate a proper RG
flow, this is inconsequential if we are only interested in the final Hamiltonian 
$\HO(\infty)$, because all unitary transformations which suppress $\HO_{od}$ 
must be equivalent up to truncation effects \cite{Hergert:2016jk}.

A benefit of the White generator is that its matrix elements are defined 
as ratios of energies, and therefore the Hamiltonian only contributes linearly 
to the magnitude of the right-hand side of the flow equations \eqref{eq:imsrg2_m0b}--
\eqref{eq:imsrg2_m2b}. This leads to a significant reduction of the 
ODE system's stiffness compared to the other generators discussed here or in
Ref.~\cite{Hergert:2016jk}, and greatly reduces the numerical effort for the
ODE solver. However, the dependence of $\etaO^\text{W}$ on energy denominators
can also be a drawback if $\Delta_{ph}$ and/or $\Delta_{pp'hh'}$ become
small, which would cause the generator's matrix elements to diverge. This can be
mitigated by using an alternative ansatz that is also inspired by White's work 
\cite{White:2002fk}:
\begin{align}
  \etaO^\text{W'}(s)
  &\equiv\frac{1}{2}\sum_{ph}\arctan\frac{2 f_{ph}(s)}{\Delta_{ph}(s)}\nord{\aaO_{p}\aO_{h}}
  +\frac{1}{8}\sum_{pp'hh'}\arctan\frac{2\Gamma_{pp'hh'}(s)}{\Delta_{pp'hh'}(s)}
    \nord{\aaO_{p}\aaO_{p'}\aO_{h'}\aO_{h}}-\;\text{H.c.}\label{eq:eta_whiteatan}\,.
\end{align}
This form emphasizes that the unitary transformation can be thought of as an
abstract rotation of the Hamiltonian. The matrix elements of $\eta^\text{W'}$
are regularized by the $\arctan$ function, and explicitly limited to the 
interval $]-\tfrac{\pi}{4},\tfrac{\pi}{4}[$. Expanding the function for 
small arguments, we recover our initial ansatz for the White generator, 
equation \eqref{eq:eta_white}. 

%
%
\subsubsection{\label{sec:imsrg_generator_imtime}The Imaginary-Time Generator}
Using $G_{pp'hh'}$ to ensure that the energy denominator is always
positive, we obtain the so-called \emph{imaginary-time generator}
\cite{Morris:2015ve,Hergert:2014vn,Hergert:2016jk},
which is inspired by imaginary-time evolution techniques that are 
frequently used in Quantum Monte Carlo methods, for instance (see
chapter 9, \cite{Carlson:2015lq} and references therein). Explicitly 
indicating the flow parameter dependence of all quantities, we
define
\begin{align}
  \etaO^\text{IT}(s)&\equiv\sum_{ph} \sgn\!\left(\Delta_{ph}(s)\right) f_{ph}(s)\nord{\aaO_{p}\aO_{h}}\notag\\
  &\hphantom{=}
   +\frac{1}{4}\sum_{pp'hh'}\sgn\!\left(\Delta_{pp'hh'}(s)\right)\Gamma_{pp'hh'}(s)
    \nord{\aaO_{p}\aaO_{p'}\aO_{h'}\aO_{h}}-\text{H.c.}\,,\label{eq:eta_imtime}
\end{align}
where
\begin{align}
  \Delta_{ph}&\equiv
    f_{pp}-f_{hh} + \Gamma_{phph} 
    =\dmatrixe{\Phi}{\nord{\aaO_{h}\aO_{p}}\HO\nord{\aaO_{p}\aO_{h}}}-\dmatrixe{\Phi}{\HO}\,.\label{eq:def_epsteinnesbet_1b}
\end{align}

For this generator, the perturbative analysis of the offdiagonal 
two-body matrix elements yields
\begin{equation}
  \Gamma_{pp'hh'}(s) = \Gamma_{pp'hh'}(0) e^{-|\Delta_{pp'hh'}|s}\,,
\end{equation}
ensuring that they are driven to zero by the evolution. We also
note that the energy difference $\Delta_{pp'hh'}$ controls
the scales of the decay. Matrix elements between states with large 
energy differences are suppressed more rapidly than those which
couple states that are close in energy. This means that $\eta^\text{IT}$ 
generates a proper renormalization group flow \cite{Kehrein:2006kx,Hergert:2016jk}.

%
%
\subsubsection{\label{sec:imsrg_generator_wegner}Wegner's Generator}
Last but not least, we want to discuss Wegner's original ansatz 
\cite{Wegner:1994dk}, which we have used in the free-space SRG applications 
in Sec.~\ref{sec:srg}:
\begin{equation}\label{eq:eta_wegner}
  \etaO^\text{WE}(s) \equiv \comm{\HO_d(s)}{\HO_{od}(s)}\,.
\end{equation}
Truncating $\HO_d(s)$ and $\HO_{od}(s)$ at the two-body level and using
the commutators from the appendix, it is straightforward to derive the 
one- and two-body matrix elements of $\etaO(s)$; the operator has no
zero-body component because of its anti-Hermiticity. We obtain
\begin{align}
  \eta_{ij} &= 
  \sum_{a}(1-P_{ij})f^d_{ia}f^{od}_{aj} +\sum_{ab}(n_a-n_b)(f^d_{ab}\Gamma^{od}_{biaj}-f^{od}_{ab}\Gamma^d_{biaj}) \notag\\ 
  &\quad +\frac{1}{2} \sum_{abc}(n_an_b\bar{n}_c+\bar{n}_a\bar{n}_bn_c) (1-P_{ij})\Gamma^d_{ciab}\Gamma^{od}_{abcj}
  \label{eq:eta_wegner_m1b}\,,\\[5pt]
  \eta_{ijkl}&= 
  \sum_{a}\left\{ 
    (1-P_{ij})(f^d_{ia}\Gamma^{od}_{ajkl}-f^{od}_{ia}\Gamma^{d}_{ajkl} )
    -(1-P_{kl})(f^{d}_{ak}\Gamma^{od}_{ijal}-f^{od}_{ak}\Gamma^d_{ijal} )
    \right\}\notag \\
  &\quad+ \frac{1}{2}\sum_{ab}(1-n_a-n_b)(\Gamma^d_{ijab}\Gamma^{od}_{abkl}-\Gamma^{od}_{ijab}\Gamma^d_{abkl})\notag\\
  &\quad+ \sum_{ab}(n_a-n_b) (1-P_{ij})(1-P_{kl})\Gamma^{d}_{aibk}\Gamma^{od}_{bjal}
    \label{eq:eta_wegner_m2b}\,.
\end{align}
Structurally, Eqs.~\eqref{eq:eta_wegner_m1b} and \eqref{eq:eta_wegner_m2b} 
are identical to the flow IMSRG(2) equations except for signs stemming from 
the anti-Hermiticity of the generator. 

Superficially, the Wegner generator is quite different from the imaginary-time 
and White generators, but we can uncover commonalities by carrying out a 
perturbative analysis along the lines of the previous sections. For a HF 
Slater determinant, the one-body part of the off-diagonal Hamiltonian vanishes 
at $s=0$, and corrections that are induced by the flow start at $\OC(g^2)$
(see Eq. ~\eqref{eq:flow1b_pert}). This means that the one-body part of
the Wegner generator has the form
\begin{equation}
  \eta_{ij} = \frac{1}{2} 
      \sum_{abc}(n_an_b\bar{n}_c+\bar{n}_a\bar{n}_bn_c) (1-P_{ij})\Gamma^d_{ciab}\Gamma^{od}_{abcj}
      + \OC(g^2)\,.
\end{equation}
In the minimal decoupling scheme, the matrix elements appearing here are
counted as follows:
\begin{equation}
  \Gamma^d_{ijij} = -\Gamma^d_{jiij} = \OC(1), \quad \Gamma^d_{ijkl} = \OC(g)\; \text{for}\; (ij)\neq(kl), 
  \quad \Gamma^{od}_{ijkl} = \OC(g)\,.
\end{equation}
To obtain a $\OC(g)$ contribution to the one-body generator, we need
either $a=c$ and $b=i$, or $a=i$ and $b=c$, but then the occupation number
factor becomes
\begin{equation}
  n_{i}n_{c}\nn_c + \nn_{i} \nn_{c} n_c = 0\,.
\end{equation}
This implies that the leading contributions to $\eta_{ij}$ are of order $\OC(g^2)$.

A similar analysis for the two-body part of $\etaO^\text{WE}$ (see problem
\ref{problem:wegner_pert}) shows that
\begin{align}
  \eta_{ijkl}
    &=\left(f^d_{ii} + f^d_{jj} - f^d_{kk} + f^d_{ll} + (1-n_i-n_k)\Gamma^d_{ijij} - (1-n_k-n_l)\Gamma^d_{klkl} 
    \right.\notag\\
    &\hphantom{=}\left.\quad
    + (n_i-n_k) \Gamma^d_{ikik} + (n_j-n_k) \Gamma^d_{jkjk} + (n_i-n_l) \Gamma^d_{ilil} + (n_j-n_l) \Gamma^d_{jljl}
  \right)\Gamma^{od}_{ijkl}
  +\OC(g^2)\,.
  \label{eq:wegner2b_pert}
\end{align}
Since $\Gamma^{od}_{ijkl}$ is restricted to $pphh$ matrix elements, we immediately
obtain
\begin{equation}
  \eta_{pp'hh'} = \Delta_{pp'hh'}\Gamma_{pp'hh'}\,,
\end{equation}
and plugging this into Eq.~\eqref{eq:flow2b_pert}, we have
\begin{equation}
    \totd{}{s}\Gamma_{pp'hh'} 
  =-\left(\Delta_{pp'hh'}\right)^2G_{pp'hh'}\Gamma_{pp'hh'} + \OC(g^2)\,.
\end{equation}
Neglecting the $\OC(g^2)$ terms, we can integrate the flow equation
and find that the Wegner generator suppresses off-diagonal matrix
elements with a Gaussian exponential function,
\begin{equation}
  \Gamma_{pp'hh'}(s) = \Gamma_{pp'hh'}(0) e^{-(\Delta_{pp'hh'})^2s}\,,
\end{equation}
and therefore generates a proper RG flow. Note that this result matches 
our findings for the SRG flows of the 
$2\times2$ matrix toy model (Sec.~\ref{sec:srg_toy}) and the pairing 
Hamiltonian (Sec.~\ref{sec:srg_pairing}), which were using matrix 
versions of the Wegner generator.

In numerical applications, Wegner generators are less efficient than 
our other choices. The cost for constructing $\etaO^\text{WE}$ is of 
order $\OC(N^4_pN^2_h)$, compared to $\OC(N^2_pN^2_h)$ for the White
and imaginary-time generators. More importantly, $\etaO^\text{WE}$ 
generates very stiff flow equations because the RHS terms are cubic in 
the Hamiltonian. This forces us to use ODE solvers that are appropriate
for stiff systems, which have higher storage requirements and need more
computing time than solvers for non-stiff systems, which can be used
for imaginary-time and White IMSRG flows.


\subsection{\label{sec:imsrg_implementation}Implementation}
Now that we have all the necessary ingredients, it is time to discuss the
numerical implementation of IMSRG flows. As an example, we use a Python
code that is designed for solving the pairing Hamiltonian (see Sec.~
\ref{sec:imsrg_pairing}) but easily adaptable to other problems. The
IMSRG solver and tools for visualizing the flow can be found at
\url{https://github.com/ManyBodyPhysics/LectureNotesPhysics/tree/master/Programs/Chapter10-programs/python/imsrg_pairing}.

\subsubsection*{Basis and Matrix Element Handling}
The code \texttt{imsrg\_pairing.py} uses \texttt{NumPy} arrays to store
the one- and two-body matrix elements of the normal-ordered operators.
The underlying one- and two-nucleon states are indexed as integers and
pairs of integers, respectively. The complete lists of states are stored
in the variables \texttt{bas1B} and \texttt{bas2B}, respectively. We
also define an additional set of two-nucleon states in the list \texttt{basph2B},
for reasons that will be explained shortly. The lists are all initialized
at the beginning of the program's main routine:
\begin{lstlisting}
def main():
...
    particles  = 4

    # setup shared data
    dim1B     = 8

    # this defines the reference state
    holes     = [0,1,2,3]
    particles = [4,5,6,7]

    # basis definitions
    bas1B     = range(dim1B)
    bas2B     = construct_basis_2B(holes, particles)
    basph2B   = construct_basis_ph2B(holes, particles)

    idx2B     = construct_index_2B(bas2B)
    idxph2B   = construct_index_2B(basph2B)

  ...
\end{lstlisting}
Aside from a distinction between occupied (hole) and unoccupied (particle)
single-particle states, we do not impose any constraints on the one- and
two-nucleon bases. All permutations of two-body matrix elements $\Gamma_{ijkl}$
are stored explicitly, and so are the vanishing matrix elements like 
$\Gamma_{iikl}$ or $\Gamma_{ijkk}$ that are forbidden by the Pauli principle.
The benefit of using this ``naive'' basis construction is that we do not need
to worry whether certain combinations of quantum numbers are allowed or 
forbidden by symmetries. This is relevant for flow equation terms that cannot 
be expressed as matrix products, i.e., the contractions of one- and two-body 
operators that appear in Eqs.~\eqref{eq:imsrg2_m1b} and \eqref{eq:imsrg2_m2b}. 
To implement such terms, we also need to translate pairs of single-particle 
indices into collective two-nucleon state indices and back, which is achieved 
with the help of the lookup arrays \texttt{idx2B} and \texttt{idxph2B}.

The state and lookup lists \texttt{basph2B} and \texttt{idxph2B} are used
to work with matrices in the so-called \emph{particle-hole representation}. 
This representation allows us to write the particle-hole contributions in
the third line of Eq.~\eqref{eq:imsrg2_m2b} as matrix products. Superficially,
these terms look like they require explicit loop summations:
\begin{equation}\label{eq:imsrg2_m2b_ph}
  \totd{}{s}\Gamma^{(ph)}_{ijkl}=(1-P_{ij})(1-P_{kl})\sum_{ab}(n_a-n_b)\eta_{aibk}\Gamma_{bjal}\,.
\end{equation}
Since we are working with a Slater determinant reference state and therefore
able to distinguish particle and hole states in the single-particle basis,
we can also define hole creation and annihilation operators (see, e.g., \cite{Suhonen:2007wo}):
\begin{equation}
  \hhO_i \equiv \aO_i\,,\quad \hO_i \equiv \aaO_i\,, 
\end{equation}
that satisfy the same anticommutation relations as the regular creators
and annihilators. In addition, we also have
\begin{align}
  \acomm{\hhO_i}{\aO_j}  &= \acomm{\hO_i}{\aaO_j} = 0\,,\notag\\
  \acomm{\hhO_i}{\aaO_j} &= \acomm{\hO_i}{\aO_j} = \delta_{ij}\,.
\end{align}
Using Eq.~\eqref{eq:nord_acomm} for the hole operators, we can rewrite a 
generic normal-ordered two-body operator in the following way:
\begin{equation}
  \AO \equiv \frac{1}{4}\sum_{ijkl}A_{ijkl}\nord{\aaO_{i}\aaO_{j}\aO_{l}\aO_{k}}
    = \frac{1}{4}\sum_{ijkl}A_{ijkl}\nord{\aaO_{i}\hO_{j}\hhO_{l}\aO_{k}}
    = -\frac{1}{4}\sum_{ijkl}A_{ijkl}\nord{\aaO_{i}\hhO_{l}\hO_{j}\aO_{k}}\,.
\end{equation}
We can also define $\AO$ directly in the particle-hole representation, 
\begin{equation}
   \AO \equiv \frac{1}{4}\sum_{ijkl}A_{i\overline{j}k\overline{l}}\nord{\aaO_{i}\hhO_{j}\hO_{l}\aO_{k}}\,,
\end{equation}
where we have indicated the hole states by lines over the indices. Thus,
the operator's matrix elements in the regular particle representation and 
the particle-hole representation are related by the following expression:
\begin{equation}
  A_{i\overline{j}k\overline{l}} = - A_{ilkj}\,.
\end{equation}
We see that the switch to particle-hole representation is achieved 
by a simple rearrangement of matrix elements. The situation is more 
complicated if one works with angular-momentum coupled states, because 
then the angular momenta must be recoupled in a different order, giving 
rise to a so-called \emph{Pandya transformation} 
\cite{Pandya:1956zv,Suhonen:2007wo,Kuo:1981fk}. 

Using particle-hole matrix elements, the right-hand side of Eq.~\eqref{eq:imsrg2_m2b_ph} 
can be written as
\begin{equation}
  \sum_{ab}(n_a-n_b)\eta_{iabk}\Gamma_{bjla}
  = \sum_{ab}(n_a-n_b) \eta_{i\overline{k}b\overline{a}}\Gamma_{b\overline{a}l\overline{j}}
  \equiv M_{i\overline{k}l\overline{j}}\,,
\end{equation}
which makes it possible to evaluate the term using matrix product routines,
treating $(n_a-n_b)$ as a diagonal matrix. The resulting product matrix $M$
can then be transformed back into the particle representation, where it will 
be completely antisymmetrized by the permutation symbols in Eq.~\eqref{eq:imsrg2_m2b_ph}. 
The necessary transformations between the particle and particle-hole 
representations are implemented in the routines \texttt{ph\_transform\_2B}
and \texttt{inverse\_ph\_transform\_2B}, respectively.

\subsubsection*{Reference States}
After the basis initialization, we need to define the reference Slater
determinant \ket{\Phi} for the subsequent normal ordering. To this end, 
we simply create lists of the hole (occupied) and particle (unoccupied)
single-particle states:
\begin{lstlisting}
def main():
...

    particles  = 4

    # setup shared data
    dim1B     = 8

    # this defines the reference state
    holes     = [0,1,2,3]
    particles = [4,5,6,7]

    # basis definitions
...

    # occupation number matrices
    occ1B     = construct_occupation_1B(bas1B, holes, particles)
    occA_2B   = construct_occupationA_2B(bas2B, occ1B)
    occB_2B   = construct_occupationB_2B(bas2B, occ1B)
    occC_2B   = construct_occupationC_2B(bas2B, occ1B)

    occphA_2B = construct_occupationA_2B(basph2B, occ1B)

...
\end{lstlisting}
In addition to the elementary lists, we set up diagonal matrices for
the various occupation number factors that appear in the IMSRG(2)
flow equations \eqref{eq:imsrg2_m0b}--\eqref{eq:imsrg2_m2b},
i.e., $(n_a - n_b)$, which is required both in particle and particle-hole
representation (\texttt{construct\_occupationA\_2B}), $(1-n_a-n_b)$
(\texttt{construct\_occupationB\_2B}), and $n_a n_b$ (\texttt{construct\_occupationC\_2B}).
The latter appears when we rewrite the last occupation factor in 
the one-body flow equation \eqref{eq:imsrg2_m1b}:
\begin{equation}
  n_a n_b \nn_c + \nn_a \nn_b n_c = n_a n_b - n_a n_b n_c + (1 - n_a - n_b + n_a n_b) n_c
  = n_a n_b + (1-n_a - n_b)n_c\,.
\end{equation}

\subsubsection*{Sharing Data and Settings}
Since the routines for normal ordering the Hamiltonian and calculating 
the generator and derivatives need to access the bases, index lookups,
and occupation number matrices, we store this shared data in a Python
dictionary:

\begin{lstlisting}
...
    user_data  = {
      "dim1B":      dim1B, 
      "holes":      holes,
      "particles":  particles,
      "bas1B":      bas1B,
      "bas2B":      bas2B,
      "basph2B":    basph2B,
      "idx2B":      idx2B,
      "idxph2B":    idxph2B,
      "occ1B":      occ1B,
      "occA_2B":    occA_2B,
      "occB_2B":    occB_2B,
      "occC_2B":    occC_2B,
      "occphA_2B":  occphA_2B,

      "eta_norm":   0.0,                # variables for sharing data between ODE solver
      "dE":         0.0,                # and main routine
      
      "calc_eta":   eta_white,          # specify the generator (function object)
      "calc_rhs":   flow_imsrg2         # specify the right-hand side and truncation
    }
...
\end{lstlisting}

Rather than passing all of the data structures as separate parameters, we
can then pass \texttt{user\_data} as a parameter. By passing the data as a
dictionary, we can also avoid the creation of global variables, and make
it easier to reuse individual routines in other projects.

We also want to direct our readers' attention to the last two entries of
the dictionary. These are function objects that are used to define which
generator and flow equation routines the ODE solver will call (see below).
Through this abstraction, users can easily add additional generators, or
implement different truncations of the flow equations. The current version
of \texttt{imsrg\_pairing.py} implements all of the generators discussed
in Sec.~\ref{sec:imsrg_generator}, and the standard IMSRG(2) truncation.

\subsubsection*{Normal Ordering}
The next task of the main routine is the normal ordering of the initial
Hamiltonian: 
\begin{lstlisting}
def main():
...
    # set up initial Hamiltonian
    H1B, H2B = pairing_hamiltonian(delta, g, user_data)

    E, f, Gamma = normal_order(H1B, H2B, user_data) 
...
\end{lstlisting}
In order to facilitate the reuse of our code, we proceed in two
steps: First, we set up the Hamiltonian in the vacuum, in this case the
pairing Hamiltonian \eqref{eq:def_pairing_hamiltonian} with single-particle
spacing $\delta$ and pairing strength $g$. The one- and two-body matrix
elements are then passed to a generic routine that performs the normal ordering:
\begin{lstlisting}
def normal_order(H1B, H2B, user_data):
    bas1B     = user_data["bas1B"]
    bas2B     = user_data["bas2B"]
    idx2B     = user_data["idx2B"]
    particles = user_data["particles"]
    holes     = user_data["holes"]

    # 0B part
    E = 0.0
    for i in holes:
      E += H1B[i,i]

    for i in holes:
      for j in holes:
        E += 0.5*H2B[idx2B[(i,j)],idx2B[(i,j)]]  

    # 1B part
    f = H1B
    for i in bas1B:
      for j in bas1B:
        for h in holes:
          f[i,j] += H2B[idx2B[(i,h)],idx2B[(j,h)]]  

    # 2B part
    Gamma = H2B

    return E, f, Gamma
\end{lstlisting}

\subsubsection*{Integration}
Once the initial Hamiltonian is set up, we use the \texttt{SciPy}
\texttt{ode} class to integrate the flow equations, so that we can
switch between non-stiff and stiff solvers, and give the user as 
much control over the solver as possible.

The ODE solver calls the following derivative wrapper function:
\begin{lstlisting}
def derivative_wrapper(t, y, user_data):
...
    calc_eta  = user_data["calc_eta"]
    calc_rhs  = user_data["calc_rhs"]

    # extract operator pieces from solution vector
    E, f, Gamma = get_operator_from_y(y, dim1B, dim2B)


    # calculate the generator
    eta1B, eta2B = calc_eta(f, Gamma, user_data)

    # calculate the right-hand side
    dE, df, dGamma = calc_rhs(eta1B, eta2B, f, Gamma, user_data)

    # convert derivatives into linear array
    dy   = np.append([dE], np.append(reshape(df, -1), reshape(dGamma, -1)))

    # share data
    user_data["dE"]       = dE
    user_data["eta_norm"] = np.linalg.norm(eta1B,ord='fro')+np.linalg.norm(eta2B,ord='fro')
    
    return dy
\end{lstlisting}
This routine is very similar to the ones we used in the SRG
codes in Secs.~\ref{sec:srg_pairing_implementation} and \ref{sec:srg_nn_flow}. It extracts
$E(s), f(s),$ and $\Gamma(s)$ from the solution vector, and calls 
appropriate routines to construct the generator and the derivatives.
This is where the function object entries of the \texttt{user\_data}
dictionary come into play: We use them as an abstract interface
to call the routines we assigned to the dictionary in our \texttt{main} 
routine. This is much more elegant than selecting the generator and
flow equation truncation scheme via \texttt{if...elif} clauses. Most 
importantly, we do not need to modify the derivative routine at all 
if we want to add new generators and truncation schemes, but only
need to assign the new functions to \texttt{calc\_eta} and 
\texttt{calc\_rhs}.

In the ODE loop, we do not require output at an externally chosen
value of $s$, but check and process the intermediate solution after
each accepted (not attempted!) internal step of the solver. This is
achieved by setting the option \texttt{step=True}. We extract 
$E(s), f(s),$ and $\Gamma(s)$, and use it to calculate diagnostic
quantities like the second- and third-order energy corrections, 
$\Delta E^{(2)}(s)$ and $\Delta E^{(3)}(s)$, as well as the norms 
of $f_{od}(s)$ and $\Gamma_{od}(s)$:  
\begin{lstlisting}
def main():
...
    while solver.successful() and solver.t < sfinal:
        ys = solver.integrate(sfinal, step=True)
        
        dim2B = dim1B*dim1B
        E, f, Gamma = get_operator_from_y(ys, dim1B, dim2B)

        DE2 = calc_mbpt2(f, Gamma, user_data)
        DE3 = calc_mbpt3(f, Gamma, user_data)

        norm_fod     = calc_fod_norm(f, user_data)
        norm_Gammaod = calc_Gammaod_norm(Gamma, user_data)

        print("%8.5f %14.8f   %14.8f   %14.8f   %14.8f   %14.8f   %14.8f   %14.8f   %14.8f"%(
          solver.t, E , DE2, DE3, E+DE2+DE3, user_data["dE"], user_data["eta_norm"], norm_fod, norm_Gammaod))

        if abs(DE2/E) < 10e-8: break

\end{lstlisting}
As discussed in earlier sections, the off-diagonal matrix elements 
create $1p1h$ and $2p2h$ admixtures to the reference state wave function 
which give rise to the energy corrections. These admixtures are suppressed
as we decouple, hence the size of the energy corrections must decrease.
In the limit $s\to\infty$, we expect them to be completely absorbed into
the RG-improved Hamiltonian (see Sec.\ref{sec:imsrg_pairing}). It is
by this reasoning that we use the relative size of the second-order correction
to the flowing energy, $\Delta E^2(s)/E(s)$, as the stopping criterion for
the flow. Once this quantity falls below $10^{-8}$, we terminate the evolution.

\subsubsection*{Optimizations}
The Python implementation of the IMSRG that we describe here can solve
problems with small single-particle basis sizes in reasonable time. To 
tackle large-scale calculations, we implement the IMSRG in 
languages like C/C++ or Fortran that are closer to the hardware.

As mentioned in the course of the discussion, we have favored simplicity 
in the design of the Python code, which leaves significant room for optimization.
For instance, we can exploit that nucleons are Fermions, and 
antisymmetrize the two-nucleon basis states. This reduces the storage for
matrix elements involving identical particles by a factor of four, because
we only need to store one of the antisymmetrized matrix elements 
\begin{equation}
  \Gamma_{ijkl} = -\Gamma_{jikl} = - \Gamma_{ijlk} = \Gamma_{jilk}\,.
\end{equation}
The storage can be reduced even further if we also exploit the Hermiticity
and anti-Hermiticity of $\HO$ and $\etaO$, respectively.

Another important tool for optimization are symmetries of the Hamiltonian.
Nuclear Hamiltonians conserve the total angular momentum, parity, and the 
isospin projection of quantum states, which implies that $f$ and $\Gamma$ 
are block-diagonal in the corresponding quantum numbers. Since the generator 
and the derivatives are constructed from the Hamiltonian, they have the
same symmetries, and are block-diagonal as well. Exploiting this block
structure, we can reduce the storage requirements, which are naively $\OC(N^4)$
for each operator, by about one to two orders of magnitude. We can 
also explicitly work on the blocks instead of the full matrices when 
we evaluate the right-hand sides of the IMSRG flow equations.  The complex 
couplings between blocks prevents us from evolving individual blocks or
small groups separately, in contrast to the free-space SRG case, where
this was possible (see Sec.\ref{sec:srg_nn_flow}).

\subsection{\label{sec:imsrg_pairing}IMSRG Solution of the Pairing Hamiltonian}
Let us now use the code from the previous section to solve the 
Schr\"odinger equation for four particles that interact via the pairing
Hamiltonian \eqref{eq:def_pairing_hamiltonian}.

%
%
\subsubsection{Ground-State Calculations}

\begin{figure}[t]
  \setlength{\unitlength}{\textwidth}
  \begin{picture}(1.0000,0.4000)
    \put(0.0000,0.0000){\includegraphics[width=0.48\unitlength]{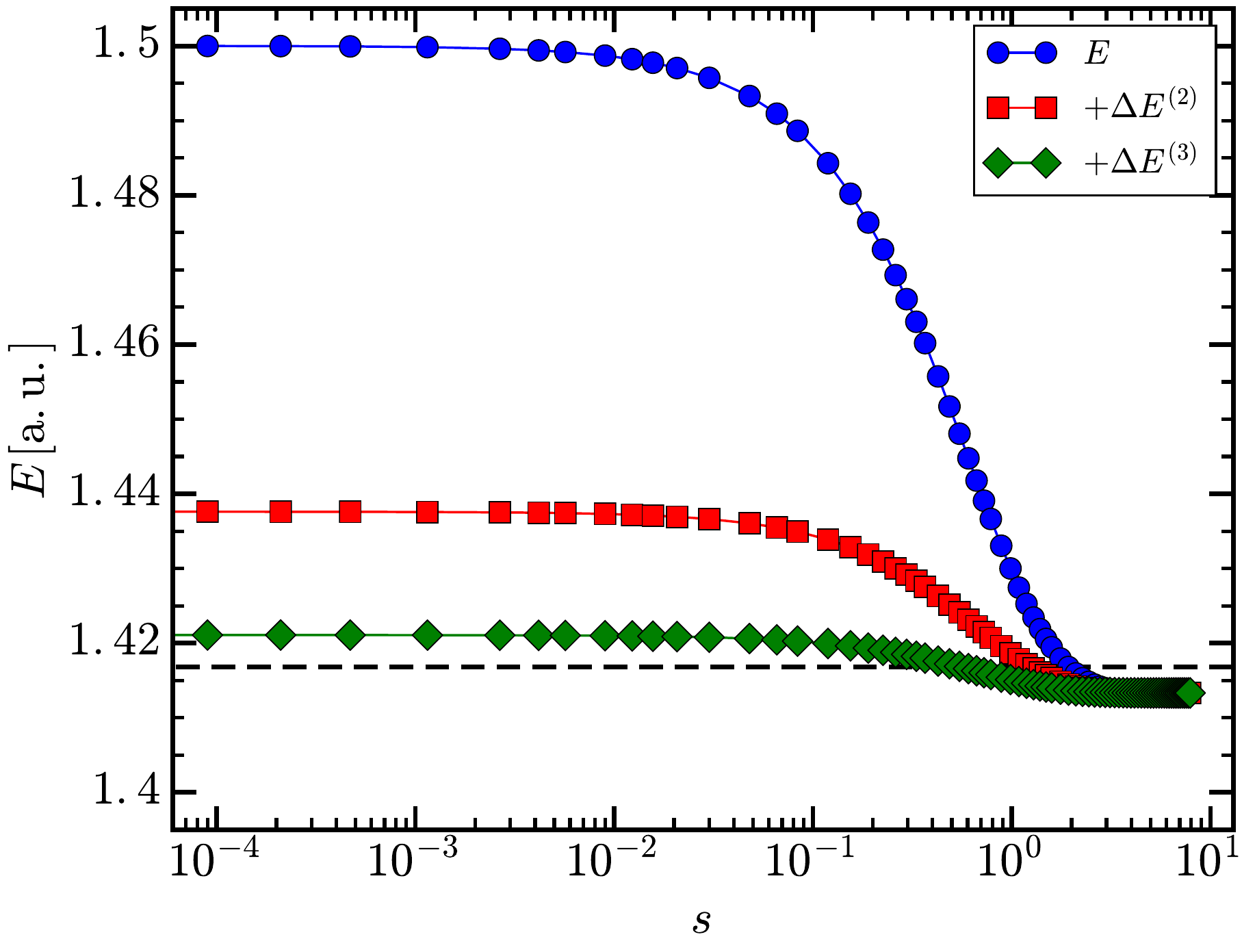}}
    \put(0.5000,0.0000){\includegraphics[width=0.49\unitlength]{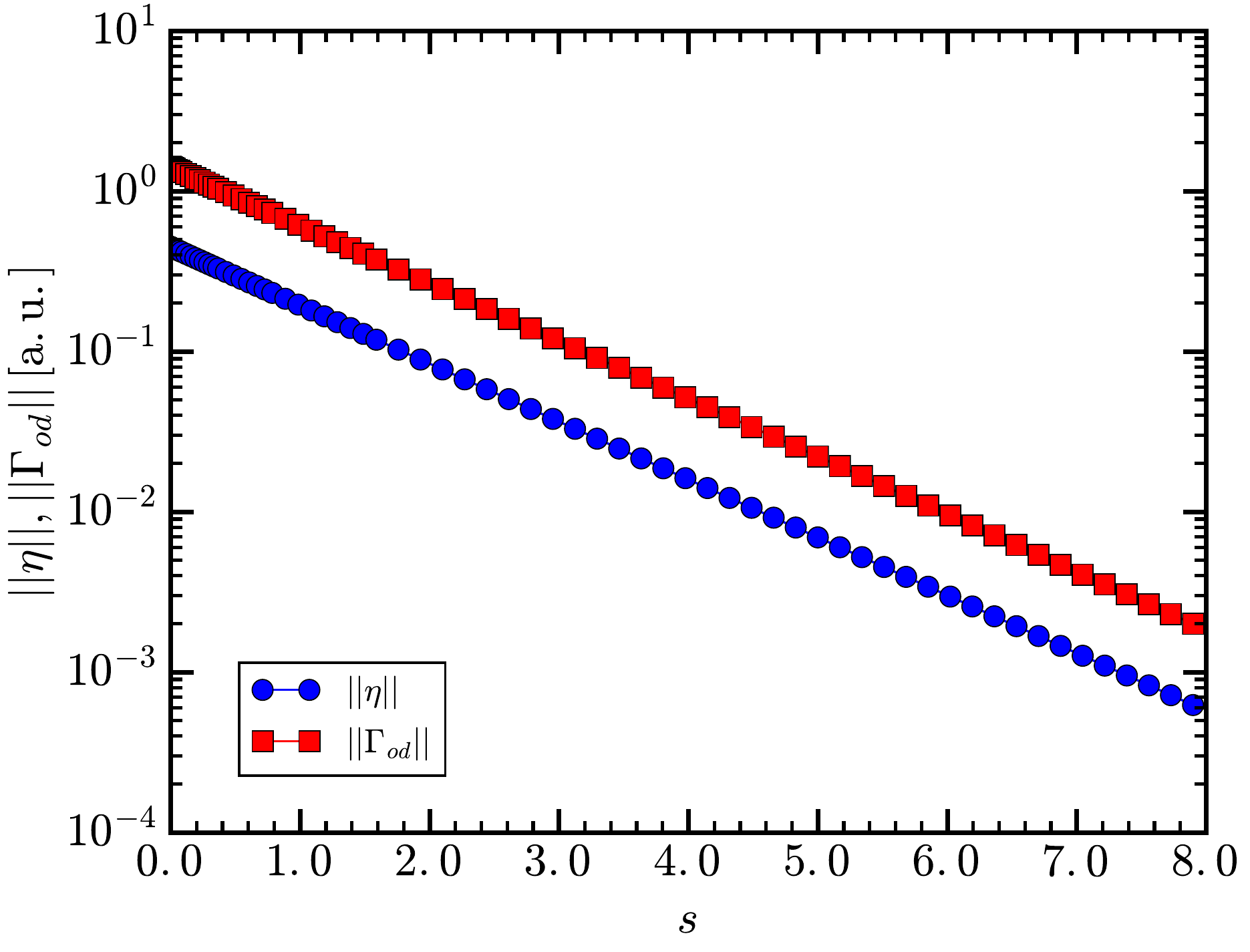}}
  \end{picture}

  \caption{\label{fig:imsrg_gs}IMSRG(2) flow for the ground state of the pairing 
  Hamiltonian with $\delta=1.0, g=0.5$ (cf.~Sec.~\ref{sec:srg_pairing_example}).
  Calculations were performed with the White generator, Eq.~\eqref{eq:eta_white}.
  \emph{Left panel:} Flowing ground-state energy $E(s)$ plus perturbative second and
  third-order energy corrections for $\HO(s)$. The exact ground-state energy is 
  indicated by the dashed line. \emph{Right panel:} Norm of the White generator and
  the off-diagonal Hamiltonian $\HO_{od}(s)$ (note that $||f_{od}(s)||=0$\,).}
\end{figure}

As a first application, we calculate the ground-state energy for the 
pairing Hamiltonian with $\delta=1.0$ and $g=0.5$, which we studied
using SRG matrix flows in Sec.~\ref{sec:srg_pairing_example}. In
the left panel of Fig.~\ref{fig:imsrg_gs}, we show the flowing 
ground-state energy $E(s)$. Starting from the energy of the uncorrelated
reference state, which is $E(0)=2\delta-g=1.5$, we obtain a final 
energy $E(\infty)=1.4133$, which is slightly below the exact result
$1.4168$.

The mechanism by which the flowing ground-state energy is absorbing
correlation energy can be understood by considering the zero-body flow 
equation \eqref{eq:imsrg2_m0b} in the perturbative approach we introduced 
in Sec.~\ref{sec:imsrg_generator}. We note that
\begin{align}
  \totd{E}{s} &=     
    \underbrace{\sum_{ab}(n_{a}-n_{b})\eta^{a}_{b}f^{b}_{a}}_{\OC(g^4)}
    +\underbrace{\frac{1}{4}\sum_{abcd}
        \left(\eta^{ab}_{cd}\Gamma^{cd}_{ab}-\Gamma^{ab}_{cd}\eta^{cd}_{ab}\right)
        n_{a}n_{b}\bar{n}_{c}\bar{n}_{d}}_{\OC(g^2)}\,.
\end{align}
For the White generator \eqref{eq:eta_white},
\begin{equation}
    \Gamma^{pp'}_{hh'}(s)=\Gamma^{pp'}_{hh'}(0)e^{-s}\,,\quad 
    \Gamma^{hh'}_{pp'}(s)=\Gamma^{hh'}_{pp'}(0)e^{-s}\,.
\end{equation}
As we can see in the left panel of Fig.~\ref{fig:imsrg_gs}, the
off-diagonal matrix elements and the generator indeed decay 
exponentially with a single, state-independent scale. Plugging the matrix
elements into the energy flow equations to $\OC(g^2)$, we have  
\begin{equation}
  \totd{E}{s}=\frac{1}{2}\sum_{pp'hh'}\frac{|\Gamma^{pp'}_{hh'}(0)|^2}{\Delta^{pp'}_{hh'}(0)}e^{-2s}\,.
\end{equation}
Integrating over the flow parameter, we obtain
\begin{equation}
  E(s) = E(0) - \frac{1}{4}\sum_{pp'hh'}\frac{|\Gamma^{pp'}_{hh'}(0)|^2}{|\Delta^{pp'}_{hh'}(0)|}
      \left(1 - e^{-2s}\right)\,.\label{eq:flow_E_pert}
\end{equation}
We recognize the second-order energy correction, evaluated with the
initial Hamiltonian, and see that $E(s)$ will decrease 
with $s$ (i.e., the binding energy increases). In the limit $s\to\infty$,
the entire correction is shuffled into the zero-body piece of the evolved 
Hamiltonian. As discussed in Sec.~\ref{sec:imsrg_flow}, the complete 
IMSRG(2) flow performs a more complex re-summation of correlations, but we 
can see from Fig.~\ref{fig:imsrg_gs} that it certainly encompasses the complete 
second order. In fact, we see that the third-order correction is completely 
absorbed into the final $E(\infty)$ as well. Readers who are interested in
more details are referred to the extensive discussion in Ref.~\cite{Hergert:2016jk} 
(also see \cite{Morris:2015ve}).

\begin{figure}[t]
  \setlength{\unitlength}{\textwidth}
  \begin{picture}(1.0000,0.4000)
    \put(0.0000,0.0000){\includegraphics[width=0.48\unitlength]{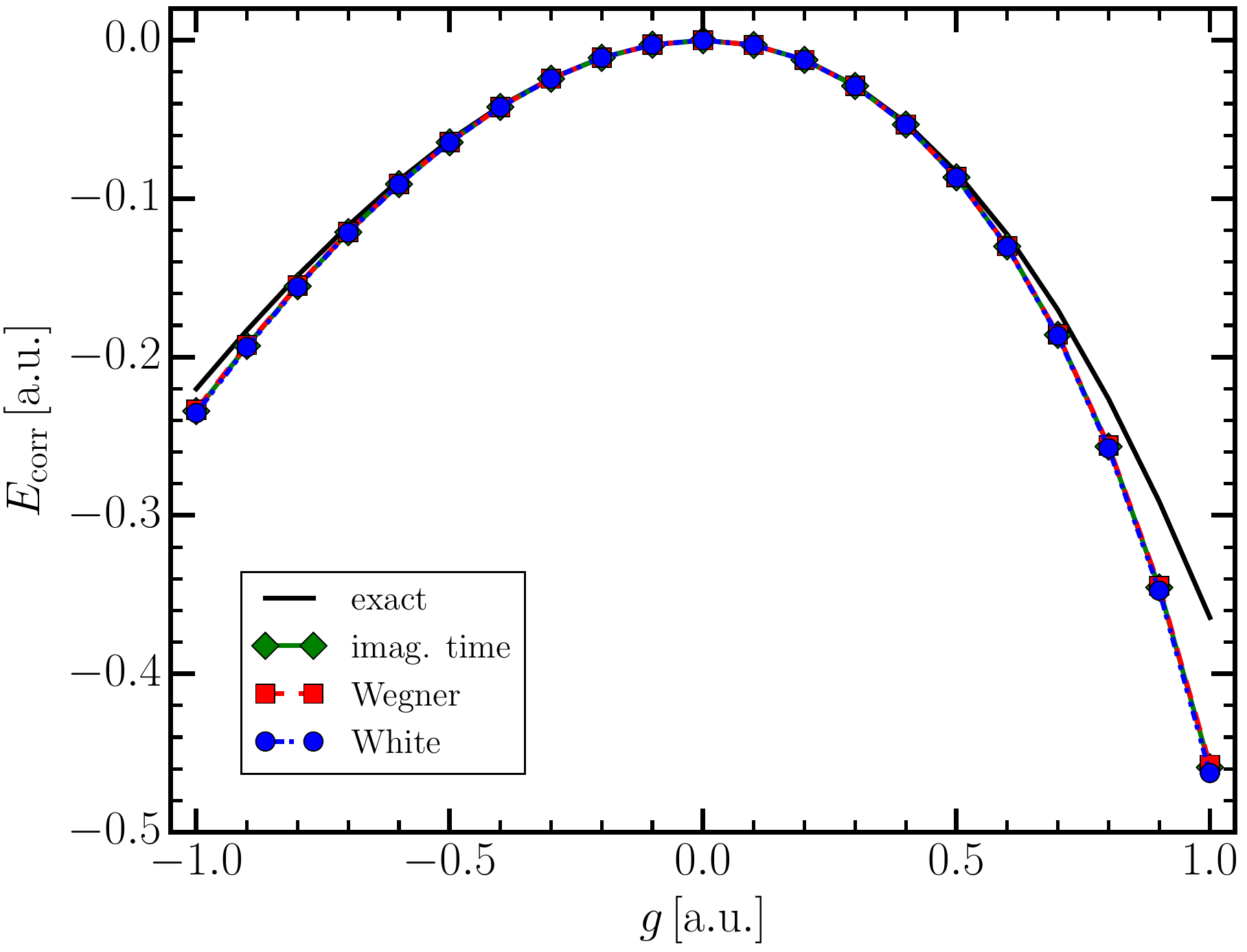}}
    \put(0.5000,0.0000){\includegraphics[width=0.48\unitlength]{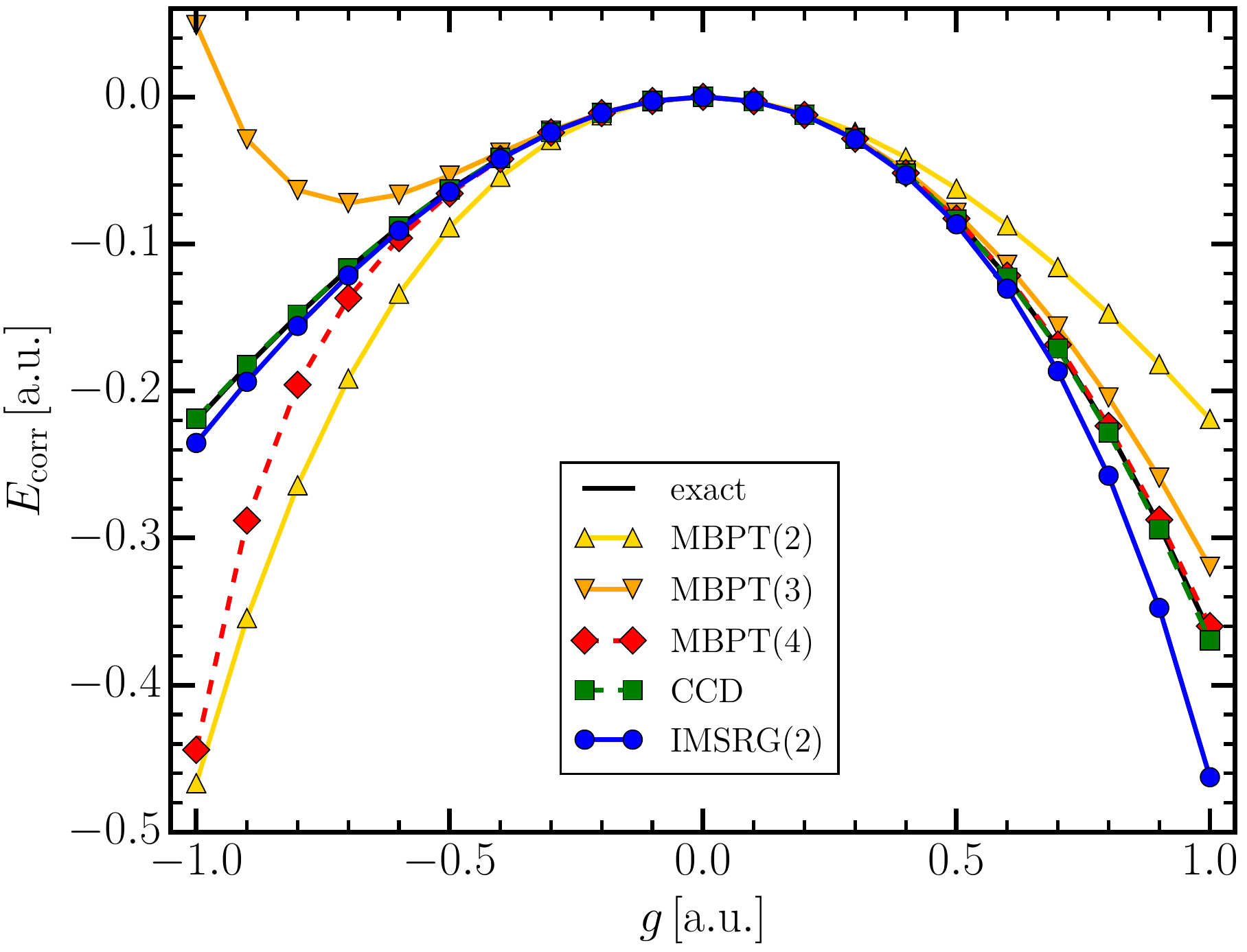}}
  \end{picture}

  \caption{\label{fig:imsrg_corr}
  Ground-state correlation energies as a function of the pairing strength, 
  for $\delta=1.0$.
  \emph{Left panel:} IMSRG(2) correlation energies for different flow
  generators.
  \emph{Right panel:} Comparison of IMSRG(2) and other many-body methods. 
  }
\end{figure}

In Fig.~ \ref{fig:imsrg_corr}, we show the IMSRG(2) correlation energy
\begin{equation}
  E_\text{corr} = E(\infty) - E_\text{HF}
\end{equation}
as a function of the pairing strength $g$, holding the single-particle 
level spacing constant at $\delta=1.0$. As we see in the left panel, 
the IMSRG(2) results for the White (Eq.~\eqref{eq:eta_white}), imaginary
time (Eq.~\eqref{eq:eta_imtime}) and Wegner generators (Eq.~\eqref{eq:eta_wegner})
are practically identical as we evolve to $s\to\infty$. This behavior
is expected, because all three generators are based on the same off-diagonal
Hamiltonian, and we have explained in the previous sections that the 
specific choice of the generator then only affects the numerical aspects
of the flow, with the potential exception of accumulated truncation
errors. Such errors clearly do not matter for the present case.

In the range $-0.5\lesssim g \lesssim 0.5$, the IMSRG(2) is in excellent 
agreement with the exact diagonalization. This range corresponds to 
a region of weak correlations between the four particles that we 
consider in our system, because the ratio between the characteristic excitation
scale of the uncorrelated many-body states (diagonal matrix elements of 
Eq.~\eqref{eq:def_h_matrix}) and the pairing strength (off-diagonal matrix 
elements) is small,
\begin{equation}
  \left|\frac{g}{2\Delta E}\right| = \left|\frac{g}{4\delta}\right| \leq 0.125\,.
\end{equation}
Going beyond $g\geq 0.5$, the correlations grow stronger, and the IMSRG(2) 
starts to overestimate the size of the correlation energy. The absolute 
deviation is about $0.1$ at $g=1.0$, where the exact eigenvalue of the pairing 
Hamiltonian is $0.6355$. 

Interestingly, the deviations are much smaller in the opposite case, $g=-1.0$. The 
right-hand panel of Fig.~ \ref{fig:imsrg_corr} sheds further light on this matter.
There, we compare the IMSRG(2) correlation energy to results from finite-order
many-body perturbation theory (MBPT) and Coupled Cluster with Doubles Excitations
(CCD) --- see chapter 8 for details. We see that the correlation
energies from finite-order MBPT alternate in sign for a repulsive interaction,
which suggests that the agreement between the exact solution, IMSRG(2), and CCD
is due to cancellations at all orders that these methods take into account.
Looking back at the attractive pairing force for $g\sim1.0$, we observe that 
the MBPT results have converged to the exact solution once fourth-order 
corrections are taken into account, and CCD gives the same result. The
IMSRG(2) overestimates the correlation energy in this region because it 
undercounts a set of four repulsive fourth-order diagrams by a factor $1/2$,
while CCD takes them into account completely 
(see \cite{Hergert:2016jk,Parzuchowski:2016pi}).

\begin{figure}[t]
  \setlength{\unitlength}{\textwidth}
  \begin{picture}(1.0000,0.4000)
    \put(0.0000,0.0000){\includegraphics[width=0.48\unitlength]{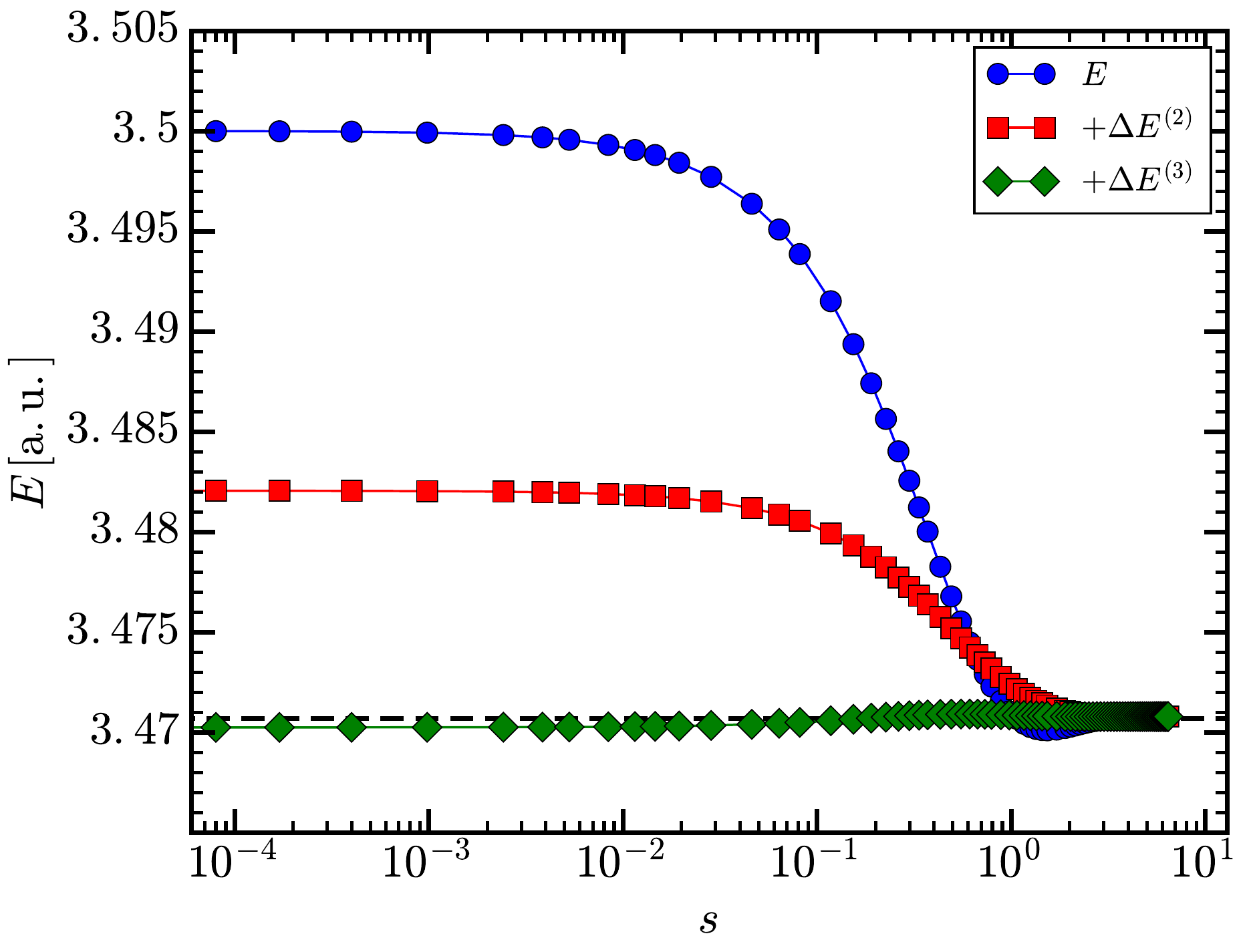}}
    \put(0.5000,0.0000){\includegraphics[width=0.48\unitlength]{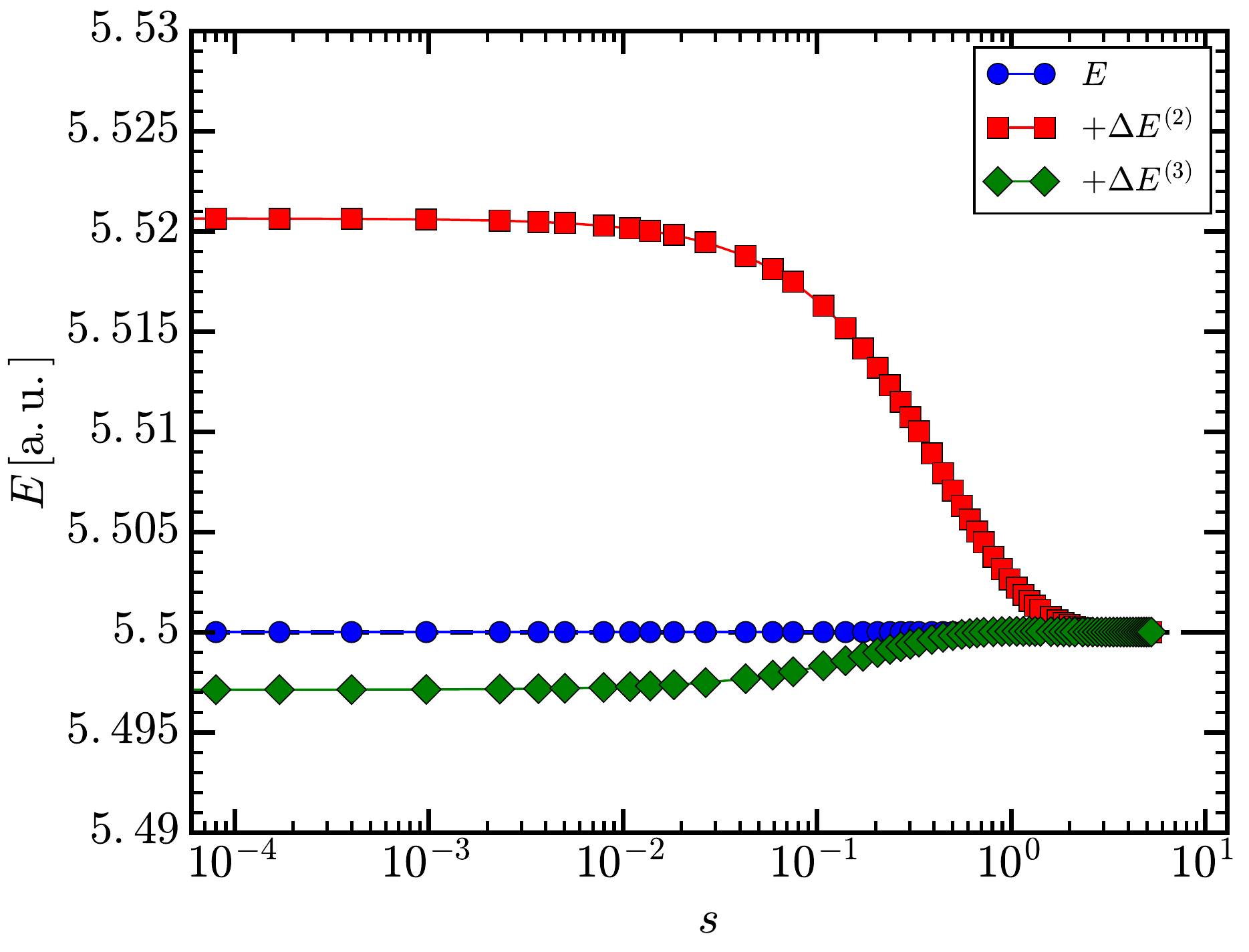}}
  \end{picture}

  \caption{\label{fig:imsrg_excited}
    IMSRG(2) flow for different reference states, using the pairing 
    Hamiltonian with $\delta=1.0, g=0.5$.
    Calculations were performed with the White generator, Eq.~\eqref{eq:eta_white}.
    For Slater determinants with uncorrelated energies $4-g=3.5$
    (\emph{left panel}) and $6-g=5.5$ (\emph{right panel}), the 
    IMSRG(2) flow targets the 2nd and 3rd excited states,
    see Sec.~\ref{sec:srg_pairing_example}.
  }
\end{figure}

%
%
\subsubsection{Targeting Excited States}
In Sec.~\ref{sec:imsrg_offidag}, we mentioned that the choice of
reference state will affect which eigenstate of the Hamiltonian the 
IMSRG evolution is targeting. To illustrate this, Fig.~ \ref{fig:imsrg_excited}
shows the IMSRG(2) flows starting from uncorrelated states with
energies $4-g$ (with nucleons occupying the $p=1$ and $p=2$ 
single-particle states, see Tab.~\ref{tab:srg_pairing_sp}) and $6-g$
(with $p=0$ and $p=2$ occupied), respectively. For $g=0.5$ and $\delta=1.0$,
the exact eigenvalues are $3.4707$ and $5.5$. 

As we can see in the left panel, the IMSRG(2) flow towards the
correct energy. Between $s=1$ and $s=3$, it overshoots the exact
energy. As discussed in Sec.~\ref{sec:imsrg_variational}, the
IMSRG(2) is not a variational method, so this is unproblematic. In
the limit $s\to\infty$, the flow converges to $3.4708$, matching
the energy of the first excited state. The flow shown in the right
panel of Fig.~\ref{fig:imsrg_excited} is particularly interesting.
Recall from Sec.~\ref{sec:srg_pairing_example} that the state with 
energy $5.5$ is already a degenerate eigenstate of the Hamiltonian, and 
therefore supposed to be invariant under a unitary flow. This is indeed
the case, at least to the eight digits recorded in the flow data file 
\texttt{imsrg-white\_d1.0\_g+0.5\_N4\_ev3.flow}. It also means that
all perturbative corrections in the IMSRG(2) summation must cancel out.
While the second- and third-order energy corrections are of opposite
sign, they are not of the same size. Thus, contributions from fourth
and higher orders are involved in the cancellation, and the invariance
of $E(s)$ demonstrates that they are indeed generated by the IMSRG(2)
flow. We conclude our discussion of the IMSRG treatment of the pairing 
Hamiltonian here.

\subsection{\label{sec:imsrg_neutron_matter}Infinite Neutron Matter}

After discussing the pairing Hamiltonian, we now want to apply the IMSRG(2) 
to a large-scale problem, namely the calculation of the equation of state
for pure neutron matter. Python scripts and data are available at \url{https://github.com/ManyBodyPhysics/LectureNotesPhysics/tree/master/Programs/Chapter10-programs/python/imsrg_pnm}. The C++ program is available from \url{https://github.com/ManyBodyPhysics/LectureNotesPhysics/tree/master/Programs/Chapter10-programs/cpp/imsrg_pnm}.

We work in a basis of plane wave states, which is
set up just like in the CC case discussed in Sec.~8.7.
We work in a spherical periodic cell $L=(N/\rho)^{1/3}$, where $N$ is the number 
of neutrons in the cell, and $\rho$ the neutron matter density. Because of 
the periodic boundary conditions, the momenta $p_x,p_y,p_z$ are discretized,
and we can write the single-particle states as $\ket{n_x,n_y,n_z,s_z}$ 
($s_z$ is the spin projection of the neutron). We impose the truncation
\begin{equation}
  n_x^2 + n_y^2 + n_z^2 \leq \Nmax\,.
\end{equation}

\begin{figure}[t]
  \setlength{\unitlength}{\textwidth}
  \begin{picture}(1.0000,0.4000)
    \put(0.0000,0.0000){\includegraphics[width=0.5\unitlength]{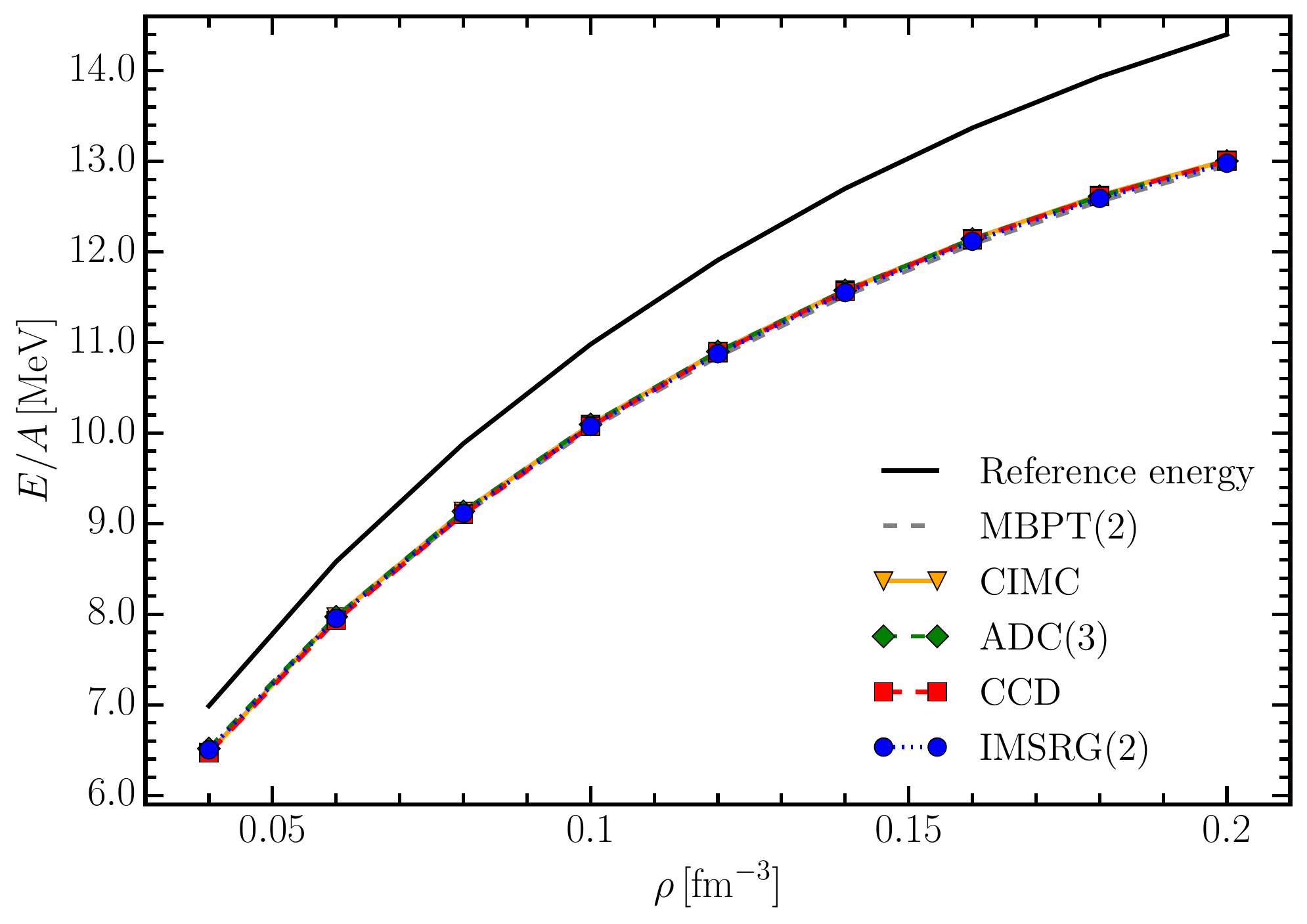}}
    \put(0.5000,0.0000){\includegraphics[width=0.5\unitlength]{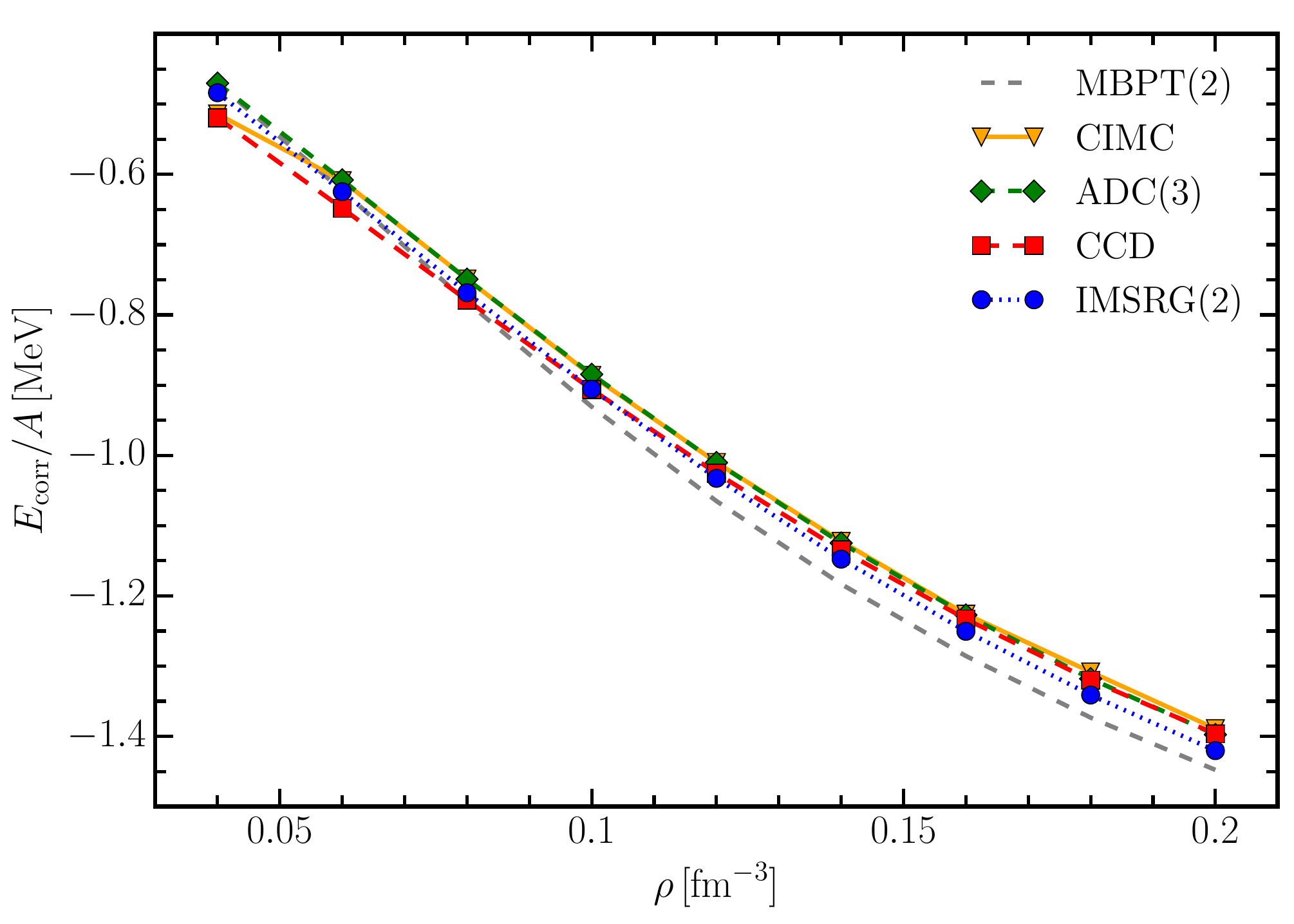}}
  \end{picture}

  \caption{\label{fig:imsrg_pnm}
    Equation of state for pure neutron matter from IMSRG(2) and other many-body
    methods, based on the Minnesota $NN$ potential \cite{Thompson:1977fk}.
    The left panel shows the energy per particle obtained with IMSRG(2), CCD (
    Chap.~8), CIMC (Chap.~9), and the ADC(3) 
    Self-Consistent Green's Function scheme (Chap.~11), the
    right panel the correlation energy per particle from the same methods.
  }
\end{figure}

In Fig.~\ref{fig:imsrg_pnm}, we show the IMSRG(2) results for neutron matter
for the semi-realistic Minnesota $NN$ potential \cite{Thompson:1977fk},
using $N=66$ neutrons in an $\Nmax=36$ basis. The left panel shows the 
IMSRG(2) equation of state (EOS) which is essentially the same as that of 
methods with comparable correlation content, namely CCD, the Configuration 
Interaction Monte Carlo based on CCD wave functions (CIMC, see Sec.~9.6.1) 
and the Self-Consistent Green's Functions in the ADC(3) scheme 
(Sec.~11.3.1). We also include the MBPT(2) EOS, which
is in very good agreement with the more sophisticated methods.
This shows that pure neutron matter is only weakly correlated, and
the many-body expansion is clearly converging rapidly. The IMSRG(2)
(and other methods) gain an additional 10\% additional energy compared
to the uncorrelated HF EOS, all the way from dilute matter at $\rho=0.05\fm^{-3}$
to $\rho=0.2\,\fm^3$, which is more than twice the neutron density of 
typical stable nuclei. 

In the right panel of Fig.~\ref{fig:imsrg_pnm}, we show the correlation 
energy per particle, which reveals some differences between the various
methods. As the neutron matter density comes and correlations are expected
to become increasingly important, MBPT(2) gains the highest amount of
correlation energy, just as in our results for the pairing Hamiltonian
(cf.~Fig.~\ref{fig:imsrg_corr}). Curiously, CCD gives the most binding
of all methods in dilute neutron matter, but eventually, the CCD correlation
energy is very similar to that of CIMC and ADC(3), which should be superior
approximations to the exact ground state (see Chaps.~9 and
11). As the density increases the IMSRG(2) starts to gain
more binding from correlations than CCD, its closest cousin among the 
considered methods, but not as much additional binding energy as MBPT(2).
This reflects our findings for the pairing Hamiltonian, where we 
observed the same phenomenon (see Fig.~\ref{fig:imsrg_corr}). As explained
in Sec.~\ref{sec:imsrg_pairing}, the reason for this energy gain compared
to CCD is the under-counting of certain repulsive fourth-order diagrams
in the IMSRG(2), see \cite{Hergert:2016jk}.

\section{\label{sec:current}Current Developments}
After covering the essential concepts of the SRG and IMSRG, and discussing
both the formal and technical aspects of their applications, we want to
introduce our readers to the three major directions of current IMSRG
research: These are the use of the so-called Magnus expansion to explicitly 
construct the IMSRG transformation (Sec.~\ref{sec:current_magnus}), the
Multireference IMSRG for generalizing the method to correlated 
reference states (Sec.~\ref{sec:current_mrimsrg}), and the construction
of effective Hamiltonians for use in configuration interaction and 
Equation-of-Motion methods, which allows us to tackle excited states 
(Sec.~\ref{sec:current_hamiltonians}).

\subsection{\label{sec:current_magnus}Magnus Formulation of the IMSRG}

Despite its modest computational scaling and the flexibility to tailor 
the generator to different applications, IMSRG calculations based on 
the direct integration of Eqs.~(\eqref{eq:opflow}) are limited by memory 
demands of the ODE solver in many realistic cases. The use of a
high-order solver is essential, as the accumulation of integration-step
errors destroys the unitary equivalence between $H(s)$ and $H(0)$ even
if no truncations are made in the flow equations. State-of-the-art
solvers can require the storage of 15-20 copies of the solution vector
in memory, which is the main computational bottleneck of the method
(see, e.g., \cite{Hindmarsh:1983pd,Brown:1989qd,Hindmarsh:2005kl}).

Matters are complicated further if we also want to calculate expectation
values for observables besides the Hamiltonian. General operators have
to be evolved consistently using the flow equation 
\begin{equation} \label{eq:obsflow}
  \totd{}{s} \OO(s) = \comm{\etaO(s)}{\OO(s)}\,,
\end{equation}
but since storage of $\etaO(s)$ at every point of the flow trajectory
is prohibitively expensive, we are forced to solve Eq.~\eqref{eq:obsflow} 
simultaneously with the flow equation for the Hamiltonian. The evaluation
of $N$ observables besides the Hamiltonian implies that the dimension of
the ODE system \eqref{eq:imsrg2_m0b}--\eqref{eq:imsrg2_m2b} grows by a
factor $N+1$. In addition, generic operators can evolve with rather different 
characteristic scales than the Hamiltonian, increasing the likelihood of 
the ODEs becoming stiff.

We can now overcome these limitations by re-formulating the IMSRG using
the Magnus expansion from the theory of matrix differential equations
\cite{Magnus:1954xy,Blanes:2009fk}. Magnus proved that the path-ordered 
series defining the IMSRG transformation, Eq.~\eqref{eq:def_U_series},
can be summed into a true exponential expression if the generator $\etaO$
meets certain conditions (see \cite{Magnus:1954xy}):
\begin{equation} \label{eq:def_U_magnus}
  \UO(s) \equiv e^{\OmegaO(s)}\,.
\end{equation}
This allows us to derive a flow equation for the anti-Hermitian Magnus
operator $\OmegaO(s)$:
\begin{equation}\label{eq:magnus_flow}
  \frac{d\OmegaO}{ds}= \sum_{k=0}^\infty \frac{B_k}{k!} \mathrm{ad}_{\OmegaO}^k\left(\etaO\right)\,,
\end{equation}
where $B_k$ are the Bernoulli numbers, and 
\begin{align}
  \mathrm{ad}_{\OmegaO}^0\left(\etaO\right)&=\etaO\\
  \mathrm{ad}_{\OmegaO}^k\left(\etaO\right) &= \comm{\OmegaO}{\mathrm{ad}_{\OmegaO}^{k-1}\left(\etaO\right)}\,.
\end{align}
As in the standard IMSRG(2), we truncate $\etaO$ and $\OmegaO$ as well
as their commutator at the two-body level. We refer to the resulting calculation 
scheme as the Magnus(2) formulation of the IMSRG. 
The series of nested commutators generated by $\mathrm{ad}^k_{\OmegaO}$ 
is evaluated recursively, until satisfactory convergence of the right-hand 
side of Eq.~\eqref{eq:magnus_flow} is reached \cite{Morris:2015ve}. At each 
integration step, we use $\UO(s)$ to construct the
Hamiltonian $\HO(s)$ via the Baker-Campbell-Hausdorff (BCH) formula
\begin{equation}
  \HO(s)\equiv e^{\OmegaO(s)}\HO(0) e^{-\OmegaO(s)} = \sum_{k=0}^\infty\frac{1}{k!}\mathrm{ad}^k_{\OmegaO(s)}\left(\HO(0)\right)\,,
\end{equation}
(the flow parameter dependence is stated explicitly here for clarity).
Like $\etaO$ and $\OmegaO$, the Hamiltonian is truncated at the two-body
level.

A major advantage of the Magnus formulation stems from the fact that the 
flow equations for $\OmegaO(s)$ can be solved using a simple first-order 
Euler step method without any loss of accuracy, resulting in substantial 
memory savings and a modest reduction in CPU time.  While sizable 
integration-step errors accumulate in $\Omega(s)$ with a first-order method, 
upon exponentiation the transformation is still unitary, and the transformed 
$H(s)=\UO(s)H\UUO(s)$ is unitarily equivalent to the initial Hamiltonian aside 
from the truncations made while evaluating the BCH formula. For further details 
on the implementation of the Magnus formulation, see Ref.~\cite{Morris:2015ve}.

\begin{figure}[t]
  \begin{center}
    \includegraphics[width=.6\textwidth]{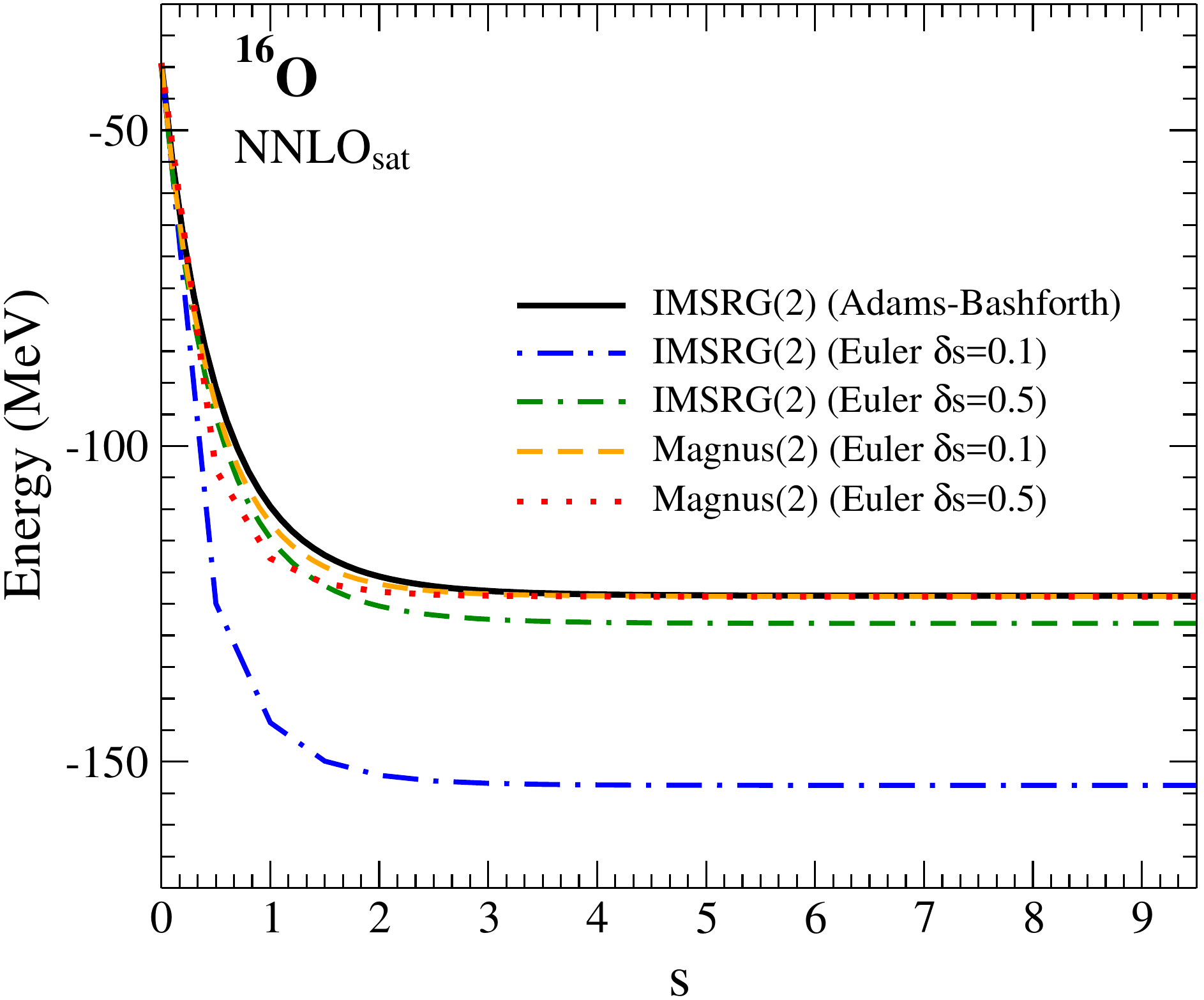}
  \end{center}

  \caption{\label{fig:timestep_O16}
    IMSRG(2) and Magnus(2) ground-state energy of $\nuc{O}{16}$, 
    starting from the \NNLOsat{} $NN+3N$ interaction \cite{Ekstrom:2015fk}. The 
    solid black line is the IMSRG(2) flow obtained with
    an adaptive solver based on the Adams-Bashforth method \cite{Radhakrishnan:1993fk}, 
    while the other lines show Magnus(2) and IMSRG(2) flows 
    obtained with a simple Forward Euler method and different 
    fixed step sizes. All calculations were done in an
    $\eMax=8$ model space, with $\hbar\omega = 24$~MeV for the
    underlying harmonic-oscillator basis.}
\end{figure}

The insensitivity of the Magnus scheme to integration-step errors is illustrated 
in Fig.~\ref{fig:timestep_O16}, which shows flowing ground-state energy for 
$^{16}$O, calculated with the chiral \NNLOsat{} $NN+3N$ 
interaction by Ekstr\"{o}m et al.~\cite{Ekstrom:2015fk}. The black solid line 
denotes the results of a standard IMSRG(2) calculation using a high-order 
predictor-corrector solver \cite{Hindmarsh:1983pd,Hindmarsh:2005kl}, while the 
other curves denote IMSRG(2) and Magnus(2) calculations using a first-order Euler 
method with different step sizes $\delta s$. Unsurprisingly, the IMSRG(2) calculations 
using a first-order Euler method are very poor, with different step sizes
converging to different large-$s$ limits. The Magnus(2) calculations,
on the other hand, converge to the same large-$s$ limit in excellent
agreement with the standard IMSRG(2) results.

A second major advantage of the Magnus formulation of the IMSRG is that
we can evaluate arbitrary observables by simply using the final 
Magnus operator $\OmegaO(\infty)$ to calculate
\begin{equation}
  \OO(\infty)\equiv e^{\OmegaO(\infty)}\OO(0) e^{-\OmegaO(\infty)}\,.
\end{equation}
This is obviously much more convenient than dealing with observables
in the direct IMSRG(2) approach, because we do not have to deal with
the doubling, tripling, \ldots of the already large system of flow
equations. In contrast to the prohibitive space required to store 
$\etaO(s)$ for all values of $s$, we can also easily archive the 
$\Omega(\infty)$ for long-term use, e.g., when we want to look at
new observables in future studies. 

\subsection{\label{sec:current_mrimsrg}The Multi-Reference IMSRG}

%
%
\subsubsection{Correlated Reference States}
Many-body bases built from a single Slater determinant and its particle-hole
excitations work best for systems with large gaps in the single-particle 
spectrum, e.g., closed-shell nuclei. If the gap is small, particle-hole 
excited basis states can be near-degenerate with the reference determinant, 
which results in strong configuration mixing. When the mixing involves 
configurations in which many nucleons are excited simultaneously, many-body 
physicists speak of \emph{static} or \emph{collective correlations} in the 
wave function, as opposed to \emph{dynamic correlations} that are caused by 
the excitations of a small number of nucleons only. 

Important examples are the emergence of nuclear superfluidity \cite{Dean:2003ei}
or diverse rotational and vibrational bands in open-shell nuclei (see, e.g., \cite{Bohr:1999vn}).
These phenomena are conveniently described by using the concept of intrinsic
wave functions that explicitly break appropriate symmetries of the Hamiltonian.
For instance, nuclear superfluidity can be treated in to leading-order in 
the self-consistent Hartree-Fock-Bogoliubov (HFB) approach, which is formulated 
in terms of anti-symmetrized product states of fermionic quasi-particles that 
are superpositions of particles and holes. Because of this, the intrinsic HFB wave 
functions are superpositions of states with different particle numbers. The
broken symmetries must eventually be restored by means of projection methods,
which have a long history in nuclear many-body theory
\cite{Peierls:1973fk,Ring:1980bb,Egido:1982sd,Robledo:1994qf,Flocard:1997fx,Sheikh:2000xx,Dobaczewski:2007hh,Bender:2009rv,Duguet:2009ph,Lacroix:2009aj,Lacroix:2012vn,Duguet:2015ye}).

The standard IMSRG framework as described in Sec.~\ref{sec:imsrg} works
with an uncorrelated reference state, and therefore puts the entire onus
of describing correlations on the transformation $\UO(s)$. The
computational cost limits us to the IMSRG(2) scheme, or an eventual
approximate IMSRG(3) that is roughly analogous to completely renormalized  
Coupled Cluster schemes with approximate triples \cite{Binder:2013fk}. An 
IMSRG(4), let alone the complete IMSRG(A) scheme, are not feasible. Thus, 
the IMSRG, like CC, is best suited to the description of dynamic correlations.
We can mitigate this shortcoming by extending the IMSRG to work with correlated
reference states, and building static correlation that would otherwise require
an IMSRG(4),\ldots scheme directly into the reference state. This leads us to
the Multi-Reference IMSRG (MR-IMSRG) \cite{Hergert:2013ij,Hergert:2014vn,Hergert:2017kx}, 
which is constructed using the generalized normal ordering and Wick's
theorem developed by Kutzelnigg and Mukherjee \cite{Kutzelnigg:1997fk,Mukherjee:1997yg}.

%
%
\subsubsection{Generalized Normal Ordering}
In Ref.~\cite{Kutzelnigg:1997fk}, Kutzelnigg and Mukherjee developed a 
generalized normal ordering for arbitrary reference states. In the brief
discussion that follows, we use the slightly different notation of 
Kong \emph{et al.}~\cite{Kong:2010kx}.

First, we introduce a pseudo-tensorial notation for strings of creation 
and annihilation operators, to facilitate book-keeping and make the formalism
more compact. A product of $k$ creators and annihilators each is written as
\begin{equation}
  \AO^{i_1\ldots i_k}_{j_1\ldots j_k}\equiv
  \aaO_{i_1}\ldots\aaO_{i_k}\aO_{j_k}\ldots\aO_{j_1}\,.
\end{equation}
We do not consider particle-number changing operators in the present work,
because they cause ambiguities in the contraction and sign rules for the
$A$ operators that are defined in the following. The anticommutation relations 
imply
\begin{equation}\label{eq:def_A_permutations}
  \AO^{\PC(i_1 \ldots i_k)}_{\PC'(j_1 \ldots j_k)} 
  = (-1)^{\pi(\PC)+\pi(\PC')} \AO^{i_1\ldots i_k}_{j_1\ldots j_k}\,,
\end{equation}
where $\pi(\PC)=\pm1$ indicates the parity (or signature) of a permutation 
$\PC$. A general $k$-body operator can now be written as
\begin{equation}
 O^{(k)} = \frac{1}{(k!)^2}\sum_{\substack{i_1\ldots i_k\\j_1\ldots j_k}}
  o^{i_1\ldots i_k}_{j_1\ldots j_k}\AO^{i_1\ldots i_k}_{j_1\ldots j_k}\,,
\end{equation}
where we assume that the coefficients $o^{i_1\ldots i_k}_{j_1\ldots j_k}$
are antisymmetrized, and therefore also obey equation \eqref{eq:def_A_permutations}
under index permutations. 

Next, we introduce \emph{irreducible $k$-body density matrices $\lambda^{(k)}$}. In
the one-body case, we have the usual density matrix
\begin{equation}
  \lambda^{i}_{j} \equiv \matrixe{\Phi}{\AO^{i}_{j}}{\Phi}\,,
\end{equation}
and for future use, we also define
\begin{equation}
  \xi^{i}_{j} \equiv \lambda^{i}_{j} - \delta^{i}_{j}\,.
\end{equation}
Up to a factor $(-1)$ that unifies the sign rules for one-body contractions
presented below, $\xi^{(1)}$ is simply the generalization of the hole 
density matrix for a correlated state (cf.~Eq.~\eqref{eq:def_hole_contraction}. 
In the natural orbital basis, both one-body density matrices are diagonal,
with fractional occupation numbers $0\leq n_i, \nn_i \leq 1$ as eigenvalues. 

For $k\geq 2$, we denote full density matrices by
\begin{align}
  \rho^{i_1 \ldots i_k}_{j_1 \ldots j_k} &= 
    \dmatrixe{\Phi}{\AO^{i_1 \ldots i_k}_{j_1 \ldots j_k}}\,,
\end{align}
and define
\begin{align}
  \lambda^{ij}_{kl} &\equiv \rho^{ij}_{kl} - \AC\{\lambda^{i}_{k}\lambda^{j}_{l}\}\,, \label{eq:def_Lambda2}\\
  \lambda^{ijk}_{lmn}&\equiv \rho^{ijk}_{lmn} - \AC\{\lambda^{i}_{l}\lambda^{jk}_{mn}\} -
                      \AC\{\lambda^{i}_{l}\lambda^{j}_{m}\lambda^{k}_{n}\}\,,\label{eq:def_Lambda3}
\end{align}
etc., where $\AC\{\ldots\}$ fully antisymmetrizes the indices of the expression
within the brackets, e.g.,
\begin{equation}
  \AC\{\lambda^{i}_{k}\lambda^{j}_{l}\} = \lambda^{i}_{k}\lambda^{j}_{l} - 
    \lambda^{i}_{l}\lambda^{j}_{k}\,.
\end{equation}
From equation \eqref{eq:def_Lambda2}, it is easy to see that $\lambda^{(2)}$ encodes the
two-nucleon correlation content of the reference state $\ket{\Phi}$. If 
the reference state is a Slater determinant, i.e., an independent-particle
state, the full two-body density matrix factorizes, and $\lambda^{(2)}$
vanishes:
\begin{equation}
  \lambda^{ij}_{kl} = \rho^{ij}_{kl} - \AC\{\lambda^{i}_{k}\lambda^{j}_{l}\}
  = \lambda^{i}_{k}\lambda^{j}_{l} -  \lambda^{i}_{k}\lambda^{j}_{l} - 
    \left(\lambda^{i}_{k}\lambda^{j}_{l} -  \lambda^{i}_{k}\lambda^{j}_{l}\right) = 0\,.
\end{equation}
Equation \eqref{eq:def_Lambda3} shows that $\lambda^{(3)}$ is constructed
by subtracting contributions from three independent particles as well as
two correlated nucleons in the presence of an independent spectator particle
from the full three-body density matrix, and therefore encodes the genuine
three-nucleon correlations. This construction and interpretation  
generalizes to irreducible density matrices of rank $k$.

Normal-ordered one-body operators are constructed in the same manner as in
the standard normal ordering of Sec.~\ref{sec:nord}:
\begin{equation}
  \nord{A^{a}_{b}} \equiv A^{a}_{b} - \matrixe{\Phi}{A^{a}_{b}}{\Phi} = A^{a}_{b} - \lambda^{a}_{b}\,.
\end{equation}
For a two-body operator, we have the expansion
\begin{align}
  A^{ab}_{cd} &= \nord{A^{ab}_{cd}} 
                  + \lambda^{a}_{c}\nord{A^{b}_{d}} 
                  - \lambda^{a}_{d}\nord{A^{b}_{c}} 
                  + \lambda^{b}_{d}\nord{A^{a}_{c}}
                  - \lambda^{b}_{c}\nord{A^{a}_{d}}
                  + \lambda^{a}_{c}\lambda^{b}_{d}
                  - \lambda^{a}_{d}\lambda^{b}_{c}
                  + \lambda^{ab}_{cd}\,.\label{eq:nord_2B}
\end{align}
As a consequence of equation \eqref{eq:def_A_permutations}, the sign of each
term is determined by the product of the parities of the permutations 
that map upper and lower indices to their ordering in the initial operator. 
Except for the last term, this expression looks like the result for the
regular normal ordering, with pairwise contractions of indices giving
rise to one-body density matrices. The last term, a contraction
of four indices, appears because we are dealing with an arbitrary, 
correlated reference state here. For a three-body operator, we 
obtain schematically
\begin{align}
  A^{abc}_{def} &= \nord{A^{abc}_{def}} 
                  + \AC\{\lambda^{a}_{d}\nord{A^{bc}_{ef}} \}
                  + \AC\{\lambda^{a}_{d}\lambda^{b}_{e}\nord{A^{c}_{f}} \}
                  + \AC\{\lambda^{ab}_{de}\nord{A^{c}_{f}} \} \notag\\
                &\hphantom{=} + \lambda^{abc}_{def}
                  + \AC\{\lambda^{a}_{d}\lambda^{bc}_{ef} \}
                  + \AC\{\lambda^{a}_{d}\lambda^{b}_{e}\lambda^{c}_{f} \}\,,
\end{align}
and the procedure can be extended to higher particle rank in an
analogous fashion.

When we work with arbitrary reference states, the regular Wick's 
theorem of Sec.~\ref{sec:nord} is extended with additional contractions:
{\setlength{\fboxsep}{1pt}
\begin{align}
  \nord{A^{a{\fbox{\scriptsize $b$}}}_{cd}}\nord{A^{ij}_{\fbox{\scriptsize $k$}l}} &= - \lambda^{\fbox{\scriptsize $b$}}_{\fbox{\scriptsize $k$}}\nord{A^{aij}_{cdl}}\,,\label{eq:contract_lambda1B}\\ 
  \nord{A^{ab}_{\fbox{\scriptsize $c$}d}}\nord{A^{i\fbox{\scriptsize $j$}}_{kl}} &= - \xi^{\fbox{\scriptsize $j$}}_{\fbox{\scriptsize $c$}}\nord{A^{bia}_{dkl}}\,,\label{eq:contract_chi1B}\\[10pt]
  \nord{A^{\fbox{\scriptsize $ab$}}_{cd}}\nord{A^{ij}_{\fbox{\scriptsize $kl$}}} &= + \lambda^{\fbox{\scriptsize $ab$}}_{\fbox{\scriptsize $kl$}}\nord{A^{ij}_{cd}}\,,\label{eq:contract_lambda2Ba}\\ 
  \nord{A^{a{\fbox{\scriptsize $b$}}}_{cd}}\nord{A^{\fbox{\scriptsize $i$}j}_{\fbox{\scriptsize $kl$}}} &= - \lambda^{\fbox{\scriptsize $ib$}}_{\fbox{\scriptsize $kl$}}\nord{A^{aj}_{cd}}\,,\label{eq:contract_lambda2Bb}\\ 
  \nord{A^{ab}_{\fbox{\scriptsize $c$}d}}\nord{A^{\fbox{\scriptsize $ij$}}_{\fbox{\scriptsize $k$}l}} &= - \lambda^{\fbox{\scriptsize $ij$}}_{\fbox{\scriptsize $ck$}}\nord{A^{ab}_{dl}}\,,\label{eq:contract_lambda2Bc}\\ 
  \nord{A^{\fbox{\scriptsize $ab$}}_{c\fbox{\scriptsize$d$}}}\nord{A^{\fbox{\scriptsize $i$}j}_{\fbox{\scriptsize $kl$}}} &= - \lambda^{\fbox{\scriptsize $abi$}}_{\fbox{\scriptsize $dkl$}}\nord{A^{j}_{c}}\,,\label{eq:contract_lambda3B}\\ 
  \nord{A^{\fbox{\scriptsize $ab$}}_{\fbox{\scriptsize$cd$}}}\nord{A^{\fbox{\scriptsize $ij$}}_{\fbox{\scriptsize $kl$}}} &= +\lambda^{\fbox{\scriptsize $abij$}}_{\fbox{\scriptsize $cdkl$}}\,.\label{eq:contract_lambda4B}
\end{align}
}
The new contractions \eqref{eq:contract_lambda2Ba}--\eqref{eq:contract_lambda4B} 
increase the number of terms when we expand operator products. Fortunately,
the overall increase in complexity is manageable.

Applying the generalized normal ordering to the intrinsic nuclear $A$-body 
Hamiltonian \eqref{eq:def_Hint} we obtain
\begin{align}
  \HO &= E 
        + \sum_{ij}f^{i}_{j}\nord{A^{i}_{j}} 
        + \frac{1}{4}\sum_{ijkl}\Gamma^{ij}_{kl}\nord{A^{ij}_{kl}}
        + \frac{1}{36}\sum_{ijklmn}W^{ijk}_{lmn}\nord{A^{ijk}_{lmn}}\,,
\end{align}
with the individual contributions 
\begin{align}
  E &\equiv \left(1-\frac{1}{A}\right)\sum_{ab}t^{a}_{b}\lambda^{a}_{b}
        + \frac{1}{4}\sum_{abcd}\left(\frac{1}{A}t^{ab}_{cd} + v^{ab}_{cd}\right)\rho^{ab}_{cd}
      + \frac{1}{36}\sum_{abcdef}v^{abc}_{def} \rho^{abc}_{def}\,,
      \label{eq:def_mr_E0}\\
  f^{i}_{j} &\equiv \left(1-\frac{1}{A}\right)t^{i}_{j} + \sum_{ab}\left(\frac{1}{A}t^{ia}_{jb} + v^{ia}_{jb}\right)\lambda^{a}_{b}
  + \frac{1}{4}\sum_{abcd}v^{iab}_{jcd}\rho^{ab}_{cd}\,,\label{eq:def_mr_f}   \\
  \Gamma^{ij}_{kl} &\equiv \frac{1}{A}t^{ij}_{kl} + v^{ij}_{kl} + \sum_{ab}v^{ija}_{klb}\lambda^{a}_{b}\,,\label{eq:def_mr_Gamma}\\
  W^{ijk}_{lmn}&\equiv v^{ijk}_{lmn}\,.
\end{align}
Here, we use the full density matrices for compactness, but it is easy to express
equations \eqref{eq:def_mr_E0}--\eqref{eq:def_mr_Gamma} completely in terms of irreducible density
matrices by using equations \eqref{eq:def_Lambda2} and \eqref{eq:def_Lambda3}. 

%
%
\subsubsection{MR-IMSRG Flow Equations}
We evaluate the operator flow equation \eqref{eq:opflow} using the generalized
Wick's theorem, truncating all operators the two-body level, and obtain the 
MR-IMSRG(2) flow equations \cite{Hergert:2013ij,Hergert:2014vn,Hergert:2017kx}:
\begin{align}
  \totd{E}{s} &=     
    \sum_{ab}(n_{a}-n_{b})\eta^{a}_{b}f^{b}_{a}
    +\frac{1}{4}\sum_{abcd}
        \left(\eta^{ab}_{cd}\Gamma^{cd}_{ab}-\Gamma^{ab}_{cd}\eta^{cd}_{ab}\right)
        n_{a}n_{b}\bar{n}_{c}\bar{n}_{d}
    \notag\\
  &\hphantom{=}
    +\frac{1}{4}\sum_{abcd}\left(\totd{}{s}\Gamma^{ab}_{cd}\right)\lambda^{ab}_{cd}
    +\frac{1}{4}\sum_{abcdklm}\left(\eta^{ab}_{cd}\Gamma^{kl}_{am}-\Gamma^{ab}_{cd}\eta^{kl}_{am}\right)\lambda^{bkl}_{cdm}\,,
    \label{eq:mr_flow_0b_tens}
  \\[10pt]
  \totd{}{s}f^{i}_{j} &=
    \sum_{a}\left(\eta^{i}_{a}f^{a}_{j}-f^{i}_{a}\eta^{a}_{j}\right)
    +\sum_{ab}\left(\eta^{a}_{b}\Gamma^{bi}_{aj}-f^{a}_{b}\eta^{bi}_{aj}\right)(n_{a}-n_{b})
  \notag\\
  &\hphantom{=}
  +\frac{1}{2}\sum_{abc}
    \left(\eta^{ia}_{bc}\Gamma^{bc}_{ja}-\Gamma^{ia}_{bc}\eta^{bc}_{ja}\right)\left(n_{a}\bar{n}_{b}\bar{n}_{c}+\bar{n}_{a}n_{b}n_{c}\right)
  \notag\\
  &\hphantom{=}
      +\frac{1}{4}\sum_{abcde}\left(\eta^{ia}_{bc}\Gamma^{de}_{ja}-\Gamma^{ia}_{bc}\eta^{de}_{ja}\right)\lambda^{de}_{bc} 
    +\sum_{abcde}\left(\eta^{ia}_{bc}\Gamma^{be}_{jd}-\Gamma^{ia}_{bc}\eta^{be}_{jd}\right)\lambda^{ae}_{cd}
  \notag\\
  &\hphantom{=}
      -\frac{1}{2}\sum_{abcde}\left(\eta^{ia}_{jb}\Gamma^{cd}_{ae}-\Gamma^{ia}_{jb}\eta^{cd}_{ae}\right)\lambda^{cd}_{be}
    +\frac{1}{2}\sum_{abcde}\left(\eta^{ia}_{jb}\Gamma^{bc}_{de}-\Gamma^{ia}_{jb}\eta^{bc}_{de}\right)\lambda^{ac}_{de}\,,
  \label{eq:mr_flow_1b_tens}\\[10pt]
  \totd{}{s}\Gamma^{ij}_{kl}&=  
  \sum_{a}\left(\eta^{i}_{a}\Gamma^{aj}_{kl}+\eta^{j}_{a}\Gamma^{ia}_{kl}-\eta^{a}_{k}\Gamma^{ij}_{al}-\eta^{a}_{l}\Gamma^{ij}_{ka}
  -f^{i}_{a}\eta^{aj}_{kl}-f^{j}_{a}\eta^{ia}_{kl}+f^{a}_{k}\eta^{ij}_{al}+f^{a}_{l}\eta^{ij}_{ka}\right)
  \notag\\
  &\hphantom{=}
    +\frac{1}{2}\sum_{ab}\left(\eta^{ij}_{ab}\Gamma^{ab}_{kl}-\Gamma^{ij}_{ab}\eta^{ab}_{kl}\right)
     \left(1-n_{a}-n_{b}\right)
  \notag\\
  &\hphantom{=}
    +\sum_{ab}(n_{a}-n_{b})\left(\left(\eta^{ia}_{kb}\Gamma^{jb}_{la}-\Gamma^{ia}_{kb}\eta^{jb}_{la}\right)-\left(\eta^{ja}_{kb}\Gamma^{ib}_{la}-\Gamma^{ja}_{kb}\eta^{ib}_{la}\right)\right)\,.
  \label{eq:mr_flow_2b_tens}
\end{align}
All single-particle indices and occupation numbers (cf.~Sec.~\ref{sec:nord}) 
refer to natural orbitals, and the $s$-dependence has been suppressed for 
brevity. Because we use general reference states, the MR-IMSRG flow equations 
also include couplings to correlated pairs and triples of nucleons in that
state through the irreducible density matrices $\lambda^{(2)}$ and 
$\lambda^{(3)}$. The single-reference limit 
(Eqs.~\eqref{eq:imsrg2_m0b}--\eqref{eq:imsrg2_m2b}) 
can be obtained by setting the irreducible density matrices $\lambda^{(2)}$ 
and $\lambda^{(3)}$ to zero in the previous expressions.

Superficially, the computational cost for the evaluation of the MR-IMSRG(2) 
flow equations is dominated by the final term of Eq.~\eqref{eq:mr_flow_0b_tens},
which is of $\OC(N^7)$. However, since storage of the complete $\lambda^{(3)}$ 
is prohibitive in large-scale calculations, we impose certain constraints on
the reference state, which in turn restrict the non-zero matrix elements to
small subsets of the entire matrix. For example, for particle-number projected
HFB reference states, $\lambda^{(3)}$ is almost diagonal, which reduces the
effort for the zero-body flow equation to $\OC(N^4)$. We have also explored
reference states from No-Core Shell Model calculations in a small model space 
\cite{Gebrerufael:2016rp},
which limit the indices of $\lambda^{(3)}$ to 5-10 single-particle states
out of a complete single-particle basis that is one to two orders of magnitude
larger. In a similar scenario, we have used reference states consisting of 
a valence space (or active space, in chemistry parlance) on top of an inert
core, as in the traditional nuclear Shell model (cf.~Sec.~\ref{sec:current_hamiltonians}). 
In that case, the correlations are restricted to this valence space, and 
$\lambda^{(3)}$ is only non-zero if all indices refer to valence space
(active space) single-particle states. Thus, the main driver of the computational
effort is still the two-body flow equation, at $\OC(N^6)$, just like in the
regular IMSRG(2). Equation \eqref{eq:mr_flow_2b_tens} actually has exactly
the same for as its single-reference counterpart, Eq.~\eqref{eq:imsrg2_m2b},
except that the occupation numbers can now have arbitrary values between 0 and 1.

%
%
\subsubsection{Decoupling and Generators}
In the multireference case, we choose a suitable correlated reference 
state, and construct its excitations by applying all possible one- 
and two-body operators:
\begin{equation}
  \ket{\Phi},\,\nord{\AO^{i}_{j}}\ket{\Phi},\,\nord{\AO^{ij}_{kl}}\ket{\Phi},\ldots\,.
\end{equation}
The properties of the normal ordering ensure that the excited states are
orthogonal to the reference state, but they are in general not orthogonal 
to each other: for instance,
\begin{equation}
  \dmatrixe{\Phi}{\nord{A^{i}_{j}}\nord{A^{k}_{l}}}
  =-\lambda^{i}_{l}\xi^{k}_{j} + \lambda^{ij}_{kl}
  =n_i\nn_j \delta^{i}_{l}\delta^{k}_{j} + \lambda^{ij}_{kl}\,,
\end{equation}
where $0\leq n_i, \nn_i \leq 1$. Moreover, there can be linear dependencies
between the excitations of the correlated reference state, so the matrix
representations of the Hamiltonian and other operators in this basis can 
be rank deficient. While the rank deficiency poses a major challenge for 
multireference CC methods (see, e.g., \cite{Lyakh:2012zr}), for us it only
means that we are implementing the MR-IMSRG flow on a matrix that has
spurious zero eigenvalues that are typically far removed from the low-lying
part of the spectrum in which we are most interested.

To identify the off-diagonal Hamiltonian, we can proceed like in the
single-reference IMSRG, and try to satisfy the decoupling conditions
\begin{align}
  \matrixe{\Phi}{\HO(\infty)\nord{\AO^{i}_{j}}}{\Phi}&=0\,,\label{eq:mr_decoupling1B}\\
  \matrixe{\Phi}{\HO(\infty)\nord{\AO^{ij}_{kl}}}{\Phi}&=0\,,\label{eq:mr_decoupling2B}\\
  \ldots& \notag
\end{align}
and corresponding conditions for the conjugate matrix elements. The 
matrix elements can be evaluated with the generalized Wick's theorem, 
e.g.,
\begin{align}
  \dmatrixe{\Phi}{H\nord{\AO^{i}_{j}}}
  =\nn_i n_j f^{j}_{i}+\sum_{ab}f^{a}_{b}\lambda^{ai}_{bj}
   +\frac{1}{2}\sum_{abc} 
    \left( \nn_i \lambda^{bc}_{ja}\Gamma^{bc}_{ia} - n_j \Gamma^{ja}_{bc} \lambda^{ia}_{bc} \right)
   +\frac{1}{4}\sum_{abcd}\Gamma^{ab}_{cd}\lambda^{iab}_{jcd}\,.\label{eq:mr_coupling_1b}
\end{align}
The first term is merely the generalization of the one-body
particle-hole matrix element from Sec.~\ref{sec:imsrg_decoupling}: In the 
single-reference limit, the occupation number prefactor is nonzero if $i$ and
$j$ are particle and hole indices, respectively. In addition, the matrix
element depends on the irreducible densities $\lambda^{(2)}$ and $\lambda^{(3)}$
due to the coupling of the Hamiltonian to correlated pairs and triples of nucleons 
in the reference state. The coupling condition to two-nucleon excitations, 
Eq.~\eqref{eq:mr_decoupling2B}, not only has a much more complicated 
structure than its single-reference counterpart, but even depends
on $\lambda^{(4)}$ (see \cite{Hergert:2017kx} for details).
Constructing and storing $\lambda^{(4)}$ is essentially out
of the question in general MR-IMSRG applications, hence we 
are forced to introduce truncations to evaluate Eq.~\eqref{eq:mr_coupling_1b} 
and similar matrix elements. This implies that we can only
achieve approximate decoupling in general.

In recent applications, we have found the variational perspective
introduced in Sec.~\ref{sec:imsrg_variational} to be useful. We
can write the decoupling conditions as
\begin{align}
  \dmatrixe{\Phi}{H\nord{\AO^{i}_{j}}} 
    &= \frac{1}{2}\dmatrixe{\Phi}{\acomm{H}{\nord{\AO^{i}_{j}}}} 
    + \frac{1}{2}\dmatrixe{\Phi}{\comm{H}{\nord{\AO^{i}_{j}}}}\,,\\
  \dmatrixe{\Phi}{H\nord{\AO^{ij}_{kl}}} 
    &= \frac{1}{2}\dmatrixe{\Phi}{\acomm{H}{\nord{\AO^{ij}_{kl}}}} 
    + \frac{1}{2}\dmatrixe{\Phi}{\comm{H}{\nord{\AO^{ij}_{kl}}}}\,,
\end{align}
and at least suppress the second terms in both equations through what 
amounts to a minimization of the ground-state energy under unitary variation. 
This means that we aim to satisfy the IBCs introduced in Sec.~\ref{sec:imsrg_variational}, 
(also see \cite{Mukherjee:2001uq,Kutzelnigg:2002kx,Kutzelnigg:2004vn,Kutzelnigg:2004ys}).
Evaluating the commutators, we obtain
\begin{align}
  \dmatrixe{\Phi}{\comm{H}{\nord{\AO^{i}_{j}}}}
    &=(n_j-n_i)f^{j}_{i}-\frac{1}{2}\sum_{abc}\left(  
     \Gamma^{ja}_{bc}\lambda^{ia}_{bc}
    -\Gamma^{ab}_{ic}\lambda^{ab}_{jc}
    \right)\,,
\end{align}
\begin{align}
  \dmatrixe{\Phi}{\comm{H}{\nord{\AO^{ij}_{kl}}}}
  &=
    \Gamma^{kl}_{ij}(\nn_i\nn_jn_kn_l - n_in_j\nn_k\nn_l)
  +\sum_{a}
     \left(
     (1-P_{ij})f^{a}_{i}\lambda^{aj}_{kl}
    -(1-P_{kl})f^{k}_{a}\lambda^{ij}_{al}
    \right)
    \notag\\[3pt]
  &\hphantom{=}
    +\frac{1}{2}\left(
      (\lambda\Gamma)^{kl}_{ij}\left(1-n_i-n_j\right)
      -(\Gamma\lambda)^{kl}_{ij}\left(1-n_k-n_l\right)
    \right)
  \notag\\[3pt]
  &\hphantom{=}
    +(1-P_{ij})(1-P_{kl})\sum_{ac}
       \left(n_j-n_k\right)\Gamma^{ak}_{cj}\lambda^{ai}_{cl}
    \notag\\[3pt]
  &\hphantom{=}
    +\frac{1}{2}\sum_{abc}\left(
       (1-P_{kl})\Gamma^{ka}_{bc}\lambda^{aij}_{bcl} 
      -(1-P_{ij})\Gamma^{ab}_{ic}\lambda^{abj}_{ckl} 
    \right)\,.
\end{align}
Like the MR-IMSRG(2) flow equations \eqref{eq:mr_flow_0b_tens}--\eqref{eq:mr_flow_2b_tens},
these expressions only depend linearly on $\lambda^{(2)}$ and $\lambda^{(3)}$,
which makes untruncated implementations feasible.

We use the IBCs to define the so-called Brillouin generator as
\begin{align}
  \eta^{i}_{j}&\equiv\dmatrixe{\Phi}{\comm{H}{:\AO^{i}_{j}:}}\\
  \eta^{ij}_{kl}&\equiv\dmatrixe{\Phi}{\comm{H}{:\AO^{ij}_{kl}:}}
\end{align}
Because the matrix elements of $\etaO$ are directly given
by the residuals of the IBCs, it can be interpreted as the \emph{gradient}
of the energy with respect to the parameters of the unitary transformation 
at each step of the flow. At the fixed point of the flow, $\etaO=0$,
and the flowing zero-body part of the Hamiltonian, $E(\infty)$, will be an extremum 
of the energy. Indeed, $\etaO$ has behaved in this manner in all 
numerical applications to date, generating a monotonic flow of the  
energy towards the converged results \cite{Hergert:2017kx}.

%
%
\subsubsection{Example: The Oxygen Isotopic Chain}
\begin{figure}[t]
  \begin{center}
    \includegraphics[width=\textwidth]{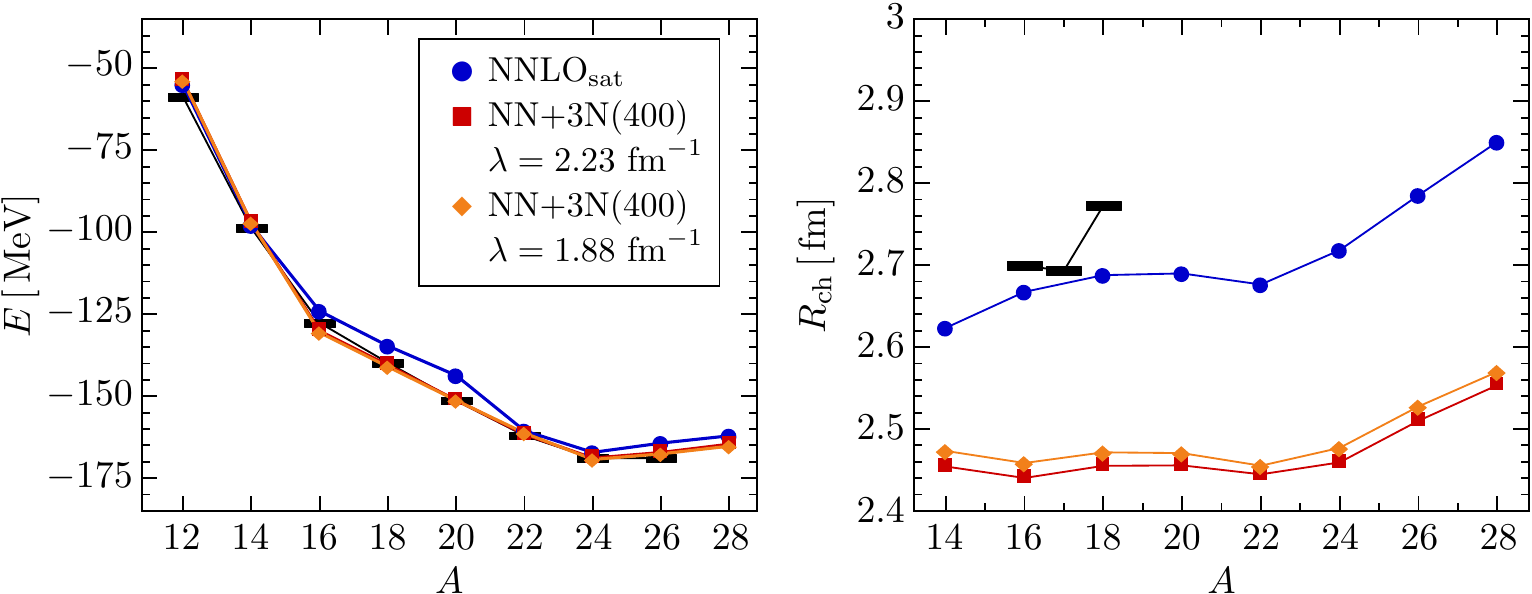}
  \end{center}

  \caption{\label{fig:OXX} MR-IMSRG(2) ground-state energies and charge radii of the oxygen 
    isotopes for \NNLOsat{} and $NN\!+\!3N(400)$ at $\lambdaSRG=1.88,\ldots,2.24\fmi$ 
    ($\eMax=14,\EMax=14$, and optimal $\hw$). Black bars indicate experimental 
    data \cite{Wang:2012uq,Angeli:2013rz}.
  }
\end{figure}

As a sample application of the MR-IMSRG(2), we use spherical,
particle-number projected HFB vacua (see, e.g., \cite{Ring:1980bb,Hergert:2009zn}
to compute the ground-state energies and radii of the even oxygen 
isotopes (odd isotopes have irreducible densities that are non-scalar 
under rotation, which requires a future extension of our framework) 
\cite{Hergert:2013ij,Lapoux:2016xu}. Our results are shown in Fig.~
\ref{fig:OXX}. We use various chiral $NN+3N$ interactions.

The $NN+3N(400)$ Hamiltonian consists of the \NNNLO{} interaction
by Entem and Machleidt \cite{Entem:2003th,Machleidt:2011bh}, with
cutoff $\Lambda_{NN}=500\,\MeV$, and a local \NNLO{} interaction
with a reduced cutoff $\Lambda_{NN}=400\,\MeV$ \cite{Roth:2011kx,Gazit:2009qf}.
The low-energy constants (LECs), i.e., the parameters of the chiral
Hamiltonian, are entirely fixed by fitting data in the $A=2,3,4$ systems, 
and it is evolved to lower resolution scales $\lambdaSRG$ via free-space 
SRG, as discussed in Sec.~\ref{sec:srg_nn}. In contrast, the LECs of \NNLOsat{} 
are also optimized with respect to selected many-body data \cite{Ekstrom:2015fk},
and it is sufficiently soft that we use it as is. 

While $NN+3N(400)$ gives a good reproduction of the oxygen ground-state
energies, an issue with the Hamiltonian's saturation properties is revealed
by inspecting the oxygen charge radii (see Fig.~\ref{fig:OXX}). 
The theoretical charge radii are about 10\% smaller
than the experimental charge radius of $\nuc{O}{16}$, $\Rch=2.70\,\fm$ 
\cite{Angeli:2013rz}, and the sharp increase for $\nuc{O}{18}$ is missing 
entirely. The variation of $\lambdaSRG$ produces only a 0.2\%
change in the ground-state energies, but this is the result of
a fine-tuned cancellation between induced $4N$ forces that are
generated by the $NN$ and $3N$ pieces of the Hamiltonian, and
should not be seen as representative for chiral interactions in
general. The charge radii grow larger as $\lambdaSRG$ decreases,
which is consistent with a study for light nuclei 
by Schuster \emph{et al.}\cite{Schuster:2014oq}. The authors found 
that two- and three-body terms that are induced by consistently evolving
the charge radius operator to lower $\lambdaSRG$ have the opposite
effect and \emph{reduce} its expectation value. These terms have not 
been included here, but need to be considered for complete consistency
in the future.

The MR-IMSRG(2) ground-state energies obtained with \NNLOsat{} are 
slightly \emph{lower} than those for $NN\!+\!3N(400)$ in the proton-rich
isotopes $\nuc{O}{12,14}$, and above the $NN\!+\!3N(400)$ energies in 
$\nuc{O}{16-28}$. From $\nuc{O}{16-22}$, the \NNLOsat{} ground-state energies
exhibit a parabolic behavior as opposed to the essentially linear trend
we find for $NN\!+\!3N(400)$. A possible cause is the inclusion of the $\nuc{O}{22,24}$
ground-state energies in the optimization protocol, which constrains
the deviation of the energies from experimental data in these nuclei. 
\NNLOsat{} predicts the drip line at $\nuc{O}{24}$, and the trend for 
the $\nuc{O}{26,28}$ resonance energies is similar to the $NN\!+\!3N(400)$ 
case. For \NNLOsat{}, the charge radii for the bound oxygen isotopes are 
about 10\% larger than for $NN\!+\!3N(400)$, which is expected given 
the use of the $\nuc{O}{16}$ charge radius in the optimization of 
the LECs (also see Ref.~\cite{Lapoux:2016xu}). For the resonant states, 
the increase is even larger, but continuum effects must be considered to 
make a meaningful comparison. We note that \NNLOsat{} also fails to describe 
the sharp jump in $\Rch$ at $\nuc{O}{18}$.


\subsection{\label{sec:current_hamiltonians}Effective Hamiltonians}

\begin{figure}[t]
  \setlength{\unitlength}{0.5\textwidth}
  \begin{center}
  \begin{picture}(1.0000,0.8000)
    \put(0.0000,0.0000){\includegraphics[width=\unitlength]{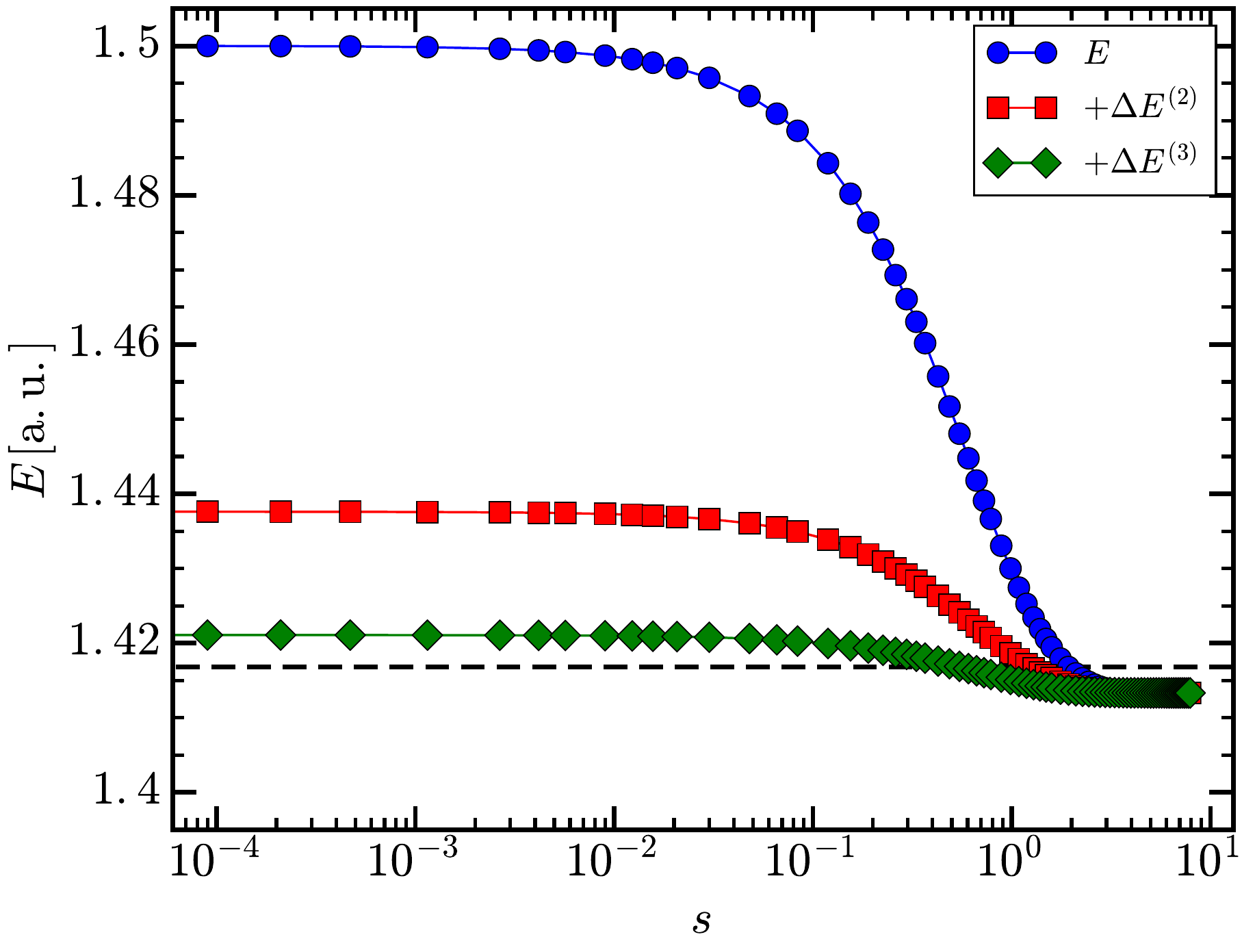}}
  \end{picture}
  \end{center}
  \vspace{-5pt}

  \caption{\label{fig:imsrg_mbpt}IMSRG(2) flow for the ground state of the pairing 
  Hamiltonian with $\delta=1.0, g=0.5$ (cf.~Sec.~\ref{sec:srg_pairing_example}).
  Calculations were performed with the White generator, Eq.~\eqref{eq:eta_white}.
  The figures shows the flowing ground-state energy $E(s)$ plus perturbative second and
  third-order energy corrections for $\HO(s)$. The exact ground-state energy is 
  indicated by the dashed line. }
\end{figure}

A recurring theme of this chapter has been the transformation of 
nuclear Hamiltonians to a shape that facilitates their subsequent application
in many-body calculations. We have stressed this point in our discussion
of the free-space SRG, in particular (see Sec.~\ref{sec:srg_nn}), but 
it applies to the IMSRG (or the MR-IMSRG) as well. Recall our application 
of the IMSRG to the pairing Hamiltonian in Sec.~\ref{sec:imsrg_pairing},
where we primarily focused on how correlations that are usually probed by 
perturbative corrections are shuffled into the flowing ground-state energy 
$E(s)$ in the limit $s\to\infty$ (see Fig.~\ref{fig:imsrg_mbpt}). 
We can also interpret the results shown in this figure in a slightly different way: At 
any given value of $s$, $E(s)$ would result from a simple HF
calculation with the Hamiltonian $\HO(s)$, which has absorbed correlations
because of an RG improvement. The result of the simple HF calculation approaches
the exact result as we evolve, aside from truncation errors, of course.
The same is true for the MBPT(2) and MBPT(3) calculations, summing $E(s)$ 
plus perturbative corrections through the indicated order. Of course, these
approaches are already closer to the exact result in the first place. 

This example illustrates the potential benefits of using Hamiltonians that
have been improved through IMSRG evolution as input for other many-body 
methods. In this section, we will briefly discuss applications in the 
traditional nuclear Shell model, which will give us access to a wealth 
of spectroscopic observables like excitation energies and transition rates. 
We will also look at the use of IMSRG Hamiltonians in Equation-of-Motion 
methods, which are an alternative approach to the computation of excited-state 
properties.

%
%
\subsubsection{Non-Empirical Interactions for the Nuclear Shell Model}
In IMSRG ground-state applications, we use the RG flows to decouple a 
suitable reference state from $npnh$ excitations (see Sec.~\ref{sec:imsrg_decoupling}). 
From a more general perspective, we can view this as a decoupling of
different \emph{sectors} of the many-body Hilbert space by driving the 
couplings of these sectors to zero. We are not forced to restrict the
decoupling to a single state, but could target multiple states at once
\cite{Tsukiyama:2012fk,Parzuchowski:2016pi,Hergert:2017kx} --- all we
need to do is tailor our definition of the off-diagonal Hamiltonian
to the problem, as in all SRG and IMSRG applications!

\begin{figure}[t]
  \setlength{\unitlength}{0.3\textwidth}
  \begin{center}
  \begin{picture}(1.2000,0.9000)
    \put(0.1000,0.0000){\includegraphics[width=\unitlength]{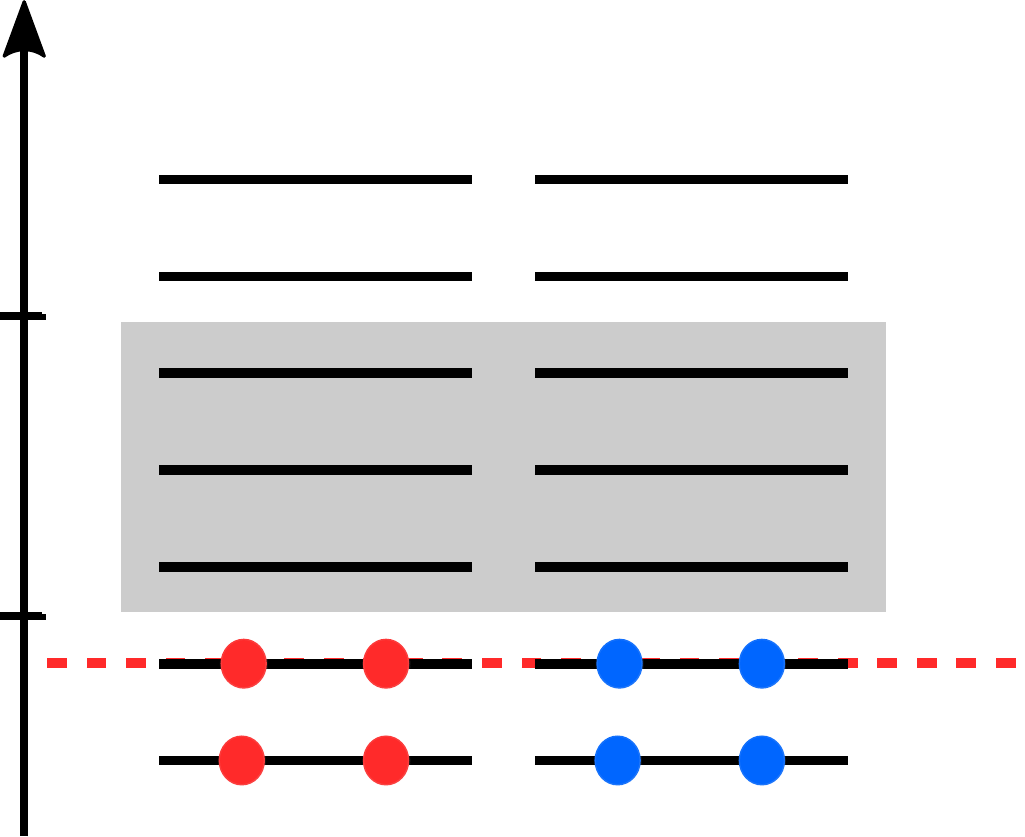}}
    \put(0.0000,0.0900){$h$}
    \put(0.0000,0.3400){$v$}
    \put(0.0000,0.5900){$q$}
    \put(1.1200,0.1500){$\varepsilon_F$}
  \end{picture}
  \end{center}

  \caption{\label{fig:valence_space}
    Separation of the single-particle basis into hole (h), valence particle (v)
    and non-valence particle (q) states. The Fermi energy of the fully occupied
    core, $\varepsilon_F$, is indicated by the red dashed line.
  }
\end{figure}
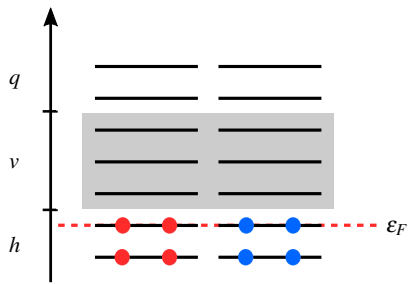

In the nuclear Shell model, we split the single-particle basis in our 
calculation into core or hole ($h$), valence particle ($v$) and non-valence 
particle ($q$) orbitals (see Fig.~\ref{fig:valence_space})\footnote{In
quantum chemistry, what we call a valence space is usually refereed to as
the \emph{active space}.}. The actual many-body calculation for a nucleus
with $A$ nucleons is an exact diagonalization of the Hamiltonian matrix in a 
subspace of the Hilbert space that is spanned by configurations 
of the form
\begin{equation}\label{eq:def_configurations}
  \ket{\aaO_{v_1}\ldots\aaO_{v_{A_v}}} \equiv \aaO_{v_1}\ldots\aaO_{v_{A_v}}\ket{\Phi}\,,
\end{equation}
where $\ket{\Phi}$ is the wave function for a suitable core with $A_c$
nucleons, and the $A_v$ valence nucleons are distributed over the valence
orbitals $v_i$ in all possible ways. Since the core is assumed to be inert, 
it can be viewed as a vacuum state for the valence configurations. The matrix 
representation of the Hamiltonian in the space spanned by these configurations 
is
\begin{equation}
  \matrixe{v'_{1}\ldots v'_{A_v}}{\HO}{v_{1}\ldots v_{A_v}}
  = \matrixe{\Phi}{\aO_{v'_{A_v}}\ldots\aO_{v'_1}\HO\aaO_{v_1}\ldots\aaO_{v_{A_v}}}{\Phi}\,.
\end{equation}
This expression suggests that we normal order the Hamiltonian and other 
operators with respect to the core wave function $\ket{\Phi}$, which can
be obtained from a simple spherical HF calculation. The state $\ket{\Phi}$
takes on the role of the reference state for the IMSRG flow, but recent
studies have shown that choosing either individual configurations for the 
target nucleus or ensembles of configurations as references will reduce 
truncation errors due to omitted induced terms, see 
\cite{Stroberg:2016fk,Stroberg:2016th}. When the resulting valence-space
interactions are used to calculate nuclear ground-state energies, we find
excellent agreement with direct IMSRG ground-state calculations, which 
indicates that the introduction of the inert core is justified, at least
for the used $NN+3N$ forces with low resolution scales $\lambdaSRG$.

\begin{figure}[t]
  \begin{center}
    \includegraphics[width=0.78\textwidth]{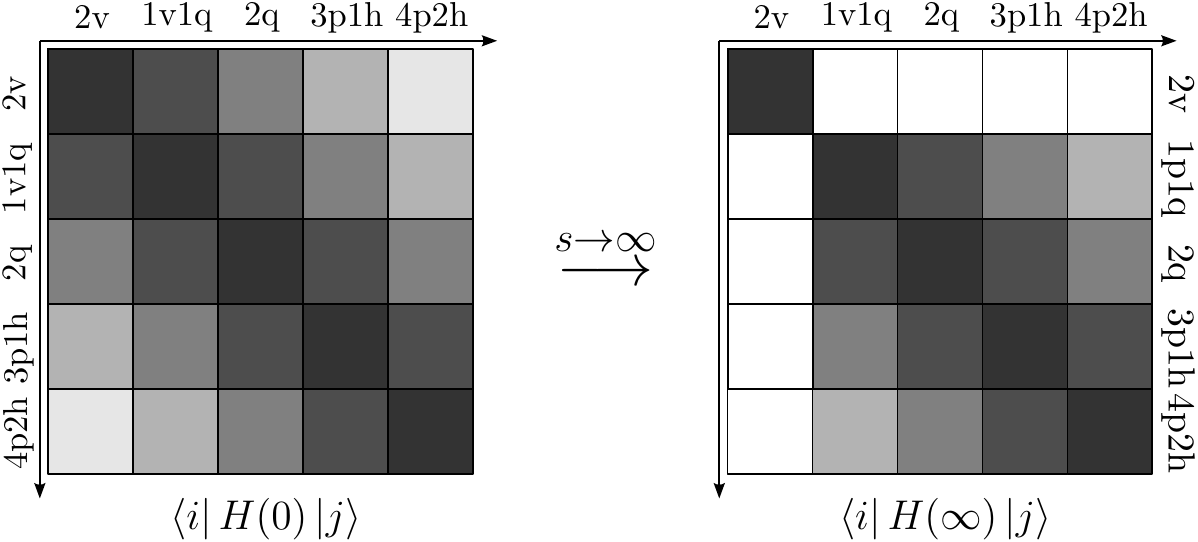}
  \end{center}

  \caption{\label{fig:valence_decoupling}
    Schematic view of IMSRG valence-space decoupling for two valence nucleons (p$\,=\,$v,\,q).
  }
\end{figure}

We want to use the IMSRG evolution to decouple the configurations
\eqref{eq:def_configurations} from states that involve excitations 
of the core, just as in the ground-state calculations. In addition, 
we need to decouple them from states containing nucleons in non-valence 
particle states (see Fig.~\ref{fig:valence_decoupling}). Working in 
IMSRG(2) truncation, i.e., assuming up to two-body terms in $\HO(s)$, 
we can identify the matrix elements that couple pairs of valence-space 
particles to $1q1v$, $2q$, $3p1h$, and $4p2h$ excitations, respectively, 
where $p=v,q$. For each type of matrix element, we show the antisymmetrized 
Goldstone diagrams \cite{Shavitt:2009} that represent the excitation process. 
Additional diagrams due to permutations of the nucleons or Hermitian 
adjoints are suppressed for brevity.

Diagrams (I) and (II) are eliminated if matrix elements of $f$ and 
$\Gamma$ that contain at least one q index are chosen to be off-diagonal. 
Diagrams (III) and (V) are eliminated by decoupling the reference-state, i.e.,
the core, which requires $f^{p}_{h}$ and $\Gamma^{pp'}_{hh'}$ to be off-diagonal 
(cf.~Sec.~\ref{sec:imsrg_decoupling}). This only leaves diagram (IV), 
which vanishes if matrix elements of the type $\Gamma^{pp'}_{vh}$ 
vanish. Thus, we define \cite{Tsukiyama:2012fk,Hergert:2017kx}
\begin{equation}\label{eq:def_Hod}
  H_{od} \equiv \sum_{i\neq i'}f^{i}_{i'}\nord{\AO^{i}_{i'}}
           + \frac{1}{4}\left(\sum_{pp'hh'}\Gamma^{pp'}_{hh'}\nord{\AO^{pp'}_{hh'}}
           + \sum_{pp'vh}\Gamma^{pp'}_{vh}\nord{\AO^{pp'}_{vh}}
           + \sum_{pqvv'}\Gamma^{pq}_{vv'}\nord{\AO^{pq}_{vv'}}\right)
           + \text{H.c.}\,.
\end{equation}
This definition of the off-diagonal Hamiltonian holds for an arbitrary
number of valence particles $A_v$. For $A_v=1$, diagram (II) vanishes, 
while diagrams (I) and (III)-(V) have the same topology, but one 
less spectator nucleons. Analogously, diagrams (I)-(V) merely contain 
additional spectator nucleons for $A_v>2$.

\begin{table*}[t]
  \setlength{\unitlength}{0.1\textwidth}
  \begin{tabular*}{\textwidth}{lm{0.16\textwidth}m{0.2\textwidth}m{0.45\textwidth}}
    \hline\hline
     no. & type & diagram & energy difference $\Delta$ \\
    \hline\\[-5pt]
     I & $\matrixe{2p}{\HO}{2p}$ & 
    \begin{picture}(2.0000,1.3000)
      \put(0.0000,0.1500){\includegraphics[height=\unitlength]{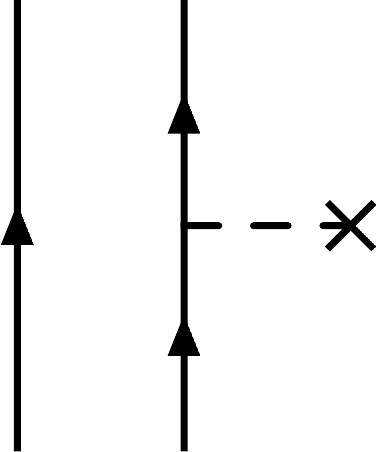}}
      \put(0.3750,1.2000){\footnotesize$p$}
      \put(0.3750,0.0000){\footnotesize$p'$}
    \end{picture}
    & 
    $f^{p}_{p} - f^{p'}_{p'}$\\
    II & $\matrixe{2p}{\HO}{2p}$ & 
    \begin{picture}(2.0000,1.4000)
      \put(0.0000,0.1500){\includegraphics[height=\unitlength]{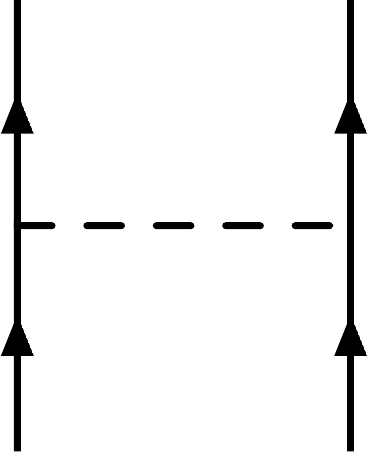}}
      \put(0.0150,1.2000){\footnotesize$p$}
      \put(0.7400,1.2000){\footnotesize$p'$}
      \put(0.0150,0.0000){\footnotesize$p''$}
      \put(0.7400,0.0000){\footnotesize$p'''$}
    \end{picture}
     & 
    $f^{p}_{p} + f^{p'}_{p'} - f^{p''}_{p''} - f^{p'''}_{p'''} + \Gamma^{pp'}_{pp'} - \Gamma^{p''p'''}_{p''p'''}$\\
    III & $\matrixe{3p1h}{\HO}{2p}$ &
    \begin{picture}(2.0000,1.4000)
      \put(0.0000,0.1500){\includegraphics[height=\unitlength]{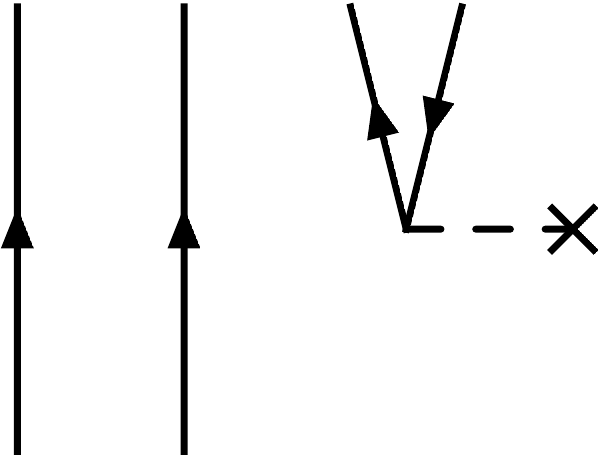}}
      \put(0.7400,1.2000){\footnotesize$p$}
      \put(0.9600,1.2000){\footnotesize$h$}
  \end{picture}
    & 
    $f^{p}_{p} - f^{h}_{h} - \Gamma^{ph}_{ph}$
    \\
    IV & $\matrixe{3p1h}{\HO}{2p}$ &
    \begin{picture}(2.000,1.4000)
      \put(0.0000,0.1500){\includegraphics[height=\unitlength]{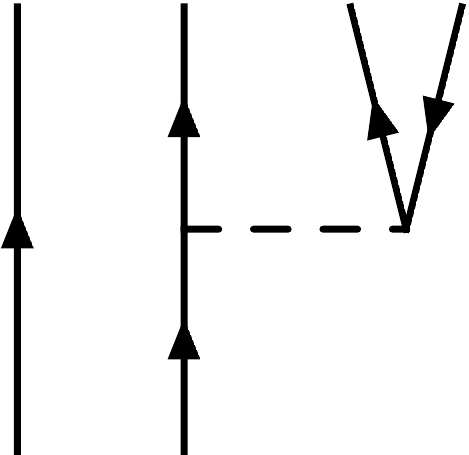}}
      \put(0.3750,1.2000){\footnotesize$p$}
      \put(0.7400,1.2000){\footnotesize$p'$}
      \put(0.9600,1.2000){\footnotesize$h$}
      \put(0.3750,0.0000){\footnotesize$p''$}
    \end{picture}
    & 
    $f^{p}_{p} + f^{p'}_{p'} - f^{p''}_{p''}-f^{h}_{h} + \Gamma^{pp'}_{pp'}-\Gamma^{ph}_{ph}-\Gamma^{p'h}_{p'h}$ 
    \\
    V & $\matrixe{4p2h}{\HO}{2p}$ &
    \begin{picture}(2.000,1.4000)
      \put(0.0000,0.1500){\includegraphics[height=\unitlength]{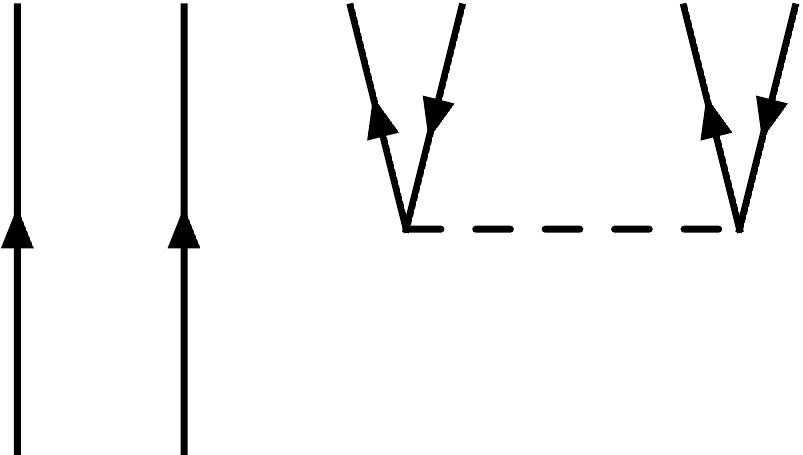}}
      \put(0.7400,1.2000){\footnotesize$p$}
      \put(0.9600,1.2000){\footnotesize$h$}
      \put(1.4800,1.2000){\footnotesize$p'$}
      \put(1.7000,1.2000){\footnotesize$h'$}
    \end{picture}
    &
    $f^{p}_{p} + f^{p'}_{p'} - f^{h}_{h} - f^{h'}_{h'} 
    + \Gamma^{pp'}_{pp'} + \Gamma^{hh'}_{hh'} - \Gamma^{ph}_{ph} 
    - \Gamma^{p'h'}_{p'h'} - \Gamma^{ph'}_{ph'}-\Gamma^{p'h}_{p'h}$
    \\ 
    \hline\hline
  \end{tabular*}
  \caption{\label{tab:diagrams} 
    Classification of matrix elements of the many-body Hamiltonian 
    in the many-body Hilbert space spanned by $(n+2)$p$n$h excitations 
    of the reference state (cf.~Fig.~\ref{fig:valence_decoupling}). For 
    each matrix element, we show the corresponding antisymmetrized 
    Goldstone diagrams \cite{Shavitt:2009} involving the one- and 
    two-body parts of $\HO$ (permutations involving spectator particles 
    which are required by antisymmetry are implied), as well as the 
    energy differences appearing in the matrix elements for $\eta(s)$ 
    in each case (see text).
  }
\end{table*}

Using $\HO_{od}$ in the construction of generators, we evolve the 
Hamiltonian by solving the IMSRG(2) flow equations \eqref{eq:imsrg2_m0b}--\eqref{eq:imsrg2_m2b}. 
The evolved Hamiltonian is given by
\begin{equation}
 \HO(\infty) = E + \sum_{v}f^{v}_{v}\nord{\AO^{v}_{v}} + \frac{1}{4}\sum_{v_i,v_j,v_k,v_l}
  \Gamma^{v_iv_j}_{v_kv_l}\nord{\AO^{v_iv_j}_{v_kv_l}} + \ldots\,,
\end{equation}
where the explicitly shown terms are the core energy, single-particle
energies, and two-body matrix elements that are used as input for 
a subsequent Shell model diagonalization. 

A possible subtlety is associated with the treatment of the mass-number 
dependence of the intrinsic Hamiltonian \eqref{eq:def_Hint}. We interpret 
it as a dependence on the mass-number \emph{operator} $\AO$, which acts 
on the many-body states on which we are operating, i.e., configurations 
in the \emph{target} nucleus, and therefore the mass number of the target
should be used in the intrinsic Hamiltonian at all stages of a calculation
\cite{Hergert:2009wh,Stroberg:2016fk,Stroberg:2016th}. This is appropriate 
because the combined IMSRG + Shell model calculation is supposed to approximate 
the results of an exact diagonalization for that particular nucleus.

The naive computational scaling for the valence-decoupling procedure 
described here is $\OC(N^6)$, just like that of IMSRG(2) ground-state 
calculations. On 2015/16 computing hardware, typical evolutions require
about 100-1000 core hours, assuming a single major shell as the valence
space. The Shell model calculation is typically less expensive in that
case. However, it will start to dominate the computational scaling as 
soon as we have to consider extended valence spaces consisting of two
or more major shells, because of the factorial growth of 
the Shell model basis (cf.~Sec.~\ref{sec:srg_pairing}).

\begin{figure}[t]
  \begin{center}
    \includegraphics[width=0.98\textwidth]{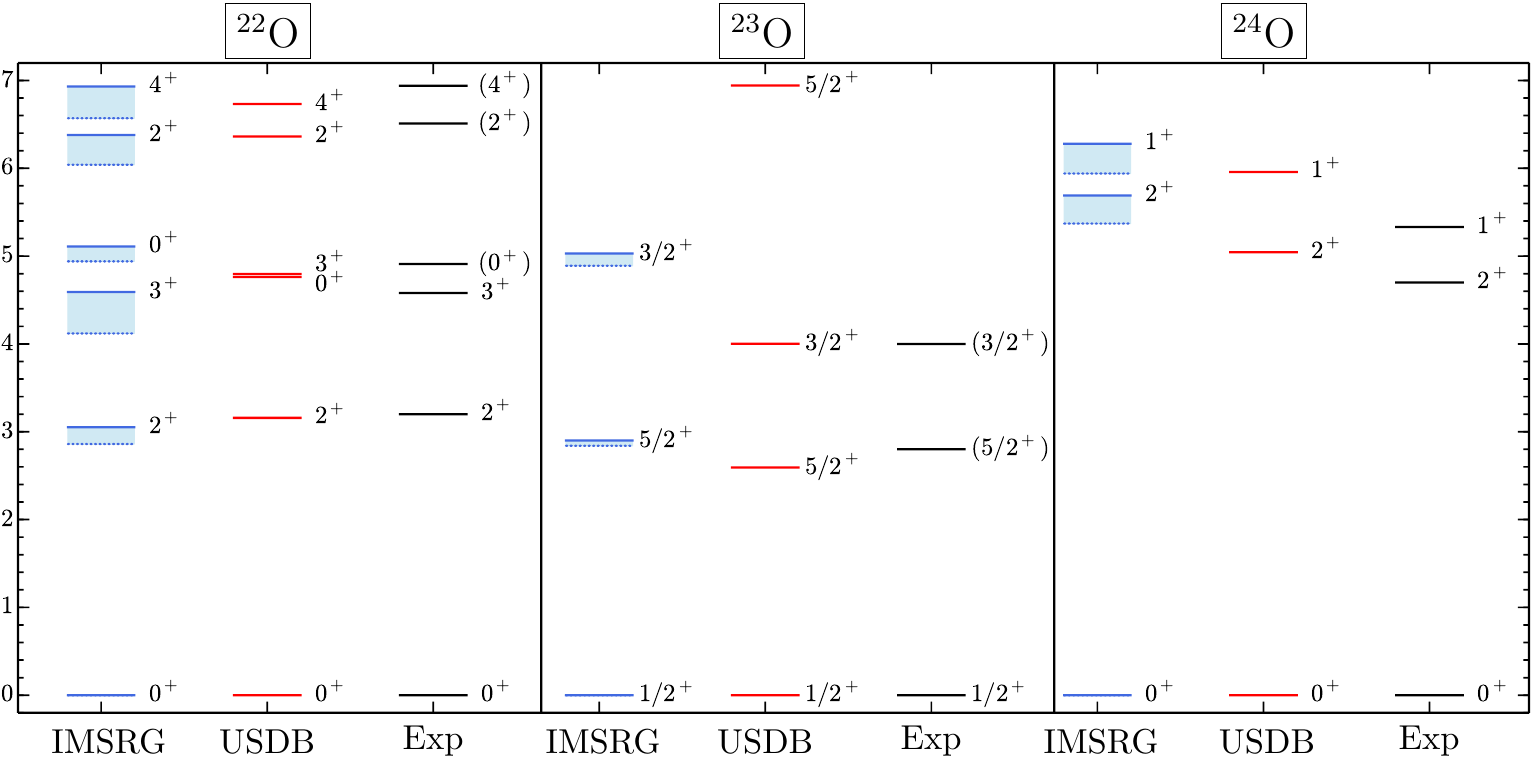}
  \end{center}

	\caption{Excited-state spectra of $^{22,23,24}$O based on the chiral
	$NN+3N(400)$ interaction with $\lambdaSRG=1.88\,\fmi$, compared to 
	results with the phenomenological USDB interaction \cite{Brown:2006fk}
	and experimental data, see \cite{Bogner:2014tg,Hergert:2017kx} for 
	full details. Dotted and solid lines are results for $\hbar \omega=20\,\MeV$
	and $\hbar \omega=24\,\MeV$, respectively, which are shown as an indicator
	of convergence. \label{fig:Ospectra}}
\end{figure}

As an example, Fig.~\ref{fig:Ospectra} shows the low-lying excitation 
spectra of $\nuc{O}{22-24}$ from Shell model calculations with IMSRG-derived
valence-space interactions. These interactions were generated from an 
underlying chiral $NN+3N$ Hamiltonian consisting of the
\NNNLO{} $NN$ interaction by Entem and Machleidt with $\Lambda_{NN}=500\,\MeV$,
and an \NNLO{} $3N$ interaction with $\Lambda_{3N}=400\,\MeV$, which has
been evolved to $\lambdaSRG=1.88\,\fmi$ \cite{Entem:2003th,Machleidt:2011bh,Gazit:2009qf,Roth:2011kx}
(also cf.~Secs.~\ref{sec:srg_nn},\ref{sec:current_mrimsrg}). We compare
our results to the gold-standard phenomenological USDB interaction by
Brown and Richter, which describes more than 600
ground-state and excitation energies in $sd-$shell ($8\leq Z,N\leq 20$) 
nuclei with an rms deviation of merely $\sim130\,\keV$ \cite{Brown:2006fk},
and to experimental data. In the nuclei shown, the agreement is quite
satisfactory given that the chiral input Hamiltonian is entirely fixed
by $A\leq4$ data. The IMSRG interactions turn out to perform quite well 
in the entire lower $sd-$shell, achieving an rms deviation of $\sim580\,\keV$
in about 150 states \cite{Stroberg:2016fk,Hergert:2017kx}. The chiral 
$3N$ interactions are found to be of crucial importance for the correct
reproduction of level orderings and spacings. 

%
%
\subsubsection{Equation-of-Motion Methods}

Equation-of-Motion (EOM) methods \cite{Rowe:1968eq} are a useful alternative
to the Shell model when it comes to the calculation of excited states,
in particular when extended valence spaces lead to prohibitively large
Shell model basis dimensions. In these approaches, the Schr\"odinger
equation is rewritten in terms of ladder operators
that create excited eigenstates from the exact ground state:
\begin{equation}\label{eq:schroedinger_eom}
  \HO\ket{\Psi_n} = E_n\ket{\Psi} \qquad\longrightarrow\qquad \HO\XXO_n\ket{\Psi_0} = E_n\XXO_n\ket{\Psi_0}\,.
\end{equation}
Formally, $\QQO_n$ is given by the dyadic product $\ket{\Psi_n}\bra{\Psi_0}$,
and by thinking of the exact eigenstates in a CI sense, it is easy to see that 
they can be expresed as a linear combination of up to $A$-body excitation and 
de-excitation operators acting on the ground state. We can further rewrite
Eq.~\eqref{eq:schroedinger_eom} as the Equation of Motion
\begin{equation}\label{eq:eom}
  \comm{\HO}{\XXO_n}\ket{\Psi_0} = (E_n-E_0)\XXO_n\ket{\Psi_0} \equiv \omega_n \XXO_n\ket{\Psi_0}\,,
\end{equation}
and introduce systematic approximations to the $\XXO_n$ and the ground-state 
$\ket{\Psi_0}$. For example, by replacing $\ket{\Psi_0}$ with a simple Slater 
determinant and using the ansatz 
\begin{equation}
  \XXO_n = \sum_k X_{nk}\aaO_k\,, 
\end{equation}
we obtain Hartree-Fock theory, for
\begin{equation}
  \XXO_n = \sum_{ph} X^{(n)}_{ph}\aaO_p\aO_h\,, 
\end{equation}
we have the Tamm-Dancoff Approximation (TDA) for excited states, and
\begin{equation}\label{eq:rpa_ladder}
  \XXO_n = \sum_{ph} X^{(n)}_{ph}\aaO_p\aO_h - Y^{(n)}_{ph}\aaO_h\aO_p\,, 
\end{equation}
yields the Random Phase Approximation (RPA) in quasi-boson approximation
\cite{Ring:1980bb,Suhonen:2007wo}. Plugging the Slater determinant reference
state and the ansatz for the ladder operators into the Eq.~ 
\eqref{eq:eom}, we end up with a regular or a generalized eigenvalue problem, 
which we solve for the amplitudes appearing in the $\XXO_n$ operators. Since 
the computed amplitudes can be used to improve the ground-state ansatz, it is
usually possible to construct self-consistent solutions of the EoM \eqref{eq:eom}
in a given truncation.

Since we have casually referred to the approximate ground state as a 
reference state already, it will not come as a surprise to our readers
that we can quite naturally combine EOM methods with the IMSRG. Per construction,
the reference state $\ket{\Phi_0}$ will be the ground state of the final IMSRG Hamiltonian 
\begin{equation}
  \Hfinal\equiv\UO(\infty)\HO(0)\UUO(\infty)
\end{equation}. Multiplying Eq.~\eqref{eq:eom} by $\UO(\infty)$ and recalling that 
\begin{equation}
  \UO(\infty)\ket{\Psi_0} = \ket{\Phi_0}\,,
\end{equation}
we obtain the unitarily transformed EOM
\begin{equation}\label{eq:eom_imsrg}
  \comm{\Hfinal}{\overline{X}^\dag_n}\ket{\Phi_0} = \omega_n \overline{X}^\dag_n\ket{\Phi_0}\,.
\end{equation}
The solutions $\overline{X}^\dag_n$ can be used to obtain the eigenstates 
of the unevolved Hamiltonian via
\begin{equation}
  \ket{\Psi_n} = \UUO(\infty)\overline{X}^\dag_n\ket{\Phi_0}\,.
\end{equation}

In current applications, we include up to $2p2h$ excitations in the ladder 
operator \cite{Parzuchowski:2016pi}:
\begin{equation}\label{eq:eom_imsrg_ladder}
  \overline{X}^\dagger_n = \sum_{ph} \bar{X}^{(n)}_{ph} \nord{\aaO_{p}\aO_{h}} + \frac{1}{4}\sum_{pp'hh'} \bar{X}^{(n)}_{pp'hh'} \nord{\aaO_{p}\aaO_{p'}\aO_{h'}\aO_{h}} \,.
\end{equation}
Note that the operator only contains excitation operators because de-excitation 
operators annihilate the reference state $\ket{\Phi_0}$ and therefore do not 
contribute in the EOM:
\begin{equation}
   \comm{\nord{\aaO_{h}\aaO_{h'}\aO_{p'}\aO_{p}}}{\overline{H}}\ket{\Phi_0}
   = (\overline{H} - E)\nord{\aaO_{h}\aaO_{h'}\aO_{p'}\aO_{p}}\ket{\Phi_0} = 0\,.
\end{equation}
The ladder operator $\overline{X}^\dag$ can be systematically improved by including higher
particle-hole excitations, until we reach the $ApAh$ level which would amount 
to an exact diagonalization of $\Hfinal$. Denoting the operator rank of the 
ladder operator by $m$ and the IMSRG truncation by $n$, we refer to a specific
combined scheme as EOM-IMSRG($m$,$n$). 

Like TDA and RPA, the EOM \eqref{eq:eom_imsrg} can be implemented as an eigenvalue 
problem that can be tackled with the Lanczos-Arnoldi or Davidson algorithms 
\cite{Lanczos:1950sp,Arnoldi:1951kk,Davidson:1989pi} that are also used in CI
approaches. As mentioned throughout this chapter, these algorithms only require
knowledge of matrix-vector products. In the EOM-IMSRG, the required product is
identified once we realize that the commutator in Eq.~\eqref{eq:eom_imsrg} can
be rewritten as the connected product of $\Hfinal$ and $\overline{X}^\dag$:
\begin{equation}\label{eq:mat_vec_prod}
  \comm{\Hfinal}{\overline{X}^\dag_n} = \{\Hfinal\overline{X}^\dag_n\}_C\,.
\end{equation}
Thus, the matrix-vector product can be computed with the same commutator routines 
that are used in the evaluation of the IMSRG flow equations. Thus, the scaling of
EOM-IMSRG($m$,$n$) is the same as that of the IMSRG($\max{m,n}$), e.g., $\OC(N^6)$
for the EOM-IMSRG(2,2) scheme that is the method we primarily use at this point 
\cite{Parzuchowski:2016pi}. 

\begin{figure}[t]
  \setlength{\unitlength}{\textwidth}
  \begin{picture}(1.0000,0.5000)
    \put(0.1500,0.0000){\includegraphics[height=0.51\unitlength]{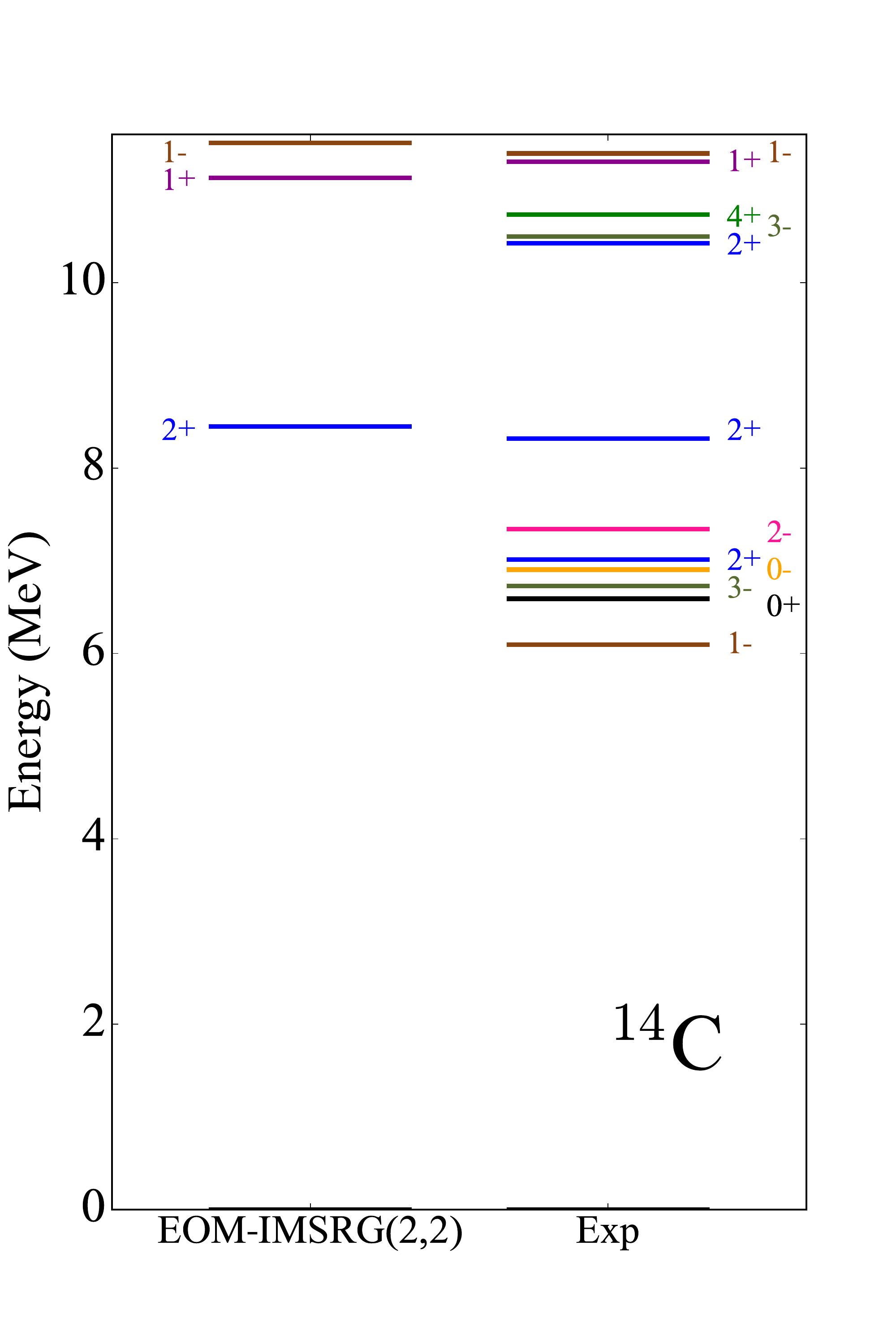}}
    \put(0.5000,0.0000){\includegraphics[height=0.51\unitlength]{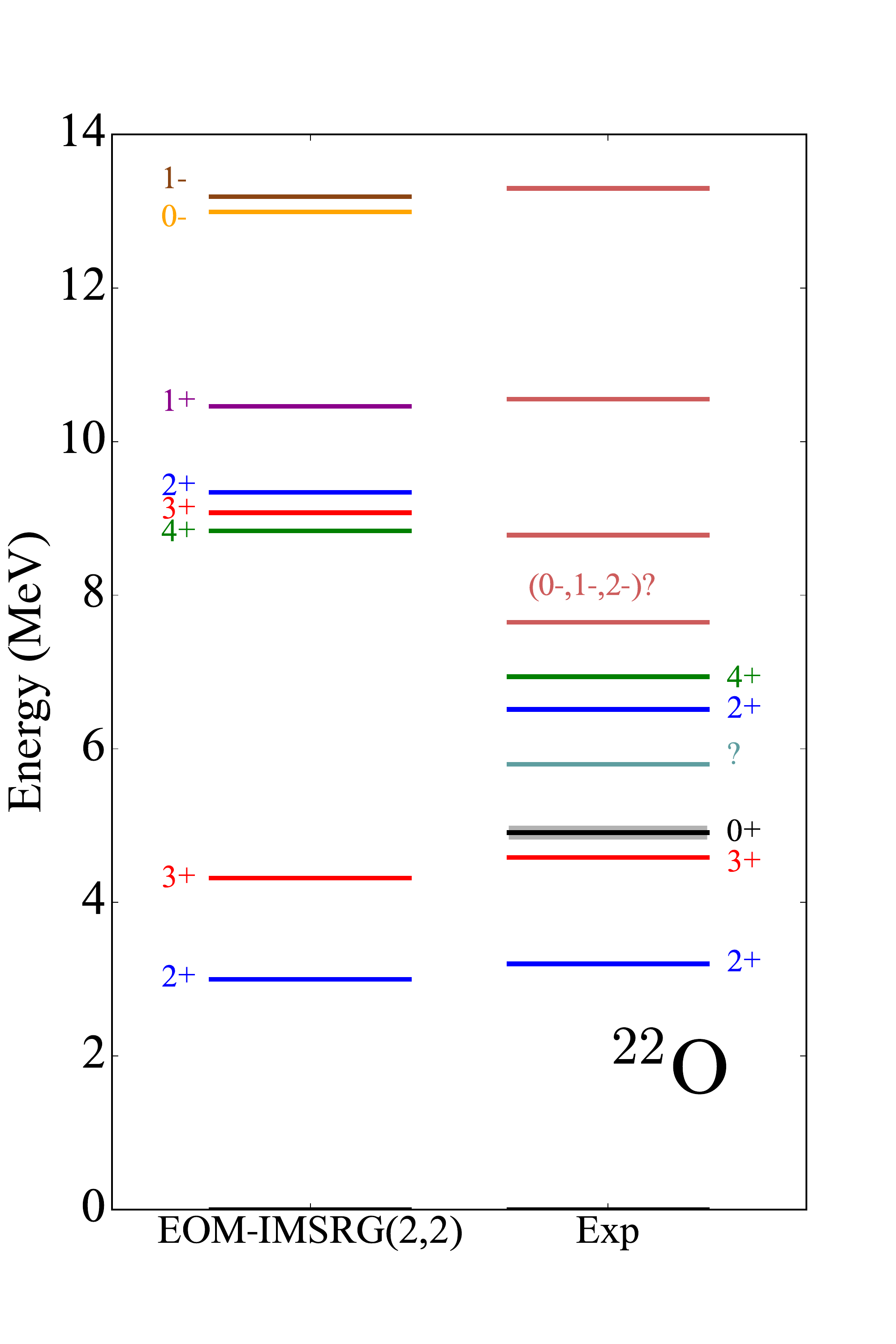}}
  \end{picture}

  \caption{\label{fig:eom}EOM-IMSRG(2,2) excitation spectra of $\nuc{C}{14}$ and $\nuc{O}{22}$,
  calculated with the chiral $NN+3N(400)$ Hamiltonian for $\lambdaSRG=1.88\,\fmi$ (
  cf.~Fig.~\ref{fig:Ospectra}), compared with experimental data.}
\end{figure}

In Fig.~\ref{fig:eom}, we show sample spectra of $\nuc{C}{14}$ and $\nuc{O}{22}$
from EOM-IMSRG(2,2) calculations with the chiral $NN+3N(400)$ Hamiltonian 
($\lambdaSRG=1.88\,\fmi$). We see that certain levels seem to be reproduced
quite well, while experimentally observed states below $8\,\MeV$ are either 
missing in the calculation in the case of $\nuc{C}{14}$), or found at higher 
excitation theoretical excitation energy in the case of $\nuc{O}{22}$. In the
latter case, we can compare the EOM-IMSRG(2,2) calculation to the IMSRG+SM
spectrum shown in Fig.~\ref{fig:Ospectra}, in which the group of $2^+,3^+,4^+$
states is found much closer to their experimental counterparts. This suggests that
the overestimation of the excitation energy in the EOM approach is caused by the
truncation of the ladder operator at the $2p2h$ level, while the Shell model
contains all allowed $npnh$ excitations in the valence space. Conversely, we can
conclude that the states that the EOM-IMSRG(2,2) is reproducing well are dominated
by $1p1h$ and $2p2h$ excitations \cite{Parzuchowski:2016pi}. 

An advantage of the EOM-IMSRG(2,2) over the IMSRG+SM is that
it allows us to compute negative parity states, to the level of accuracy
that the truncation allows. In the Shell model, the description of such
states requires a valence space consisting of two major shells, which 
usually makes the exact diagonalization prohibitively expensive, especially
in $sd-$shell nuclei and beyond.

\subsection{\label{sec:current_remarks}Final Remarks}
Superficially, the discussion in Sec.~\ref{sec:current} focused on 
three different subject areas, namely Magnus methods, MR-IMSRG, and the 
construction of effective Hamiltonians, but we expect that our readers
can already tell that these directions are heavily intertwined. 

The Magnus formulation of the IMSRG will greatly facilitate the evaluation 
of general observables not just in the regular single-reference version of
the method, but also in the MR-IMSRG. Magnus methods can also be readily 
adapted to the construction of effective valence-space operators besides 
the nuclear interaction, e.g., radii and electroweak transitions. In addition, 
it allows us to construct systematic and computationally tractable approximations 
to the full IMSRG(3) \cite{Morris:2016xp,Parzuchowski:2016pi}, similar to 
the non-iterative treatment of triples in Coupled Cluster methods 
\cite{Taube:2008kx,Taube:2008vn,Piecuch:2005dp,Binder:2013fk}.

While the current version of our EOM technology is based on the single-reference
IMSRG, we will formulate a multireference EOM scheme based on final 
Hamiltonians from the MR-IMSRG evolution next. Such a MR-EOM-IMSRG
will hold great potential for the description of excitations in 
deformed or weakly bound nuclei, which would require excessively 
large valence spaces in traditional CI approaches, and at the same
time require an explicit treatment of static correlations that the
uncorrelated reference wave functions used in the IMSRG and EOM-IMSRG
cannot provide. 

The MR-IMSRG has also recently been used to pre-diagonalize Hamiltonians
that serve as input for the No-Core Shell Model (NCSM), merging the two 
approaches into an iterative scheme that we call the In-Medium NCSM (IM-NCSM 
for short) \cite{Gebrerufael:2016rp}. In this combined approach, we can
converge NCSM results in model spaces whose dimensions are orders of
magnitude smaller than those of the regular NCSM or its importance-truncated
variant \cite{Roth:2007fk,Roth:2009eu,Roth:2014fk}, which should significantly
expand its range of applicability. A similar idea might be used fruitfully
in the traditional Shell model with a core as well, where the use of 
statically correlated reference states and an MR-IMSRG evolution might
help to overcome shortcomings in the Shell model interactions for specific
nuclei.
\section{\label{sec:conclusions}Conclusions}
In this chapter, we have presented a pedagogical introduction to the
SRG and IMSRG, and discussed their applications in the context of the
nuclear many-body problem. The former has become maybe the most popular
tool for pre-processing nuclear interactions and operators, leading to
vast improvements in the rate of convergence of many-body calculations,
and extending the range of nuclei that can be tackled in \emph{ab initio}
approaches \cite{Barrett:2013oq,Hagen:2014ve,Binder:2014fk,Soma:2014eu,
Hergert:2016jk,Hergert:2017kx}. The IMSRG implements SRG concepts directly
in the $A$-body system, relying on normal-ordering methods to control
the size of induced operators and make systematic truncations feasible.
As we have demonstrated through a variety of applications, the IMSRG
is an extremely versatile and powerful addition to the canon of 
quantum many-body methods. In Sec.~\ref{sec:current}, we have given
an overview of the main thrusts of current IMSRG research, and we
hope that our readers will be inspired to contribute to these developments,
or find ways in which the IMSRG framework can be useful to their
own research programs. 

The explicit RG aspect of the IMSRG framework is a unique feature
that sets it apart from most other many-body methods on the market. 
When the comparison with 
those other methods and experimental data is our first and foremost 
concern, we are primarily interested in the $s\to\infty$ limit of the 
IMSRG or MR-IMSRG evolution, but the flow trajectory is an enormous 
source of additional insight. By studying the flows, not just the
final fixed points, we can gain a new understanding of how many-body 
correlations are reshuffled between the wave function and the Hamiltonian, 
or different pieces of 
the Hamiltonian, making transparent what is only implicitly assumed 
in other methods. Like in the free-space SRG (or other RG methods), we 
have the freedom to work at intermediate values of $s$ if this is more 
practical than working at $s=0$ or in the limit 
$s\to\infty$, especially if either of these extremes would lead to the 
accrual of unacceptable numerical errors in our results (see, e.g., 
\cite{Gebrerufael:2016rp,Li:2015nq,Li:2016rm}). 
This is the inherent power of a framework that integrates many-body and 
renormalization group techniques, and the reason why we consider the IMSRG 
to be an extremely valuable tool for quantum many-body theory.


\begin{acknowledgement}
The authors are indebted to a multitude of colleagues for many stimulating 
discussions of the SRG and IMSRG that are reflected in this work. We are
particularly grateful to Angelo Calci, Thomas Duguet, Dick Furnstahl, 
Kai Hebeler, Morten Hjorth-Jensen, Jason Holt, Robert Roth, Achim Schwenk, 
Ragnar Stroberg, and Kyle Wendt.

The preparation of this chapter was supported in part by NSF Grant No.~PHY-1404159 
and the NUCLEI SciDAC Collaboration under the U.S.~Department of Energy Grant 
No.~DE-SC0008511. H.~H. gratefully acknowledges the National Superconducting 
Cyclotron Laboratory (NSCL)/Facility for Rare Isotope Beams (FRIB) and Michigan 
State University (MSU) for startup support during the preparation of this work.
Computing resources were provided by the MSU High-Performance Computing Center 
(HPCC)/Institute for Cyber-Enabled Research (iCER).
\end{acknowledgement}

\section{\label{sec:exercises_projects}Exercises and Projects}

\begin{prob}\label{problem:tint}
  \item[a)] Prove that the two forms of the intrinsic kinetic energy operator given in 
  Eqs.~\eqref{eq:def_Tint_1B2B} and \eqref{eq:def_Tint_2B} are equivalent.

  \item[b)] Now consider the expectation values of the two forms of $\Tint$
  in a state that does not have a fixed particle number, e.g., as in 
  Bardeen-Cooper-Schrieffer (BCS) \cite{Bardeen:1957bz,Bardeen:1957tv}
  or Hartree-Fock-Bogoliubov (HFB) theory (see, e.g., \cite{Ring:1980bb}). Expand
  the $\frac{1}{\AO}$ dependence of Eqs.~\eqref{eq:def_Tint_1B2B} and \eqref{eq:def_Tint_2B} 
  into series around $\expect{\AO}$ by introducing $\Delta\AO=\AO-\expect{\AO}$, 
  and compare the series expansions order by order.
  (A thorough discussion of the issue can be found in \cite{Hergert:2009wh}.)
\end{prob}

\begin{prob}\label{problem:nord}
  \item[a)] Prove that the expectation value of a normal-ordered operator in the
  reference state vanishes (Eq.~\eqref{eq:nord_ex}):
    \begin{equation}
      \dmatrixe{\Phi}{\nord{\aaO_{i_1}\ldots\aO_{j_1}}} = 0\,.
    \end{equation}
  Start by considering a one-body operator, and extend your result to the general
  case by induction.
  \item[b)] Show that $\aaO_i$ and $\aO_j$ anticommute freely in a normal-ordered
  string (Eq.~\eqref{eq:nord_acomm}). 
  \begin{equation}
    \nord{\ldots \aaO_i\aO_j\ldots} = -\nord{\ldots\aO_j\aaO_i\ldots}\,.
  \end{equation}
  Consider the one-body case first, as in problem \ref{problem:nord}(a).
  \item[c)] Prove the following schematic expression for products of normal-ordered
  operators:
    \begin{equation}
      \AO^{[M]}\BO^{[N]} = \sum_{k=|M-N|}^{M+N}\CO^{[k]}\,.
    \end{equation}
  \item[d)] Show that the following rule applies for commutators of normal-ordered
    operators:
    \begin{equation}
      \comm{\AO^{[M]}}{\BO^{[N]}} = \sum_{k=|M-N|}^{M+N-1}\CO^{[k]}\,.
    \end{equation}
    Thus, the largest particle rank appearing in the expansion of the commutator of 
    normal-ordered $M-$ and $N-$body operators is $M+N-1$.
  \item[e)] We can view the free-space operators as being normal-ordered with respect
    to the vacuum state. How are the expansion formulas for products and commutators
    modified in that case?
\end{prob}

\begin{prob}\label{problem:ph_orthogonality}
  Use Wick's theorem to show that the basis consisting of a Slater determinant $\ket{\Phi}$ 
  and its particle-hole excitations, 
  \begin{equation}
    \ket{\Phi},\,\nord{\aaO_{p}\aO_h}\ket{\Phi},\,\nord{\aaO_{p}\aaO_{p'}\aO_{h'}\aO_{h}}\ket{\Phi},\ldots\,,
  \end{equation}
  is orthogonal if the underlying single-particle basis is orthonormal.
\end{prob}

\begin{prob}\label{problem:wegner_pert}
  \item[a)]
    Validate the leading-order perturbative expression for the Wegner generator, Eq.~\eqref{eq:wegner2b_pert}.
  \item[b)]
    Let us now assume that we have used a Slater determinant reference state that 
    has \emph{not} been optimized by performing a Hartree-Fock calculation. Using
    Epstein-Nesbet partioning, the one-body part of the off-diagonal Hamiltonian
    is then counted as $f^{od}_{ij}=\OC(g)$ instead of $\OC(g^2)$ during the flow.
    Show that the one-body part of the Wegner generator has the following perturbative
    expansion in this case: 
    \begin{equation}\label{eq:wegner1b_pert}
      \eta_{ij} = \left(f^d_{ii} - f^d_{jj} - (n_i-n_j)\Gamma^d_{ijij}\right)f^{od}_{ij} + \OC(g^2)\,.
    \end{equation}
    Interpret the expression in the parentheses.
\end{prob}

\begin{prob}\textbf{Project: Optimization of the IMSRG(2) Code} \\
  In Sec.~\ref{sec:imsrg_implementation}, we mention several ways of optimizing 
  the performance of the Python code, like taking into account antisymmetry of
  two-body (and three-body) states, or exploiting symmetries and the resulting
  block structures.

  \item[a)] Optimize the storage requirements and speed of the Python code by taking 
  the antisymmetry of states as well as the Hermiticity (anti-Hermiticity) of $\HO$ ($\eta$) 
  into account.

  \item[b)] Identify the symmetries of the pairing Hamiltonian, and construct a
  variant of \texttt{imsrg\_pairing.py} that is explicitly block diagonal in
  the irreducible representations of the corresponding symmetry group.

  \item[c)] A significant portion of the code \texttt{imsrg\_pairing.py} consists of
  infrastructure routines that are used to convert between one- and two-body bases.
  We could avoid this inversion if we treat $\Gamma$ and the two-body part of the 
  generator as \emph{rank-four tensors} instead of matrices. \texttt{NumPy}
  offers tensor routines that can be used to evaluate tensor contractions and
  products, in particular \texttt{numpy.tensordot()}. Rewrite \texttt{imsrg\_pairing.py}
  in terms of tensors, and compare the performance of your new code to the
  original version.
\end{prob}

\begin{prob}
  \textbf{Project: IMSRG(3) for the Pairing Hamiltonian} \\
  Throughout this chapter, we used a single-particle basis consisting of only 
  8 states in our discussions of the pairing Hamiltonian. For such a small basis 
  size, it is possible to implement to include explicit 3N operators in the IMSRG
  evoltion, that is, to work in the IMSRG(3) scheme.

  \item[a)] Derive the IMSRG(3) flow equations. (Note: compare your results with
  Ref.~\cite{Hergert:2016jk}).

  \item[b)] Implement the IMSRG(3). The Python code discussed in Sec.~\ref{sec:imsrg_implementation}
  provides a good foundation, but you may find it necessary to switch to a 
  language like C/C++ or Fortran for performance reasons. 

  \item[c)] ``Interpolate'' between IMSRG(2) and IMSRG(3) by selectively activating
  flow equation terms, and document the impact of these intermediate steps.

\end{prob}

\section*{\label{app:products}Appendix: Products and Commutators of Normal-Ordered Operators}
\addcontentsline{toc}{section}{Appendix}
In this appendix, we collect the basic expressions for products and commutators of
normal-ordered one- and two-body operators. All single-particle indices refer to 
the natural orbital basis, where the one-body density matrix is diagonal
\begin{equation}
  \rho_{kl} = \matrixe{\Phi}{\aaO_{l}\aO_{k}}{\Phi}  = n_k \delta_{kl}\,,\quad n_k\in\{0,1\}\,,
\end{equation}
(notice the convention for the indices of $\rho$, cf.~\cite{Ring:1980bb}). We also
define the hole density matrix
\begin{equation}
  \bar\rho_{kl} \equiv \matrixe{\Phi}{\aO_{k}\aaO_{l}}{\Phi}  = \delta_{kl} - \rho_{kl} \equiv \nn_k\delta_{kl}
\end{equation}
whose eigenvalues are 
\begin{equation}
  \nn_{k}  = 1 - n_{k}\,,
\end{equation}
i.e., 0 for occupied and 1 for unoccupied single-particle states. Finally, 
we will again use the permutation symbol $\PO_{ij}$ to interchange the indices 
in any expression, 
\begin{equation}
  \PO_{ij} g(\ldots,i,\ldots,j) \equiv g(\ldots,j,\ldots,i)\,
\end{equation}
(see Sec.~\ref{sec:imsrg} and chapter 8).

\subsection*{Operator Products}
\begin{align}
  \nord{\aaO_{a}\aO_{b}}\nord{\aaO_{k}\aO_{l}}
  &=\nord{\aaO_{a}\aaO_{k}\aO_{l}\aO_{b}}
    -n_a\delta_{al}\nord{\aaO_{k}\aO_{b}}
    +\nn_b\delta_{bk}\nord{\aaO_{a}\aO_{l}}
    +n_a\nn_b\delta_{al}\delta_{bk}
\end{align}

\begin{align}
  \nord{\aaO_{a}\aO_{b}}\nord{\aaO_{k}\aaO_{l}\aO_{n}\aO_{m}}
  &=\nord{\aaO_{a}\aaO_{k}\aaO_{l}\aO_{n}\aO_{m}\aO_{b}}
    +\left(1-\PO_{mn}\right)n_a\delta_{an}\nord{\aaO_{k}\aaO_{l}\aO_{m}\aO_{b}}
    -\left(1-\PO_{kl}\right)\nn_b\delta_{bl}\nord{\aaO_{a}\aaO_{k}\aO_{n}\aO_{m}}\notag\\
  &\hphantom{=}  
    -\left(1-\PO_{kl}\right)\left(1-\PO_{mn}\right)n_{a}\nn_{b}\delta_{am}\delta_{bl}\nord{\aaO_{k}\aO_{n}}
\end{align}

\begin{align}
  &\nord{\aaO_{a}\aaO_{b}\aO_{d}\aO_{c}}\nord{\aaO_{k}\aaO_{l}\aO_{n}\aO_{m}}
  \notag\\
  &=\nord{\aaO_{a}\aaO_{b}\aaO_{k}\aaO_{l}\aO_{n}\aO_{m}\aO_{d}\aO_{c}}
    \notag\\
  &\hphantom{=}
    +(1-\PO_{ab})(1-\PO_{mn})n_a\delta_{am}\nord{\aaO_{b}\aaO_{k}\aaO_{l}\aO_{n}\aO_{d}\aO_{c}}
    -(1-\PO_{cd})(1-\PO_{kl})\nn_c\delta_{ck}\nord{\aaO_{a}\aaO_{b}\aaO_{l}\aO_{n}\aO_{m}\aO_{d}}
  \notag\\
  &\hphantom{=}
    +(1-\PO_{mn})n_an_b\delta_{am}\delta_{bn}\nord{\aaO_{k}\aaO_{l}\aO_{d}\aO_{c}}
    +(1-\PO_{cd})\nn_c\nn_d\delta_{ck}\delta_{dl}\nord{\aaO_{a}\aaO_{b}\aO_{n}\aO_{m}}
  \notag\\
  &\hphantom{=}
    +(1-\PO_{ab})(1-\PO_{cd})(1-\PO_{kl})(1-\PO_{mn})n_a\nn_c\delta_{am}\delta_{ck}\nord{\aaO_{b}\aaO_{l}\aO_{n}\aO_{d}}
  \notag\\
  &\hphantom{=}
  +(1-\PO_{ab})(1-\PO_{kl})(1-\PO_{mn})
    n_b \nn_c \nn_d \delta_{bn}\delta_{ck}\delta_{dl}
    \nord{\aaO_{a}\aO_{m}}
  \notag\\
  &\hphantom{=}
   +(1-\PO_{cd})(1-\PO_{kl})(1-\PO_{mn})
     \nn_d n_a n_b \delta_{dl}\delta_{an}\delta_{bm}
    \nord{\aaO_{k}\aO_{c}}
  \notag\\
  &\hphantom{=}
    + (1-\PO_{kl})(1-\PO_{mn})n_a n_b \nn_c \nn_d  \delta_{am}\delta_{bn}\delta_{ck}\delta_{dl}
\end{align}

\subsection*{Commutators}

\begin{align}
  \comm{\nord{\aaO_{a}\aO_{b}}}{\nord{\aaO_{k}\aO_{l}}}
  &=\delta_{bk}\nord{\aaO_{a}\aO_{l}}-\delta_{al}\nord{\aaO_{k}\aO_{b}}+(n_{a}-n_{b})\delta_{al}\delta_{bk}
\end{align}

\begin{align}
  \comm{\nord{\aaO_{a}\aO_{b}}}{\nord{\aaO_{k}\aaO_{l}\aO_{n}\aO_{m}}}
  &
    =\left(1-\PO_{kl}\right)\delta_{bk}\nord{\aaO_{a}\aaO_{l}\aO_{n}\aO_{m}}
    -\left(1-\PO_{mn}\right)\delta_{am}\nord{\aaO_{k}\aaO_{l}\aO_{n}\aO_{b}}
    \notag\\
  &\hphantom{=}  
    +\left(1-\PO_{kl}\right)\left(1-\PO_{mn}\right)(n_a-n_b)\delta_{an}\delta_{bl}\nord{\aaO_{k}\aO_{m}}
\end{align}

\begin{align}
  &\comm{\nord{\aaO_{a}\aaO_{b}\aO_{d}\aO_{c}}}{\nord{\aaO_{k}\aaO_{l}\aO_{n}\aO_{m}}}
  \notag\\
  &=
    \left(1-\PO_{ab}\right)\left(1-\PO_{mn}\right)\delta_{am}\nord{\aaO_{b}\aaO_{k}\aaO_{l}\aO_{n}\aO_{d}\aO_{c}}
    -\left(1-\PO_{cd}\right)\left(1-\PO_{kl}\right)\delta_{kc}\nord{\aaO_{a}\aaO_{b}\aaO_{l}\aO_{n}\aO_{m}\aO_{d}}
  \notag\\
  &\hphantom{=}
    +(1-\PO_{cd})\left(\nn_{c}\nn_{d} - n_c n_d \right)\delta_{ck}\delta_{dl}\nord{\aaO_{a}\aaO_{b}\aO_{n}\aO_{m}}
    +(1-\PO_{ab})\left(n_a n_b -\nn_{a} \nn_{b} \right)\delta_{am}\delta_{bn}\nord{\aaO_{k}\aaO_{l}\aO_{d}\aO_{c}}
  \notag\\
  &\hphantom{=}
    +(1-\PO_{ab})(1-\PO_{cd})(1-\PO_{kl})(1-\PO_{mn})\left(n_{b} - n_{d}\right)\delta_{bn}\delta_{dl}
    \nord{\aaO_{a}\aaO_{k}\aO_{m}\aO_{c}}
  \notag\\
  &\hphantom{=}
   +(1-\PO_{ab})(1-\PO_{mn})\left(n_{b}\nn_{c}\nn_{d} - \nn_{b}n_{c}n_{d}\right)\delta_{bn}\delta_{ck}\delta_{dl}
   \nord{\aaO_{a}\aO_{m}}
  \notag\\
  &\hphantom{=}
   -(1-\PO_{cd})(1-\PO_{kl})\left(n_{d}\nn_{a}\nn_{b} - \nn_{d}n_{a}n_{b}\right)\delta_{dl}\delta_{am}\delta_{bn}
   \nord{\aaO_{k}\aO_{c}}
  \notag\\
  &\hphantom{=}
  +(1-\PO_{ab})(1-\PO_{cd})
    \left(n_{a} n_{b} \nn_{c} \nn_{d} - \nn_{a} \nn_{b} n_{c} n_{d}\right)\delta_{am}\delta_{bn}\delta_{ck}\delta_{dl}
\end{align}

\bibliographystyle{spphys}
\bibliography{chapter10}

\end{document}